\documentclass[rmp,aps,amssymb,twocolumn,nofootinbib,floatfix]{revtex4}

\usepackage{amsmath,amssymb,bm}
\usepackage{graphics}
\usepackage{epsfig}
\usepackage{graphicx}

\def\inbar{\,\vrule height1.5ex width.4pt depth0pt}
\def\IR{\relax{\rm I\kern-.18em R}}
\def\IC{\relax\hbox{$\inbar\kern-.3em{\rm C}$}}

\def\be{\begin{equation}}
\def\ee{\end{equation}}

\begin{document}
\title{Statistical physics of social dynamics}

\author{Claudio Castellano}
\email{claudio.castellano@roma1.infn.it}
\affiliation{SMC, INFM-CNR and Dipartimento di Fisica, ``Sapienza''
  Universit\`a di Roma, Piazzale A. Moro 2, 00185 Roma I-ITALY}
\author{Santo Fortunato}
\email{fortunato@isi.it}
\affiliation{Complex Networks Lagrange Laboratory, ISI Foundation,
  Viale S. Severo 65, 10133, Torino, I-ITALY}
\author{Vittorio Loreto}
\email{vittorio.loreto@roma1.infn.it}
\affiliation{Dipartimento di Fisica, ``Sapienza'' Universit\`a di Roma
  and SMC, INFM-CNR, Piazzale A. Moro 2, 00185 Roma I-ITALY and
  Complex Networks Lagrange Laboratory, ISI Foundation, Viale
  S. Severo 65, 10133, Torino, I-ITALY}

\begin{abstract}  
  Statistical physics has proven to be a very fruitful framework to
  describe phenomena outside the realm of traditional physics. The
  last years have witnessed the attempt by physicists to study
  collective phenomena emerging from the interactions of individuals
  as elementary units in social structures.  Here we review the state
  of the art by focusing on a wide list of topics ranging from
  opinion, cultural and language dynamics to crowd behavior, 
  hierarchy formation, human dynamics, social
  spreading. We highlight the connections between these problems and
  other, more traditional, topics of statistical physics. We also
  emphasize the comparison of model results with empirical data from
  social systems.
\end{abstract}                                                                 

\pacs{05.00.00 Statistical physics, thermodynamics, and nonlinear
  dynamical systems; 89.20.-a Interdisciplinary applications of
  physics; 89.75.-k Complex systems}
\maketitle
\tableofcontents

\section{INTRODUCTION}
\label{sec:intro}

The concept that many laws of nature are of statistical origin is so
firmly grounded in virtually all fields of modern physics, that
statistical physics has acquired the status of a discipline on its
own. Given its success and its very general conceptual framework, in
recent years there has been a trend toward applications of statistical
physics to interdisciplinary fields as diverse as biology, medicine,
information technology, computer science, etc..  In this context,
physicists have shown a rapidly growing interest for a statistical
physical modeling of fields patently very far from their
traditional domain of
investigations~\cite{stauffer06e,chakrabarti06}. In social phenomena
the basic constituents are not particles but humans and every
individual interacts with a limited number of peers, usually
negligible compared to the total number of people in the system. In
spite of that, human societies are characterized by stunning global
regularities~\cite{buchanan07}.  There are transitions from disorder
to order, like the spontaneous formation of a common language/culture
or the emergence of consensus about a specific issue. There are
examples of scaling and universality.  These macroscopic phenomena
naturally call for a statistical physics approach to social behavior,
i.e., the attempt to understand regularities at large scale as
collective effects of the interaction among single individuals,
considered as relatively simple entities.

It may be surprising, but the idea of a physical modeling of social
phenomena is in some sense older than the idea of statistical modeling
of physical phenomena.  The discovery of quantitative laws in the
collective properties of a large number of people, as revealed for
example by birth and death rates or crime statistics, was one of the
factors pushing for the development of statistics and led many
scientists and philosophers to call for some quantitative
understanding (in the sense of physics) on how such precise
regularities arise out of the apparently erratic behavior of single
individuals. Hobbes, Laplace, Comte, Stuart Mill and many others
shared, to a different extent, this line of thought~\cite{ball04}.
This point of view was well known to Maxwell and Boltzmann and
probably played a role when they abandoned the idea of describing the
trajectory of single particles and introduced a statistical
description for gases, laying the foundations of modern statistical
physics. The value of statistical laws for social sciences has been
foreseen also by Majorana in his famous tenth
article~\cite{majorana42,majorana42eng}. But it is only in the past
few years that the idea of approaching society within the framework of
statistical physics has transformed from a philosophical declaration
of principles to a concrete research effort involving a critical mass
of physicists.  The availability of new large databases as well as the
appearance of brand new social phenomena (mostly related to the
Internet) and the tendency of social scientists to move toward the
formulation of simplified models and their quantitative analysis, have
been instrumental for this change.

In this review we mostly discuss several different aspects of a single
basic question of social dynamics: how do the interactions between
social agents create order out of an initial disordered situation?
Order is a translation in the language of physics of what is denoted
in social sciences as consensus, agreement, uniformity, while disorder
stands for fragmentation or disagreement.  It is reasonable to assume
that without interactions heterogeneity dominates: left alone, each
agent would choose a personal response to a political question, a
unique set of cultural features, his own special correspondence
between objects and words.  Still it is common experience that shared
opinions, cultures, languages do exist. The focus of the statistical
physics approach to social dynamics is to understand how this comes
about.  The key factor is that agents interact and this generally
tends to make people more similar (although many counterexamples
exist).  Repeated interactions in time lead to higher degrees of
homogeneity, that can be partial or complete depending on the temporal
or spatial scales.  The investigation of this phenomenon is
intrinsically dynamic in nature.

A conceptual difficulty immediately arises when trying to approach
social dynamics from the point of view of statistical physics.  In
usual applications, the elementary components of the systems
investigated, atoms and molecules, are relatively simple objects,
whose behavior is very well known: the macroscopic phenomena are not
due to a complex behavior of single entities, rather to nontrivial
collective effects resulting from the interaction of a large number of
'simple' elements.

Humans are exactly the opposite of such simple entities: the detailed
behavior of each of them is already the complex outcome of many
physiological and psychological processes, still largely unknown. No
one knows precisely the dynamics of a single individual, nor the way
he interacts with others.  Moreover, even if one knew the very nature
of such dynamics and such interactions, they would be much more
complicated than, say, the forces that atoms exert on each other. It
would be impossible to describe them precisely with simple laws and
few parameters.  Therefore any modeling of social agents inevitably
involves a huge and unwarranted simplification of the real problem.
It is then clear that any investigation of models of social dynamics
involves two levels of difficulty. The first is in the very definition
of sensible and realistic microscopic models; the second is the usual
problem of inferring the macroscopic phenomenology out of the
microscopic dynamics of such models. Obtaining useful results out of
these models may seem a hopeless task.

The critique that models used by physicists to describe social systems
are too simplified to describe any real situation is most of the times
very well grounded. This applies also to highly acclaimed models
introduced by social scientists, as Schelling's model for urban
segregation~\cite{schelling71} and Axelrod's model~\cite{axelrod97}
for cultural dissemination.  But in this respect, statistical physics
brings an important added value. In most situations qualitative (and
even some quantitative) properties of large scale phenomena do not
depend on the microscopic details of the process. Only higher level
features, as symmetries, dimensionality or conservation laws, are
relevant for the global behavior. With this concept of {\em
  universality} in mind one can then approach the modelization of
social systems, trying to include only the simplest and most important
properties of single individuals and looking for qualitative features
exhibited by models.  A crucial step in this perspective is the
comparison with empirical data which should be primarily intended as
an investigation on whether the trends seen in real data are
compatible with plausible microscopic modeling of the individuals, are
self-consistent or require additional ingredients.

The statistical physics approach to social dynamics is currently
attracting a great deal of interest, as indicated by the large and
rapidly increasing number of papers devoted to it.  The newcomer can
easily feel overwhelmed and get lost in the steadily growing flow of
new publications.  Even for scholars working in this area it is
difficult to keep up on the new results that appear at an impressive
pace.  In this survey we try to present, in a coherent and structured
way, the state of the art in a wide subset of the vast field of social
dynamics, pointing out motivations, connections and open problems.
Specific review articles already exist for some of the topics we
consider and we will mention them where appropriate.  We aim at
providing an up-to-date and -- as much as possible -- unified
description of the published material.  Our hope is that it will be
useful both as an introduction to the field and as a reference.

When writing a review on a broad, interdisciplinary and active field,
completeness is, needless to say, a goal out of reach.  For this
reason we spell out explicitly what is in the review and what is not.
We focus on some conceptually homogeneous topics, where the common
thread is that individuals are viewed as adaptive instead of rational
agents, the emphasis being on communication rather than strategy.  A
large part of the review is devoted to the dynamics of opinions
(Sec.~\ref{sec:opinions}) and to the related field of cultural
dissemination (Sec.~\ref{sec:culture}). Another large section
describes language dynamics (Sec.~\ref{sec:language}), intended both
as the formation and evolution of a language and as the competition
between different languages.  In addition we discuss some other
interesting issues as crowd dynamics (Sec.~\ref{sec:collective}), the
emergence of hierarchies (Sec.~\ref{sec:bonabeau}), social spreading
phenomena (Sec.~\ref{sec:rumor}), coevolution of states and
topology (Sec.~\ref{sec:coevolution}) and what is becoming established
as 'human dynamics' (Sec.~\ref{sec:humandynamics}). Although it is
often very difficult to draw clear borders between disciplines, we
have in general neglected works belonging to the field of econophysics
as well as to evolutionary theory~\cite{blythe07} and evolutionary
game theory, except for what concerns the problem of language
formation. On such topics there are excellent books and
reviews~\cite{mantegna99, bouchaud00,lux07} to which we refer the
interested reader.  We leave out also the physical investigation of
vehicular traffic, a rather well established and successful
field~\cite{chowdhury00,helbing01,nagatani02}, though akin to
pedestrian behavior in crowd dynamics. The hot topic of complex
networks has a big relevance from the social point of view, since many
nontrivial topological structures emerge from the self-organization of
human agents. Nevertheless, for lack of space, we do not to discuss
such theme, for which we refer to~\cite{barabasi99b, newman03b,
  dorogovtsev02, boccaletti06}.  Networks will be considered but only
as substrates where the social dynamics may take place. Similarly, we
do not review the intense recent activity on epidemics
spreading~\cite{anderson91,lloyd01} on
networks~\cite{pastor-satorras01b,may06}, though we devote a section
to social spreading phenomena.  It is worth remarking that, even if we
have done our best to mention relevant social science literature and
highlight connections to it, the main focus of this work remains the
description of the statistical physics approach to social dynamics.

The reader will realize that, with the exception of some specific
sections, there is a striking unbalance between empirical evidence and
theoretical modelization, in favor of the latter.  This does not
correspond to our personal choice: it is a rather objective reflection
of a great disproportion in the literature on social dynamics. In this
respect things are very different from other related fields like
complex networks or econophysics.  In those cases the great bursts of
activity occurred recently have been essentially data-driven: the
availability of unprecedented huge new datasets has spurred first
their thorough empirical characterization and, based on that, an
intense theoretical activity followed for their modeling, well rooted
in the comparison between theory and data.  For some of the fields that will
be reviewed below, things have gone the other way around. The
introduction of a profusion of theoretical models has been mainly
justified by vague plausibility arguments, with no direct connection
to measurable facts.  Very little attention has been paid to a
stringent quantitative validation of models and theoretical
results. The contribution of physicists in establishing social
dynamics as a sound discipline grounded on empirical evidence has been
so far insufficient. We hope that the present review will be
instrumental in stimulating the statistical physics community towards
this goal. The latest developments of the field, especially those
related to so-called Social Web, hint at a change of perspective in
this direction.

\section{GENERAL FRAMEWORK: CONCEPTS AND TOOLS}
\label{sec:general}

Despite their apparent diversity, most research lines we
shall review are actually closely connected from the point of view of
both the methodologies employed and, more importantly, of the general
phenomenology observed. Opinions, cultural and linguistic traits, social status,
kinematic features are always modeled in terms of a small set of variables whose dynamics is
determined by social interactions.  The
interpretation of such variables will be different in the various
cases: a binary variable will indicate yes/no to a political question
in opinion dynamics, two synonyms for a certain object in language
evolution, two languages in competition, whether somebody has been reached 
by a rumor or not, etc..  Other details may
differ, but often results obtained in one case can immediately be
translated in the context of other sub-fields. In all cases the
dynamics tends to reduce the variability of the initial state and this
may lead to an ordered state, with all the agents sharing the
same features (opinion, cultural or linguistic traits, velocity) or to a
fragmented (disordered) state. The way in which those systems evolve
can thus be addressed in a unitary way using well known tools and
concepts from statistical physics.  In this spirit some of the
relevant general questions we will consider in the review include:
What are the fundamental interaction mechanisms that allow for the
emergence of consensus on an issue, a shared culture, a common
language, collective motion, a hierarchy?  
What favors the homogenization process? What hinders it?

Generally speaking the drive toward order is provided by the tendency
of interacting agents to become more alike. This effect is often
termed 'social influence' in the social science
literature~\cite{festinger50} and can be seen as a counterpart of
ferromagnetic interaction in magnets.  Couplings of anti-ferromagnetic
type, i.e., pushing people to adopt a state different from the state of their
neighbors, are also in some cases important and will be considered.

Any modelization of social agents inevitably neglects a huge number of
details. One can often take into account in an effective form such
unknown additional ingredients assuming the presence of noise.  A
time-independent noise in the model parameters often represents the
variability in the nature of single individuals.  On the other hand a
time-dependent noise may generate spontaneous transitions of agents
from one state to another. A crucial question has then to do with the
stability of the model behavior with respect to such perturbations.
Do spontaneous fluctuations slow down or even stop the ordering
process?  Does diversity of agents' properties strongly affect the
model behavior?

An additional relevant feature is the topology of the interaction
network.  Traditional statistical physics usually deals with
structures whose elements are located regularly in space (lattices) or
considers the simplifying hypothesis that the interaction pattern is
all-to-all, thus guaranteeing that the mean field approximation is
correct.  This assumption, often also termed homogeneous mixing,
generally permits analytical treatment, but it is hardly realistic in
a social context.  Much more plausible interaction patterns are those
denoted as complex networks (see Sec.~\ref{sec:topology}).  The
study of the effect of their nontrivial topological properties on
models for social dynamics is a very hot topic.

One concept playing a special role in many social dynamic models and
having no equally common counterpart in traditional statistical
physics is 'bounded confidence', i.e., the idea that in order to
interact two individuals must be not too different. This parallels
somewhat the range of interaction in physics: if two particles are too
far apart they do not exert any influence on each other. However let
us stress that the distance involved in bounded confidence is not
spatial, rather being defined in a sort of opinion space.  We will
discuss in the review several instances of this general principle.

Let us finally clarify some problems with nomenclature.  Being a
strongly interdisciplinary field, in social dynamics there is a
natural tendency towards a rather free (or sloppy) use of terms.  This
heterogeneity is in some cases very confusing as it happens for some
words (like polarization) that have been used with opposite meaning.
For the sake of clarity we specify that in this review consensus
indicates a configuration where all agents share the same state. When
many possible states are possible but only two of them survive in the
population we speak of polarization. Fragmentation indicates instead a
configuration with more than two surviving states.

\subsection{Order and disorder: the Ising paradigm}
\label{subsec:order-disorder}

In the previous section we have seen that the common theme of social
dynamics is the understanding of the transition from an initial
disordered state to a configuration that displays order (at least
partially).  Such type of transitions abound in traditional
statistical physics~\cite{huang87,kubo85}. It is worth summarizing
some important concepts and tools used in that context, as they are
relevant also for the investigation of social dynamics. We will
illustrate them using a paradigmatic example of order-disorder
transitions in physics, the one exhibited by the Ising model for
ferromagnets~\cite{binney92}. Beyond its relevance as a physics model,
the Ising ferromagnet can be seen as a very simple model for opinion
dynamics, with agents being influenced by the state of the majority of
their interacting partners.

Consider a collection of $N$ spins (agents) $s_i$ that can assume two
values $\pm 1$. Each spin is energetically pushed to be aligned with
its nearest neighbors.  The total energy is
\be
H = - \frac{1}{2} \sum_{<i,j>} s_i s_j,
\label{HIsing}
\ee
where the sum runs on the pairs of nearest-neighbors spins. Among the
possible types of dynamics, the most common
(Metropolis)~\cite{landau05} takes as elementary move a single
spin flip that is accepted with probability $\exp(-\Delta E/k_B T)$,
where $\Delta E$ is the change in energy and $T$ is the temperature.
Ferromagnetic interactions in Eq.~(\ref{HIsing}) drive the system
towards one the two possible ordered states, with all positive or all
negative spins.  At the same time thermal noise injects fluctuations
that tend to destroy order.  For low temperature $T$ the ordering
tendency wins and long-range order is established in the system, while
above a critical temperature $T_c$ the system remains macroscopically
disordered.  The transition point is characterized by the average
magnetization $m=1/N \sum_i \langle s_i \rangle$ passing from $0$ for
$T>T_c$ to a value $m(T)>0$ for $T<T_c$. The brackets denote the
average over different realizations of the dynamics. This kind of
transitions is exhibited by a wealth of systems. Let us simply
mention, for its similarity with many of the social dynamic models
discussed in the review, the Potts model~\cite{wu82}, where each spin
can assume one out of $q$ values and equal nearest neighbor values are
energetically favored.  The Ising model corresponds to the special
case $q=2$.

\begin{figure}
  \includegraphics[width=\columnwidth]{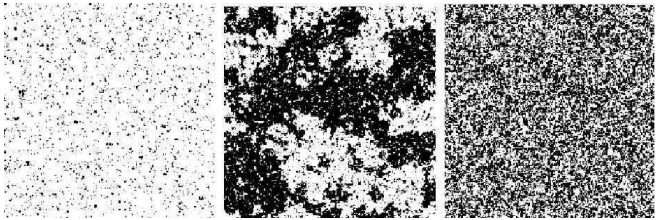}
  \caption{Snapshots of equilibrium configurations of the Ising model
    (from left to right) below, at and above $T_c$.}
\label{equilibrium}
\end{figure}

It is important to stress that above $T_c$ no infinite-range order
is established, but on short spatial scales spins are correlated: there
are domains of $+1$ spins (and others of $-1$ spins) extended over
regions of finite size. Below $T_c$ instead these ordered regions
extend to infinity (they span the whole system), although at finite
temperature some disordered fluctuations are present on short scales
(Fig.~\ref{equilibrium}).

Not only the equilibrium properties just described, that are attained
in the long run, are interesting.  A much investigated and nontrivial
issue~\cite{bray94} is the way the final ordered state at $T<T_c$ is
reached, when the system is initially prepared in a fully disordered
state.  This ordering dynamics is a prototype for the analogous
processes occurring in many models of social dynamics.  On short time
scales, coexisting ordered domains of small size (both positive and
negative) are formed.  The subsequent evolution occurs through a {\em
  coarsening} process of such domains, which 
grow larger and larger while their global statistical features remain
unchanged over time. This is the dynamic scaling phenomenon: the
morphology remains statistically the same if rescaled by the typical
domain size, which is the only relevant length in the system and grows
over time as a power-law.

Macroscopically, the dynamic driving force towards order is surface
tension.  Interfaces between domains of opposite magnetization cost in
terms of energy and their contribution can be minimized by making them
as straight as possible.  This type of ordering is often referred to
as curvature-driven and occurs in many of the social systems described
in this review.  The presence of surface tension is a consequence of
the tendency of each spin to become aligned with the majority of its
neighbors.  When the majority does not play a role, the qualitative
features of the ordering process change.

The dynamic aspect of the study of social models requires the
monitoring of suitable quantities, able to properly identify the
buildup of order.  The magnetization of the system is not one of such
suitable quantities.  It is not sensitive to the size of single
ordered domains, while it measures their cumulative extension, which
is more or less the same during most of the evolution.  The
appropriate quantity to monitor the ordering process is the
correlation function between pairs of spins at distance $r$ from each
other, $C(r,t)= \langle s_i(t) s_{i+r}(t)\rangle-\langle
s_i(t)\rangle^2$, where brackets denote averaging over dynamic
realizations and an additional average over $i$ is implicit. The
temporal variable $t$ is measured as the average number of attempted
updates per spin. The dynamic scaling property implies that $C(r,t)$
is a function only of the ratio between the distance and the typical
domain size $L(t)$: $C(r,t) = L(t)^d F[r/L(t)]$.  $L(t)$ grows in time
as a power-law $t^{1/z}$. The dynamic exponent $z$ is universal,
independent of microscopic details, possibly depending only on
qualitative features as conservation of the magnetization or space
dimensionality. In the Glauber/Metropolis case $z=2$ in any dimension.
Another quantity often used is the density of interfaces
$n_a(t)=N_a(t)/N_p$, where $N_p$ is the total number of nearest
neighbor pairs and $N_a$ the number of such pairs where the two
neighbors are in different states: $n_a=1/2$ means that disorder is
complete, while $n_a=0$ indicates full consensus.

Finally, a word about finite size effects. The very concept of
order-disorder phase-transitions is rigorously defined only in the
limit of a system with an infinite number of particles (thermodynamic
limit), because only in that limit truly singular behavior can
arise. Social systems are generally composed by a large number $N$ of
agents, but by far smaller than the number of atoms or molecules in a
physical system.  The finiteness of $N$ must play therefore a crucial
role in the analysis of models of social dynamics~\cite{toral07}.
Studying what happens when $N$ changes and even considering the
large-$N$ limit is generally very useful, because it helps
characterizing well qualitative behaviors, understanding which
features are robust, and filtering out non-universal microscopical
details.

\subsection{Role of topology}
\label{sec:topology}

An important aspect always present in social dynamics is topology,
i.e., the structure of the interaction network describing who is
interacting with whom, how frequently and with which intensity. Agents
are thus supposed to sit on vertices (nodes) of a network, and the
edges (links) define the possible interaction patterns.

The prototype of homogeneous networks is the uncorrelated random graph
model proposed by Erd\"os and R\'enyi (ER
model)~\cite{erdos59,erdos60}, whose construction consists in drawing
an (undirected) edge with a fixed probability $p$ between each
possible pair out of $N$ given vertices. The resulting graph shows a
binomial degree distribution, the degree of a node being the number of
its connections, with average $\langle k\rangle \simeq Np$. The degree
distribution converges to a Poissonian for large $N$. If $p$ is
sufficiently small (order $1/N$), the graph is sparse and presents
locally tree-like structures. In order to account for degree
heterogeneity, other constructions have been proposed for random
graphs with arbitrary degree
distributions~\cite{molloy_reed95,molloy_reed98,aiello01,goh01,catanzaro05}.

A well-known paradigm, especially for social sciences, is that of
``small-world'' networks, in which, on the one hand, the average
distance between two agents is small~\cite{milgram67}, growing only
logarithmically with the network size, and, on the other hand, many
triangles are present, unlike ER graphs. In order to reconcile both
properties, Watts and Strogatz have introduced the small-world network
model~\cite{watts98}, which allows to interpolate between regular
low-dimensional lattices and random networks, by introducing a certain
fraction $p$ of random long-range connections into an initially
regular lattice~\cite{newman99}. In~\cite{watts98} two main quantities
have been considered: the characteristic path length $L(p)$, defined
as the number of edges in the shortest path between two vertices,
averaged over all pairs of vertices, and the clustering coefficient
$C(p)$, defined as follows. If a node $i$ has $k$ connections, then at
most $k(k-1)/2$ edges can exist between its neighbors (this occurs
when every neighbor of $i$ is connected to every other neighbor).
The clustering coefficient $C(p)$ denotes the fraction of these
allowable edges that actually exist, averaged over all nodes.
Small-world networks feature high values of $C(p)$ and low values of
$L(p)$.

Since many real networks are not static but evolving, with new nodes
entering and establishing connections to already existing nodes, many
models of growing networks have also been introduced.  The Barab\'asi
and Albert model (BA)~\cite{barabasi99}, has become one of the most
famous models for complex heterogeneous networks, and is constructed
as follows: starting from a small set of $m$ fully interconnected
nodes, new nodes are introduced one by one. Each new node selects $m$
older nodes according to the {\em preferential attachment} rule, i.e.,
with probability proportional to their degree, and creates links with
them. The procedure stops when the required network size $N$ is
reached. The obtained network has average degree $\langle k
\rangle=2m$, small clustering coefficient (of order $1/N$) and a power
law degree distribution $P(k) \sim k^{-\gamma}$, with $\gamma =
3$. Graphs with power law degree distributions with $\gamma \le 3$
are referred to as {\it scale-free networks}.

An extensive analysis of the existing network models is out of the
scope of the present review and we refer the reader to the huge
literature on the so-called complex
networks~\cite{boccaletti06,barabasi99b,dorogovtsev03,newman03b,pastor-satorras04,caldarelli07}. It
is nevertheless important to mention that real networks often differ
in many respects from artificial networks. People have used the social
network metaphor for over a century to represent complex sets of
relationships between members of social systems at all scales, from
interpersonal to international. A huge amount of work has been carried
out about the so-called social network analysis (SNA), especially in
the social science
literature~\cite{moreno34,granovetter73,granovetter83,scott00,wasserman94,freeman04}. Recently
the interest of physicists triggered the investigation of many
different networks: from the network of scientific
collaborations~\cite{newman01,newman01b,barabasi02,newman04} to that
of sexual contacts~\cite{lilijeros01} and the ongoing social
relationships~\cite{holme03}, from email exchanges
networks~\cite{ebel02,newman02,eckmann04} to the dating community
network~\cite{holme04} and to mobile communication
networks~\cite{onnela07,palla07}, just to quote a few examples. From
this experimental work a set of features characterizing social
networks have been identified. It has been shown~\cite{newman03c} how
social networks differ substantially from other types of networks,
namely technological or biological. The origin of the difference is
twofold. On the one hand they exhibit a positive correlation between
adjacent vertices (also called assortativity), while most non-social
networks~\cite{pastor-satorras01,newman03} are disassortative. A
network is said to show assortative mixing if nodes with many
connections tend to be linked to other nodes with high degree. On
the other hand social networks show clustering coefficients well above
those of the corresponding random models. These results opened the way
to a modeling activity aimed at reproducing in an artificial and
controlled way the same features observed in real social
networks~\cite{jin01}. We cannot review here all these attempts but we
have quoted some relevant references all along the review when
discussing specific modeling schemes. It is important to keep in mind
that future investigations on social dynamics will be forced to take
into account in a more stringent way structural and dynamic properties
of real social networks~\cite{roehner07}.

When applying models of social dynamics on specific topologies several
non-trivial effects may arise, potentially leading to important biases
for the dynamics. For instance, on networks with strongly heterogeneous
degree distributions the dynamics of models with binary
asymmetric interaction rules, i.e., where the two selected agents have
different roles, may be affected by the order in which the interaction
partners are selected.
This is a consequence of the fact that, in a
network with degree distribution $P(k)$,  if one picks a random
node and then one of its neighbors, it is likely that the second
node has a higher degree than the first. More precisely, the
degree distribution of the second is
$kP(k)/\langle k \rangle$~\cite{pastor-satorras04}. 
For scale-free degree distributions this may have strong effects (as
is the case of the voter model, as seen in
Sec.~\ref{sec:voter}, and of the Naming Game, as seen in
Sec.~\ref{sec:semiotic-dynamics}).


\subsection{Dynamical systems approach}
\label{sec:dyn-sys}

One of the early contribution of physicists to the study of social
systems has been the introduction of methods and tools coming from the
theory of dynamical systems and non-linear dynamics. This development
goes under the name of {\em sociodynamics}~\cite{weidlich91,weidlich02,helbing91}. The term
sociodynamics has been introduced to refer to a systematic approach to
mathematical modeling in the framework of social sciences.

Sociodynamics is a branch of synergetics~\cite{haken77} devoted to
social systems, featuring a few important differences. In synergetics
one typically starts with a large set of microscopic equations for the
elementary components and performs a reduction of the degrees of
freedom. This is not the case for social systems, for which
equations at the microscopic level are not available. In this case one has
to identify relevant macro-variables and construct directly equations
for them, based on reasonable and realistic social hypotheses, i.e.,
informed by social driving forces. The typical procedure consists in
defining probabilistic transition rates per unit of time for the jumps
between different configurations of the system corresponding to different
values of the macro-variables. The transition rates are used as
building blocks for setting up the equations of motion for the
probabilistic evolution of the set of macro-variables. The central
evolution equation in sociodynamics is the master equation, a
phenomenological first-order differential equation describing the time
evolution of the probability $P({\mathbf m},t)$ for a system to occupy
each one of a discrete set of states, defined through the set of
macro-variables ${\mathbf m}$:

\begin{equation}
  \frac{dP({\mathbf m},t)}{dt} = \sum_{{\mathbf m}^{\prime}} [W_{{\mathbf m}^{\prime},{\mathbf m}}
  P({\mathbf m}^{\prime},t) - W_{{\mathbf m},{\mathbf m}^{\prime}}
  P({\mathbf m},t)],
  \label{master-equation}
\end{equation}

\noindent where $W_{{\mathbf m},{\mathbf m}^{\prime}}$ represents the
transition rate from the state ${\mathbf m}$ to the state ${\mathbf
  m}^{\prime}$. The master equation is a gain-loss equation for the
probability of each state ${\mathbf m}$. The first term is the gain
due to transitions from other states ${\mathbf{m}}^{\prime}$, and the
second term is the loss due to transitions into other states
${\mathbf{m}}^{\prime}$.

While it is relatively easy to write down a master equation, it is
quite another matter to solve it. It is usually highly non-linear and
some clever simplifications are often needed to extract a solution. In
general only numerical solutions are available.  Moreover, typically
the master equation contains too much information in comparison to
available empirical data. For all these reasons it is highly desirable
to derive from the master equation simpler equations of motion for
simpler variables. One straightforward possibility is to consider the
equations of motion for the average values of the macro-variables
${\mathbf m}$, defined as:

\begin{equation}
  {\overline{m}}_k(t)= \sum_{{\mathbf m}} m_k P({\mathbf m},t).
  \label{meanvalues}
\end{equation}

The exact expression for the equations of motion for
${\overline{m}}_k(t)$ does not lead to simplifications because one
should already know the full probability distribution $P({\mathbf
  m},t)$.  On the other hand, under the assumption that the
distribution remains unimodal and sharply peaked for the period of
time under consideration, one has:

\begin{equation}
 \overline{P({\mathbf m},t)} \simeq P(\overline{\mathbf m(t)}),
  \label{meanvalues2}
\end{equation}

\noindent yielding the approximate equations of motions for
${\overline{m}}_k(t)$, which are now a closed system of coupled
differential equations. We refer to~\cite{weidlich02} for a complete
derivation of these equations as well as for the discussion of several
applications. The approach has also been applied to model behavioral
changes~\cite{helbing93,helbing93a,helbing94}.

\subsection{Agent-based modeling}
\label{sec:agents}

Computer simulations play an important role in the study of social
dynamics since they parallel more traditional approaches of
theoretical physics aiming at describing a system in terms of a set of
equations, to be later solved numerically and/or, whenever possible,
analytically. One of the most successful methodologies used in social
dynamics is {\em agent-based} modeling. The idea is to construct the
computational devices (known as agents with some properties) and then
simulate them in parallel to model the real phenomena. In physics this
technique can be traced back to molecular
dynamics~\cite{alder57,alder59} and Metropolis and Monte
Carlo~\cite{metropolis53} simulations. The goal is to address the
problem of the emergence from the lower (micro) level of the social
system to the higher (macro) level. The origin of agent-based modeling
can be traced back to the 1940s, to the introduction by Von Neumann
and Ulam of the notion of cellular
automaton~\cite{vonneumann66,ulam60}, e.g., a machine composed of a
collection of cells on a grid. Each cell can be found in a discrete
set of states and its update occurs on discrete time steps according
to the state of the neighboring cells. A well-known example is
Conway's Game of Life, defined in terms of simple rules in a virtual
world shaped as a 2-dimensional checkerboard. This kind of algorithms
became very popular in population biology~\cite{matsuda92}.

The notion of agent has been very important in the development of the
concept of Artificial Intelligence~\cite{mccarthy59,minsky61}, which
traditionally focuses on the individual and on rule-based paradigms
inspired by psychology. In this framework the term {\em actors} was
used to indicate interactive objects characterized by a certain number
of internal states, acting in parallel and exchanging
messages~\cite{hewitt70}. In computer science the notion of actor
turned in that of agent and more emphasis has been put on the
interaction level instead of autonomous actions.

Agent-based models were primarily used for social systems by Reynolds,
who tried to model the reality of living biological agents, known as
artificial life, a term coined in~\cite{langton96}.  Reynolds
introduced the notion of individual-based models, in which one
investigates the global consequences of local interactions of members
of a population (e.g. plants and animals in ecosystems, vehicles in
traffic, people in crowds, or autonomous characters in animation and
games). In these models individual agents (possibly heterogeneous)
interact in a given environment according to procedural rules tuned by
characteristic parameters. One thus focuses on the features of each
individual instead of looking at some global quantity averaged over
the whole population.

The Artificial Life community has been the first in developing
agent-based models~\cite{maes91,meyer90,varela92,steels95,weiss99},
but since then agent-based simulations have become an important tool
in other scientific fields and in particular in the study of social
systems~\cite{conte97,wooldridge02,macy02,schweitzer03,axelrod06b}.
In~\cite{epstein96}, by focusing on a bottom-up
approach, the first large scale agent model, the Sugarscape, has been
introduced to simulate and explore the role of social phenomena such
as seasonal migrations, pollution, sexual reproduction, combat, trade
and transmission of disease and culture. In
this context it is worth mentioning the concept of {\em Brownian
  agent}~\cite{schweitzer03} which generalizes that of Brownian
particle from statistical mechanics. A Brownian agent is an active
particle which possesses internal states, can store energy and
information and interacts with other agents through the environment.
Again the emphasis is on the parsimony in the agent definition as well
as on the interactions, rather than on the autonomous actions. Agents
interact either directly or in an indirect way through the external
environment, which provides a feedback about the activities of the
other agents. Direct interactions are typically local in time and
ruled by the underlying topology of the interaction network (see also
Sec.~\ref{sec:topology}). Populations can be homogeneous (i.e., all
agents being identical) or heterogeneous. Differently from physical
systems, the interactions are usually asymmetrical since the role of
the interacting agents can be different both for the actions performed
and for the rules to change their internal states. Agent-based
simulations have now acquired a central role in modeling complex
systems and a large literature has been rapidly developing in the last
few years about the internal structure of the agents, their activities
and the multi-agent features.  An exhaustive discussion of agent-based
models is out of the scope of the present review, but we refer
to~\cite{schweitzer03} where the role of active particles is
thoroughly discussed with many examples of applications, ranging from
structure formation in biological systems and pedestrian traffic to
the simulation of urban aggregation or opinion formation processes.

\section{OPINION DYNAMICS}
\label{sec:opinions}

\subsection{Introduction}
\label{sec:history}
           
Agreement is one of the most important aspects of social group
dynamics. Everyday life presents many situations in which it is
necessary for a group to reach shared decisions. Agreement makes a
position stronger, and amplifies its impact on society.

The dynamics of agreement/disagreement among individuals is complex,
because the individuals are. Statistical physicists working on opinion
dynamics aim at defining the opinion states of a population, and the
elementary processes that determine transitions between such
states. The main question is whether this is possible and whether this
approach can shed new light on the process of opinion formation.

In any mathematical model, opinion has to be a variable, or a set of
variables, i.e., a collection of numbers.  This may appear too
reductive, thinking about the complexity of a person and of each
individual position. Everyday life, on the contrary, indicates that
people are sometimes confronted with a limited number of positions on a
specific issue, which often are as few as two: right/left,
Windows/Linux, buying/selling, etc. If opinions can be represented by
numbers, the challenge is to find an adequate set of mathematical
rules to describe the mechanisms responsible for the evolution and
changes of them.

The development of opinion dynamics so far has been uncoordinated and
based on individual attempts, where social mechanisms considered
reasonable by the authors turned into mathematical rules, without a
general shared framework and often with no reference to real
sociological studies.  The first opinion dynamics designed by a
physicist was a model proposed in~\cite{weidlich71}. The model is
based on the probabilistic framework of sociodynamics, which was
discussed in Sec.~\ref{sec:dyn-sys}. Later on the Ising model made
its first appearance in opinion dynamics~\cite{galam82,galam91}. The
spin-spin coupling represents the pairwise interaction between agents,
the magnetic field the cultural majority or propaganda. Moreover
individual fields are introduced that determine personal preferences
toward either orientation. Depending on the strength of the individual
fields, the system may reach total consensus toward one of the two
possible opinions, or a state where both opinions coexist.

In the last decade, physicists have started to work actively in
opinion dynamics, and many models have been designed. We focus on the
models that have received more attention in the physics literature,
pointing out analogies as well as differences between them: the voter
model (Sec.~\ref{sec:voter}), majority rule models
(Sec.~\ref{sec:galam}), models based on social impact theory
(Sec.~\ref{sec:impact}), Sznajd (Sec.~\ref{sec:sznajd}) and bounded
confidence models (Sec.~\ref{sec:BCM}).  In
Sec.~\ref{sec:othermodels_opinions} other models are briefly
discussed.  Finally, in Sec.~\ref{sec:elections}, we review recent
work that aims at an empirical validation of opinion dynamics from the
analysis of data referring to large scale social phenomena.

\subsection{Voter model}
\label{sec:voter}

\subsubsection{Regular lattices}

The voter model has been named in this way for the very natural
interpretation of its rules in terms of opinion dynamics; for its
extremely simple definition, however, the model has been thoroughly
investigated also in fields quite far from social dynamics, like
probability theory and population genetics.  Voter dynamics was first
considered in~\cite{clifford73} as a model for the competition of
species and named ``voter model'' in~\cite{holley75}.  It has soon
become popular because, despite being a rather crude description of any
real process, it is one of the very few non-equilibrium stochastic
processes that can be solved exactly in any dimension~\cite{redner01}.
It can also be seen as a model for dimer-dimer heterogeneous catalysis
in the reaction controlled limit~\cite{evans93}.

\begin{figure}
\includegraphics[width=\columnwidth]{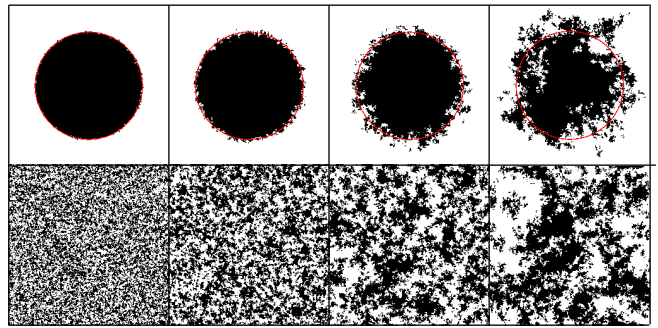}
\caption{Evolution of a two-dimensional voter model starting from a
  droplet (top) or a fully disordered configuration (bottom).
  From~\cite{dornic01}.}
\label{dornic01_1}
\end{figure}

The definition is extremely simple: each agent is endowed with a
binary variable $s = \pm 1$.  At each time step an agent $i$ is
selected along with one of its neighbors $j$ and $s_i = s_j$, i.e.,
the agent takes the opinion of the neighbor.  This update rule implies
that agents imitate their neighbors. They feel the pressure of the
majority of their peers only in an average sense: the state of the
majority does not play a direct role and more fluctuations may be
expected with respect to the zero-temperature Glauber dynamics.  Bulk
noise is absent in the model, so the states with all sites equal
(consensus) are absorbing.  Starting from a disordered initial
condition, voter dynamics tends to increase the order of the system, as in
usual coarsening processes~\cite{scheucher89}. The question is whether
full consensus is reached in a system of infinite size. In
one-dimensional lattices the dynamics is exactly the same of the
zero-temperature Glauber dynamics.  A look at the patterns generated
in two-dimensional lattices (Fig.~\ref{dornic01_1}) indicates that
domains grow but interfaces are very rough, at odds with usual
coarsening systems~\cite{bray94}.

Early studies, performed by probabilists~\cite{clifford73, holley75,
  liggett85, cox86}, have exploited the fact that the model can be
exactly mapped on a model of random walkers that coalesce upon
encounter. This duality property allows to use the powerful machinery
of random walk theory~\cite{liggett85,liggett99}.
We prefer to follow another derivation of the general solution on
lattices~\cite{frachebourg96}, based on
earlier work~\cite{krapivsky92}.  Considering a $d$-dimensional hypercubic
lattice and denoting with $S=\{s_i \}$ the state of the system, the
transition rate for a spin $k$ to flip is 
\be W_k(S) \equiv W(s_k \to -s_k) = \frac{d}{4}\left( 1 - \frac{1}{2d}
  s_k \sum_{j} s_{j} \right),
\label{voter_rates}
\ee 
where $j$ runs over all $2d$ nearest neighbors and the prefactor,
setting the overall temporal scale, is chosen for convenience.  The
probability distribution function $P(S,t)$ obeys the master equation
\be \frac{d}{dt}P(S,t) = \sum_k \left[ W_k(S^k) P(S^k,t) - W_k(S)
  P(S,t) \right], \ee
where $S^k$ is equal to $S$ except for the flipped spin $s_k$. The
linear structure of the rates~(\ref{voter_rates}) has the nice
consequence that the equations for correlation functions of any order
$\langle s_k \cdots s_l \rangle \equiv \sum_S P(S,t) s_k \cdots s_l$
can be closed, i.e., they do not depend on higher-order functions and
hence can be solved~\cite{scheucher89}.

The equation for the one-body correlation function is
\be
\frac{d}{dt}\langle s_k \rangle = \Delta_k \langle s_k \rangle,
\ee
where $\Delta_k$ is the discrete Laplace operator.  Summing over $k$
one sees that the global magnetization $\langle s \rangle =1/N \sum_k
\langle s_k\rangle$ is conserved.  This conservation immediately
allows to determine the probability that a finite system will end up
with all spins up or down (exit probability), depending on the initial
density of up spins $\rho(0)=(\langle s \rangle +1)/2$. This gives
$P_{up}(\rho(0))=\rho(0)$ in any dimension.

The two-body correlation function obeys
\be
\frac{d}{dt}\langle s_k s_l \rangle = (\Delta_k + \Delta_l)
\langle s_k s_l \rangle.
\label{voter_twobody}
\ee
The structure of this equation, as well as of those for higher-order
correlation functions, is similar {\em in any dimension} to the
equations for correlators of the one-dimensional Ising model with
zero-temperature Glauber dynamics~\cite{glauber63} and can be solved
analogously, via Laplace transform.  In this way the asymptotic
behavior of the density of active interfaces $n_a(t)=(1-\langle s_k
s_{k+1}\rangle)/2$ is derived~\cite{frachebourg96}
\be n_a(t) \sim
\left\{
\begin{array}{ll}
t^{-(2-d)/2} & d<2 \\
1/\ln(t) & d=2 \\
a-bt^{-d/2} & d>2 .
\end{array}
\right.
\label{voter_active}
\ee
Eq.~(\ref{voter_active}) shows that for $d \le 2$ the voter model
undergoes a coarsening process leading to complete consensus. For
$d>2$ instead, it exhibits asymptotically a finite density of
interfaces, i.e., no consensus is reached (in an infinite system) and
domains of opposite opinions coexist indefinitely in time.  In terms
of duality the lack of order in high dimensions is a consequence of the
transient nature of random walks in $d>2$: diffusing active interfaces
have a finite probability to meet and annihilate.  For $d=2$ the exact
expression of the density of active interfaces for large times is
\be n_a(t)=\frac{\pi}{2 \ln(t) +
  \ln(256)} + O\left(\frac{\ln t}{t}\right).
\label{voter_active_d=2}
\ee
The large constant value in the denominator of
Eq.~(\ref{voter_active_d=2}) makes the approach to the asymptotic
logarithmic decay very slow, and explains why different laws were
hypothesized, based on numerical evidence~\cite{meakin87,evans93}.

Beyond the expression for the density $n_a(t)$, the solution of
Eq.~(\ref{voter_twobody}) allows to write down a scaling form for the
correlation function $C(r,t)$~\cite{scheucher89,dornicthese}.  In
$d=2$ the solution violates logarithmically the standard scaling
form~(see Sec.~\ref{subsec:order-disorder}) holding for usual
coarsening phenomena~\cite{bray94}. This violation reflects the fact
that there is no single domain length in the system.  On the contrary,
domains of all sizes $t^{\alpha}$, with $0<\alpha \le 1/2$, are
simultaneously present~\cite{cox86, scheucher89}.

While in an infinite system consensus is reached only if $d \le 2$, in
any dimension consensus is invariably reached asymptotically if the
system is finite.  The time $T_N$ needed depends on the system size
$N$~\cite{cox89}: $T_N \sim N^2$ for $d=1$, $T_N \sim N \ln N$ for
$d=2$, while $T_N \sim N$ for $d>2$. It is worth remarking that the
way consensus is reached on finite systems has a completely different
nature for $d \leq 2$ (where the system coherently tends towards order
by coarsening) and for $d>2$ (where consensus is reached only because
of a large random fluctuation).

Expression~(\ref{voter_rates}) for the spin-flip rates is rather
special.  How much of the voter behavior is retained if rates are
modified?  A natural generalization~\cite{deoliveira93,drouffe99}
considers transition rates of the form $W_k(S) = 1/2 [1-s_k f_k(S)]$,
where $f_k(S)$ is a local function with $|f_k(S)| \leq 1$.  A local
dynamics that is spatially symmetric and preserves the up-down
symmetry requires $f_k(S)$ to be an odd function of the sum of the
nearest neighbors. In a square lattice, the local field can assume
five values, but only two of them are independent,
$f(2)=-f(-2)=x$, and $f(4)= -f(-4)=y$. Voter dynamics corresponds to
$x=1/2$ and $y=1$, while $x=y$ corresponds to the majority-vote model
(Sec.~\ref{sec:galam}), $y=2x/(1+x^2)$ gives the transition rates of
Glauber dynamics, and the case $y=2x$ corresponds to the noisy voter
model (see below).  The significance of the two parameters is
straightforward: $y$ gauges bulk noise, i.e., the possibility that a
spin fully surrounded by equal spins flips to the opposite position;
The value $y=1$ implies absence of such noise.  The parameter $x$
instead measures the amount of interfacial noise.  Simulations and a
pair approximation treatment show that the phase-diagram of this
generalized model is divided in a ferromagnetic region around the
$x=1, y=1$ point (zero-temperature Glauber dynamics) and a
paramagnetic phase, separated by a line of continuous
phase-transitions terminating at the voter model point. Changing the
interfacial noise parameter $x$, while keeping $y=1$, one finds a jump
of the order parameter, indicating a first-order transition.  Hence
the voter point is critical, sitting exactly at the transition between
order and disorder driven by purely interfacial noise.

More physical insight is provided by considering a droplet of up spins
surrounded by negative spins~\cite{dornic01}. The Cahn-Allen theory
for curvature-driven coarsening~\cite{bray94} predicts in $d=2$ a
linear decay in time of the droplet area, the rate being proportional
to surface tension.  In the voter model instead, the interface of the
droplet roughens but its radius remains statistically
unchanged~\cite{dornic01,dallasta07}, showing that no surface tension
is present (Fig.~\ref{dornic01_1}).

The phase diagram of the generalized model~\cite{deoliveira93}
described above seems to suggest that the voter model is rather
peculiar, being only a point in the line of continuous transitions
between the ferromagnetic and the paramagnetic phase.  However,
voter-like behavior (characterized by the absence of surface tension
leading to logarithmic ordering in $d=2$) can be found in other
models. It has been argued~\cite{dornic01} that voter behavior is
generically observed at order-disorder non-equilibrium transitions,
driven by interfacial noise, between dynamically symmetric absorbing
states. This symmetry may be enforced either by an up-down symmetry of
the local rules or by global conservation of the magnetization.  The
universal exponents associated to the transition are $\beta=0$, and
$\nu=1/2$ in all dimensions, while $\gamma=1/2$ for $d=1$, and
$\gamma=1$ for $d >2$ with logarithmic corrections at the upper
critical dimension $d=2$~\cite{dornic01,deoliveira03}.

The original voter dynamics does not include the possibility for a
spin to flip spontaneously when equal to all its neighbors. The noisy
voter model~\cite{scheucher89, granovsky95}, also called linear
Glauber model~\cite{deoliveira03} includes this possibility, via a
modification of the rates~(\ref{voter_rates}) that keeps the model
exactly solvable.  The effect of bulk noise is to destroy long-range
order: the noisy voter model is always in the paramagnetic phase of
the generalized model of~\cite{deoliveira93}, so that domains form
only up to a finite correlation length.  As the strength of bulk noise
is decreased, the length grows and the voter first-order transition
occurs for zero noise.

The investigation of the generalized voter universality class and its
connections with other classes of non-equilibrium phase transitions is
a complicated and open issue, approached also via
field-theoretical methods~\cite{dickman95, droz03, alhammal05}.

\subsubsection{Modifications and applications}

Being a very simple non-equilibrium dynamics with a nontrivial behavior,
the voter model has been investigated with respect to many properties
in recent years, including persistence, aging and correlated
percolation.  Furthermore, many modifications of the original dynamics have
been proposed in order to model various types of phenomena or to test
the robustness of the voter phenomenology.  A natural extension is a
voter dynamics for Potts variables (multitype voter model), where many
of the results obtained for the Ising case are easily
generalizable~\cite{sire95}.

One possible modification is the presence of quenched disorder, in the
form of one ``zealot'', i.e., an individual that does not change its
opinion~\cite{mobilia03}.  This modification breaks the conservation
of magnetization: in $d \leq 2$ the zealot influences all, inducing
general consensus with its opinion. In higher dimensions consensus is
still not reached, but in the neighborhood of the zealot the
stationary state is biased toward his opinion. The case of many
zealots has also been addressed~\cite{mobilia05, mobilia07}.

Another variant is the constrained voter model~\cite{vazquez03}, where
agents can be in three states (leftists, rightists, or centrists) but
interactions involve only centrists, while extremists do not talk to
each other. In this way a discrete analogue of bounded confidence is
implemented. Detailed analytical results give the
probabilities, as a function of the initial conditions, of ending up
with full consensus in one of the three states or with a mixture of
the extremists, with little change between $d=1$~\cite{vazquez03} and
mean field~\cite{vazquez04}.  A similar model with three states is the
AB-model~\cite{castello06}.  Here the state of an agent evolves
according to the following rules.  At each time step one randomly
chooses an agent $i$ and updates its state according to the following
transition probabilities:
\begin{eqnarray}
  p_{A\rightarrow AB} & = & 1/2 \sigma_B, \;\;\;\;\;\;\;\;\;\; p_{B\rightarrow AB} = 1/2 \sigma_A,\\
  p_{AB\rightarrow B} & = & 1/2 (1-\sigma_A), \;\; p_{AB\rightarrow A} = 1/2 (1-\sigma_B),
\label{AB-model}
\end{eqnarray}
where $\sigma_l$ ($l$=A,B,AB) are the local densities of each state in
the neighborhood of $i$. The idea here is that, in order to go from A
to B one has to pass through the intermediate state AB.  At odds with
the constrained voter model, however, here extremes do interact, since
the rate to go from state A to AB is proportional to the density of
neighbors in state B. This implies that consensus on the AB state or a
frozen mixture of A and B is not possible, the only two possible
absorbing states being those of consensus of A or B type.

In the original voter model the decision on the possible flip of a
spin depends on one neighbor. In the ``vacillating'' voter
model~\cite{lambiotte07b} a site checks the state of two of its neighbors
and flips if either is different from himself. This leads to a bias
towards the zero magnetization state, anti-coarsening behavior and
consensus time scaling exponentially with the system size $N$.
A related one-dimensional model is~\cite{lambiotte07c}.

Another modification is the introduction of memory in the form of
noise reduction~\cite{dallasta07}. Each spin has associated two
counters.  When an interaction takes place with a positive (negative)
neighbor, instead of modifying the spin the positive (negative)
counter is increased by one. The spin is updated only when one of the
counters reaches a threshold $r$. This change induces an effective
surface tension, leading to curvature-driven coarsening dynamics.

A counterintuitive effect is reported in~\cite{stark07} where a voter
model with ``inertia'' is presented: the local rate of change of a
site decreases linearly with the time since the last flip,. up to a
finite saturation value. On two-dimensional lattices, Watts-Strogatz
and fully connected networks, a sufficiently weak inertia makes global
dynamics faster: consensus time is smaller than with fixed rate.  

Other variants of the original voter model have been
devised for studying ecological problems. Recent publications
in the physics literature are the study of diversity in plant
communities (voter model with speciation~\cite{zillio05}), or the
investigation of fixation in the evolution of competing species
(biased voter model~\cite{antal06}).

\subsubsection{The voter model on networks}

Non-regular topologies have nontrivial effects on the ordering dynamics
of the voter model.

On a complete graph the Fokker-Planck equation for the probability
density of the magnetization has the form of a one-dimensional
diffusion equation with a position-dependent diffusion constant, and
can be solved analytically~\cite{slanina03}.  The lack of a drift term
is the effect of the lack of surface tension in the model
dynamics. The average time needed to reach consensus in a finite
system can be computed exactly for any value of the initial
magnetization and scales as the size of the system $N$.  The tail of
the distribution can also be computed and has an exponential decay
$\exp(-t/N)$.

When considering disordered topologies different ways of
defining the voter dynamics, that are perfectly equivalent on regular
lattices, give rise to nonequivalent generalizations of the voter
model.  When the degree distribution is heterogeneous, the order in
which a site and the neighbor to be copied are selected does matter,
because high-degree nodes are more easily chosen as neighbors than
low-degree vertices.  

The most natural generalization ({\em direct voter} model) is to pick
up a site and make it equal to one of its neighbors. In this way one
of the fundamental properties of the voter model, conservation of the
global magnetization, is violated~\cite{suchecki05, wu04}. To restore
conservation a {\em link-update} dynamics must be
considered~\cite{suchecki05}: a link is selected at random and then
one node located at a randomly chosen end is set equal to the
other. If instead one chooses first a node and copies its variable to
a randomly selected neighbor one obtains the {\em reverse (or
  invasion) voter} dynamics~\cite{castellano05}.

On highly heterogeneous substrates these different definitions result
in different behaviors.  The mean consensus time $T_N$ has been
computed in~\cite{sood05, sood07} for the direct voter dynamics on a
generic graph, by exploiting the conservation of a suitably defined
degree-weighted density $\omega$ of up spins,
\be T_N(\omega) = - N \frac{\mu_1^2}{\mu_2}
\left[ (1-\omega) \ln(1-\omega) + \omega \ln \omega) \right],
\ee 
where $\mu_k$ is the $k$-th moment of the degree-distribution.  For
networks with scale-free distributed degree (with exponent $\gamma$),
$T_N$ scales then as $N$ for $\gamma>3$ and sublinearly for $\gamma
\leq 3$, in good agreement with numerical
simulations~\cite{sood05,suchecki05,castellano05b}.  The same approach
gives, for the other versions of voter dynamics on graphs, a linear
dependence of the consensus time on $N$ for link-update dynamics
(independent of the degree distribution) and $T_N \sim N$ for any
$\gamma>2$ for the reverse-voter dynamics, again in good agreement
with simulations~\cite{castellano05, sood07}.  A general analysis of
voter-like dynamics on generic topologies is presented
in~\cite{baxter08}, with particular reference to applications in
population genetics and biodiversity studies.

\begin{figure}
\includegraphics[width=7cm,angle=-90]{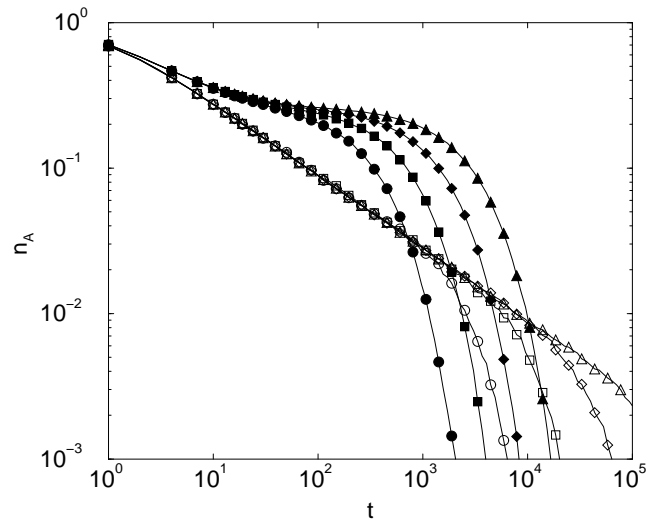}
\caption{Log-log plot of the fraction $n_A$ of active bonds between
  nodes with different opinions. Empty symbols are for the
  one-dimensional case ($p=0$).  Filled symbols are for rewiring
  probability $p=0.05$.  Data are for $N=200$ (circles), $N=400$
  (squares), $N=800$ (diamonds), $N=1600$ (triangles up) and $N=3200$
  (triangles left).  Reprinted figure with permission
  from~\cite{castellano03}. Copyright 2003 from EDP Sciences.}
\label{castellano03_1}
\end{figure}

Another interesting effect of the topology occurs when voter dynamics
is considered on small-world networks~\cite{watts98}.  After an
initial regime equal to the one-dimensional behavior, the density of
active interfaces forms a plateau (Fig.~\ref{castellano03_1}), because
shortcuts hinder their diffusive motion. The system remains trapped in
a metastable state with coexisting domains of opposite opinions, whose
typical length scales as $1/p$~\cite{castellano03,vilone04}, $p$ being
the fraction of long-range connections.

The lifetime of the metastable state scales with the linear system
size $L$ so that for finite systems consensus is eventually reached on
a temporal scale shorter than on a regular one-dimensional lattice
($L^2$).  For infinite systems instead, the state with coexisting
opinions is actually stable, leading to the conclusion that long-range
connections prevent the complete ordering of the voter model, in a way
similar to what occurs for Glauber dynamics~\cite{boyer03}.  A general
discussion of the interplay between topology and dynamics for the
voter model is presented in~\cite{suchecki05b}. A comparison between
the behavior of the voter dynamics and the AB-model on modular
networks is in~\cite{castello07}.

\subsection{Majority rule model}
\label{sec:galam}

\begin{figure}
\includegraphics[width=4cm]{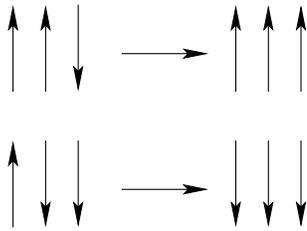}
\caption {\label{figop1} MR model. The majority opinion inside a
  discussion group (here of size three) is taken by all agents.}
\end{figure}

In a population of $N$ agents, endowed with binary opinions, a
fraction $p_+$ of agents has opinion $+1$ while a fraction $p_-=1-p_+$
opinion $-1$.  For simplicity, suppose that all agents can communicate
with each other, so that the social network of contacts is a complete
graph.  At each iteration, a group of $r$ agents is selected at random
(discussion group): as a consequence of the interaction, all agents
take the majority opinion inside the group (Fig.~\ref{figop1}). 
This
is the basic principle of the majority rule (MR) model, which was
proposed to describe public debates~\cite{galam02}.  

The group size $r$ is not fixed, but is selected at each step from a
given distribution.  If $r$ is odd, there is always a majority in
favor of either opinion. If $r$ is even, instead, there is the
possibility of a tie, i.e., that either opinion is supported by
exactly $r/2$ agents. In this case, one introduces a bias in favor of
one of the opinions, say $+1$, and that opinion prevails in the group.
This prescription is inspired by the principle of social inertia, for
which people are reluctant to accept a reform if there is no clear
majority in its favor~\cite{friedman84}. Majority rule with opinion
bias was originally applied within a simple model describing
hierarchical voting in a
society~\cite{galam86,galam90,galam99,galam00}.

Defined as $p_+^0$ the initial fraction of agents with the opinion
$+1$, the dynamics is characterized by a threshold $p_c$ such that,
for $p_+^0>p_c$ ($p_+^0<p_c$), all agents will have opinion $+1$
($-1$) in the long run. The time to reach consensus (in number of
updates per spin) scales like $\log N$~\cite{tessone04}.  If the group
sizes are odd, $p_c(r)=1/2$, due to the symmetry of the two opinions.
If there are groups with $r$ even, $p_c<1/2$, i.e., the favored
opinion will eventually be the dominant one, even if it is initially
shared by a minority of agents.

The MR model\footnote{The name Majority Rule model was actually coined
  in~\cite{krapivsky03}. Since this model is just a special case of
  the one introduced in~\cite{galam02}, we adopt this name since the
  beginning of the section.} with a fixed group size $r$ was
analytically solved in the mean field
limit~\cite{krapivsky03}.  The group size $r$ is odd, to keep the
symmetry of the two opinions.  The solution can be derived both for a
finite population of $N$ agents and in the continuum limit of
$N\rightarrow\infty$. The latter derivation is simpler~\cite{chen05a},
and is sketched here.

Let $s_k=\pm 1$ be the opinion of agent $k$; the average opinion
(magnetization) of the system is $m=1/N\sum_ks_k=p_+-p_-$.  The size of
each discussion group is $3$. At each update step, the number $N_+$ of
agents in state $+$ increases by one unit if the group state is $++-$,
while it decreases by one unit if the group state is $+--$.  One thus
has:
\begin{equation}
  dN_+=3(p_+^2p_--p_+p_-^2)=-6p_+(p_+-\frac{1}{2})(p_+-1),
\label{op_eq2}
\end{equation}
where the factor of $3$ is due to the different permutations of the
configurations $++-$ and $+--$. Eq.~(\ref{op_eq2}) can be rewritten as:
\begin{equation}
  \frac{dN_+}{N}\frac{N}{3}=\dot{p}_+=-2p_+(p_+-\frac{1}{2})(p_+-1),
\label{op_eq3}
\end{equation}
with the time increment $dt=3/N$, so that each agent is updated once
per unit of time.  The fixed points are determined by the condition
$\dot{p}_+=0$ and from Eq.~(\ref{op_eq3}) we see that this happens
when $p_+=0$, $1/2$ and $1$, respectively.  The point $p_+=1/2$ is
unstable, whereas the others are stable: starting from any $p_+\neq
1/2$, all agents will converge to the state of initial majority,
recovering Galam's result. The integration of Eq.~(\ref{op_eq3})
yields that the consensus time grows as $\log N$.

On a $d$-dimensional lattice, the discussion group is localized around 
a randomly chosen lattice site. 
In one dimension, the model is not analytically solvable.  Since the
average magnetization is not conserved by the MR dynamics, the exit
probability, i.e., the probability that the final magnetization is
$+1$, has a non-trivial dependence on the initial magnetization in the
thermodynamic limit and a minority can actually win the
contest. Consensus time grows as $N^2$.  In
higher dimensions~\cite{chen05}, the dynamics is characterized by
diffusive coarsening.  When the initial magnetization is zero, the
system may be trapped in metastable states (stripes in $2d$, slabs in
$3d$), which evolve only very slowly. This leads to the existence of
two distinct temporal scales: the most probable consensus time is
short but, when metastable states appear, the time needed is
exceedingly longer. As a consequence, the average consensus time grows
as a power of $N$, with a dimension-dependent exponent. When the
initial magnetization is non-zero, metastable states quickly
disappear.  A crude coarse-graining argument reproduces qualitatively
the occurrences of metastable configurations for any $d$. Numerical
simulations show that the MR model in four dimensions does not
reproduce the results of the mean field limit, so the upper critical
dimension of the MR model is larger than four.  The MR dynamics was
also investigated on networks with strong degree
heterogeneities~\cite{lambiotte07} and on networks with community
structure, i.e., graphs consisting of groups of nodes with a
comparatively large density of internal links with respect to the
density of connections between different
groups~\cite{lambiotte06,lambiotte07a}.  The MR model was
studied on small world lattices~\cite{li06} as well.

The MR model has been extended to multi-state opinions and plurality
rule~\cite{chen05a}.  The number of opinion states and the size of the
interaction groups are denoted with $s$ and $G$, respectively.  In the
mean field limit, the system reaches consensus for any choice of $s$
and $G$, in a time that scales like $\log N$, as in the $2$-state MR
model. On a square lattice, if the number of states $s$ is too large,
there are no groups with a majority, so the system does not evolve,
otherwise the evolution is based on diffusive coarsening, similarly to
that of the $2$-state MR model. Again, two different timescales
emerge when $s$ is small, due to the existence of metastable
states. When $s$ and $G$ approach a threshold, there is only one
domain that grows and invades all sites, so there is only one time
scale. The plurality rule is a special extension of the MR rule when
there are more than two opinion states: in this case, all agents of a
group take the opinion with the most representatives in the group.
The evolution leads to consensus for any $s$ and $G$, because all
interaction groups are active (there is always a relative majority);
when the opinions reduce to two, the dynamics becomes identical to
that of the $2$-state MR model, so there will be metastable states and
two different timescales.

Modifications of the MR model include: a model where agents can 
move in space~\cite{stauffer02b,galam02a}; a dynamics where each agent interacts with a 
variable number of neighbors~\cite{tessone04};
an extension to three opinions~\cite{gekle05}; the introduction of a probability to favor a
particular opinion, that could vary among different individuals and/or
social groups~\cite{galam05}; the presence of ``contrarians'', i.e., agents that initially take the
majority opinion in a group discussion, but that right after the
discussion switch to the opposite opinion~\cite{galam04,stauffer04a};
the presence of one-sided contrarians and unsettled
agents~\cite{borghesi06}; the presence of inflexible agents, that
always stay by their side~\cite{galam07}.

We now discuss some variants of the majority rule.  In the
majority-minority (MM) model~\cite{mobilia03b}, one accounts for the
possibility that minorities take over: in a discussion group the
majority opinion prevails with a probability $p$, whereas with a
probability $1-p$ it is the minority opinion that dominates. For a
discussion group of three agents, the magnetization $m$ changes by an
amount $2p-4(1-p)$ at each interaction, which means that, for
$p=p_c=2/3$, $m$ does not change on average, like in the voter
model. In the mean field limit, the model can be solved analytically:
the exit probability turns out to be a step function for $p>p_c$,
(i.e., the system will evolve towards consensus around the initial
majority opinion), whereas it equals $1/2$ for $p<p_c$, which means
that the system is driven towards zero magnetization.

Another interesting model based on majority rule is the majority-vote
model~\cite{liggett85}. At each update step, with a probability $1-q$
a spin takes the sign of the majority of its neighbors, with a
probability $q$ it takes the minority spin state. If there is a tie,
the spin is flipped with probability $1/2$.  The parameter $q$ is the
so-called noise parameter.  We stress that a single spin is updated at
each time step, at variance with the MR model.  For $q=0$ the model
coincides with the Ising model with zero-temperature Glauber
kinetics~\cite{glauber63}.  On a regular lattice, the majority-vote
model presents a phase transition from an ordered to a disordered
state at a critical value $q_c$ of the noise
parameter~\cite{oliveira92}. The critical exponents of the transition
are in the Ising universality class. Recent studies showed that the
majority-vote model also generates an order-disorder phase transition
on small-world lattices~\cite{campos03} and on random
graphs~\cite{pereira05,lima08}.

In a recent model, an agent is convinced if there is at least a
fraction $p$ of its neighbors sharing the same
opinion~\cite{klimek07}. This model interpolates between the rule
where majority dominates ($p=1/2$) and the unanimity rule ($p=1$),
where an agent is
influenced by its neighbors only if they all have the same
opinion~\cite{lambiotte07d}.
Another model based similar to majority-vote is presented
and studied on directed networks in~\cite{sanchez02}.

\subsection{Social impact theory}
\label{sec:impact}

The psychological theory of social impact~\cite{latane81} describes
how individuals feel the presence of their peers and how they in turn
influence other individuals.  The impact of a social group on a
subject depends on the number of the individuals in the group, on
their convincing power, and on the distance from the subject, where
the distance may refer both to spatial proximity or to the closeness
in an abstract space of personal relationships.  The original cellular
automata introduced in~\cite{latane81} and refined in~\cite{nowak90}
represent a class of dynamic models of statistical mechanics, that
are exactly solvable in the mean field limit~\cite{lewenstein92}.

The starting point is a population of $N$ individuals.  Each
individual $i$ is characterized by an opinion $\sigma_i=\pm 1$ and by
two real-valued parameters, that estimate the strength of its action
on the others: persuasiveness $p_i$ and supportiveness $s_i$, that
describe the capability to convince someone to change or to keep its
opinion, respectively. These parameters are assumed to be random
numbers, and introduce a disorder that is responsible for the complex
dynamics of the model.  The distance of a pair of agents $i$ and $j$
is $d_{ij}$.  In the simplest version, the total impact $I_i$ that an
individual $i$ experiences from his/her social environment is
\begin{equation}
I_i=\Big[\sum_{j=1}^{N}\frac{p_j}{d^\alpha_{ij}}(1-\sigma_i\sigma_j)\Big]-
\Big[\sum_{j=1}^{N}\frac{s_j}{d^\alpha_{ij}}(1+\sigma_i\sigma_j)\Big],
\label{op_eq4}
\end{equation}
where $\alpha>2$ expresses how fast the impact decreases with the distance $d_{ij}$ between two individuals.
The first term of Eq.~(\ref{op_eq4}) expresses the persuasive impact, i.e. the pressure exerted by
the agents with opposite opinions, which tend to enforce an opinion change; the second term instead
is the supportive impact, i.e. the pressure exerted by the agents with the same opinion of $i$, which
favor the status quo. In both cases, the impact of each agent on $i$ 
is proportional to its persuasiveness/supportiveness.

The opinion dynamics is expressed by the rule
\begin{equation}
\sigma_i(t+1)=-\textstyle{sgn}[\sigma_i(t)I_i(t)+h_i],
\label{op_eq5}
\end{equation}
where $h_i$ is a random field representing all sources other than
social impact that may affect the opinion (e.g. mass media).  According to
Eq.~(\ref{op_eq5}), a spin flips if the pressure in favor of the
opinion change overcomes the pressure to keep the current opinion
($I_i>0$ for vanishing $h_i$).

For a system of fully connected agents, and without individual fields,
the model presents infinitely many stationary
states~\cite{lewenstein92}. The order parameter of the dynamics is a
complex function of one variable, like in spin
glasses~\cite{mezard87}.

In general, in the absence of individual fields, the dynamics leads to
the dominance of one opinion on the other, but not to complete
consensus. If the initial magnetization is about zero, the system
converges to configurations characterized by a large majority of spins
in the same opinion state, and by stable domains of spins in the
minority opinion state.  In the presence of individual fields, these
minority domains become metastable: they remain stationary for a very
long time, then they suddenly shrink to smaller clusters, which again
persist for a very long time, before shrinking again, and so on
(``staircase dynamics'').

The dynamics can be modified to account for other processes related to
social behavior, such as learning~\cite{kohring96}, the response of a
population to the simultaneous action of a strong leader and external
influence ~\cite{kacperski96,
  kacperski97,kacperski99,kacperski00,holyst00} and the mitigation of
social impact due to the coexistence of different individuals in a
group~\cite{bordogna07}.  For a review of statistical mechanical
models of social impact, see~\cite{holyst01}.

Social impact theory neglects a number of realistic features of social
interaction: the existence of a memory of the individuals, which
reflects the past experience; a finite velocity for the exchange of
information between agents; a physical space, where agents have the
possibility to move.  An important extension of social impact theory
that includes those features is based on \textit{active Brownian
  particles}~\cite{schweitzer00,schweitzer03}, that are Brownian
particles endowed with some internal energy depot that allows them to
move and to perform several tasks as well.  The interaction is due to
a scalar opinion field, expressing the social impact of all agents/opinions
at each point in space; 
the particles/agents act as sources of the field and are in turn affected by it, both in
opinion and in space. Each agent $i$ is labeled by its opinion
$\sigma_i=\pm 1$ and its personal strength $s_i$. The field of opinion
$\sigma$, at position ${\bf r}$ and time $t$, is indicated with
$h_{\sigma}({\bf r},t)$. The transition probability rates
$w(\sigma^{\prime}|\sigma)$, for an agent to pass from opinion
$\sigma$ to opinion $\sigma^{\prime}$, with
$\sigma\neq\sigma^{\prime}$, are defined as:
\begin{equation}
w(\sigma^{\prime}|\sigma)=\eta\exp\{[h_{\sigma^{\prime}}({\bf r}, t)-h_{\sigma}({\bf r}, t)]/T\}, 
\label{op_eq8}
\end{equation}
where $T$ is a social temperature.
The dynamics is expressed by two sets of
equations: one set describes the spatio-temporal change of the
opinion field
\begin{equation}
\frac{\partial}{\partial t}h_{\sigma}({\bf r}, t)=\displaystyle{\sum_{i=1}^{N}s_i
\delta_{\sigma,\sigma_i}\delta({\bf r}-{\bf r}_i)-\gamma h_{\sigma}({\bf r}, t)+D_h\Delta h_{\sigma}({\bf r}, t)},
\label{op_eq6}
\end{equation}
the other set presents reaction-diffusion equations for the density
$n_{\sigma}({\bf r}, t)$ of individuals with opinion $\sigma$, at
position ${\bf r}$ and time $t$
\begin{eqnarray}
  \lefteqn{\displaystyle{\frac{\partial}{\partial t}n_{\sigma}({\bf r}, t)=-\nabla[n_{\sigma}({\bf r}, t)\alpha\nabla h_{\sigma}({\bf r}, t)]
      +D_n\Delta n_{\sigma}({\bf r}, t)}}\nonumber\\
  &&-
  \displaystyle{\sum_{\sigma^{\prime}\neq \sigma}[w(\sigma^{\prime}|\sigma)n_{\sigma}({\bf r}, t)-w(\sigma|\sigma^{\prime})n_{\sigma^{\prime}}({\bf r}, t)]}.
\label{op_eq7}
\end{eqnarray}
In the equations above, $N$ is the number of agents, $1/\gamma$ is the
average lifetime of the field, $D_h$ is the diffusion
constant for information exchange, $D_n$ the spatial diffusion
coefficient of the individuals, $\alpha$ measures the agents' response
to the field. The three terms on the right-hand side of Eq.~(\ref{op_eq6}) represent the
microscopic density of the agents' strength, the relaxation of the field (modeling the memory effects)
and its diffusion in space, respectively. The first two terms on the right-hand side of Eq.~(\ref{op_eq7}) 
indicate the change in the agents' density due to their motion in space, whereas the sum
expresses the balance between the gain and loss of individuals due to opinion changes.

Eqs.~(\ref{op_eq6}) and
~(\ref{op_eq7}) are coupled: depending on the local intensity of the
field supporting either opinion, an agent can change its opinion, or
migrate towards locations where its opinion has a larger
support. Opinion changes and migrations have a non-linear feedback on
the communication field, which in turn affects the agents, and so on.
The model presents three phases, depending on the values of the
parameters: a paramagnetic phase, where both opinions have the same
probability ($1/2$) of being selected at every place
(high-temperature, high-diffusion), a ferromagnetic phase, with more
agents in favor of one opinion over the other (low-temperature,
low-diffusion), and a phase in which either opinion prevails in
spatially separated domains (segregation).

\subsection{Sznajd model}
\label{sec:sznajd}

In the previous section we have seen that the impact exerted by a
social group on an individual increases with the size of the group. We
would not pay attention to a single guy staring at a blank wall;
instead, if a group of people stares at that wall, we may be tempted
to do the same.  Convincing somebody is easier for two or more people
than for a single individual. This is the basic principle behind the
Sznajd model~\cite{stauffer03a,sznajd05a}. In its original
version~\cite{sznajd00}, that we call Sznajd B, agents occupy the
sites of a linear chain, and have binary opinions, denoted by Ising
spin variables. A pair of neighboring agents $i$ and $i+1$ determines
the opinions of their two nearest neighbors $i-1$ and $i+2$, according
to these rules:
\begin{equation}
{\textrm{if }}s_i=s_{i+1}\textrm{, then }s_{i-1}=s_i=s_{i+1}=s_{i+2};
\label{rule1}
\end{equation}
\begin{equation}
\textrm{if }s_i\neq s_{i+1}\textrm{, then }s_{i-1}=s_{i+1} \textrm{ and }s_{i+2}=s_{i}.
\label{rule2}
\end{equation}
So, if the agents of the pair share the same opinion, they
successfully impose their opinion on their neighbors.  If, instead,
the two agents disagree, each agent imposes its opinion on the other
agent's neighbor.  

Opinions are updated in a random sequential order.  Starting from a
totally random initial configuration, where both opinions are equally
distributed, two types of stationary states are found, corresponding
to consensus, with all spins up ($m=1$) or all spins down ($m=-1$),
and to a stalemate, with the same number of up and down spins in
antiferromagnetic order ($m=0$).  The latter state is a consequence of
rule~(\ref{rule2}), that favors antiferromagnetic configurations, and has a
probability $1/2$ to be reached. Each of the two (ferromagnetic)
consensus states occurs with a probability $1/4$. The values of the
probability can be easily deduced from the up-down symmetry of the
model. The relaxation time of the system into one of the possible
attractors has a log-normal distribution~\cite{behera03}. The number of
agents that never changed opinion first decays as a power law of time,
and then it reaches a constant but finite value, at odds with the
Ising model~\cite{stauffer02c}. The exit probability
has been calculated analytically for both random and correlated 
initial conditions~\cite{slanina07,lambiotte07c}.

Since the very introduction of the Sznajd model, it has been argued
that a distinctive feature of its dynamics is the fact that the
information flows from the initial pair of agents to their neighbors,
at variance with the other opinion dynamics models, in which instead
agents are influenced by their neighbors. Because of that the Sznajd
model was supposed to describe how opinions spread in a society. On
the other hand, in~\cite{behera03} it has been shown that, in one
dimension, the direction of the information flow is actually irrelevant,
and that the Sznajd B dynamics is equivalent to a voter dynamics.
The only difference with the classic voter model is that an agent is
not influenced by its nearest neighbors but by its next-to-nearest
neighbors. Indeed, the dynamics of Sznajd B on a linear chain can be
summarized by the simple sentence ``just follow your next-to-nearest
neighbor''.  The fact that in Sznajd a pair of agents is updated at a
time, whereas in the voter model the dynamics affects a single spin,
introduces a factor of two in the average relaxation time of the
equivalent voter dynamics; all other features are exactly the same,
from the probability to hit the attractors to the distributions of
decision and relaxation times.  Therefore, Sznajd B does not respect
the principle of social validation which motivated its introduction,
as each spin is influenced only by a single spin, not by a pair.

\begin{figure}
\includegraphics[width=4cm]{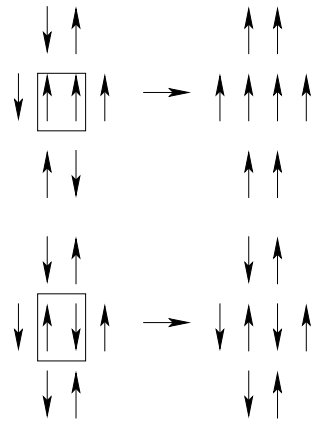}
\caption {\label{sznajd1} Sznajd model. In the most common
  version of the model (Sznajd A), a pair of
  neighboring agents with the same opinion convince all their
  neighbors (top), while they have no influence if they disagree
  (bottom).}
\end{figure}

Sznajd rule~(\ref{rule2}) is unrealistic and was soon replaced by alternative
recipes in subsequent studies.  In the most popular alternative, that
we call Sznajd A, only the ferromagnetic rule~(\ref{rule1}) holds, so the neighbors
of a disagreeing agents' pair maintain their opinions. Extensions of
the Sznajd model to different substrates usually adopt this
prescription and we shall stick to it unless stated otherwise.  On the
square lattice, for instance, a pair of neighboring agents affect the
opinions of their six neighbors only if they agree
(Fig.~\ref{sznajd1}).  
In this case, the exit probability is a step
function with threshold at $m=0$: if the initial magnetization $m<0$,
the system always attains consensus with $m=-1$; if $m>0$ initially,
the steady state is consensus with $m=1$.  The distribution of the
times required to reach complete consensus is broad, but not a
log-normal like for Sznajd B in one dimension~\cite{stauffer00}.  We
stress that Sznajd B in one dimension has no phase transition in the
exit probability, due to
the coexistence of ferro- and antiferromagnetic stationary states.

The fixed points of Sznajd A dynamics hold if one changes the size of
the pool of persuading agents.  The only exception is represented by
the so-called Ochrombel simplification of the Sznajd
model~\cite{ochrombel01}, in which a single agent imposes its opinion
on all its neighbors.

The results mentioned above were derived by computer
simulations. In~\cite{slanina03} an exact solution for a Sznajd-like
dynamics on a complete graph has been given. Here a pair of randomly
selected agents $i$ and $j$ interacts with a third agent $k$, also
taken at random.  If $s_i=s_j$, then $s_k=s_i=s_j$, otherwise nothing
happens. The evolution equation for the probability density
$P(m,t)$ that the system has magnetization $m$ at time $t$
reads:
\begin{equation}
  \frac{\partial}{\partial t}P(m,t)=-\frac{\partial}{\partial m}[(1-m^2)mP(m,t)].
\label{op_eq9}
\end{equation}
Eq.~(\ref{op_eq9}) is derived in the thermodynamic limit and it
represents a pure drift of the magnetization. The general solution is:
\begin{equation}
P(m,t)=[(1-m^2)m]^{-1}f\Big(e^{-t}\frac{m}{\sqrt{1-m^2}}\Big),
\label{op_eq11}
\end{equation}
where the function $f$ depends on the initial conditions.  If
$P(m,t=0)=\delta(m-m_0)$, i.e., the system starts with a fixed value
$m_0$ of the magnetization, $P(m,t)$ is a $\delta$-function at any
moment of the evolution; the center is pushed by the drift towards the
extremes $+1$ if $m_0>0$ or $-1$ if $m_0<0$, which are reached
asymptotically.  So, the initial magnetization $m_0$ determines the
final state of the system, which is consensus, and there is a phase
transition when $m_0$ changes sign.  Eq.~(\ref{op_eq9}) also allows to
derive the behavior of the tail of the distribution of the times to
reach the stationary states of the dynamics, which turns out to be
exponential.

Some effort has been devoted to find a proper Hamiltonian formulation
of Sznajd dynamics~\cite{sznajd02a,sznajd04,sznajd05b}.  It turns out
that the rules of the model are equivalent to the minimization of a
local function of spin-spin interactions, the so-called
\textit{disagreement function}.  On a linear chain of spins, the
disagreement function for spin $i$ reads:
\begin{equation}
E_i=-J_1s_is_{i+1}-J_2s_is_{i+2},
\label{op_eq12}
\end{equation}
where $J_1$ and $J_2$ are coupling constants, whose values determine
the type of dynamics, and $i+1$, $i+2$ are the right nearest and
next-to-nearest neighbors of $i$.  Here, spin $i$ takes the value that
minimizes $E_i$. The function $E_i$ and its minimization defines the
Two-Component (TC) model~\cite{sznajd02a}.  We remark that, when
$J_1J_2>0$, the two terms of $E_i$ are equivalent, so only one can be
kept.  Sznajd B dynamics is recovered for $-J_2<J_1<J_2$, $J_2>0$, but
the model has a much richer behavior. Based on the values of the pair
of parameters $J_1$ and $J_2$, one distinguishes four phases,
delimited by the bisectors $J_1\pm J_2=0$. Besides the known ferro-
and antiferromagnetic attractors, a new stationary configuration
emerges, with pairs of aligned spins whose signs alternate
($...++--++--...$). The TC model has been extended to the square
lattice~\cite{sznajd04}, and can be exactly solved in the mean field
limit~\cite{sznajd05b}.  In general, we stress that the model is not
equivalent to a Hamiltonian model at zero temperature, because it is
not possible to define a global energy for the system. The sum of the
disagreement function $E_i$ over all spins does not play the role of
the energy: the local minimization of $E_i$ can lead to an increase of
its global value~\cite{sznajd04}.

Sznajd dynamics turns out to be a special case of the general
sequential probabilistic model (GPM)~\cite{galam05b}.  Here, opinions
are Ising spins: the proportions of both opinions at time $t$ are
$p(t)$ (+) and $1-p(t)$ ($-$). In the mean field limit, a random group
of $k$ agents is selected, with $j$ agents with opinion $+$ and $k-j$
with opinion $-$.  The opinion dynamics of the GPM enforces consensus
among the agents of the group, which adopt opinion $+$ with a suitably
defined probability $m_{k,j}$ and opinion $-$ with probability
$1-m_{k,j}$. The probability $p(t+1)$ to find an agent sharing opinion
$+$ after the update is
\begin{equation}
p(t+1)=\sum_{j=0}^k m_{k,j}p(t)^j[1-p(t)]^{k-j}\frac{k!}{j!(k-j)!}.
\label{op_eq13}
\end{equation}
The size $k$ of the random group along with the local probabilities
$\{m_{k,j}\}$ completely define the dynamics of the GPM.  A phase
diagram can be derived as a function of the local probabilities. Only
two different phases are obtained, corresponding to consensus and
coexistence of the two opinions in equal proportions.  The phase
transition occurs at those values of the $\{m_{k,j}\}$ for which
magnetization is on average conserved: here the model has a voter
dynamics.  With suitable choices of the set $\{m_{k,j}\}$ the GPM
reproduces the MF behavior of all known models with binary opinions:
voter, majority rule, Sznajd, the majority-minority model, etc..

We now briefly review the modifications of the Sznajd model. The
dynamics has been studied on many different topologies: regular
lattices~\cite{stauffer00,chang01}, complete graphs~\cite{slanina03},
random graphs~\cite{rodrigues05}, small-world
networks~\cite{elgazzar03,he04} and scale-free
networks~\cite{bernardes02,bonnekoh03,sousa05,rodrigues05,sousa06}.
The Sznajd model on scale-free networks was recently
studied~\cite{gonzalez06} within a real space renormalization
framework. On any graph, if only Sznajd's ferromagnetic rule~(\ref{rule1}) holds,
the system undergoes a sharp dynamic phase transition from a state
with all spins down to a state will all spins up.  If the graph is not
fixed, but in evolution, like a growing network, the transition
becomes a smooth crossover between the two phases~\cite{gonzalez04}.
The phase transition holds as well if one introduces
dilution~\cite{moreira01}, if the number of opinion states is larger
than two~\cite{slanina03}, if the influence of the active pair of
agents extends beyond their neighborhood~\cite{schulze03a}, so it is a
very robust feature of the Sznajd model, although it disappears when
one includes noise~\cite{stauffer00} or antiferromagnetic
rules~\cite{chang01,sznajd04}. 

If the random sequential updating so far adopted is replaced by
synchronous updating, i.e.,  if at each iteration all agents of the
configurations are paired off and act simultaneously on their
neighbors, it may happen that an agent is induced to choose opposite
opinions by different neighboring pairs.  In this case the agent is
``frustrated'' and maintains its opinion. Such frustration hinders
consensus~\cite{stauffer02d,yu-song05}, due to the emergence of stable
clusters where both opinions coexist.  This problem can be limited if
noise is introduced~\cite{sabatelli04}, or if agents have memory, so
that, in case of conflicting advice, they follow the most frequent
opinion they had in the past~\cite{sabatelli03}.

When the possible opinion states are $q>2$, one can introduce bounded
confidence, i.e., the realistic principle that only people with similar
opinions can have an influence on each other. If we assume that two
opinions are similar if their values differ by at most one unit, and
that a pair of agents with the same opinion can convince only
neighbors of similar opinions, the Sznajd dynamics always leads to
complete consensus for $q=3$, whereas for $q>3$ it is very likely that
at least two opinions survive in the final stationary
state~\cite{stauffer02a}.  Bounded confidence allows for an extension of the Sznajd model
to real-valued opinions~\cite{fortunato05b}.  Other
studies focused on the dynamics of clusters of agents with regular opinion patterns,
ferromagnetic and/or antiferromagnetic~\cite{schneider05a},
damage spreading~\cite{roehner04,klietsch05}, the combination of
Sznajd with other convincing strategies~\cite{sousa06}, 
contrarian behavior~\cite{delalama05, wio06},
the effect on the dynamics of agents biased 
towards the global majority and/or minority opinion~\cite{schneider04,schneider05}. Recently 
extensions of Sznajd B to higher dimensions have been considered as well~\cite{kondrat08}.

The Sznajd model has found applications in different areas.  In
politics, it has been used to describe voting behavior in
elections~\cite{bernardes02,gonzalez04}; we shall discuss this issue
in Sec.~\ref{sec:elections}. Moreover, it was applied to study the
interaction of economic and personal attitudes of individuals, which
evolve according to different rules but in a coupled
manner~\cite{sznajd05}.  Sznajd dynamics has also been adopted to
model the competition of different products in an open
market~\cite{sznajd03}.  The effects of aging, diffusion and a
multi-layered society have been considered as
well~\cite{schulze03,schulze04}.  Sznajd dynamics has been adapted in
a model that describes the spread of opinions among a group of
traders~\cite{sznajd02}. Finally, Sznajd-like rules have been employed
to generate a new class of complex networks~\cite{costa05}.

\subsection{Bounded confidence models}
\label{sec:BCM}
           
\subsubsection{Continuous opinions}
\label{sec:cont_op}

In the models we have so far investigated opinion is a discrete
variable, which represents a reasonable description in several
instances.  However, there are cases in which the position of an
individual can vary smoothly from one extreme to the other of the
range of possible choices. As an example, one could think of the
political orientation of an individual, that is not restricted to the
choices of extreme Right/Left, but it includes all the options in
between, which may be indicated by the geometric position of the seat
of a deputy in the Parliament.

Continuous opinions invalidate some of the concepts adopted in models
with discrete choices, like the concepts of majority of an opinion and
equality of opinions, so they require a different framework. Indeed,
continuous opinion dynamics has historically followed an alternative
path. The first studies were carried out by applied mathematicians and
aimed at identifying the conditions under which a panel of experts
would reach a common decision~\cite{stone61,chatterjee77,cohen86}.

The initial state is usually a population of $N$ agents with randomly
assigned opinions, represented by real numbers within some
interval. In contrast to discrete opinion dynamics, here all agents
usually start with different opinions, and the possible scenarios are
more complex, with opinion clusters emerging in the final stationary
state.  The opinion clusters could be one (consensus), two
(polarization) or more (fragmentation). In principle, each agent can
interact with every other agent, no matter what their opinions are. In
practice, there is a real discussion only if the opinions of the
people involved are sufficiently close to each other. This realistic
aspect of human communications is called {\it bounded confidence}
(BC); in the literature it is expressed by introducing a real number
$\epsilon$, the {\it uncertainty} or {\it tolerance}, such that an
agent, with opinion $x$, only interacts with those of its peers whose
opinion lies in the interval $]x-\epsilon,x+\epsilon[$.

In this section we discuss the most popular BC models, i.e., the
Deffuant model~\cite{deffuant00} and that of
Hegselmann-Krause~\cite{hegselmann02}. BC models have been recently
reviewed in~\cite{lorenz07}.

\subsubsection{Deffuant model}
\label{sec:deffuant}

Let us consider a population of $N$ agents, represented by the nodes
of a graph, where agents may discuss with each other if the
corresponding nodes are connected.  Each agent $i$ is initially given
an opinion $x_i$, randomly chosen in the interval $[0,1]$.  The
dynamics is based on random binary encounters, i.e., at each time
step, a randomly selected agent discusses with one of its neighbors on
the social graph, also chosen at random.  Let $i$ and $j$ be the pair
of interacting agents at time $t$, with opinions $x_i(t)$ and
$x_j(t)$, respectively.  Deffuant dynamics is summarized as follows:
if the difference of the opinions $x_i(t)$ and $x_j(t)$ exceeds the
threshold $\epsilon$, nothing happens; if, instead,
$|x_i(t)-x_j(t)|<\epsilon$, then:
\begin{equation}
x_i(t+1)=x_i(t)+\mu[x_j(t)-x_i(t)],
\label{op_eq14}
\end{equation}
\begin{equation}
x_j(t+1)=x_j(t)+\mu[x_i(t)-x_j(t)].
\label{op_eq15}
\end{equation}

The parameter $\mu$ is the so-called convergence parameter, and its
value lies in the interval $[0,1/2]$.  Deffuant model is based on a
compromise strategy: after a constructive debate, the positions of the
interacting agents get closer to each other, by the relative amount
$\mu$. If $\mu=1/2$, the two agents will converge to the average of
their opinions before the discussion.  For any value of $\epsilon$ and
$\mu$, the average opinion of the agents' pair is the same before and
after the interaction, so the global average opinion ($1/2$) of the
population is an invariant of Deffuant dynamics.

The evolution is due to the instability of the initial uniform
configuration near the boundary of the opinion space.  Such
instability propagates towards the middle of the opinion space, giving
rise to patches with an increasing density of agents, that will become
the final opinion clusters.  Once each cluster is sufficiently far
from the others, so that the difference of opinions for agents in
distinct clusters exceeds the threshold, only agents inside the same
cluster may interact, and the dynamics leads to the convergence of the
opinions of all agents in the cluster to the same value. Therefore,
the final opinion configuration is a succession of Dirac's delta
functions.  In general, the number and size of the clusters depend on
the threshold $\epsilon$, whereas the parameter $\mu$ affects the
convergence time of the dynamics. However, when $\mu$ is small, the
final cluster configuration also depends on
$\mu$~\cite{laguna04,porfiri07}.

\begin{figure}[t]
\includegraphics[width=\columnwidth]{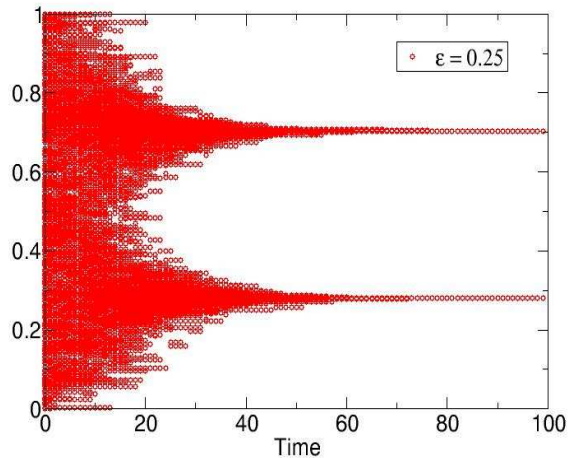}
\caption {\label{deffuant} Deffuant model. Opinion profile of a
  population of $500$ agents during its time evolution, for
  $\epsilon=0.25$. The population is fully mixed, i.e., everyone may
  interact with everybody else. The dynamics leads to a polarization
  of the population in two factions.}
\end{figure}

On complete graphs, regular lattices, random graphs and scale-free
networks, for $\epsilon>\epsilon_c=1/2$, all agents share the same
opinion $1/2$, so there is complete consensus~\cite{fortunato05a,lorenz07d}.
This may be a general property of Deffuant model, independently of the
underlying social graph.  If $\epsilon$ is small, more clusters emerge
(Fig.~\ref{deffuant}).

Monte Carlo simulations reveal that the number $n_c$ of clusters in
the final configuration can be approximated by the expression
$1/(2\epsilon)$.  This can be understood if we consider that, at
stationarity, agents belonging to different opinion clusters cannot
interact with each other, which means that the opinion of each cluster
must differ by at least $\epsilon$ from the opinions of its
neighboring clusters. In this way, within an interval of length
$2\epsilon$ centered at a cluster, there cannot be other clusters, and
the ratio $1/(2\epsilon)$ is a fair estimate for $n_c$.
 
Most results on Deffuant dynamics are derived through numerical
simulations, as the model is not analytically solvable. However, in
the special case of a fully mixed population, where everybody
interacts with everybody else, it is possible to write the general
rate equation governing the opinion dynamics~\cite{bennaim03a}. For this purpose, one
neglects individual agents and focuses on the evolution
of the opinion population $P(x,t)$, where $P(x,t)dx$ is the
probability that an agent has opinion in the interval $[x,x+dx]$. The
interaction threshold is $\epsilon=1$, but the opinion range is the
interval $[-\Delta, \Delta]$; this choice is equivalent to the usual
setting of the Deffuant model, if $\epsilon=1/{2\Delta}$. For
simplicity, $\mu=1/2$. The rate equation then reads:
\begin{eqnarray}
\frac{\partial}{\partial t}P(x,t)&=&\int_{|x_1-x_2|<1}\int dx_1dx_2P(x_1,t)P(x_2,t)\nonumber\\
&&\times \Big[\delta\Big(x-\frac{x_1+x_2}{2}\Big)-\delta(x-x_1)\Big].
\label{op_eq16}
\end{eqnarray}
Eq.~(\ref{op_eq16}) conserves the norm
$M_0=\int_{-\Delta}^{+\Delta}P(x,t)dx$ and the average opinion. The
question is to find the asymptotic state
$P_{\infty}(x)=P(x,t\rightarrow\infty)$, starting from the flat
initial distribution $P(x,t=0)=1$, for $x \in [-\Delta,\Delta]$. If
$\Delta<1/2$, all agents interact and Eq.~(\ref{op_eq16}) is
integrable.  In this case, it is possible to show that all agents
approach the central opinion $0$ and $P_{\infty}(x)=M_0\delta(x)$.

If $\Delta>1/2$, the equation is no longer analytically solvable.  The
asymptotic distribution is a linear combination of delta functions,
i.e.,
\begin{equation}
P_{\infty}(x)=\sum_{i=1}^{p}m_i\delta(x-x_i).
\label{op_eq18}
\end{equation}
The cluster masses $m_i$ must obey the conditions $\sum_im_i=M_0$ and 
$\sum_im_ix_i=0$; the latter comes from the conservation of the average opinion.
Numerical solutions of Eq.~(\ref{op_eq16}) reveal that there are
only three types of clusters: major (mass $>1$), minor (mass $<
10^{-2}$) and a central cluster located at $x=0$. These clusters are
generated by a periodic sequence of bifurcations, consisting in the
nucleation and annihilation of clusters. 

On any graph topology, Deffuant dynamics is always characterized by a
stationary state with one or more opinion clusters.  However, as the
interaction range of an agent is restricted to its topological
neighborhood, more opinion clusters emerge for low values of the
uncertainty.  Opinion homogenization involves only agents in the same
cluster: in this way, if two clusters are geometrically separated,
there will be no communication between the corresponding agents and
the final opinions will be in general different in each cluster, even
if their opinions are compatible, so that they would converge to
the same opinion on a complete graph.  The result is an increased
fragmentation of the agents' population. On scale-free networks, the
number of surviving opinions in the stationary state is proportional
to the number of agents of the population, for fixed
$\epsilon$~\cite{stauffer04d}. In particular, nodes with few
connections have a sizeable probability to be excluded from the
dynamics and to keep their opinion forever~\cite{weisbuch04}. The
result holds for both static and evolving networks~\cite{sousa04}.

Deffuant model can be defined as well if opinions are not continuous
but discretized~\cite{stauffer04}.  Here the opinion $s$ of any agent
can take one of $Q$ values, $s=1,2,...,Q$.  Opinions $s_i$ and
$s_j\neq s_i$ are compatible if $|s_i-s_j|\leq L$, where $L$ is an
integer.  The rules are the same as in Eqs.~(\ref{op_eq14}) and
~(\ref{op_eq15}), still with a real-valued convergence parameter
$\mu$, but the shift of the opinions is rounded to the nearest
integer.  In the limit $L\rightarrow\infty$ and $Q\rightarrow\infty$,
with the ratio $\epsilon=L/Q$ kept constant, one recovers the results
of the original model with continuous opinions.  If $L=1$, on a
complete graph consensus is the only stationary state if $Q=2$, i.e.,
for binary opinions\footnote{We remark that, for $L=1$, it is
impossible for the two interacting opinions to shift towards each
other, as only integer opinion values are allowed; so, as a result
of the discussion, one agent takes the opinion of the other.}.
Instead, for $Q>2$, complete consensus is never attained, but there
are at least two opinion clusters in the final configuration. On
scale-free networks the number of surviving opinions in the stationary
state is a scaling function of $Q$ and the population size $N$.

Simple modifications of Deffuant model yield rich dynamics. If agents
have individual values of $\epsilon$~\cite{weisbuch02,lorenz08}, the dynamics
is dominated by the agents with large uncertainties. In several
papers~\cite{deffuant02,amblard04,weisbuch05,deffuant04,deffuant06,aldashev07},
the uncertainties are also affected by the dynamics. In addition, they
are also coupled to the opinions, based on the principle that a small
uncertainty also implies more confidence and a higher probability to
affect other opinions. These models are able to explain how extremal
positions, initially shared by a minority of people, may eventually
dominate in society. In~\cite{bennaim05} a model in which Deffuant
compromise strategy is combined with spontaneous changes of the
agents' opinions, has been studied.  The latter phenomenon is
described as a diffusion process in the opinion space, which affects
cluster formation and evolution, with large clusters steadily
overtaking small ones.  Other extensions of Deffuant dynamics include
the introduction of an external periodic perturbation affecting all
agents at once, to simulate the effect of
propaganda~\cite{carletti06}, and the study of a more complex opinion
dynamics where the interaction of pairs of agents depends not only on
the compatibility of their opinions, but also on the coevolving mutual
affinity of the agents~\cite{bagnoli07,carletti07}.  This coupling provides a
natural and endogenous way of determining the number of opinion
clusters and their positions.

\subsubsection{Hegselmann-Krause model}
\label{sec:hk}

The model proposed in~\cite{hegselmann02} (HK) is quite similar to
that of Deffuant. Opinions take real values in an interval, say
$[0,1]$, and an agent $i$, with opinion $x_i$, interacts with
neighboring agents whose opinions lie in the range
$[x_i-\epsilon,x_i+\epsilon]$, where $\epsilon$ is the
uncertainty. The difference is given by the update rule: agent $i$
does not interact with one of its compatible neighbors, like in
Deffuant, but with all its compatible neighbors at once.  Deffuant's
prescription is suitable to describe the opinion dynamics of large
populations, where people meet in small groups, like pairs.  In
contrast, HK rule is intended to describe formal meetings, where
there is an effective interaction involving many people at the same
time.

On a generic graph, HK update rule for the opinion of agent $i$ at
time $t$ reads:
\begin{equation}
  x_i(t+1)=\frac{\sum_{j:|x_i(t)-x_j(t)|<\epsilon}a_{ij}x_j(t)}{\sum_{j:|x_i(t)-x_j(t)|<\epsilon}a_{ij}},
\label{op_eq19}
\end{equation}
where $a_{ij}$ is the adjacency matrix of the graph. So, agent $i$
takes the average opinion of its compatible neighbors. The model is
fully determined by the uncertainty $\epsilon$, unlike Deffuant
dynamics, for which one needs to specify as well the convergence
parameter $\mu$.  The need to calculate opinion averages of groups of
agents that may be rather large makes computer simulations of the HK
model rather lengthy as compared to Deffuant's. This may explain why
the HK model has not been studied by many authors.
 
The dynamics develops just like in Deffuant, and leads to the same
pattern of stationary states, with the number of final opinion
clusters decreasing if $\epsilon$ increases. In particular, for
$\epsilon$ above some threshold $\epsilon_c$, there can only be one
cluster.  On a complete graph, the final configurations are symmetric
with respect to the central opinion $1/2$, because the average opinion
of the system is conserved by the dynamics~\cite{fortunato05d}, as in
Deffuant. The time to reach the stationary state diverges in correspondence to the
bifurcation thresholds of opinion clusters, due to the presence of isolated agents lying between consecutive
clusters~\cite{fortunato05d}.

The threshold for complete consensus $\epsilon_c$ can only take one of
two values, depending on the behavior of the average degree $\langle
k\rangle$ of the underlying social graph when the number of nodes $N$
grows large~\cite{fortunato05c}.  If $\langle k\rangle$ is constant in
the limit of large $N$, as for example in lattices,
$\epsilon_c=\epsilon_1=1/2$.  Instead, if $\langle
k\rangle\rightarrow\infty$ when $N\rightarrow\infty$, as for example
in complete graphs, $\epsilon_c=\epsilon_2\sim 0.2$.  We have seen
instead that, for Deffuant, $\epsilon_c=1/2$ on any graph.

The extension of HK to discretized opinions~\cite{fortunato05a} is
essentially a voter model with bounded confidence: an agent picks at
random the opinion of a compatible neighbor. For three opinion values
and uncertainty one, the model reduces to the constrained voter
model~\cite{vazquez03}.

Other developments include: the use of alternative recipes to average
the opinions in Eq.~(\ref{op_eq19})~\cite{hegselmann05}; an analysis
of damage spreading~\cite{fortunato05}; the introduction of a general
framework where the size of the groups of interacting agents varies
from $2$ (Deffuant) to $N$ (HK)~\cite{urbig07}; the reformulation of
Deffuant and HK dynamics as interactive Markov
chains~\cite{lorenz06,lorenz07c}; analytical results on the stability
of BC opinion dynamics~\cite{lorenz05} and their ability to preserve
the relative ordering of the opinions~\cite{hendrickx07}.

\subsection{Other models}
\label{sec:othermodels_opinions}

The opinion dynamics models described so far are based on elementary
mechanisms, which explain their success and the many investigations
they have stimulated.  Such models, however, do not exhaust the wide
field of opinion dynamics.  The last years witnessed a real explosion
of new models, based on similar concepts as the classical models or on
entirely new principles.  Here we briefly survey these alternative
models.

The basic models we have seen are essentially deterministic, i.e., the
final state of the system after an interaction is always well
defined. Randomness can be introduced, in the form of a social
temperature or pure noise, but it is not a fundamental feature. Most
models of last generation, instead, focus on the importance of
randomness in the process of opinion formation.  Randomness is a
necessary ingredient of social interactions: both our individual
attitudes and the social influence of our peers may vary in a
non-predictable way.  Besides, the influence of external factors like
mass media, propaganda, etc., is also hardly predictable.  In this
respect, opinion dynamics is a stochastic process.
 
An interesting variant of Ising dynamics has been recently explored in~\cite{jiang07,jiang08}.
Here an agent surrounded by a majority of neighbors with equal opinion will flip its opinion with 
the usual Metropolis probability $\exp(-\Delta E/T)$, where $\Delta E$ is the increase of the Ising energy
due to the flip and $T$ the (social) temperature. When the majority of the neighbors disagrees
with the agent's opinion, the latter flips with a probability $q$, 
which accounts for the possibility that agents keep their opinion in spite of social pressure 
(``inflexible agents''). For $q=1$ one recovers the usual Ising dynamics. For $q<1$ instead, the model 
obeys a non-equilibrium dynamics, since detailed balance is violated. Simulations of the model 
on the lattice reveal a non-trivial phase diagram:
the ferromagnetic transition is continuous for $q$ larger than a critical value $q_c$, 
and discontinuous for $q<q_c$. The critical exponents of the continuous transitions differ both from
the mean field and the Ising exponents. On small-world lattices with a large density of shortcuts
there is no magnetization transition if $q$ is sufficiently low~\cite{jiang08}.

In~\cite{bartolozzi05} a model with binary opinions, evolving according
to a heat bath dynamics, is proposed.  The opinion field acting on a
spin is given by a linear combination with random weights of a term
proportional to the average opinion of its nearest neighbors on the
social network, with a term proportional to the average opinion of the
whole network.  When the stochastic noise exceeds a threshold, the
time evolution of the average opinion of the system is characterized
by large intermittent fluctuations; a comparison with the time series
of the Dow-Jones index at New York's Stock Exchange reveals striking
similarities.

In a recent model~\cite{kuperman02}, opinions are affected by three
processes: social imitation, occurring via majority rule, fashion,
expressed by an external modulation acting on all agents, and
individual uncertainty, expressed by random noise.  Stochastic
resonance~\cite{gammaitoni98} was observed: a suitable amount of noise
leads to a strong amplification of the response of the system to the
external modulation. The phenomenon occurs as well if one varies the
size of the system for a fixed amount of noise~\cite{tessone05}: here
the best response to the external solicitation is achieved for an
optimal population size (system size stochastic resonance).

Kinetic models of opinion dynamics were proposed
in~\cite{toscani06}. Interactions are binary, and the opinions of the
interacting pair of agents vary according to a compromise strategy
{\`a} la Deffuant, combined with the possibility of opinion diffusion,
following the original idea~\cite{bennaim05} discussed in
Sec.~\ref{sec:deffuant}.  The importance of diffusion in the process
is expressed by a random weight. The dynamics can be easily
reformulated in terms of Fokker-Planck equations, from which it is
possible to deduce the asymptotic opinion configurations of the
model. Fokker-Planck equations have also been employed to study a
dynamics similar to that of the constrained voter
model~\cite{vazquez03}, but in the presence of a social temperature,
inducing spontaneous opinion changes~\cite{delalama06}.

In~\cite{martins07,martins08} binary and continuous opinions are
combined.  The model is based on the simple idea that declared
opinions, or actions, are only a projection of the actual opinion of
the agents. Two persons can express the same preference but their
individual certitudes towards that preference may be different. So
one has to distinguish between the internal opinion, which is
expressed by a probability, and the external opinion, or action, which
is a binary variable. The internal opinion is a measure of how extreme
a position is.  Agents vary both their actions and their internal
opinions based on the observation of the actions of their peers.  In
this way one can monitor how the convictions of individuals are
coupled to their actions.  Clusters of agents with the same external
opinions display a characteristic pattern, where the agents are very
convinced of their choices (extremists) if they are well inside the
cluster, whereas they are more open-minded if they lie at the
boundary.

Synchronization has also been used to explain consensus formation.  A
variant of the Kuramoto model~\cite{kuramoto75}, where the phases of
the oscillators are replaced by unbounded real numbers, representing
the opinions, displays a phase transition from an incoherent phase
(anarchy), to a synchronized phase (consensus)
~\cite{pluchino05,pluchino06}. In~\cite{dimare06} it was shown that
several opinion dynamics models can be reformulated in the context of
strategic game theory.

Some models focus on specific aspects of opinion dynamics.
In~\cite{indekeu04} it has been pointed out that the influence of
network hubs in opinion dynamics is overestimated, because it is
unlikely that a hub-agent devotes much time to all its social
contacts. If each agent puts the same time in its social
relationships, this time will be distributed among all its social
contacts; so the effective strength of the interaction between two
neighboring agents will be the smaller, the larger the degrees of the
agents.  If the spin-spin couplings are renormalized according to this
principle, the Ising model on scale-free networks always has a
ferromagnetic threshold, whereas it is known that, with uniform
couplings, networks with infinite degree variance are magnetized at
any temperature~\cite{aleksiejuk02,leone02}.  The issue of how opinion
dynamics is influenced by the hierarchical structure in
societies/organizations has also been
investigated~\cite{laguna05,grabowski06a}. Other authors investigated
fashion~\cite{nakayama04}, the interplay between opinions and personal
taste~\cite{bagnoli04} and the effect of opinion surveys on the
outcome of elections~\cite{alves02}.

It is worth mentioning how the close formal similarities between the
fields of opinion and language dynamics leads to the idea that models
proposed in the framework of language dynamics could suitably apply
also in modeling opinion formation. One example is represented by a
variant of the naming game~\cite{baronchelli07}, as defined in
Sec.~\ref{sec:language}.

\subsection{Empirical data}
\label{sec:elections}

One of the main contributions of the physical approach to opinion
dynamics should be to focus on the quantitative aspects of the
phenomenon of consensus formation, besides addressing the mere
qualitative question of when and how people agree/disagree.  What is
needed is then a quantitative phenomenology of opinion dynamics, to
define the phenomenon in a more objective way, posing severe
constraints on models.  Sociological investigations have been so far
strongly limited by the impossibility of studying processes involving
large groups of individuals. However, the current availability of
large datasets and of computers able to handle them makes for the
first time such empirical analysis possible.

Elections are among the largest scale phenomena involving people and
their opinions. The number of voters is of the order of millions for
most countries, and it can easily reach hundreds of millions in
countries like Brazil, India and the USA. A great deal of data is
nowadays publicly available in electronic form.  The first empirical
investigations carried out by physicists concerned Brazilian
elections~\cite{costafilho99}.  The study focused on the distribution
of the fraction $\nu$ of votes received by a candidate. Datasets
referring to the federal elections in 1998 revealed the existence of a
characteristic pattern for the histogram $P(\nu)$, with a central
portion following the hyperbolic decay $1/\nu$, and an exponential
cutoff for large values of $\nu$. Interestingly, datasets
corresponding to candidates to the office of state deputy in several
Brazilian states revealed an analogous pattern.  A successive analysis
on data referring to state and federal elections in $2002$ confirmed
the results for the elections in $1998$, in spite of a change in the
political rules that constrained alliances between
parties~\cite{costafilho03}.  Indian data displayed a similar pattern
for $P(\nu)$ across different states, although discrepancies were also
found~\cite{gonzalez04}.  Data on Indonesian elections are consistent
with a power law decay of $P(\nu)$, with exponent close to one, but
are too noisy to be reliable~\cite{situngkir04}.  Claims that Mexican
elections also obey a similar pattern are not clearly supported by the
data~\cite{morales06}.

The peculiar pattern of $P(\nu)$ was interpreted as the result of a
multiplicative process, which yields a log-normal distribution for
$\nu$, due to the Central Limit Theorem~\cite{costafilho99}.  The
$1/\nu$ behavior can indeed be reproduced by a log-normal function, in
the limit where the latter has a large variance.  A microscopic model
based on Sznajd opinion dynamics was proposed
in~\cite{bernardes02}. Here, the graph of personal contacts between
voters is a scale-free network {\em \`a la} Barab\'asi-Albert;
candidates are initially the only nodes of the network in a definite
opinion state, a suitably modified Sznajd dynamics spreads the
candidates' opinions to all nodes of the network. The model reproduces
the empirical curve $P(\nu)$ derived from Brazilian elections. The
same mechanism yields, on different social graphs, like pseudo-fractal
networks~\cite{gonzalez04} and a modified Barab\'asi-Albert network
with high clustering~\cite{sousa05}, a good agreement with empirical
data.  The weakness of this model, however, is that a non-trivial
distribution is only a transient in the evolution of the system. For
long times the population will always converge to the only stable
state of Sznajd dynamics, where every voter picks the same candidate,
and the corresponding distribution is a $\delta$-function. All studies
stopped the modified Sznajd dynamics after a certain, carefully
chosen, time.  A recent model based on simple opinion spreading yields
a distribution similar to the Brazilian curve, if the underlying
social graph is an Erd\"os-R\'enyi network, whereas on scale-free
networks the same dynamics fails to reproduce the
data~\cite{travieso06}.

\begin{figure}
\includegraphics[width=\columnwidth]{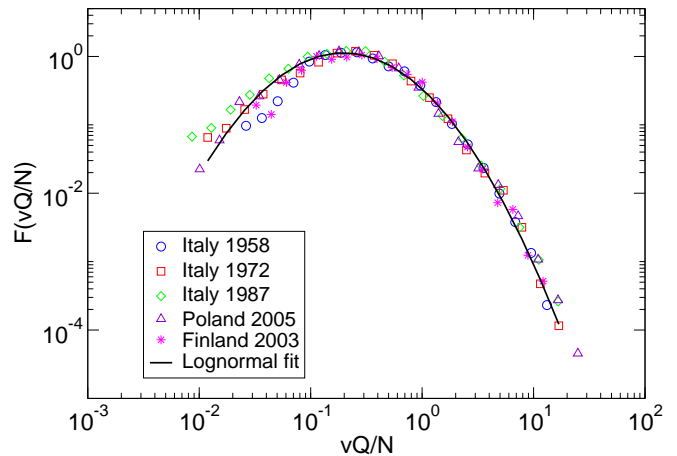}
\caption {\label{universality} Distribution of electoral performance
  for candidates in proportional elections held in Italy, Poland and
  Finland. The remarkable overlap shows that the curve is a universal
  feature of the voting process. From~\cite{fortunato07}.}
\end{figure}

The power law decay in the central region of $P(\nu)$, observed in
data sets relative to different countries and years, could suggest
that this pattern is a universal feature of the distribution. But this
is unlikely because candidates' scores strongly depend on the
performance of their parties, which is determined by a much more
complex dynamics. Indeed, municipal election data display a different
pattern~\cite{lyra03}.  Instead, the performances of candidates of the
same party can be objectively compared. This can be done in
proportional elections with open lists~\cite{fortunato07}. In this
case the country is divided into constituencies, and each party
presents a list of candidates in each constituency.  There are three
relevant variables: the number of votes $v$ received by a candidate,
the number $Q$ of candidates presented by the party in the
corresponding list and the total number $N$ of votes received by the
party list. Therefore, the distribution of the number of votes
received by a candidate should be a function of three variables,
$P(v,Q,N)$.  It turns out instead that $P(v,Q,N)$ is a scaling
function of the single variable $vQ/N$, with a log-normal shape, and,
remarkably, this function is the same in different countries and years
(Fig.~\ref{universality}). 
This finding justifies a simple microscopic
description of voting behavior, using the tools and methods of
statistical physics.  A model based on word-of-mouth spreading,
similar to that of ~\cite{travieso06}, is able to reproduce the data.

Other studies disclose a correlation between the scores of a party and
the number of its members in German elections~\cite{schneider05b} and
a polarization of the distribution of votes around two main candidates
in Brazilian elections for mayor~\cite{araripe06}. An empirical
investigation of the relation between party size and temporal
correlations in electoral results is described in~\cite{andresen07}.

\begin{figure}[t]
\includegraphics[width=6.8cm,angle=-90]{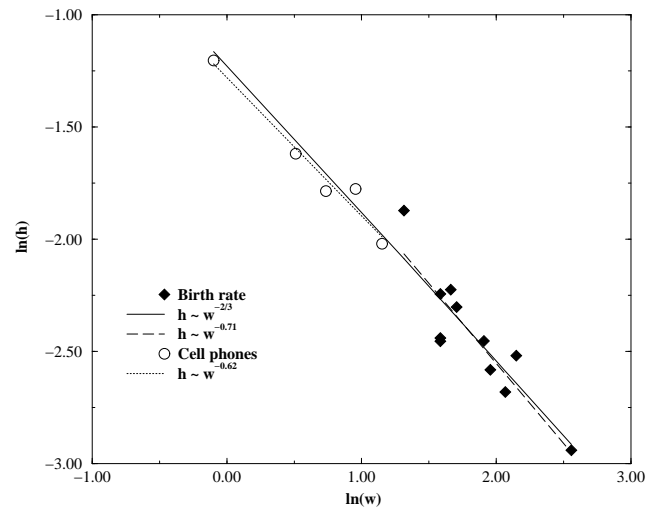}
\caption {\label{univbouchaud} Relation between the maximal speed of
  change and the duration of the change for birth rates and the number
  of mobile phones in several European countries. The linear
  regression of the data points in double logarithmic scale is
  consistent with the universal behavior predicted by the Random
  Field Ising Model at zero temperature. From~\cite{michard05}.}
\end{figure}

In~\cite{michard05} it was suggested that extreme events like booms
of products/fashions, financial crashes, crowd panic, etc., are
determined by a combination of effects, including the personal
attitude of the agents, the public information, which affects all
agents, and social pressure, represented by the mutual interaction
between the agents. This can be formally described within the
framework of the Random Field Ising Model at zero temperature, which
successfully describes hysteresis in random magnets and other physical
phenomena, like the occurrence of crackling
noise~\cite{sethna01}. Here, opinions are binary, attitudes are
real-valued numbers in $]-\infty,+\infty[$, corresponding to the
random fields, the public information is a global field $F(t)$, slowly
increasing with the time $t$, and the interaction term is the sum of
Ising-like couplings between pairs of agents. The order parameter $O$
of the system is the average opinion of the population.  By increasing
the field $F$, $O$ displays a sharp variation, due to large groups of
agents that simultaneously switch opinion.  The evolution of the speed
of change $dO/dF$ as a function of $F$ follows a universal bell-shaped
curve in the transition region, with a characteristic relation between
the height $h$ of the peak and its width $w$: $h\sim w^{-2/3}$.  This
relation was indeed observed in empirical data on extreme events, such
as the dramatic drop of birth rates in different European countries in
the last decades, the rapid diffusion of mobile phones in Europe in
the late '90s, and the decrease of the clapping intensity at the end
of applauses (Fig.~\ref{univbouchaud}).

For the future, more data are needed. Several phenomena of consensus
formation could be empirically analyzed, for instance spreading of
fads and innovations, sales dynamics, etc..

\section{CULTURAL  DYNAMICS}
\label{sec:culture}

In the previous section we have reviewed the active field of opinion
dynamics. In parallel, there has been in recent years a growing
interest for the related field of cultural dynamics. The border
between the two fields is not sharp and the distinction is not
clear-cut.  The general attitude is to consider opinion as a scalar
variable, while the more faceted culture of an individual is modeled
as a vector of variables, whose dynamics is inextricably coupled.
This definition is largely arbitrary, but we will adopt it in the
review.

The typical questions asked with respect to cultural influence are
similar to those related to the dynamics of opinions: what are the
microscopic mechanisms that drive the formation of cultural domains?
What is the ultimate fate of diversity? Is it bound to persist or all
differences eventually disappear in the long run?  What is the role of
the social network structure?

\subsection{Axelrod model}
\label{sec:axelrod}

A prominent role in the investigation of cultural dynamics has been
played by a model introduced by Axelrod in~\cite{axelrod97}, that has attracted a
lot of interest from both social scientists and physicists.

The origin of its success among social scientists is in the inclusion
of two mechanisms that are believed to be fundamental in the
understanding of the dynamics of cultural assimilation (and
diversity): social influence and homophily.  The first is the tendency
of individuals to become more similar when they interact. The second
is the tendency of likes to attract each other, so that they interact
more frequently.  These two ingredients were generally expected by
social scientists to generate a self-reinforcing dynamics leading to a
global convergence to a single culture.  It turns out instead that the
model predicts in some cases the persistence of diversity.

From the point of view of statistical physicists, the Axelrod model is a
simple and natural ``vectorial'' generalization of models of opinion
dynamics that gives rise to a very rich and nontrivial phenomenology,
with some genuinely novel behavior. The model is defined as
follows. Individuals are located on the nodes of a network (or on the
sites of a regular lattice) and are endowed with $F$ integer variables
$(\sigma_1,\ldots,\sigma_F)$ that can assume $q$ values, $\sigma_f=0,1,...,q-1$. 
The variables are called cultural {\em features}
and $q$ is the number of the possible {\em traits} allowed per
feature.  They are supposed to model the different ``beliefs,
attitudes and behavior'' of individuals.  In an elementary dynamic
step, an individual $i$ and one of his neighbors $j$ are selected and
the overlap between them
\be \omega_{i,j} =
\frac{1}{F}\sum_{f=1}^{F} \delta_{\sigma_f(i), \sigma_f(j)},
\ee 
is computed, where $\delta_{i,j}$ is Kronecker's delta. 
With probability $\omega_{i,j}$ the interaction takes
place: one of the features for which traits are different
$(\sigma_f(i) \neq \sigma_f(j))$ is selected and the trait of the
neighbor is set equal to $\sigma_f(i)$. Otherwise nothing happens.  It
is immediately clear that the dynamics tends to make interacting
individuals more similar, but the interaction is more likely for
neighbors already sharing many traits (homophily) and it becomes
impossible when no trait is the same.  There are two stable
configurations for a pair of neighbors: when they are exactly equal,
so that they belong to the same cultural region or when they are
completely different, i.e., they sit at the border between cultural
regions.
 
Starting from a disordered initial condition (for example with uniform
random distribution of the traits) the evolution on any finite system
leads unavoidably to one of the many absorbing states, which belong to
two classes: the $q^F$ ordered states, in which all individuals have
the same set of variables, or the other, more numerous, frozen states
with coexistence of different cultural regions.

\begin{figure}
\includegraphics[width=8cm,angle=-90]{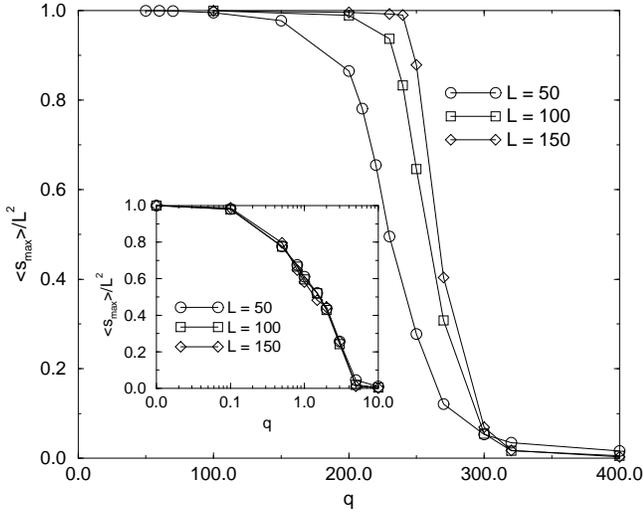}
\caption{Axelrod model. Behavior of the order parameter $\langle
  S_{max}\rangle/L^2$ vs. $q$ for three different system sizes and
  $F=10$. In the inset the same quantity is reported for
  $F=2$. From~\cite{castellano00}.}
\label{castellano00_1}
\end{figure}

It turns out that which of the two classes is reached depends on the
number of possible traits $q$ in the initial
condition~\cite{castellano00}.  For small $q$ individuals share many
traits with their neighbors, interactions are possible and quickly
full consensus is achieved.  For large $q$ instead, very few
individuals share traits.  Few interactions occur, leading to the
formation of small cultural domains that are not able to grow: a
disordered frozen state.  On regular lattices, the two regimes are
separated by a phase transition at a critical value $q_c$, depending
on $F$ (Fig.~\ref{castellano00_1}).

Several order parameters can be defined to characterize the
transition.  One of them is the average fraction $\langle S_{max}
\rangle/N$ of the system occupied by the largest cultural region.
$N$ is the number of individuals in the system.
In the ordered phase this fraction is finite (in the limit $N \to
\infty$), while in the disordered phase cultural domains are of finite
size, so that $\langle S_{max}\rangle /N \sim 1/N$. Another (dis)order
parameter often used~\cite{gonzalez-avella05} is $g=\langle N_g
\rangle / N$, where $N_g$ is the number of different domains in the
final state.  In the ordered phase $g \to 0$, while it is finite in
the disordered phase.

In two dimensions the nature of the transition depends on the value of
$F$.  For $F=2$ there is a continuous change in the order parameter
at $q_c$, while for $F>2$ the transition is discontinuous
(Fig.~\ref{castellano00_1})\footnote{Since $q$ is discrete calling the
  transition ``continuous'' is a slight abuse of language. We will
  adopt it because the transition is associated with the divergence of
  a length, as in usual transitions.}.  Correspondingly the
distribution of the size $s$ of cultural domains at the transition
is a power law with $P(s) \sim s^{-\tau}$
exponent smaller than $2$ ($\tau \approx 1.6$) for $F=2$ while the
exponent is larger than $2$ ($\tau \approx 2.6$) for any $F>2$.  In
one-dimensional systems instead~\cite{klemm03c}, the transition is
continuous for all values of $F$.

\begin{figure}[t]
\includegraphics[width=8cm,angle=-90]{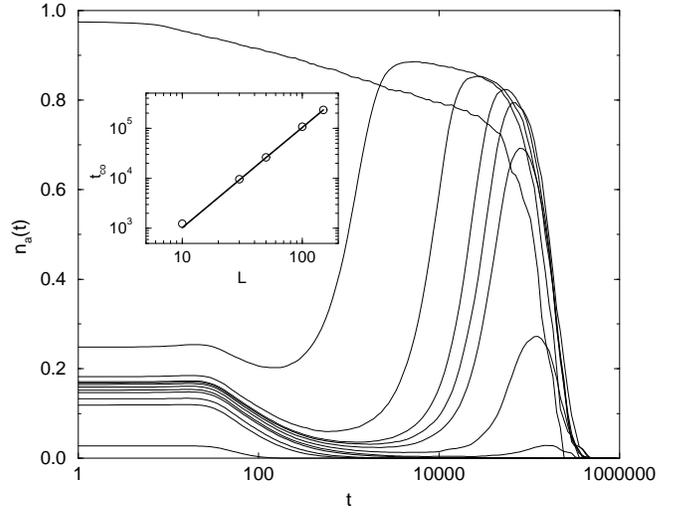}
\caption{Plot of the density of active links $n_a(t)$ for $F=10$,
  $L=150$ and (top to bottom)
  $q=1,100,200,230,240,250,270,300,320,400,500,10000$.  The inset
  reports the dependence of the freezing time $t_{co}$ on $L$ for
  $F=10$ and $q = 100<q_c$. The bold line has slope
  $2$. From~\cite{castellano00}.}
\label{castellano00_3}
\end{figure}

It is worth remarking that, upon interaction, the overlap between two
neighbors always increases by $1/F$, but the change of a trait in an
individual can make it more dissimilar with respect to his other
neighbors.  Hence, when the number of neighbors is larger than $2$, each
interaction can, somewhat paradoxically, result in an increase of the
general level of disorder in the system.  This competition is at the
origin of the nontrivial temporal behavior of the model in $d=2$,
illustrated in Fig.~\ref{castellano00_3}: below the transition but
close to it ($q \lesssim q_c$) the density of active links (connecting
sites with overlap different from $0$ and $1$) has a highly non monotonic
behavior.

Most investigations of the Axelrod model are based on
numerical simulations of the model dynamics.  Analytical approaches
are just a few.  A simple mean field
treatment~\cite{castellano00,vilone03,vazquez07} consists in writing
down rate equations for the densities $P_m$ of bonds of type $m$, i.e.,
connecting individuals with $m$ equal and $F-m$ different features.
The natural order parameter in this case is the steady state number of
active links $n_a = \sum_{m=1}^{F-1} P_m$, that is zero in the
disordered phase, while it is finite in the ordered phase.  This
approach gives a discontinuous transition for any $F$.  In the
particular case of $F=2$ the mean field equations can be studied
analytically in detail~\cite{vazquez07}, providing insight into the
non-monotonic dynamic behavior for $q \lesssim q_c$ and showing that
the approach to the steady state is governed by a timescale diverging
as $|q-q_c|^{-1/2}$.  Some information about the behavior of the Axelrod
model for $F=2$ and $q=2$ is obtained also by a mapping to the
constrained voter model~\cite{vazquez04} discussed in
Sec.~\ref{sec:voter}.

\subsection{Variants of Axelrod model}
\label{sec:sub-axelrod}

In his seminal paper, Axelrod himself mentioned many possible variants
of his model, to be studied in order to investigate the effect of
additional ingredients as the topology of the interactions, random
noise, the effect of mass media and many others.  Over the years this
program has been followed by many researchers.

\begin{figure}
\includegraphics[width=\columnwidth]{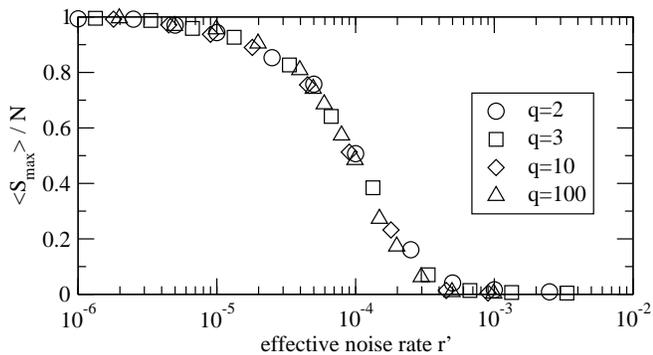}
\caption{Order parameter $\langle S_{\rm max}\rangle/N$ as a function
  of the effective noise rate $r'=r(1-1/q)$ for different values of
  $q$.  Simulations have been run in systems of size $N=50^2$ with
  $F=10$.  From~\cite{klemm03}.}
\label{klemm03_3}
\end{figure}

The possibility of one individual to change spontaneously one of his
traits, independently of his neighborhood, is denoted as ``cultural
drift'' in social science and corresponds to the addition of flipping
events driven by random noise.  In~\cite{klemm03} it is demonstrated
that the inclusion of noise at rate $r$ has a profound influence on
the model, resulting in a noise-induced order-disorder transition,
practically independent of the value of the parameter $q$
(Fig.~\ref{klemm03_3}).

For small noise the state of the system is monocultural for any $q$,
because disordered configurations are unstable with respect to the
perturbation introduced by the noise: the random variation of a trait
unfreezes in some cases the boundary between two domains leading to
the disappearance of one of them in favor of the other.  However, when
the noise rate is large, the disappearance of domains is compensated by
the rapid formation of new ones, so that the steady state is
disordered.  The threshold between the two behaviors is set by the
inverse of the average relaxation time for a perturbation $T(N)$, so
that the transition occurs for $r_c T(N) = O(1)$. An approximate
evaluation of the relaxation in $d=2$ gives $T=N \ln(N)$, in good
agreement with simulations, while $T \sim N^2$ on one
dimension~\cite{klemm05}. 
Since $T(N)$ diverges with $N$, the conclusion is that, no matter how small
the rate of cultural drift is, in the thermodynamic limit the system
remains always disordered for any $q$.

The discovery of the fragility of the Axelrod model with respect to
the presence of noise immediately raises the question ``What is the
simplest modification of the original model that preserves the
existence of a transition in the presence of noise?''.
In~\cite{kuperman06} two modified Axelrod-like dynamics have been
introduced, where the interaction between individuals is also
influenced by which trait is adopted by the majority of agents in the
local neighborhood.  Similar ingredients are present in two other
variants of the Axelrod model recently proposed~\cite{flache07}. A
convincing illustration that these modifications lead to a robust
phenomenology with respect to the addition of (at least weak) noise is
still lacking.

Another variant of the original definition of the model is the
introduction of a threshold such that, if the overlap is smaller than
a certain value $\theta$, no interaction takes
place~\cite{flache07}. Unsurprisingly no qualitative change occurs,
except for a reduction of the ordered region of the
phase diagram~\cite{desanctis07}. Another possibility named
``interaction noise'', is that for $\omega$ smaller than the
threshold, the interaction takes place with probability $\delta$.  This
kind of noise favors ordering but again does not lead to drastic
changes of the model behavior~\cite{desanctis07}.

In order to understand the effect of complex interaction topologies on
its behavior, the Axelrod model has been studied on small-world and
scale-free graphs~\cite{klemm03b}.  In the first case, the transition
between consensus and a disordered multicultural phase is still
observed, for values of the control parameter $q_c$ that grow as a
function of the rewiring parameter $p$. Since the WS network for $p=1$
is a random network (and then practically an infinite-dimensional
system) this is consistent with the observation of the transition also
in the mean field approaches~\cite{castellano00, vazquez07}.  The
scale-free nature of the BA network dramatically changes the picture.
For a given network of size $N$ a somewhat smeared-out transition is
found for a value $q_c$, with bistability of the order parameter, the
signature of a first-order transition.  However numerical simulations
show that the transition threshold grows with $N$ as $q_c \sim
N^{0.39}$, so that in the thermodynamic limit the transition
disappears and only ordered states are possible. This is similar to
what occurs for the Ising model on scale-free networks, where the
transition temperature diverges with system size~\cite{leone02}.

Another natural modification of the original Axelrod model concerns
the effect of media, represented by some external field or global
coupling in the system. One possible way to implement an external
field consists in defining a mass media cultural message as a set of
fixed variables $M=(\mu_1,\mu_2,\ldots,
\mu_F)$~\cite{gonzalez-avella05}.  With probability $B$ the selected
individual interacts with the external field $M$ exactly as if it were
a neighbor. With probability $1-B$ the individual selects instead one
of his actual neighbors.  Rather unexpectedly the external field turns
out to favor the multicultural phase, in agreement with early
findings~\cite{shibanai01}.  The order-disorder transition point is
shifted to smaller values of the control parameter $q_c(B)$. For $B$
larger than a threshold such that $q_c(B^*)=0$ only the disordered
phase is present: a strong external field favors the alignment of some
individuals with it, but it simultaneously induces a decoupling from
individuals too far from it.

Similar conclusions are drawn when a global coupling or a local
non-uniform coupling are considered~\cite{gonzalez-avella06}.  In all
cases the ordered region of the phase diagram is reduced with respect
to the case of zero field and it shrinks to zero beyond a certain
field strength $B^*$. Interestingly, for $q>q_c(B=0)$ a vanishing
field has the opposite effect, leading to an ordered monocultural
state. The limit $B \to 0$ is therefore discontinuous.  The same type
of behavior is also found for indirect mass-media feedback, i.e., when
sites accept the change of a trait only with probability $R$, if the
new value of the trait is not the same of the
majority~\cite{gonzalez-avella07}.

In the Axelrod model the numerical value of traits is just a label:
nothing changes if two neighbors have traits that differ by $1$ or by
$q-1$.  In order to model situations where this difference actually
matters, it has been proposed~\cite{flache06} to consider some
features to be ``metric'', i.e., such that the contribution to the
overlap of a given feature is $[1-\Delta \sigma_f/(q-1)]/F$, where
$\Delta \sigma_f$ is the difference between the trait values.  In this
way the Axelrod model becomes similar to the vectorial version of the Deffuant
model. Although a systematic investigation has not been performed, it
is clear that this variation favors the reaching of consensus, because
only maximal trait difference ($\Delta \sigma_f = q-1$) totally
forbids the interaction.  A related variation with ``metric'' features
is described in~\cite{desanctis07}.

Other recent works deal with a version of the Axelrod model with both
an external field and noise~\cite{mazzitello07} and one where
individuals above a fixed threshold do not
interact~\cite{parravano06}.

\subsection{Other multidimensional models}

At odds with the detailed exploration of the behavior of the Axelrod
model, much less attention has been paid to other types of dynamics
for vectors of opinions.

In the original paper on the Deffuant model~\cite{deffuant00}, a
generalization to vectorial opinions is introduced, considering in
this case binary variables instead of continuous ones. This gives a
model similar to the Axelrod model with $q=2$ traits per feature, with
the difference that the probability of interaction between two agents
as a function of their overlap is a step function at a bounded
confidence threshold $d$.  In mean field a transition between full
consensus for large threshold and fragmentation for small $d$ is
found.

A similar model is studied in~\cite{laguna03}. In this case, when two
agents are sufficiently close to interact, each pair of different variables
may become equal with a probability $\mu$.  Again a transition between
consensus and fragmentation is found as a function of the bounded
confidence threshold, but its properties change depending on whether
$\mu=1$ or $\mu<1$.

A generalization of continuous opinions (the HK model) to the
vectorial (2-dimensional) case is reported in~\cite{fortunato05d} for
a square opinion space, with both opinions ranging between $0$ and
$1$, and square or circular confidence ranges. Assuming homogeneous
mixing and solving the rate equations, it turns out that no drastic
change occurs with respect to the ordinary HK model. The consensus
threshold is practically the same. When there is no consensus the
position of coexisting clusters is determined by the shape of the
opinion space. An extension of Deffuant and HK models to vectorial
opinions has been proposed in~\cite{lorenz07a,lorenz08a}. Here
opinions sit on a hypercubic space or on a simplex, i.e., the
components of the opinion vectors sum up to one. It turns out that
consensus is easier to attain if the opinion space is a simplex rather
than hypercubic.

Other vectorial models are considered in Section~\ref{sec:coevolution}
on the coevolution of networks and states

\section{LANGUAGE DYNAMICS}
\label{sec:language}

Language dynamics is an emerging field~\cite{wichmann08} that focuses
on all processes related to emergence, change, evolution, interactions
and extinction of languages. In this section we shall review some of
the main directions in which this field is evolving.

Models for language dynamics and evolution can be roughly divided in
two main categories: {\em sociobiological} and {\em sociocultural}
approaches. This distinction somehow parallels the debate {\em nature
  versus nurture}~\cite{galton1874,ridley03} which concerns the
relative importance of an individual's innate qualities ("nature")
with respect to personal experiences ("nurture") in determining or
causing individual differences in physical and behavioral traits.

The sociobiological approach~\cite{hurford89,pinker90} postulates that
successful communicators, enjoying a selective advantage, are more
likely to reproduce than worse communicators.  Successful
communication contributes thus to biological fitness: i.e., good
communicators leave more offspring. The most developed branch of
research in this area is represented by the evolutionary
approaches. Here the main hypothesis is that communication strategies
(which are model-dependent) are innate, in the spirit of the nativist
approach~\cite{chomsky1965}, and transmitted genetically across
generations.  Thus if one of them is better than the others, in an
evolutionary time span it will displace all rivals, possibly
becoming the unique strategy of the population. The term strategy
acquires a precise meaning in the context of each particular
model. For instance, it can be a strategy for acquiring the lexicon of
a language, i.e., a function from samplings of observed behaviors to
acquired communicative behavior patterns~\cite{hurford89, oliphant97a,
  oliphant96b, nowak99b}, or it can simply coincide with the lexicon
of the parents~\cite{nowak99}, but other possibilities
exist~\cite{steels05h}.

On the other hand, in sociocultural approaches language is seen as a
complex dynamical system that evolves and self-organizes, continuously
shaped and reshaped by its users~\cite{steels03b}. Here good
strategies do not provide higher reproductive success but only better
communication abilities. Agents can select better strategies
exploiting cultural choice and direct feedback in
communications. Moreover, innovations can be introduced due to the
inventing ability of the agents. Thus, the study of the
self-organization and evolution of language and meaning has led to the
idea that a community of language users can be seen as a complex
dynamical system which collectively solves the problem of developing a
shared communication system. In this perspective, which has been
adopted by the novel field of semiotic dynamics, the theoretical tools
developed in statistical physics and complex systems science acquire a
central role for the study of the self-generated structures of
language systems.

\subsection{Evolutionary approaches}
\label{sec:evolutionary}

According to the sociobiological approach~\cite{hurford89,
  oliphant97a, oliphant96b,nowak99b,nowak06}, evolution is the main
responsible both for the origin and the emergence of natural language
in humans~\cite{pinker90}. Consequently, natural selection is the
fundamental driving force to be introduced in models.  Evolutionary
game theory~\cite{smith82} was formulated with the aim of adapting
classical game theory~\cite{vonneumann47,osborne94} to deal with
evolutionary issues, such as the possibility for agents to adapt,
learn and evolve. The approach is phenotypic, and the fitness of a
certain phenotype is, roughly speaking, proportional to its diffusion
in the population. Strategies of classical game theory are substituted
by traits (genetic or cultural), that are inherited, possibly with
mutations. The search for Nash equilibria~\cite{nash50} becomes the
quest for evolutionary stable strategies. A strategy is stable if a
group adopting it cannot be invaded by another group adopting a
different strategy. Finally, a fundamental assumption is that the
payoff from a game is interpreted as the fitness of the agents
involved in the game.  The Evolutionary Language Game (ELG)
~\cite{nowak99b, nowak99} aims at modeling the emergence of language
resorting to evolutionary game theory and to the concept of language
game~\cite{wittgenstein53german,wittgenstein53english}. For a recent
experimental paper we refer to~\cite{lieberman07}.

\subsubsection{Evolutionary language game}
\label{sec:evolutionary_language_game}

In this section we analyze in some detail how the problem of the
evolution of a common vocabulary [or more generally a common set of
conventions~\cite{lewis69}, syntactic or grammatical rules] is
addressed in the framework of evolutionary game theory. The formalism
we use is mutuated by~\cite{nowak99b}, but the basic structure of the
game was already included in the seminal paper~\cite{hurford89} about
the evolution of Saussurean signs~\cite{saussure1916}.

A population of agents (possibly early hominids) lives
in an environment with $n$ objects. Each individual is able of produce
a repertoire of $m$ words (sounds or signals, in the original
terminology) to be associated with objects. Individuals are
characterized by two matrices $P$ and $Q$, which together form a
language $L$. The {\em production matrix} $P$ is a $n\times m$ matrix
whose entry, $p_{ij}$, denotes the probability of using word $j$ when
seeing object $i$, while the {\em comprehension matrix} $Q$ is a $m
\times n$ matrix, whose entry, $q_{ji}$, denotes the probability for
a hearer to associate sound $j$ with object $i$, with the following
normalization conditions on the rows $\sum_{j=1}^{m} p_{ij}=1$ and
$\sum_{i=1}^{n} q_{ji}=1$.

A pair of matrices $P$ and $Q$ identifies a language $L$. Imagine then
two individuals $I_1$ and $I_2$ speaking languages $L_1$ (defined by
$P_1$ and $Q_1$) and $L_2$ (defined by $P_2$ and $Q_2$). The typical
communication between the two involves the speaker, say $I_1$,
associating the signal $j$ to the object $i$ with probability
$p_{ij}$. The hearer $I_2$ infers object $i$ with probability
$\sum_{j=1}^m p_{ij}^{(1)} q_{ji}^{(2)}$.  If one sums over all the
possible objects, one gets a measure of the ability, for $I_1$, to
convey information to $I_2$: $\sum_{i=1}^n \sum_{j=1}^m p_{ij}^{(1)}
q_{ji}^{(2)}$. A symmetrized form of this expression defines the
so-called payoff function, i.e., the reward obtained by two individuals
speaking languages $L_1$ and $L_2$ when they communicate:

\begin{equation}
F(L_1,L_2) =\frac{1}{2} \sum_{i=1}^n\sum_{j=1}^m 
(p_{ij}^{(1)}q_{ji}^{(2)}+ p_{ij}^{(2)}q_{ji}^{(1)}).
\label{overall_payoff}
\end{equation}

From the definition of the payoff it is evident that each agent is
treated once as hearer and once as speaker and they both receive a
reward for successful communication.

The crucial point of the model is the definition of the matrices $P$
and $Q$ which have to be initialized in some way at the beginning of
the simulation. In principle there is no reason why $P$ and $Q$ should
be correlated. On the other hand the best possible payoff is obtained
by choosing $P$ as a binary matrix having at least one $1$ in every
column (if $n \geq m$) or in every row (if $n \leq m$) and $Q$ as
a binary matrix with $q_{ji}=1$, if $p_{ij}$ is the largest entry
in a column of $P$. If $n=m$ the maximum payoff is obtained for
$P$ having one $1$ in every row and column and $Q$ being the
transposed matrix of $P$. In general the maximum payoff is given by
$F_{max}= min\left\{m,n\right\}$.  It is also worth noting that the
presence of two completely uncorrelated matrices for the production,
$P$, and comprehension, $Q$, modes, already present
in~\cite{hurford89, oliphant97a, oliphant96b}, could lead to
pathological situations as remarked in~\cite{komarova04b}, where a
single matrix is adopted for both tasks.

In a typical situation one simulates a population of $N$ individuals
speaking $N$ different languages $L_1$ to $L_N$ (by randomly choosing
the matrices $P_k$ and $Q_k$, for $k=1,...,N$). In each round of the
game, every individual communicates with every other individual, and
the accumulated payoffs are summed up, e.g. the payoff received by
individual $k$ is given by $F_k = \sum_{l=1}^{N} F(L_k,L_l)$, with $l
\neq k$. As already mentioned, the payoff is interpreted as
fitness. In a parental learning scheme each individual will produce an
offspring (without sexual reproduction) with the probability $f_k =
F_k/\sum_{l} F_l$. In this way each individual gives rise on average
to one offspring for the next generation and the population size
remains constant. The individuals of the new generation learn the
language of their parents by constructing an association matrix $A$,
whose element $a_{ij}$ records how many times the individual has
observed its parent associating object $i$ and signal $j$ in $K$
different samplings. The production and comprehension matrices $P$ and
$Q$ are easily derived from the association matrix $A$ as:

\begin{equation}
p_{ij}= a_{ij} /\sum_{l=1}^m a_{il}\;\;\;\;q_{ji}= a_{ji} /\sum_{l=1}^m a_{lj}.
\label{association_production}
\end{equation}

The form of the matrix $A$ clearly depends on $K$. In the limit $K
\rightarrow \infty$ the offspring reproduces the production matrix of
its parent and $A=P$. For finite values of $K$, learning occurs with
incomplete information and this triggers mutations occurring in the
reproduction process.

An important observation is in order. In such a scheme the language of
an individual, i.e., the pair $(P,Q)$, determines its fitness and, as a
consequence, the reproduction rate.  On the other hand what is
inherited is not directly the language but a mechanism to learn the
language which is language specific, i.e., a language acquisition
device in the spirit of the nativist approach~\cite{chomsky1965}.
Therefore the traits transmitted to the progeny can be different from
the language itself.

This evolutionary scheme leads the population to converge to a common
language, i.e., a pair of $(P,Q)$ matrices shared by all
individuals. The common language is not necessarily optimal and
the system can often get stuck in sub-optimal absorbing states where
{\em synonymy} (two or more signals associated to the same object) or
{\em homonymy} (the same signal used for two or more objects) are
present.  The convergence properties to an absorbing state depend on
the population size $N$ as well as on $K$, but no systematic analysis
has been performed in this direction.  Another interesting direction
is related to the underlying topology of the game. What described so
far corresponds to a fully connected topology, where each agent
interacts with the whole population. It is certainly of interest
exploring different topological structures, more closely related to
the structure of social networks, as discussed
in~\cite{lieberman05,szabo07} (see also the recent~\cite{tavares07} on
the model proposed in~\cite{komarova01}).

The model can then be enriched by adding a probability of errors in
perception~\cite{nowak99}, i.e., by introducing a probability $u_{ij}$
of misinterpreting signal $i$ as signal $j$. The terms $u_{ij}$ are
possibly defined in terms of similarities between signals. The maximum
payoff for two individuals speaking the same language is now reduced,
hence the maximum capacity of information transfer.  This result is
referred to as {\em linguistic error limit}~\cite{nowak99,nowak99c}:
the number of distinguishable signals in a protolanguage, and
therefore the number of objects that can be accurately described by
this language, is limited.  Increasing the number of signals would not
increase the capacity of information transfer [it is worth mentioning
here the interesting parallel between the formalism of evolutionary
language game with that of information theory~\cite{plotkin00}]. A
possible way out is that of combining signals into
words~\cite{smith03}, opening the way to a potentially unlimited
number of objects to refer to. In this case it is shown that the
fitness function can overcome the error limit, increasing
exponentially with the length of words~\cite{nowak99,nowak99c}. This
is considered one of the possible ways in which evolution has selected
higher order structures in language, e.g. syntax and grammar.  We
refer to~\cite{nowak99,nowak06} for details about the higher stages in
the evolution of language.

\subsubsection{Quasispecies-like approach}

The model described in the previous section can be cast in the
framework of a deterministic dynamical system (see for a recent
discussion~\cite{traulsen05} and references therein). We consider
again the association matrix $A$, a $n \times m$ matrix whose entries,
$a_{ij}$, are non-zero if there is an association between the object
$i$ and the signal $j$. In this case we consider a binary matrix with
entries taking either the value $0$ or $1$. The possible number of
matrices $A$ is then $M=2^{nm}$. This matrix is also denoted as the
lexical matrix~\cite{komarova01b}. In a population of $N$ individuals
denote now with $x_k$ the fraction of individuals with the association
matrix $A_k$, with $\sum_{k=1}^M x_k =1$. One can define the evolution
of $x_k$ as given by the following equation:

\begin{equation}
{\dot x_k} = \sum_l f_l x_l Q_{lk}-\phi x_k\;\;, \; l=1,..,M=2^{nm},
\label{deterministic_lexical}
\end{equation}

\noindent where $f_l$ is the fitness of individuals with the
association matrix $A_l$ (from now on individual $l$), $f_l =
\sum_k F(A_l,A_k) x_k$, with the assumption that $x_k$ is the
probability to speak with an individual $k$; $\phi$ defines the
average fitness of the population, $\phi= \sum_l f_l x_l$, while
$Q_{lk}$ denotes the probability that someone learning from an
individual with $A_l$ will end up with $A_k$. The second term on the
right-hand side keeps the population size constant.

Eqs.~(\ref{deterministic_lexical}) represent a particular case of the
quasispecies equations~\cite{eigen71,eigen79}. The quasispecies model
is a description of the process of Darwinian evolution of
self-replicating entities within the framework of physical
chemistry. These equations provide a qualitative understanding of
the evolutionary processes of self-replicating macromolecules such as
RNA or DNA or simple asexual organisms such as bacteria or
viruses. Quantitative predictions based on this model are difficult,
because the parameters that serve as input are hard to obtain from
actual biological systems.  In the specific case in which mutation is
absent, i.e., $Q_{ij}=0$ if $i\neq j$, one recovers the so-called {\em
  replicator equations} of evolutionary game theory~\cite{smith82},
which, it is worth recalling, are equivalent to the Lotka-Volterra
equations in $M-1$ dimensions~\cite{hofbauer98}.

\subsection{Semiotic Dynamics approach}
\label{sec:semiotic-dynamics}

Semiotic dynamics looks at language as an evolving system where new
words and grammatical constructions may be invented or acquired, new
meanings may arise, the relation between language and meaning may
shift (e.g. if a word adopts a new meaning), the relation between
meanings and the world may shift (e.g. if new perceptually grounded
categories are introduced). 


\subsubsection{The Naming Game}

The {\em Naming Game} (NG) possibly represents the simplest example of
the complex processes leading progressively to the establishment of
human-like languages. It was expressly conceived to explore
the role of self-organization in the evolution of
language~\cite{steels95,steels96d} and it has acquired, since then, a
paradigmatic role in the whole field of semiotic dynamics. The
original paper~\cite{steels95}, focuses mainly on the formation of
vocabularies, i.e., a set of mappings between words and meanings (for
instance physical objects). In this context, each agent develops its
own vocabulary in a random private fashion. But agents are forced to
align their vocabularies, through successive conversation, in order to
obtain the benefit of cooperating through communication. Thus, a
globally shared vocabulary emerges, or should emerge, as the result of
local adjustments of individual word-meaning association. The
communication evolves through successive conversations, i.e., events
that involve a certain number of agents (two, in practical
implementations) and meanings. It is worth remarking that here
conversations are particular cases of language games, which, as
already pointed out
in~\cite{wittgenstein53german,wittgenstein53english}, can be used to
describe linguistic behavior, even if they can include also
non-linguistic behavior, such as pointing.

This original seminal idea triggered a series of contributions along
the same lines and many variants have been proposed over the years. It
is particularly interesting to mention the work proposed
in~\cite{ke02a}, that focuses on an imitation model which simulates
how a common vocabulary is formed by agents imitating each other,
either using a mere random strategy or a strategy in which imitation
follows the majority (which implies non-local information for the
agents). A further contribution of this paper is the introduction of
an interaction model which uses a probabilistic representation of the
vocabulary. The probabilistic scheme is formally similar to the
framework of evolutionary game theory seen in
Sec.~\ref{sec:evolutionary_language_game}, since to each agent a {\em
  production} and a {\em comprehension} matrices are associated.
Differently from the approach of ELG, here the matrices are
dynamically transformed according to the social learning process and
the cultural transmission rule. A similar approach has been proposed
in~\cite{lenaerts05}.

In the next section we shall present a {\em minimal} version of the NG
which results in a drastic simplification of the model definition,
while keeping the same overall phenomenology. This version of the NG
is suitable for massive numerical simulations and analytical
approaches. Moreover the extreme simplicity allows for a direct
comparison with other models introduced in other frameworks of
statistical physics as well as in other disciplines.

\paragraph{The Minimal Naming Game}

The simplest version of the NG~\cite{baronchelli06a} is played by a
population of $N$ agents, on a fully connected network, trying to
bootstrap a common name for a given object. Each player is
characterized by an inventory of word-object associations he/she
knows.  All agents have empty inventories at time $t=0$. At each time
step ($t=1,2,..$), two players are picked at random and one of them
plays as speaker and the other as hearer. 
\begin{figure}
\includegraphics[width=7cm]{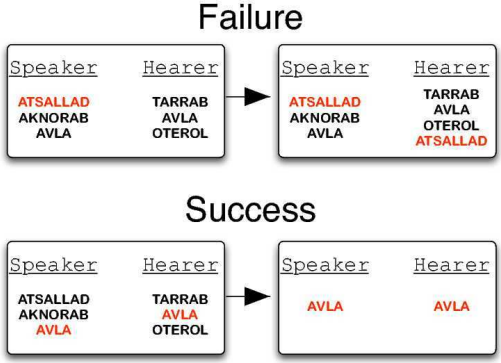}
  \caption{Naming Game. Examples of the dynamics of the
    inventories in a failed (top) and a successful (bottom) game. The
    speaker selects the word highlighted. If the hearer does
    not possess that word he includes it in his inventory
    (top). Otherwise both agents erase their inventories only keeping
    the winning word (bottom). 
    \label{rules_naming}}
\end{figure}
Their interaction obeys the
rules described in Fig.~\ref{rules_naming}.

\paragraph{Macroscopic analysis}

\begin{figure}
  \includegraphics[width=\columnwidth]{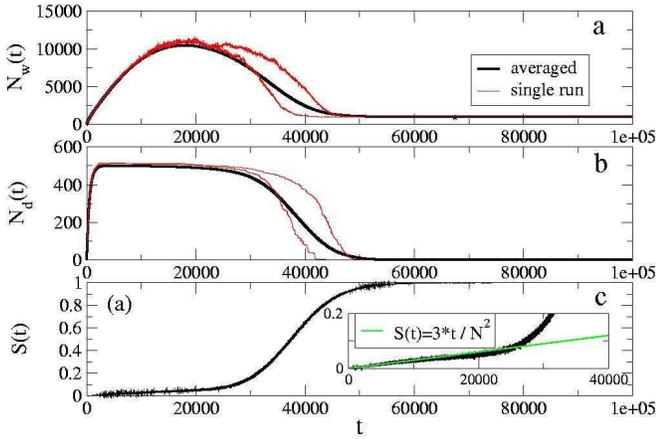}
  \caption{Naming Game. a) Total number of words present in the
    system, $N_w(t)$; b) Number of different words, $N_d(t)$; c)
    Success rate $S(t)$, i.e., probability of observing a successful
    interaction at time $t$. The inset shows the linear behavior of
    $S(t)$ at small times. The system reaches the final absorbing
    state, described by $N_w(t) = N$, $N_d(t) = 1$ and $S(t) = 1$, in
    which a global agreement has been reached.
    From~\cite{baronchelli06a}.
    \label{pheno_naming}}
\end{figure}

The first property of interest is the time evolution of the total
number of words owned by the population $N_w(t)$, of the number of
different words $N_d(t)$, and of the success rate $S(t)$
(Fig.~\ref{pheno_naming}). 
%


After a transient period, the system undergoes spontaneously a
disorder-order transition to an asymptotic state where global
consensus emerges, i.e., every agent has the same word for the same
object. It is remarkable that this happens starting from completely
empty inventories for each agent. The asymptotic state is one where a
word invented during the time evolution takes over with respect to the
other competing words and imposes itself as the leading word. In this
sense the system spontaneously selects one of the many possible
coherent asymptotic states and the transition can thus be seen as a
symmetry breaking transition. The dynamics of the Naming Game is
characterized by the following scaling behavior for the convergence
time $t_{conv}$, the time and the height of the peak of $N_w(t)$,
namely $t_{max}$ and $N_w^{max} = N_w(t_{max})$. It turns out that all
these quantities follow power law behaviors: $t_{max} \sim
N^{\alpha}$, $t_{conv} \sim N^{\beta}$, $N_{max} \sim N^{\gamma}$ and
$t_{diff} = (t_{conv} - t_{max}) \sim N^{\delta}$, with exponents
$\alpha=\beta=\gamma=\delta =1.5$ (with a subtle feature around the
disorder-order transition where an additional timescale emerges). The
values of those exponents can be understood through simple scaling
arguments~\cite{baronchelli06a}\footnote{Here the time is the number
  of binary interactions}.

\subsubsection{Symmetry breaking: a controlled case}

Consider now a simpler case in which there are only two words at the
beginning of the process, say A and B, so that the population can be
divided into three classes: the fraction of agents with only $A$,
$n_A$, the fraction of those with only the word $B$, $n_B$, and
finally the fraction of agents with both words, $n_{AB}$. Describing
the mean-field time evolution of the three species is straightforward:

\begin{eqnarray}
{\dot{n}}_A & = & - n_A n_B + n^2_{AB} + n_A n_{AB}\nonumber\\
{\dot{n}}_B & = & - n_A n_B + n^2_{AB} + n_B
n_{AB}\label{2words_naming}\\ {\dot{n}}_{AB} & = & +2 n_A n_B - 2
n^2_{AB} -(n_A + n_B) n_{AB}. \nonumber
\end{eqnarray}

The system of differential equations~(\ref{2words_naming}) is
deterministic. It presents three fixed points in which the system can
collapse depending on the initial conditions.  If $n_A(t = 0) > n_B(t
= 0)$ [$n_B(t = 0) > n_A(t = 0)$], at the end of the evolution we will
have the stable fixed point $n_A = 1$ [$n_B = 1$] and, consequently
$n_B = n_{AB} = 0$ [$n_A = n_{AB} = 0$]. If, on the other hand, we
start from $n_A(t = 0) = n_B(t = 0)$, the equations lead to $n_A
= n_B = 2n_{AB} = 0.4$. The latter situation is clearly unstable,
since any external perturbation would make the system fall in one of
the two stable fixed points.

Eqs.~(\ref{2words_naming}) however, are not only a useful example to
clarify the nature of the symmetry breaking process. In fact, they
also describe the interaction among two different populations that
converged separately to two distinct conventions. In this perspective,
Eqs.~(\ref{2words_naming}) predict that the larger population will
impose its conventions. In the absence of fluctuations, this is true
even if the difference is very small: $B$ will dominate if $n_B(t = 0)
= 0.5 + \epsilon$ and $n_A(t = 0) = 0.5 - \epsilon$, for any $0
<\epsilon \leq 0.5$ and $n_{AB}(t = 0) = 0$. Data from simulations
show that the probability of success of the convention of the minority
group $n_A$ decreases as the system size increases, going to zero in
the thermodynamic limit ($N \rightarrow \infty$). A similar approach
has been proposed to model the competition between two
languages~\cite{abrams03}.  We discuss this point in
Sec.~\ref{sec:competition}. Here it is worth remarking the formal
similarities between modeling the competition between synonyms in a NG
framework and the competition between languages: in both cases a
synonym or a language are represented by a single feature, e.g. the
characters $A$ or $B$, for instance, in
Eqs.~(\ref{2words_naming}). The similarity has been made more evident
by the subsequent variants of the model introduced in~\cite{abrams03}
to include explicitly the possibility of bilingual individuals. In
particular in~\cite{wang05,minett04->08} deterministic models for the
competition of two languages have been proposed, which include
bilingual individuals. In~\cite{castello06} a modified version of the
voter model (see Sec.~\ref{sec:voter}) including bilingual
individuals has been proposed, the so-called AB-model.  In a fully
connected network and in the limit of infinite population size, the
AB-model can be described by coupled differential equations for the
fractions of individuals speaking language $A$, $B$ or $AB$, that are,
up to a constant normalization factor in the timescale, identical to
Eqs.~(\ref{2words_naming}). 

\subsubsection{The role of the interaction topology}

As already mentioned in Sec.~\ref{sec:topology}, social networks play
an important role in determining the dynamics and outcome of language
change. The first investigation of the role of topology was proposed
in 2004, at the 5th Conference on Language evolution,
Leipzig~\cite{ke04-->08}. Since then many approaches focused on
adapting known models on topologies of increasing complexity: regular
lattices, random graphs, scale-free graphs, etc.

The NG, as described above, is not unambiguously defined on general
networks. As already observed in Sec.~\ref{sec:topology} and in
Sec.~\ref{sec:voter}, when the degree distribution is heterogeneous,
the order in which an agent and one of its neighbors are selected does
matter, because high-degree nodes are more easily chosen as neighbors
than low-degree nodes.  Several variants of the NG on generic networks
can be defined. In the {\em direct NG} ({\em reverse NG}) a randomly
chosen speaker (hearer) selects (again randomly) a hearer (speaker)
among its neighbors. In a {\em neutral} strategy one selects an edge
and assigns the role of speaker and hearer with equal probability to
the two nodes~\cite{dallasta06b}.

\begin{figure} 
  \includegraphics[width=\columnwidth]{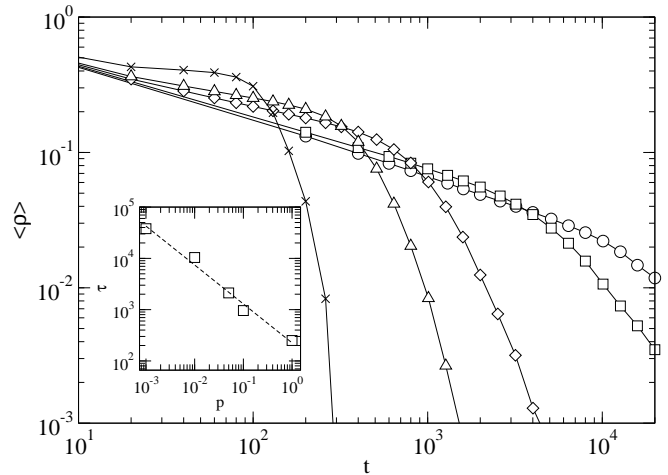}
  \caption{AB-model. Time evolution of the average density
    $\left<\rho\right>$ of bilingual individuals in small-world
    networks for different values of the rewiring parameter $p$. From
    left to right: $p=1.0, 0.1, 0.05, 0.01, 0.0$. The inset shows the
    dependence of the characteristic lifetime $\tau$ on the rewiring
    parameter $p$. The dashed line corresponds to a power law fit
    $\tau \sim
    p^{-0.76}$. From~\cite{castello06}.\label{castello06_sw}}
\end{figure}

On low-dimensional lattices consensus is reached through a coarsening
phenomenon~\cite{baronchelli06c} with a competition among the
homogeneous clusters corresponding to different conventions, driven by
the curvature of the interfaces~\cite{bray94}. A scaling of the
convergence time as ${\cal O}(N^{1+1/d})$ has been conjectured, where
$d\le4$ is the lattice dimension. Low-dimensional lattices require
more time to reach consensus compared to a fully connected graph, but
a lower use of memory. A similar analysis has been performed for the
$AB$-model~\cite{castello06}. The effect of a small-world topology has
been investigated in~\cite{dallasta06} in the framework of the NG and
in~\cite{castello06} for the $AB$-model. Two different regimes are
observed. For times shorter than a crossover time, $t_{cross} = {\cal
  O}(N/p^2)$, one observes the usual coarsening phenomena as long as
the clusters are one-dimensional, i.e., as long as the typical cluster
size is smaller than $1/p$. For times much larger than $t_{cross}$,
the dynamics is dominated by the existence of shortcuts and enters a
mean field like behavior. The convergence time is thus expected to
scale as $N^{3/2}$ and not as $N^3$ (as in $d=1$). Small-world
topology allows to combine advantages from both finite-dimensional
lattices and fully connected networks: on the one hand, only a finite
memory per node is needed, unlike the ${\cal O}(N^{1/2})$ in fully
connected graphs; on the other hand the convergence time is expected
to be much shorter than in finite dimensions. In~\cite{castello06},
the dynamics of the $AB$-model on a two-dimensional small world
network has been studied. Also in this case a dynamic stage of
coarsening is observed, followed by a fast decay to the $A$ or $B$
absorbing states caused by a finite size fluctuation
(Fig.~\ref{castello06_sw}).  
The NG has been studied on complex
networks as well. Here the convergence time $t_{conv}$ scales as
$N^{\beta}$, with $\beta \simeq 1.4 \pm 0.1$, for both Erd\"{o}s-Renyi
(ER)~\cite{erdos59,erdos60} and Barab\'asi-Albert
(BA)~\cite{barabasi99} networks. The scaling laws observed for the
convergence time are general robust features not affected by further
topological details~\cite{dallasta06b,dallasta06c}. Finally it is
worth mentioning how the Naming Games with local broadcasts on random
geometric graphs have been investigated in~\cite{lu08} as a model for
agreement dynamics in large-scale autonomously operating wireless
sensor networks.

\subsection{Other models}

A variant of the NG has been introduced with the aim of mimicking the
mechanisms leading to opinion and convention formation in a population
of individuals~\cite{baronchelli07}. In particular a new parameter,
$\beta$, has been added mimicking an {\em irresolute attitude} of the
agents in making decisions ($\beta=1$ corresponds to the NG). The
parameter $\beta$ is simply the probability that, in a successful
interaction, both the speaker and the hearer update their memories
erasing all opinions except the one involved in the interaction (see
Fig.~\ref{rules_naming}). This negotiation process displays a
non-equilibrium phase transition from an absorbing state in which all
agents reach a consensus to an active (not frozen as in the Axelrod
model~\cite{axelrod97}) stationary state characterized either by
polarization or fragmentation in clusters of agents with different
opinions. Interestingly, the model displays the non-equilibrium
phase transition also on heterogeneous networks, in contrast with
other opinion dynamics models, like for instance the Axelrod
model~\cite{klemm03b}, for which the transition disappears for
heterogeneous networks in the thermodynamic limit.

A hybrid approach, combining vertical and horizontal transmission of
cultural traits, has been proposed in~\cite{ke02a}, while an
evolutionary version of the Naming Game has been introduced
in~\cite{lipowski08}.

A very interesting approach to language change has been proposed
in~\cite{croft00,baxter06}, based on an utterance selection model.
Another interesting approach to language dynamics, the so-called
Iterated Learning Model (ILM)~\cite{kirby01}, focuses on cultural
evolution and learning~\cite{niyogi06} and explores how the mappings
from meanings to signals are transmitted from generation to
generation.  In this framework several results have been obtained
concerning the emergence of a linguistic structure,
e.g. compositionality~\cite{smith03}.

\subsection{Language competition}
\label{sec:competition}

Models of language evolution usually focus on a single population,
which is supposed to be isolated from the rest of the world.  However,
real populations are not isolated, but they keep interacting with each
other. These steady interactions between peoples play a big role in
the evolution of languages.
 
\begin{figure}[t]
  \includegraphics[width=\columnwidth]{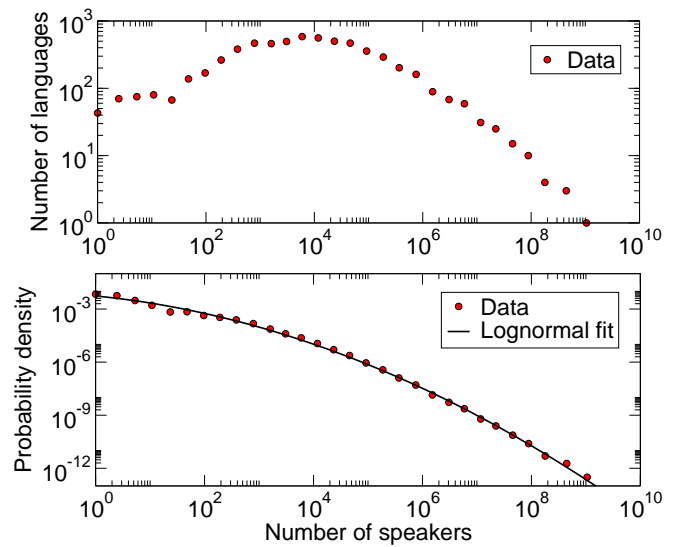}
  \caption{\label{langsize} Distribution of language sizes. The
    $x$-axis represents the number of individuals speaking a language.
    In the upper diagram, on the $y$ axis the number of languages
    spoken by $x$ individuals is reported. In the lower diagram the
    number of languages is divided by the size of the corresponding
    bins, obtaining the probability distribution, well fitted by a
    log-normal function (continuous line).  The total number of
    languages sampled is $6142$. (Data from
    {\tt http://www.ethnologue.com/}).}
\end{figure}

At present, there are about 6,500 languages in the world, with a very
uneven geographic distribution. Most of these languages have very few
speakers, and are threatened by
extinction~\cite{sutherland03}. Indeed, it is plausible that in the
future increasing numbers of people will be pushed to adopt a common
language by socio-economic factors, leading to the survival of a few
major linguistic groups, and to the extinction of all other languages.
According to some estimates, up to $90\%$ of present languages might
disappear by the end of the 21st century~\cite{krauss92}.  The
histogram of the language sizes, shown in Fig.~\ref{langsize}, has a
regular shape, which closely resembles a log-normal
distribution\footnote{Similarly, in~\cite{ausloos07}, the distribution
  of the number of adherents to religions has been investigated.}.  
Several models
of language competition have been proposed with the aim of reproducing
such distribution.  However, we stress that the observed distribution
of language sizes may not be a stable feature of language diversity,
as there is no reason to believe that it has kept its shape over the
past centuries and that it will keep it in the future. Morevoer, it
has been suggested that the current histogram may be a consequence of
pure demographic growth~\cite{zanette07}.

Modeling language competition means studying the interaction between
languages spoken by adults. Language evolution shares several features
with the evolution of biological species. Like species, a language can
split into several languages, it can mutate, by modifying
words/expressions over time, it can face extinction.  Such
similarities have fostered the application of models used to describe
biological evolution in a language competition context. The models can
be divided in two categories: macroscopic models, where only average
properties of the system are considered, are based on differential
equations; microscopic models, where the state of each individual is
monitored in time, are based on computer simulations.

\subsubsection{Macroscopic models}
\label{sec:abramsstrogatz}

\begin{figure}
\includegraphics[width=\columnwidth,angle=270]{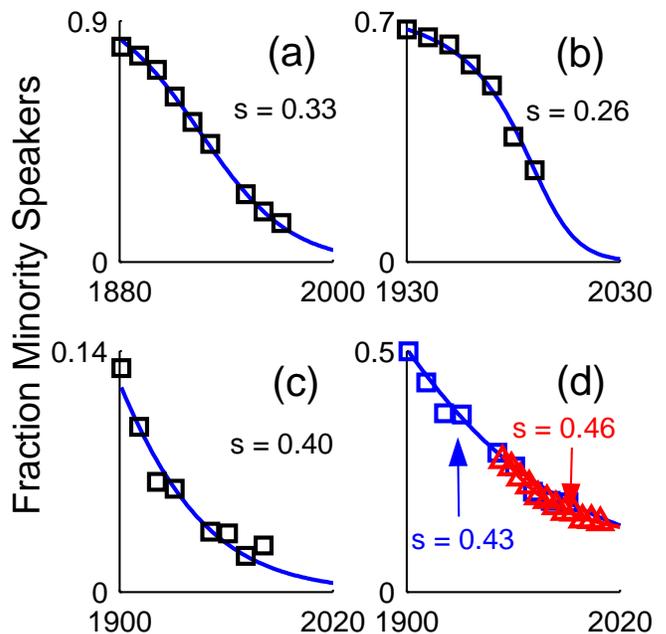}
\caption {\label{abr}. Dynamics of language extinction according to
  the model of Abrams and Strogatz.  The four panels show the
  comparison of the model with real data on the proportion of speakers
  over time for (a) Scottish Gaelic in Sutherland, Scotland, (b)
  Quechua in Huanuco, Peru, (c) Welsh in Monmouthshire, Wales and (d)
  Welsh in all of Wales. In (d) both historical data (squares) and the
  results of a recent census (triangles) are
  plotted. From~\cite{abrams03}.}
\end{figure}

The first macroscopic model of language competition was a dynamic
model proposed in~\cite{abrams03} (AS), describing how two languages,
$A$ and $B$, compete for speakers.  The languages do not evolve in
time; the attractiveness of each language increases with its number
of speakers and perceived status, which expresses the social and
economic benefits deriving from speaking that language.  We indicate
with $x$ and with $0\leq s\leq 1$ the fraction of speakers and the
status of $A$, respectively.  Accordingly, language $B$ has a fraction
$y=1-x$ of speakers, and its status is $1-s$.  The dynamics is given
by the simple rate equation
\begin{equation}
\frac{dx}{dt}=c(1-x)x^as-cx(1-x)^a(1-s),
\label{comp2}
\end{equation}
where $a$ and $c$ are parameters which, along with $s$, fix the model
dynamics\footnote{We remark that the parameter $c$ is an overall
  multiplicative constant of the right-hand side of Eq.~(\ref{comp2})
  and can be absorbed in the time unit, without affecting the
  dynamics.}. Eq.~(\ref{comp2}) expresses the balance between the rates
of people switching from language $B$ to $A$ and from $A$ to $B$. The
dynamics has only two stable fixed points, corresponding to $x=0$ and
$x=1$.  There is a third fixed point, corresponding to $x=1/2$,
$s=1/2$, when the two languages are equivalent, but it is unstable, as
confirmed by numerical simulations of a microscopic version of the AS
model on different graph topologies~\cite{stauffer07}.  Therefore, the
AS model predicts the dominance of one of the two languages and the
consequent extinction of the other. The dominant language is the one
with the initial majority of speakers and/or higher status.
Comparisons with empirical data reveal that the model is able to
reproduce the decrease in time of the number of speakers of various
endangered languages (Fig.~\ref{abr}).
The AS model is minimal and neglects important aspects of
sociolinguistic interaction.  In actual situations of language
competition, the interaction between two languages A and B often
occurs through speakers who are proficient in both languages.
In~\cite{mira05} bilingual speakers were introduced in the AS model. A
parameter $k$ expresses the similarity of the two competing languages
A and B and is related to the probability for monolingual speakers to
turn bilingual. For each choice of the AS parameters $a$, $s$, there
is a critical value $k_{min}(a,s)$ such that, for $k>k_{min}(a,s)$,
the system reaches a steady state characterized by the coexistence of
one group of monolingual speakers with a group of
bilinguals. Monolingual speakers of the endangered language are bound
to disappear, but the survival of the language is ensured by
bilingualism, provided A and B are similar enough. The model describes
well historical data on the time evolution of the proportions of
speakers of Galician and Castillian Spanish in Galicia.
In~\cite{minett04->08} a more complex modification of the AS model,
incorporating bilingualism and language transmission between adults
and from adults to children, was proposed. The model predicts the same
extinction scenario of the AS model, unless special strategies of
intervention are adopted when the number of speakers of the endangered
language decreases below a threshold.  Effective intervention
strategies are the enhancement of the status of the endangered
language and the enforcement of monolingual education of the children.

In~\cite{patriarca04} the effect of population density is introduced,
by turning the rate equation of the AS model into a reaction-diffusion
equation. Here people can move on a plane, divided in two regions; in
each regions one language has a higher status than the other one.  The
system converges to a stable configuration where both languages
survive, although they are mostly concentrated in the zones where they
are favored.  In a recent work~\cite{pinasco06} it was shown that
language coexistence in the same region is possible, if one accounts
for the population dynamics of the two linguistic communities, instead
of considering the whole population fixed, like in the AS model.  The
dynamics is now ruled by a set of generalized Lotka-Volterra
equations, and presents a non-trivial fixed point when the rate of
growth of the population of speakers of the endangered language
compensates the rate of conversion to the dominant language.

\subsubsection{Microscopic models}
\label{sec:agentbasedlanguage}

Many microscopic models of language competition represent language as
a set of $F$ independent features ($F$ usually goes from $8$ to $64$),
with each feature taking one out of $Q$ values. This is also the
representation of culture in the Axelrod model (see
Sec.~\ref{sec:axelrod}); indeed, language diversity is an aspect of
cultural diversity. If $Q=2$, language is a bit-string, a
representation used for biological species~\cite{eigen71}.  For a
recent review of language competition simulations
see~\cite{schulze08}.

In the Schulze model~\cite{schulze05b}, the language of each
individual evolves according to three mechanisms, corresponding to
random changes, transfer of words from one language to another and
learning of a new language. There are three parameters: $p$, $q$ and
$r$.  With probability $p$, a randomly chosen feature of an agent's
language is modified: with probability $q$, the new value is that of
the corresponding feature of a randomly picked individual, otherwise a
value taken at random. Finally, there is a probability $(1-x)^2r$ that
an agent switches to a language spoken by a fraction $x$ of the
population.  If agents are the nodes of a network, the language of an
individual can only be affected by its neighbors.  Simulations show
that there is a sharp transition between a phase in which most people
speak the same language (low $p$), and a phase in which no language
dominates (high $p$) and the distribution of language sizes is roughly
log-normal, like the empirical distribution (Fig.~\ref{langsize}). The
agreement with the data improves by sampling the evolving model
distribution over a large time interval~\cite{stauffer06b}. An
analytical formulation of the Schulze model has been recently
proposed~\cite{zanette07a}.

We notice that here languages have no intrinsic fitness, i.e., they
are all equivalent for the dynamics, at variance with biological
species and with the macroscopic models of the previous section, where
the different status of languages is responsible for their
survival/extinction. The possible dominance of one language is
determined by initial fluctuations, that make a linguistic community
slightly larger than the others and more likely to capture speakers
fleeing from other communities.

Several modifications of the Schulze model have been proposed.  Agents
can age, reproduce and die~\cite{schulze05b}; they can move on the
sites of a lattice, forming linguistic communities that are spatially
localized~\cite{schulze05a}; they can be bilingual~\cite{schulze08}.
To avoid the dominance of a single language, it is enough to stop the
flight from an endangered language when the number of its speakers
decreases below a threshold~\cite{schulze06}.  A linguistic taxonomy
can be introduced, by classifying languages into families based on
similarities of the corresponding bit strings~\cite{wichmann07}. This
enables to control the dynamics of both individual languages and of
their families.

The model in~\cite{viviane06} describes the colonization of a
territory by a population that eventually splits into different
linguistic communities.  Language is represented by a number, so it
has no internal structure.  The expansion starts from the central site
of a square lattice, with some initial population size.  Free sites
are occupied by a neighboring population with a probability
proportional to the number of people speaking that language, which is
a measure of the fitness of that population. The language of a group
conquering a new site mutates with a probability that is inversely
proportional to its fitness.  The simulation stops when all lattice
sites have been occupied. The resulting linguistic diversity displays
similar features as those observed in real linguistic diversity, like
the distribution of language sizes (Fig.~\ref{langsize}). The
agreement improves by introducing an upper bound for the fitness of a
population~\cite{viviane06a}, or by representing languages as bit
strings~\cite{oliveira07}.

Social Impact Theory (see Sec.~\ref{sec:impact}) was applied to
model language change~\cite{nettle99a,nettle99b}.  Here, there are two
languages and agents are induced to join the linguistic majority
because it exerts a great social pressure.  Language mixing, for which
a new language may originate from the merging of two languages, was
implemented in~\cite{kosmidis05}.  In this model, the biological
fitness of the agents may increase if they learn words of the other
language.  The model accounts for the emergence of bilingualism in a
community where people initially speak only one of two languages.
In~\cite{schwammle05} there are two languages and agents move on a
lattice, are subjected to biological aging and can reproduce.  People
may grow bilingual; bilinguals may forget one of the two languages, if
it is minoritarian in their spatial surroundings. As a result, if the
two linguistic communities are spatially separated, they can coexist
for a long time, before the dynamics will lead to the dominance of one
of them. Bilingual agents are also present in the modified version of
the voter model proposed in~\cite{castello06}, discussed in
Sec.~\ref{sec:voter}.
 
\section{CROWD BEHAVIOR}
\label{sec:collective}

Collective motion is very common in nature. Flocks of birds, fish schools, 
swarms of insects are among the most spectacular
manifestations~\cite{parrish97}.  Humans display similar behavior in
many instances: pedestrian motion, panic, vehicular traffic,
etc.. 

The origin of collective motion has represented a puzzle for many
years.  One has the impression that each individual knows exactly what
all its peers are doing in the group and acts accordingly. It is
plausible instead that an individual has a clear perception of what
happens in its neighborhood, ignoring what most of its peers are
doing. We are then faced again with a phenomenon where local
interactions determine the emergence of a global property of the
system, in this case collective motion. Therefore it is not surprising
that in the last years physicists have worked in this field. In this
section we shall give a brief account of the most important results on
crowd behavior. For a review of the studies on vehicular traffic we
refer to~\cite{helbing01,nagatani02,kerner04}.

\subsection{Flocking}
\label{sec:flocking}

To study the collective motion of large groups of organisms, the concept of 
{\it self-propelled particles} (SPP) has been introduced~\cite{vicsek95,czirok99a,czirok00}.
SPP are particles driven by an intrinsic force, produced by an energy depot
which is internal to the particles, just like it happens in real organisms.
In the original model~\cite{vicsek95}, $N$ particles move on a squared surface with periodic boundary conditions.
A particle $i$ is characterized by its position ${\bf x_i}$ and velocity ${\bf v_i}$.
For simplicity, it is assumed that the velocities of the particles equal in modulus the constant $v_0$
at any moment of the system's evolution.
This is where the self-propelled feature of the particles sets in.
The interaction is expressed by a simple rule: at each time step a particle $i$ assumes the average direction of 
motion of the particles lying within a local neighborhood $S(i)$, with a random perturbation, i.e. 
\be
\theta_i(t+\Delta t)=\langle\theta(t)\rangle_{S(i)}+\xi,
\label{crowd1}
\ee
where $\theta_i$ is the angle indicating the velocity direction of particle $i$,
$\Delta t$ the time step and the noise $\xi$ is a random variable with a uniform distribution in the range 
$[-\eta/2,\eta/2]$. Noise is a realistic ingredient of the model, relaxing   
the strict prescription of alignment to the average direction of motion of the neighboring peers.

The model resembles the classical ferromagnetic $XY$ model, where the velocities 
of the particles play the role of the spins and the noise that of 
temperature~\cite{binney92}. As in the $XY$ model, 
spins tend to align.
In the limit $v_0\rightarrow 0$ the SPP model resembles Monte Carlo simulations of diluted 
$XY$ ferromagnets.
However, when $v_0\neq 0$, it is a non-equilibrium model involving transport, as particles
continuously move on the plane. 

\begin{figure}
  \includegraphics[width=\columnwidth]{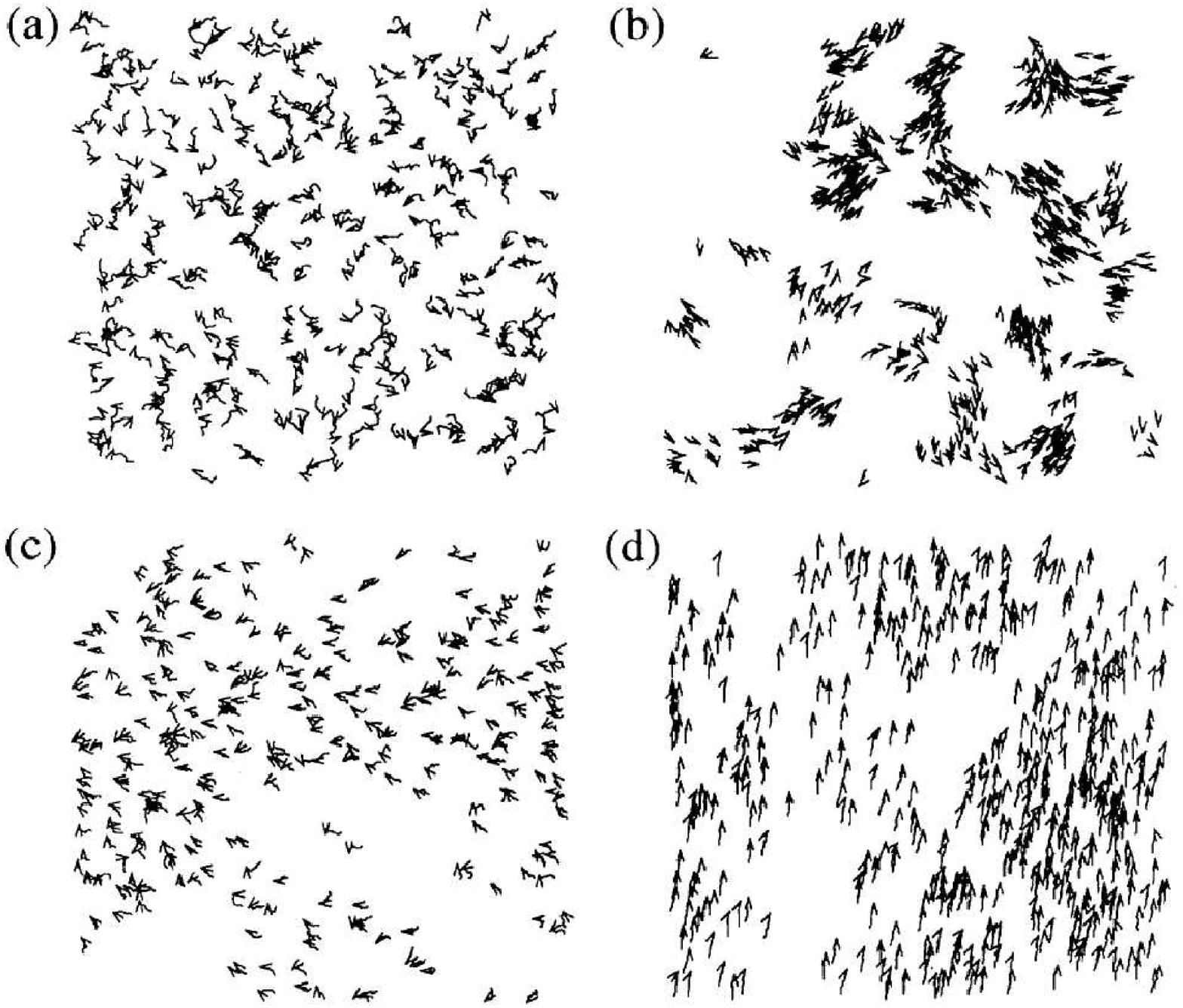}
  \caption{\label{vicsek} Velocity fields of particles in the model
    introduced in~\cite{vicsek95}. A phase transition from a
    disordered (a) to an ordered (d) state of collective
    motion is observed by decreasing the amount of noise in the
    system. From~\cite{vicsek95}.}
\end{figure}

Initially, the directions of the velocity vectors are randomly assigned, so that
there is no organized flow of particles in the system. After some relaxation time, 
the system reaches a steady state. The order parameter 
is the magnitude $\Phi$ of the average velocity of the system
\be
\Phi=\frac{1}{N}\mid\sum_{j}{\bf v_j}\mid.
\label{crowd2}
\ee
Collective motion arises when $\Phi > 0$. 
The level of noise $\eta$ is the control parameter, 
just like temperature in spin models. 
Numerical simulations indicate that, in the
thermodynamic limit of infinite system size, $\Phi$ varies with 
$\eta$ as follows
\be \Phi(\eta) \sim
\left\{
\begin{array}{ll}
\left[\frac{\eta_c(\rho)-\eta}{\eta_c(\rho)}\right]^{\beta} & {\textrm{ for }} \eta<\eta_c(\rho) \\
0 &  {\textrm{ for }} \eta>\eta_c(\rho)
\end{array}
\right.
\label{crowd3}
\ee
where $\rho$ is the density of the particles.
Eq.~(\ref{crowd3}) indicates that the system undergoes a continuous
kinetic transition from a phase,
where the initial rotational symmetry of the system is preserved, to a phase where it is broken 
due to the emergence of collective motion (Fig.~\ref{vicsek}). 
The flow is generated for values of the noise lower than
a threshold $\eta_c(\rho)$, which depends on the particle density. 
Like in equilibrium phase transitions, the fluctuations of the order parameter diverge at the critical threshold.
Numerical estimates yield
$\beta=0.42\pm 0.03$, different from the mean field value $1/2$. The critical line  
$\eta_c(\rho)\sim \rho^\kappa$, with $\kappa=0.45\pm 0.05$, and differs from the 
corresponding line for diluted ferromagnets: here, when $\eta\rightarrow 0$,  
the critical density vanishes; for diluted ferromagnets, instead, where the absence of noise
corresponds to a vanishing temperature, the critical density   
approaches the finite percolation threshold.

These results hold for any non-vanishing value of the velocity $v_0$.
It is remarkable that the model has an ordering phase transition in
two dimensions, as it is known that the equilibrium $XY$ model does
not order, due to the formation of Kosterlitz-Thouless
vortices~\cite{kosterlitz73}.  Simulations show that, for $v_0 \approx
0$, vortices indeed form in the first steps of the evolution, but they
are unstable and disappear in the long run~\cite{czirok97}.

Similar SPP models were also introduced in one and three
dimensions. In one dimension ~\cite{czirok99c} the interaction rule
has to be modified: since there is only one direction of motion,
imposing a constant particle's velocity would yield a trivial
dynamics, without contact to real situations.  In a realistic
dynamics, particles change direction of motion after slowing down.
This can be accomplished with a simple modification of rule
(\ref{crowd1}).  The resulting model displays the same kinetic
transition as in two dimensions, but with different exponents
($\beta=0.60\pm 0.05$, $\kappa \approx 1/4$). Collective motion
emerges in three dimensions as well~\cite{czirok99}. Here the kinetic
phase transition is less surprising, as in three dimensions even
equilibrium continuous spin models have a ferromagnetic phase
transition. The critical exponents are found to be consistent with the
mean field exponents in equilibrium systems. In the limit
$v_0\rightarrow 0$ the model becomes analogous to the
three-dimensional diluted Heisenberg model~\cite{binney92}.  However,
for 3D SPP collective motion can be generated for any value of the
particle density $\rho$, whereas the static model cannot order for
densities lower than the percolation threshold $\rho_{perc}\approx 1$.

Equilibrium systems cannot display an ordering phase transition in 2D
because of the Mermin-Wagner theorem~\cite{mermin66}, so the kinetic
transition observed in the SPP model is due to its intrinsic
dynamical, non-equilibrium character.  To explain the peculiar
features of the model continuum theories have been
proposed~\cite{toner95,toner98,czirok99c}, where the relevant
variables are the velocity and density fields ${\bf v}({\bf x},t)$ and
$\rho({\bf x},t)$.

In a first approach, continuum hydrodynamic equations
of motion were used~\cite{toner95,toner98}. The equation terms are dictated by the strict
requirement that rotational symmetry must be preserved
\be
\begin{array}{lll}
\partial_t{\bf v}+({\bf v}\nabla){\bf v}&=&\alpha{\bf v}-\beta|{\bf v}|^2{\bf v}-\nabla P+D_L\nabla(\nabla{\bf v})\\
&+&
D_1\nabla^2{\bf v}+D_2({\bf v}\nabla)^2{\bf v}+\xi,\\
\partial_t\rho+\nabla(\rho{\bf v})&=&0,
\end{array}
\label{crowd4}
\ee
where $P$ is the analogue of pressure, 
$\alpha$, $\beta$ are coefficients ruling the symmetry breaking, the $D_{L,1,2}$ are diffusion constants and $\xi$ 
is uncorrelated Gaussian random noise. The first equation contains some additional terms with respect
to standard Navier-Stokes equations for a simple fluid, because the SPP model does not conserve the momentum by construction,
so Galilean invariance is broken. The second equation expresses the conservation of the mass.
An analysis of these equations based on the dynamic renormalization group~\cite{forster77} 
reveals that the system orders in 2D, and that in more than four dimensions it behaves 
just like the corresponding equilibrium model, with equal exponents. 
On the other hand, there should be no phase transition 
in one dimension, in contrast to numerical findings. For this reason, a different continuum theory 
was derived directly from the master equation of the 1D SPP model~\cite{czirok99c}. Its equations read
\be
\begin{array}{lll}
\partial_t{U}&=&f(U)+\mu^2\partial_x^2U+\alpha\frac{(\partial_xU)(\partial_x\rho)}{\rho}+\zeta\\
\partial_t\rho&=&-v_0\partial_x(\rho U)+D\partial_x^2\rho,
\end{array}
\label{crowd5}
\ee
where $U(x,t)$ is the velocity field, $f(U)$ a self-propulsion term, $\zeta$ the noise.
The crucial feature of Eqs.~(\ref{crowd5}) is the existence of the non-linear term 
${(\partial_xU)(\partial_x\rho)}/{\rho}$. A linear stability analysis of the equations reveals 
that there is an ordered phase if the noise is sufficiently low,
as the domain walls separating groups of particles travelling in opposite directions
are unstable. Numerical solutions of Eqs.~(\ref{crowd5}) show that the continuum theory and the
discrete 1D SPP model belong to the same universality class.


\subsection{Pedestrian behavior}
\label{sec:pedestrians}

Pedestrian behavior has been empirically studied since the
1950s~\cite{hankin58}.  The first physical modeling  
was proposed in~\cite{henderson71} where it was conjectured
that pedestrian flows are similar to gases or fluids and
measurements of pedestrian flows were compared with Navier-Stokes
equations. However, realistic macroscopic models should account for
effects like maneuvers to avoid collisions, for which energy and
momentum are not conserved. Moreover, they should consider the
``granular'' structure of pedestrian flows, as each pedestrian
occupies a volume that cannot be penetrated by others. Therefore,
microscopic models have recently attracted much
attention~\cite{schreckenberg01,galea03}. One distinguishes two main
approaches: models based on cellular automata (CA) and the social
force model.

In CA models of pedestrian
dynamics~\cite{blue98,blue00,fukui99,muramatsu99,muramatsu00a,burstedde01,maniccam03},
time and space are discretized.  The pedestrian area is divided into
cells, which can be either empty or occupied by a single agent or an
obstacle.  A pedestrian can move to an empty neighboring cell at each
time step. The motion of a single pedestrian is a biased random walk,
where the bias is represented by a field residing on the space cells,
which determines the transition rates of the agent towards neighboring
cells, much like it happens in chemotaxis. CA models are
computationally very efficient, but they do not describe well the
complex phenomenology observed in real pedestrian dynamics, mostly
because of space discretization, which constrains the directions of
traffic flows. Therefore, models where agents can move in continuous
space are more likely to be successful. Among them, the social force
model introduced by Helbing and
coworkers~\cite{helbing94,helbing95,helbing00a,helbing02} had a big
impact: the main reason is that the actual forces between agents are
computed, which allows more quantitative predictions as compared to CA
models.

The social force model is based on the concept that behavioral changes
of individuals are driven by an external {\it social force}, which
affects the motivation of the individual and determines its actions.
Pedestrians have a particular
destination and a preferred walking speed. The motion of a
pedestrian is determined by its tendency to maintain its speed and
direction of motion and the perturbations due to the presence of other
pedestrians and physical barriers (walls).

The general equation of motion for a pedestrian $i$ is
\be
m_i\frac{d{\bf v}_i}{dt}={\bf F}_i^{(pers)}+{\bf F}_i^{(soc)}+{\bf \xi}_i(t),
\label{crowd6}
\ee
where $m_i$ is the mass of $i$ and ${\bf v}_i$ its velocity. 
The tendency to maintain the desired walking speed
${\bf v}_i^{(0)}$ is expressed by the first force term ${\bf F}_i^{(pers)}$, which reads
\be
{\bf F}_i^{(pers)}=m_i\frac{{\bf v}_i^{(0)}-{\bf v}_i}{\tau_i},
\label{crowd7}
\ee
where $\tau_i$ is the reaction time of $i$. The social force ${\bf F}_i^{(soc)}$, which describes the
influence of the environment, is a superposition of two terms, expressing 
the interaction with other pedestrians and with barriers, respectively. 
The interaction with other
pedestrians is described by a repulsive potential, expressing the need
to avoid collisions, and by an attractive potential, expressing the
tendency to come closer to persons/objects that the pedestrian finds
interesting. A typical choice for the repulsive force ${\bf F}_{ij}^{(rep)}$ 
between two pedestrians $i$ and $j$ is
\be
{\bf F}_{ij}^{(rep)}=A_i\exp\left(\frac{r_{ij}-d_{ij}}{B_i}\right){\bf n}_{ij},
\label{crowd8}
\ee
where $A_i$ and $B_i$ are constants, indicating the strength and the range of the interaction, respectively,
$r_{ij}$ is the sum of the radii of the pedestrians, which are modelled as disks, $d_{ij}$ the distance
between their centers of mass and ${\bf n}_{ij}$ the normalized vector pointing from $j$ to $i$.
Pedestrians try to keep a certain
distance from walls/obstacles as well, and the repulsion is often described by the same
term as in Eq.~(\ref{crowd8}), where one considers the distance of $i$ from the nearest point of the walls. 
The noise ${\bf \xi}_i(t)$ is added to account for
non-predictable individual behavior. 

This simple framework
predicts realistic scenarios, like the formation of ordered lanes of
pedestrians who intend to walk in the same direction and the
alternation of streams of pedestrians that try to pass a narrow door
into opposite directions. The existence of lanes reduces the risk of
collisions and represents a more efficient configuration for the
system. On the other hand, this is a spontaneously emerging property
of the system, as the agents are not explicitly instructed by the
model to behave this way. The repulsion between pedestrians moving
towards each other implies that the pedestrians shift a little aside
to avoid the collision: in this way small groups of people moving in
the same direction are formed.  These groups are stable, due to the
minimal interactions between people of each group, and attract other
pedestrians who are moving in the same direction.

In~\cite{helbing00} it is shown that a nontrivial non-equilibrium
phase transition is induced by noise: a system of particles driven in
opposite directions inside a two-dimensional periodic strip can get
jammed in crystallized configurations if the level of noise exceeds
some critical threshold ({\it freezing by heating}), in contrast to
the expectation that more noise corresponds to more disorder in the
system. This can explain how jams can arise in situations of great
collective excitation, like panic.  Surprisingly, the crystallized
state has a higher energy than the disordered state corresponding to
particles flowing along the corridor, so it is metastable.

\begin{figure}
  \includegraphics[width=\columnwidth]{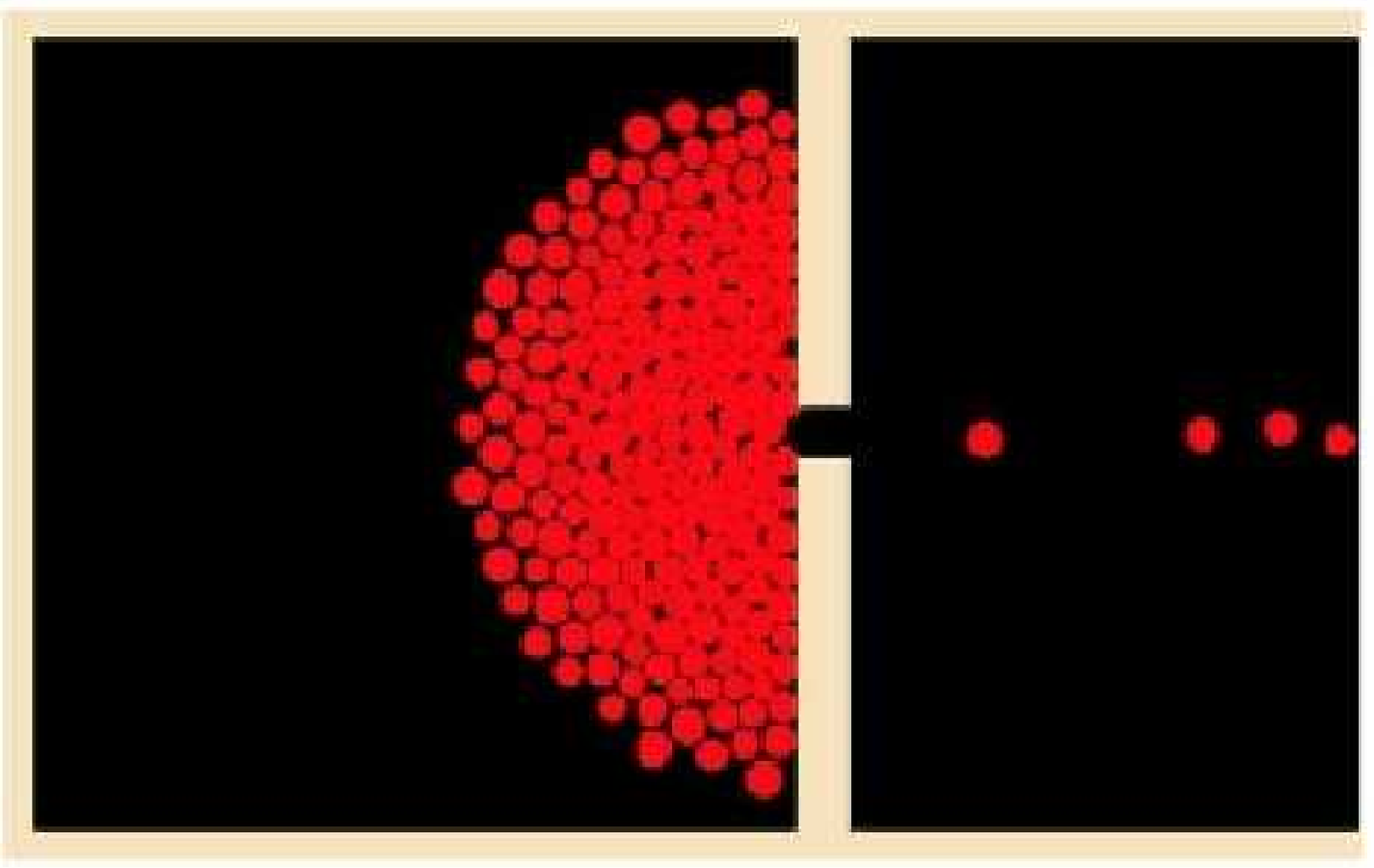}
\includegraphics[width=\columnwidth]{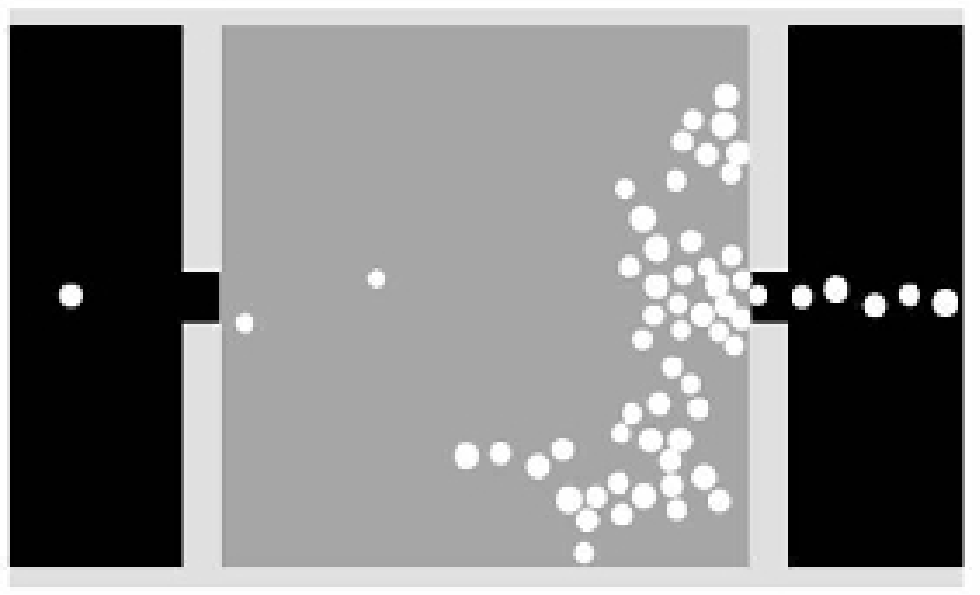}
\caption{\label{panic} Panic behavior. (Top) Escape from a room with a
  single exit. The exit is clogged by the people, who can leave the
  room only from time to time in bunches of individuals. (Bottom)
  Escape from a smoky room. The time to empty the room is minimal if
  people maintain their self-control and look at what the others are
  doing. Adapted from~\cite{helbing00a}.}
\end{figure}

The model introduced in~\cite{helbing00} has been adapted to simulate
situations in which people inside a room are reached by a sudden
alarming information (e.g. fire alarm) and try to run away through one
of the exits (escape panic)~\cite{helbing00a}. Additional force terms
are considered to account for realistic features of panicking crowds,
like the impossibility of excessive body compression and of tangential
motion of people about each other and along the walls. The model
describes phenomena observed in real panic situations: for example,
people attempting to leave a room through a single narrow exit,
generate intermittent clogging of the exit, so that people are unable
to flow continuously out of the room, but groups of individuals of
various sizes escape in an irregular succession
(Fig.~\ref{panic}a). 
This bursty behavior was actually observed in an
empirical study on mice attempting to exit out of a water
pool~\cite{saloma03}. Moreover, due to the friction of people in
contact, the time to empty the room is minimal in correspondence to
some optimal value of the individual speed: for higher speeds, the
total escape time increases ({\it faster is slower effect}). Placing
columns near the exits improves the situation, even if it seems
against intuition. Another situation deals with people trying to
escape from a smoky room, i.e., a room whose exits are not visible
unless one happens to stand close to them (Fig.~\ref{panic}b). In this
case, the agents do not have a preferential direction of motion, as
they have first to find the exits. The question is whether it is more
effective for the individuals to act on their own or to rely on the
action of the people close to them. The process is modeled by
introducing a panic parameter, that expresses the relative importance
of independent action and herding behavior [where herding is simulated
by a term analogous to the alignment rule in~\cite{vicsek95}]. It
turns out that the optimal chances of survival are attained when each
individual adopts a mixed strategy, based both on personal initiative
and on herding. In fact, through individualistic behavior some lucky
ones find quickly the exits and are followed by the others because of
imitation. A detailed account of evacuation dynamics can be found in
the recent review ~\cite{schadschneider08}.

Other studies concern the statistical features of mobility patterns of
individuals in physical space. In~\cite{brockmann06} the scaling
properties of human travels have been investigated by tracking the
worldwide dispersal of bank notes through bill-tracking websites. It
turns out that the distribution of traveling distances decays
algebraically, and is well reproduced within a two-parameter
continuous time random walk model. Moreover also in this case a
power-law distribution for the interevent times between two
consecutive sighting of the banknotes has been reported. These studies
highlight the importance of the web as a platform for social oriented
experiments (see also Sect.~\ref{sec:humandynamics}). Very recently
the mobility patterns of individuals have been investigated by
tracking the geographical location of mobile phone
users~\cite{gonzalez08}. Also in this case the distribution of
displacements over all users is well approximated by a truncated
power-law and analyzed in terms of truncated L\'evi flights,
i.e. random walks with power-law distributed step sizes.

\subsection{Applause dynamics}
\label{sec:applause}

Applause represents another remarkable example of social
self-organization.  At the end of a good performance the audience, 
after an initial uncoordinated phase,
often produces a synchronized clapping, where everybody claps at the
same time and with the same frequency. Synchronization occurs in many
biological and sociological processes~\cite{strogatz93,strogatz94}, from the flashing of
Southeast Asian fireflies to the chirping of crickets, from oscillating chemical reactions to
menstrual cycles of women living together for long times. 

Rhythmic applauses have been explored in detail, both
empirically and theoretically. In the first pioneering investigations~\cite{neda00a,neda00} 
applauses were recorded after several good theater and opera performances, with microphones placed both 
at some distance from the audience and close to randomly selected spectators. 
\begin{figure}[t]
\includegraphics[width=\columnwidth]{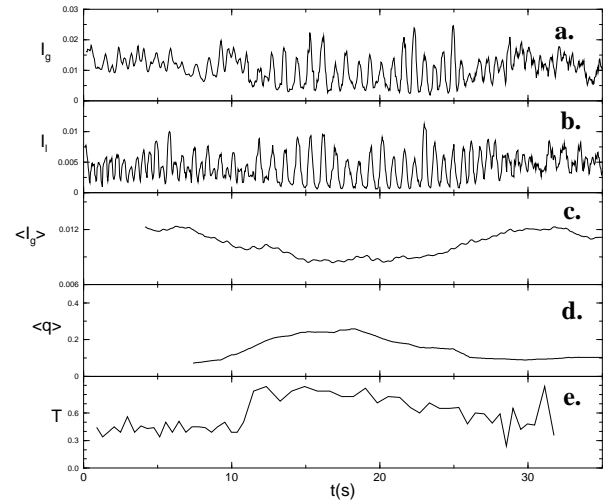}
\caption{Applause dynamics. Time evolution of the sound intensity of applauses in theaters and concert halls.
The rhythmic applause occurs in the time window between $12$s and $25$s.
a) Global noise intensity, measured by a microphone (averaged over a narrow time window of $0.2$s). 
b) Local noise intensity. 
c) Average noise intensity over a moving time window of $3$s. d) Order parameter $q$. e) Clapping period. Adapted from~\cite{neda00a}. }
\label{applause}
\end{figure}
The intensity of noise, heard at some distance from the audience or by one of the spectators (Fig.~\ref{applause}a, b) 
shows that the signal becomes periodic during the phase of the rhythmic applause, and that the average intensity 
of the sound decreases (Fig.~\ref{applause}c). Spectators usually start with a high frequency of clapping, which
reduces to approximately a half in the synchronized phase (Fig.~\ref{applause}e). Moreover, an order parameter
for synchronization was introduced and computed,
i.e. the maximum of the normalized correlation between the signal $s(t)$ and a harmonic function 
\be
r(t)={\textrm{ max }}_{\{T, \Phi\}}\left[ \frac{\int_{t-T}^{t+T}s(t)\sin(2\pi/T+\Phi)dt}{\int_{t-T}^{t+T}s(t)}\right],
\label{crowd9}
\ee
where $\Phi$ and $T$ vary in $[0, 2\pi]$ and between two convenient values $T_{min}$ and $T_{max}$, respectively.
The time evolution of the order parameter $q(t)$ clearly indicates the emergence of synchronization (Fig.~\ref{applause}d).

The dynamics of rhythmic applause has
been explained in the framework of the Kuramoto
model~\cite{kuramoto75}. The audience is described by a system of non-identical $N$ rotators. Each 
rotator $i$ is characterized by its phase $\phi_i$ and natural frequency $\omega_i$, which is assigned
according to a probability distribution $g(\omega)$. The dynamics is expressed by a system of $N$ coupled 
differential equations
\be
\frac{d\phi_i}{dt}=\omega_i+\frac{K}{N}\sum_{j=1}^{N}\sin(\phi_i-\phi_j),
\label{crowd10}
\ee
where $K$ is the interaction coupling. Each rotator interacts with all its peers, trying to
minimize the phase difference with the other rotators.  
In the thermodynamic limit $N\rightarrow\infty$ there is a critical coupling $K_c$ such that,
if $K>K_c$, there is a synchronized phase in which the rotators are partially aligned. 
If the distribution of natural frequencies $g(\omega)$ is Gaussian,
with dispersion $D$, one finds that $K_c=D\sqrt{2/\pi^3}$,
i.e. the critical coupling is proportional to the dispersion of the natural frequencies.
So, if $D$ is small, synchronization is likely to occur. The clapping
frequency of the rhythmic applause is indeed small, and so is its
dispersion, as confirmed by experiments performed on individual
spectators~\cite{neda00a}. On the other hand, the frequencies of the
enthusiastic clapping at the beginning of the applause have a much
higher dispersion, which hinders synchronization. 

From Fig.~\ref{applause} we see that
the phase of rhythmic applause is not stable, as it disappears after a while, although 
it may be reached again at a later time. This is due to the fact 
that the noise intensity decreases during the rhythmic applause, because of the lower clapping frequency 
(Fig.~\ref{applause}c). Spectators may find that the reduced intensity of the sound does not adequately represent
their appreciation of the performance and decide to increase the clapping frequency in order to produce more noise.
However, this brings the system again in the phase of enthusiastic clapping, and synchronization is lost.

In a more realistic
model, spectators are represented as two-mode stochastic oscillators,
and are only driven by the goal of producing some desired global level
of noise, with or without synchronization~\cite{neda03}.

\subsection{Mexican wave}
\label{sec:ola}

We conclude with another striking example of coherent collective
motion, i.e., the Mexican wave, also called La Ola, which is the wave
created by spectators in football stadia when they rapidly leap from
the seats with their arms up and successively sit down while a
neighboring section of people starts the same sequence.
In~\cite{farkas02,farkas06} an empirical study of this peculiar
phenomenon has been reported and simple models to describe it 
proposed.  These models were inspired by the literature on excitable
media~\cite{greenberg78,bub02}, where each unit of the system can
switch from an inactive to an active state if the density of active
units in their neighborhood exceeds a critical threshold. The
influence of a neighbor on an excitable subject decreases with its
distance from the subject and is higher if the neighbor sits on the
side where the wave comes from.  The total influence of the neighbors
is compared with the activation threshold of the spectator, which is
uniformly distributed in some range of values.  It turns out that a
group of spectators must exceed a critical mass in order to initiate
the process.  The models are able to reproduce size, form, velocity
and stability of real waves.

\section{FORMATION OF HIERARCHIES}
\label{sec:bonabeau}

Hierarchical organization is a peculiar feature of many animal
species, from insects to fishes, from birds to mammals, including
humans~\cite{wilson71,allee42,guhl68,chase80}.  Individuals usually
have a well defined rank inside their group, and the rank essentially
determines their role in the community. Highly-ranked individuals have
easier access to resources, they have better chances to reproduce,
etc..  Hierarchies also allow for an efficient distribution of
different tasks within a society, leading to a specialization of the
individuals.

The origin of hierarchical structures in animal and human societies is
still an open issue and has stimulated a lot of activity in the past
decades. The problem is to understand why and how from individuals
with initial identical status, inequalities emerge.  For instance, one
wonders how, in human societies, a strongly elitarian wealth
distribution could arise starting from a society where people
initially own an equal share of resources.  A possible explanation is
that hierarchies are produced by intrinsic attributes of the
individuals, e.g. differences in weight or size (for animals), and
talent or charisma (for humans). However, already in
1951~\cite{landau51,landau51a}, it was pointed out that intrinsic
factors alone could not be responsible for the hierarchies observed in
animal communities, and that the interactions between individuals play
a crucial role in the establishment of dominance relationships. The
hypothesis that hierarchy formation is a self-organization phenomenon
due to social dynamics has meanwhile become the most
widespread~\cite{chase82,francis88,chase02}. Here we discuss the most 
important results in this area.

\subsection{The Bonabeau model}
\label{bonabeau}

Dominance relationships seem to be determined by the outcome of fights
between individuals.  Laboratory experiments on various species hint
at the existence of a positive feedback
mechanism~\cite{hogeweg83,chase94,theraulaz95}, according to which
individuals who won more fights have an enhanced probability to win
future fights as compared to those who were less successful
(winner/loser effects). Memory effects are also important: observations on cockroaches 
show that insects removed from a group and reinserted after some time do not regain immediately their 
original rank~\cite{dugatkin94}.
Based on these empirical findings, Bonabeau et
al. proposed a simple model to explain the emergence of hierarchies
from an initial egalitarian society~\cite{bonabeau95}.

In the Bonabeau model, agents occupy the sites of a two-dimensional square lattice with
linear dimension $L$.  Each site can host only one agent and the
density of the agents on the lattice is $\rho$, which is the control
parameter of the system.  Every agent performs a random walk on the
lattice, moving to a randomly selected neighboring site at each
iteration. If the site is free, the agent occupies it.  If the site is
hosting another agent, a fight arouses between the two, and the winner
gets the right to occupy the site. In this way, if the winner is the
attacking agent, the two competitors switch their positions at
the end of the fight, otherwise they keep their original
positions. 

The outcome of the fight depends on the relative strength
of the two opponents. The strength $F$ of an agent grows with the
number of fights it wins.  Agent $i$ is stronger than agent $j$ if
$F_i>F_j$. The fight is a stochastic process, in which the stronger
agent has better chances to prevail, but it is not bound to win.  The
probability $Q_{ij}$ that agent $i$ defeats agent $j$ is expressed by
a Fermi function:
\begin{equation}
Q_{ij}=\frac{1}{1+\exp\{-\eta[F_i-F_j]\}},
\label{bona1}
\end{equation}
where $\eta$ is a sort of inverse temperature, measuring the level of randomness in the dynamics:
for $\eta\rightarrow\infty$, the stronger agent always wins, for $\eta\rightarrow 0$ both agents 
win with the same probability $1/2$. When an agent wins/loses a fight, its strength is
increased/decreased by one unit. Memory effects are expressed by a relaxation process of the
strengths of all agents, which decrease by a constant fraction $\mu$ 
at each time step.

In an egalitarian society, all agents have equal strength. A broad
distribution of strength would indicate the existence of hierarchies
in the system and is reflected in the distribution of the
winning probabilities $Q_{ij}$ in the stationary state.  So, the variance 
\begin{equation}
\sigma=\langle Q_{ij}^2\rangle - \langle Q_{ij} \rangle^2
\label{bona3}
\end{equation}
can be used as order parameter for the system~\cite{sousa00}. For an egalitarian society,
$\sigma=0$; a hierarchical society is characterized by a strictly
positive value of the variance $\sigma$.

In simulations of the Bonabeau model, agents are initially distributed
at random on the lattice, the strengths of all agents are usually
initialized to zero (egalitarian society) and one iteration consists
of one sweep over all agents, with each agent performing a
diffusion/fighting step.  The main result is that there is a critical
density $\rho_c(\eta,\mu)$ for the agents such that, for $\rho<\rho_c(\eta,\mu)$, society is
egalitarian, whereas for $\rho>\rho_c(\eta,\mu)$ a hierarchical organization is
created~\cite{bonabeau95}. This is due to the interplay between competition and relaxation: at low densities, 
fights are rare, and the dynamics is dominated by relaxation, which 
keeps the individual strengths around zero; at large densities, instead, the grow of the strengths
of some individuals is not counterbalanced by relaxation, and social differences emerge.

\subsubsection{Mean field solution}
\label{MFbonabeau}

The model can be analytically solved in the mean field limit, in which spatial correlations are absent
and the density $\rho$
expresses the probability for two agents to meet and fight. The evolution equation
for the strength of an agent $F_i$ is
\begin{equation}
\frac{dF_i}{dt}=H_i(\{F_j\})=-\mu F_i+\frac{\rho}{N}\sum_{j=1}^{N}\frac{\sinh\eta(F_i-F_j)}{1+\cosh\eta(F_i-F_j)},
\label{bona4}
\end{equation}
where $N$ is the number of agents. From Eq.~(\ref{bona4}) one derives that the mean strength $\langle F\rangle$  
decays exponentially to zero, so the stationary states are all characterized by $\langle F\rangle=0$. To check for the
stability of the special solution $F_i=0, \forall i$, corresponding to an egalitarian society, 
one computes the eigenvalues of the Jacobian matrix 
\begin{equation}
H_{ij}=\frac{\partial}{\partial F_j}R_i(0,0,...,0)=\left(-\mu+\frac{\rho\eta}{2}\right)\delta_{ij}-\frac{\rho\eta}{2N}.
\label{bona5}
\end{equation}
It turns out that there are only two different eigenvalues, namely $-\mu<0$ and $\rho\eta/2-\mu$
(in the large-$N$ limit), 
which is negative only when $\rho<\rho_c(\eta,\mu)=2\mu/\eta$. 
We conclude that, for $\rho>\rho_c(\eta,\mu)$, the egalitarian solution is linearly unstable, and a hierarchy emerges.
For $\eta<2$ the hierarchy transition is discontinuous, because a subcritical bifurcation occurs at 
$\rho_c(\eta,\mu)$. In this case there are several metastable states, along with the egalitarian one,  
so it is possible to observe hierarchical structure even for $\rho<\rho_c(\eta,\mu)$. Indeed, simulations
reveal that the stationary state is very sensitive to the choice of the initial conditions~\cite{bonabeau95}.
For $\eta\geq 2$, the bifurcation is critical, so
the transition is continuous and characterized by critical exponents
and critical slowing down.

\subsubsection{Modifications of the Bonabeau model}
\label{Mod_bonabeau}

The most popular modification of the
Bonabeau model \cite{stauffer03} consists in replacing $\eta$ in Eq.~(\ref{bona1})
with the standard deviation $\sigma$ of 
Eq.~(\ref{bona3}). By doing so, the probabilities $\{Q_{ij}\}$ are calculated using
the variance of their distribution, which changes in time, so there is
a feedback mechanism between the running hierarchical structure of
society and the dominance relationships between agents. 

For this model, analytical work in the mean field limit~\cite{lacasa06} showed
that the egalitarian fixed point is stable at all densities, at odds
with simulation results, that support the existence of a phase
transition to a hierarchical system~\cite{malarz06b}.  The apparent
discrepancy is due to the fact that, above a critical density, a
saddle-node bifurcation takes place. Both the egalitarian and the
hierarchical fixed points are stable, and represent possible endpoints
of the dynamics, depending on the initial conditions.
The model has been simulated on regular
lattices~\cite{stauffer03}, complete graphs~\cite{malarz06b} and
scale-free networks~\cite{gallos05}.  The phase transition holds in
every case, although on scale-free networks the critical density may
tend to zero in the thermodynamic limit of infinite agents.  On the
lattice, the model yields a society equally divided into leaders and
followers, which is not realistic. If the variation of the strength is
larger for a losing agent than for a winning agent, instead, the
fraction of agents that turn into leaders decreases rapidly with the
amount of the asymmetry~\cite{stauffer03c}.

Some authors proposed modifications of the moving rule for the
agents. In~\cite{odagaki06,tsujiguchi07} two particular situations
have been investigated, corresponding to what is called a timid and a
challenging society, respectively.  Timid agents do not look for
fights, but try to move to a free neighboring site. If there are none,
they pick a fight with the weakest neighbor. Two phase transitions
were observed by increasing the population density: a continuous one,
corresponding to the emergence of a middle class of agents, who are
fairly successful, and a discontinuous one, corresponding to the birth
of a class of winners, who win most of their fights.  In a challenging
society, agents look for fights, and choose the strongest neighbor as
opponent.  Hierarchies already emerge at low values of the population
density; in addition, since strong agents have a big attractiveness,
spatial correlations arise with the formation of small domains of
agents at low and intermediate densities.

\subsection{The Advancement-Decline model}
\label{advancement}

A simple model based on the interplay between advancement 
and decline, similar to the Bonabeau model,
has been proposed in~\cite{bennaim05a}. Agents have an integer-valued strength, and
interact pairwise.  The advancement dynamics is deterministic: the
strength of the stronger competitor increases by one unit. If both
agents have equal strength, both advance. The memory effect of the
Bonabeau model now consists of a declining process, in that the
strength of each individual decreases by one unit at rate $r$, as long
as it is positive. The parameter $r$ fixes the balance of advancement
and decline. 

The model is solvable in the mean field limit. We call $f_j(t)$ the fraction of agents
with strength $j$ at time $t$, and $F_k=\sum_{j=0}^{k}f_j$. The dynamics of the cumulative distribution
$F_k$ is described by the non-linear master equation
\begin{equation}
\frac{dF_k}{dt}=r(F_{k+1}-F_k)+F_{k}(F_{k-1}-F_k),
\label{bona6}
\end{equation}
where the first two terms express the contribution of the decline process, the other two that of
advancement. At the beginning all agents have zero strength, i.e. $f_0(0)=1$, corresponding to 
$F_k(0)=1, \forall k\geq 0$ (the reflecting wall at $k=0$ comes from imposing $F_{-1}(t)=0, \forall t$).

In the limit of continuous strength, Eq.~(\ref{bona6}) becomes a partial differential equation,
\begin{equation}
\frac{\partial F(k,t)}{\partial t}=[r-F(k,t)]\frac{\partial F(k,t)}{\partial k}.
\label{bona7}
\end{equation}
%
In the limit of $t\rightarrow\infty$ and finite strength, and since $F(k,\infty)\leq 1$ by definition,
Eq.~(\ref{bona7}) leaves only two possibilities:
\be
\begin{array}{lll}
F(k,\infty)&=&r,\hskip1cm{\textrm{ for }} r<1 ;\\
{\partial F(k,\infty)}/{\partial k}&=&0, \hskip1cm{\textrm{ for }} r\geq 1.
\end{array}
\label{bona8}
\ee
In this way, for $r\geq 1$, all agents have finite strength in the stationary state. For 
$r<1$, instead, we have a hierarchical society, with
a fraction $r$ of agents having finite strength (lower class), 
coexisting with a fraction $1-r$ of agents
whose strength diverges with $t$ (middle class). 

A dimensional analysis of Eq.~(\ref{bona7}) reveals that the strength of the middle class agents increases linearly with $t$.
One can then check for solutions of the form $F_k\sim\Phi(x)$, with $x=k/t$ and the boundary condition $\Phi(\infty)=1$. 
Eq.~(\ref{bona7}) becomes
\begin{equation}
x\Phi^\prime(x)=(\Phi(x)-r)\Phi^\prime(x),
\label{bona9}
\end{equation}
whose solution is
\be \Phi(x) \sim
\left\{
\begin{array}{ll}
x+r,  & \hskip1cm x<1-r\,; \\
1, &  \hskip1cm x\geq 1-r\,.
\end{array}
\right.
\label{bona10}
\ee
From Eq.~(\ref{bona10}) one deduces that the maximum strength of the system is $(1-r)t$, 
i.e. it grows linearly with time, as expected. 
In the limit of finite strength ($x\rightarrow 0$) one has $\Phi(0)=r$, which 
recovers the result on the fraction of lower class agents. 

In a later paper~\cite{bennaim06}, the
model has been generalized by introducing a stochastic advancement
dynamics, in which the stronger competitor of a pair of interacting
agents wins with a probability $p$.  This model yields a richer phase
diagram. In some region of the parameter space, a new egalitarian
class emerges, in which the strength distribution of the agents is
strongly peaked and moves with constant velocity, like a traveling
wave.  The model has been successfully applied to describe the
dynamics of sport competitions~\cite{bennaim07}.  Moreover, it has
inspired a generalization to competitive games involving more than two
players at the same time~\cite{bennaim06a}.

\section{HUMAN DYNAMICS}
\label{sec:humandynamics}

\subsection{Empirical data}

\begin{figure}
  \includegraphics[width=\columnwidth]{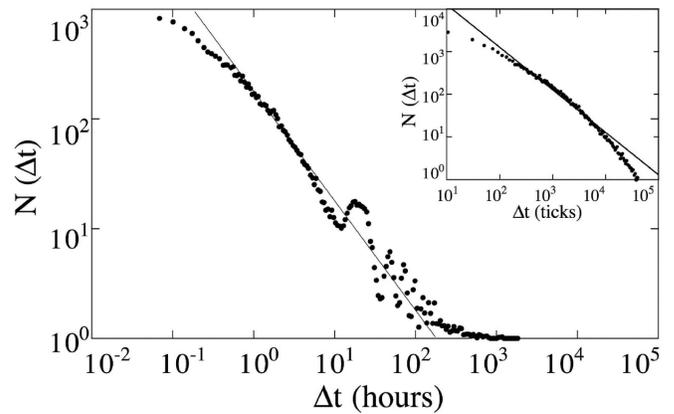}
  \caption {\label{humandynamics} Distribution of the response time
    until an email message is answered. (Inset) The same distribution
    is measured in ticks, i.e., units of messages sent in the
    system. Binning is logarithmic. The solid lines follow ${\Delta
      t}^{-1}$ and are meant as guides for the
    eye. From~\cite{eckmann04}}.
\end{figure}

One of the key questions in social dynamics concerns the behavior of
single individuals, namely how an individual chooses a convention,
takes a decision, schedules his tasks and more generally decides to
perform a given action. Most of these questions are obviously very
difficult to address, due to the psychological and social factors
involved. Nevertheless in the last few years several studies have
tried to quantitatively address these questions, mainly relying on the
availability of data through the web. A first valuable source of data
have been the logs of email exchanges. In particular the structure of
email networks has been first studied in~\cite{ebel02,newman02},
focusing on the spreading of informatic viruses. The emergence of
coherent, self-organized, structures in email traffic has been
reported in~\cite{eckmann04}, using an information-theoretic approach
based on the analysis of synchronization among triplets of users.  It has
been highlighted how nontrivial dynamic structures emerge as a
consequence of time correlations when users act in a synchronized
manner. The observed probability distribution of the response time
$\tau$ until a message is answered, features a broad distribution,
roughly approximated by a $1/\tau$ power law behavior
(Fig.~\ref{humandynamics}). 
The same kind of data have been analyzed
in~\cite{johansen04} and a generalized response time distribution
$\sim 1/\tau$ for human behavior in the absence of deadlines has
been suggested.

A different scaling, as far as the distribution of response times is
concerned, has been observed in mail correspondence
patterns~\cite{oliveira05,vazquez06}. In the correspondence of
Einstein, Darwin and Freud it has been found that the probability that
a letter will be replied in $\tau$ days is well approximated by a
power-law ${\tau}^{-\alpha}$, with $\alpha = 3/2$.

From all these empirical data a picture emerges where human dynamics
is characterized by bursts of events separated by short interevent
times followed by long periods of inactivity. This
burstiness~\cite{kleinberg02} can be quantified by looking for
instance at the distribution of interevent times that typically
displays heavy-tailed non-Poissonian statistics. A recent
study~\cite{alfi07} investigated how people react to deadlines for
conference registration.

\subsection{Queuing models}

Let us now focus on the explanation of the above mentioned phenomenology.
The very same database collected and used in~\cite{eckmann04} has been
analyzed in~\cite{barabasi05}, where the origin of bursts and heavy
tails in the probability distribution of the response time to an email
has been explained as a consequence of a decision-based queuing
process. The model is defined as follows. Each human agent has a list
with $L$ tasks, each task being assigned with an {\em a priori}
priority parameter $x_i$ (for $i\in [1,...,L]$), chosen from a
distribution $\rho(x)$. At each time step the agent selects the task
with the highest priority with probability $p$ and executes it, while
with probability $1-p$ a randomly selected task is performed. The
executed task is then removed from the list and replaced with another
one with priority again randomly extracted from $\rho(x)$. Computer
simulations of the model showed that for the deterministic protocol
($p \rightarrow 1$) the probability distribution of the times spent by
the tasks on the list features a power-law tail with exponent
$\alpha=1$.


The
exact solution of the Barab\'asi model for two tasks
($L=2$)~\cite{vazquez05} confirmed the $1/\tau$ behavior with an
exponential cut-off over a characteristic time
$(\ln{({2}/(1+p))})^{-1}$, which diverges for $p \rightarrow 1$.  The
limit $p \rightarrow 1$ is subtle because not only the characteristic
time diverges but the prefactor vanishes.

This issue has been addressed in~\cite{gabrielli07}, where an exact
probabilistic description of the Barab\'asi model for $L=2$ has been
given in the extremal limit, $p=1$, i.e., when only the most recently
added task is executed. In this limit the Barab\'asi model can be
exactly mapped into Invasion Percolation (IP)~\cite{wilkinson83} in
$d=1$. Using the so-called run time statistics (RTS)~\cite{marsili94},
originally introduced to study the IP problem in $d=2$, it has been
found that the exact waiting time distribution for a task scales as
$\tau^{-2}$, unlike the result $1/\tau$ found in~\cite{vazquez05},
which is valid for the stationary state when $0<p<1$. This behavior
disappears in the limit $p\rightarrow 1$, since the prefactor
vanishes. In summary, for $0<p<1$ finite time deviations from
stationarity relax exponentially fast and the dynamics is well
described by the stationary state. On the other hand, for $p
\rightarrow 1$ the stationary state becomes trivial and finite time
deviations relax so slowly that the task list dynamics has to be
described as an intrinsically nonstationary dynamics.

In~\cite{vazquez06} the case where limitations on the number of tasks
an individual can handle at any time is
discussed~\cite{cobham54}. Here tasks are, for instance, letters to be
replied. One assumes that letters arrive at rate $\lambda$ following a
Poisson process with exponential arrival time distribution, while
responses are written at rate $\mu$. Each letter is assigned a
discrete priority parameter $x=1,2,...,r$ upon arrival and the unanswered letter with highest
priority will be always chosen for a reply. This
model predicts a power law behavior for the waiting time distribution
of the individual tasks, with an exponent equal to $3/2$, with an
exponential cut-off characterized by a characteristic time $\tau_0$
given by $\tau_0 = 1/(\mu (1-\sqrt{\rho})^2)$, where $\rho=
\lambda/\mu$. The ratio $\rho$ is called traffic intensity and it
represents the natural control parameter for the model. For $\rho =1$
one has a critical regime where the queue length performs a $1-d$
random walk with a lower bound at zero. This allows to explain the
origin of the exponent $3/2$ in terms of the exponent of the return
time probability for a random walker.

Conditions for the emergence of scaling in the inter-event time
distribution have been addressed in~\cite{hidalgo06bis}. A
generalization of the queuing model introducing a time-dependent
priority for the tasks in the list (aging) has been discussed
in~\cite{blanchard06}. A further generalization of the queue model
with continuous valued priorities has been introduced and studied
in~\cite{grinstein06}. In this case a new task is added to the queue
with probability $\lambda$ and assigned a priority $x$ $(0 \leq x \leq
1)$ chosen from a probability distribution $\rho(x)$. With probability
$\mu$ the highest priority task is executed. Two asymptotic waiting
time distributions have been found analytically. For $\lambda = \mu <
1$ one obtains a power law (with exponent $3/2$) for the waiting time
distribution.  For $\lambda < \mu < 1$ a characteristic time $\tau_0 =
(\mu -\lambda)^2 / (4\mu (1-\lambda))>$ emerges such that, for $\tau \ll
\tau_0$, the waiting time distribution goes as $P(\tau) \sim
e^{-\tau/\tau_0} \tau^{-5/2}$, while for $1 \ll \tau \ll \tau_0$ $P(\tau)
\sim \tau^{-3/2}$. The asymptotic behaviour of $P(\tau)$ changes with
the introduction of a cost for switching between different classes of
tasks~\cite{grinstein06}.

Finally, in~\cite{bedogne07} a continuous model has been introduced,
which adopts an infinite queuing list where each element is assigned a
real non-negative number representing its priority of execution. The
priorities are drawn from a given probability distribution function
$\rho(x)$. At each time step a task is selected with a selection
probability function, which is an increasing function of the priority:

\begin{equation}
\Pi(x,t) = \frac{x^{\gamma}P(x,t)}{\int_0^{\infty}x^{\gamma}P(x,t) dx},
\end{equation}

\noindent where $\gamma$ is a non-negative real number and $P(x,t)$ is the
so-called waiting room priority distribution. The resulting waiting
time distribution $P(\tau)$ can be analytically found and it depends
explicitly on the priority distribution density function
$\rho(x)$. The scaling $P(\tau) \sim 1/\tau$ is recovered when
$\rho(x)$ is exponentially distributed.

\subsection{The social web}

Recently, a new paradigm where human dynamics plays an important role
has been quickly gaining ground on the World Wide Web: Collaborative
Tagging~\cite{golder06,cattuto07}. In web applications like
\textit{del.icio.us} (\url{http://del.icio.us}), \textit{Flickr}
(\url{http://www.flickr.com}), \textit{CiteULike}
(\url{http://www.citeulike.org}), users manage, share and browse
collections of online resources by enriching them with semantically
meaningful information in the form of freely chosen text labels
(\textit{tags}). The paradigm of collaborative tagging has been
successfully deployed in web applications designed to organize and
share diverse online resources such as bookmarks, digital photographs,
academic papers, music and more.  Web users interact with a
collaborative tagging system by posting content (\textit{resources})
into the system, and associating text strings (\textit{tags}) with
that content. 
\begin{figure}
\includegraphics[width=\columnwidth]{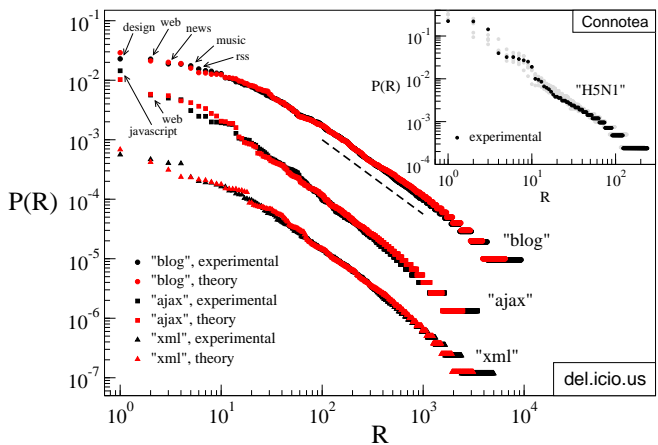}
\caption {\label{tagging} Frequency-rank plots for tags co-occurring
  with a selected tag. Experimental data (black symbols) are shown for
  \textit{del.icio.us} (circles for tags co-occurring with the popular
  tag \textit{blog}, squares for \textit{ajax} and triangles for
  \textit{xml}). For the sake of clarity, the curves for \textit{ajax}
  and \textit{xml} are shifted down by one and two decades,
  respectively.  All curves exhibit a power-law decay for high ranks
  (a dashed line corresponding to the power law $R^{-5/4}$ is provided
  as an aid for eye) and a shallower behavior for low ranks. Gray (red
  online) symbols are theoretical data obtained by computer
  simulations of the stochastic process described
  in~\cite{cattuto07,cattuto06}. (Inset) the same graph for the much
  younger system \textit{Connotea}. From~\cite{cattuto07}.}
\end{figure}
Fig.~\ref{tagging} reports the frequency-rank
distributions for the tags co-occurring with a few selected ones.  The
high-rank tail of the experimental curves displays a power-law
behavior, signature of an emergent hierarchical structure,
corresponding to a generalized Zipf's law~\cite{zipf49} with an
exponent between $1$ and $2$.  Since power laws are the standard
signature of self-organization and of human
activity~\cite{mitzenmacher03,newman05,barabasi05}, the presence of a
power-law tail is not surprising. The observed value of the exponent,
however, deserves further investigation, because the mechanisms
usually invoked to explain Zipf's law and its
generalizations~\cite{zanette05} do not look very realistic for the
case at hand, and a mechanism grounded on experimental data should be
sought. Moreover, the low-rank part of the frequency-rank curves
exhibits a flattening typically not observed in systems strictly
obeying Zipf's law.  Several aspects of the underlying complex
dynamics may be responsible for this feature: on the one hand this
behavior points to the existence of semantically equivalent and
possibly competing high-frequency tags (e.g. \textit{blog} and
\textit{blogs}).  More importantly, this flattening behavior may be
ascribed to an underlying hierarchical organization of tags
co-occurring with the one we single out: more general tags
(semantically speaking) will tend to co-occur with a larger number of
other tags.

At the global level the set of tags, though determined with no
explicit coordination, evolves in time and leads towards patterns of
terminology usage. Hence one observes the emergence of a loose
categorization system that can be effectively used to navigate through
a large and heterogeneous body of resources.  It is interesting to
investigate the way in which users interact with those systems.  Also
for this system a hyperbolic law for the user access to the system has
been observed~\cite{cattuto07}. In particular if one looks at the
temporal autocorrelation function for the sequence of tags
co-occurring with a given tag (e.g. \textit{blog}), one observes a
$1/(t+\tau)$ behavior, which suggests a heavy-tailed access to the
past state of the system, i.e., a memory kernel for the user access to
the system. On this basis a stochastic model of user
behavior~\cite{cattuto07} has been proposed, embodying two main
aspects of collaborative tagging: (i) a frequency-bias mechanism
related to the idea that users are exposed to each other's tagging
activity; (ii) a notion of memory (or aging of resources) in the form
of a heavy-tailed access to the past state of the system.  Remarkably,
this simple scheme is able to account quantitatively for the observed
experimental features, with a surprisingly high accuracy (see
Fig.~\ref{tagging}).  This points to the direction of a universal
behavior of users, who, despite the complexity of their own cognitive
processes and the uncoordinated and selfish nature of their tagging
activity, appear to follow simple activity patterns.

The dynamics of information access on the web represents another
source of data and several experiments have been performed in the last
few years. In~\cite{johansen00,johansen01} the dynamic response of the
{\em internauts} to a point-like perturbation as the announcement of a
web interview on stock market crashes has been
investigated. In~\cite{chessa04,chessa06} a cognitive model, based on
the mathematical theory of point processes, has been proposed, which
extends the results of~\cite{johansen00,johansen01} to download
relaxation dynamics.  In~\cite{dezso06} the visitation patterns of
news documents on a web portal have been considered. The problem of
collective attention has been recently addressed in~\cite{wu07} in the
framework of the website {\tt Digg.com}. Digg allows users to submit
news stories they discover from the Internet. A new submission
immediately appears on a repository webpage called {\em Upcoming
  Stories}, where other members can find the story and, if they like
it, add a digg to it. A so-called digg number is shown next to each
story's headline, which simply counts how many users have digged the
story in the past. The distribution of diggs accumulated by the news
stories has been found to be well approximated by a log-normal
curve. This behaviour has been explained in terms of a simple
stochastic model where the number of diggs acquired by a news story at
time $t$ can be expressed as:

\begin{equation}
N_t = (1+ r_t X_t) N_{t-1}, 
\end{equation}

where $X_1, X_2, ..., X_t$ are positive i.i.d. random variables
with a given mean and variance. The parameter $r_t$ is a
time-dependent novelty factor consisting of a series of decreasing
positive numbers with the property that $r_1 = 1$ and $r_t \rightarrow
0$ as $t \rightarrow \infty $. Experimentally it has been found that
$r_t$ follows a stretched-exponential relation with a characteristic
time of about one hour.

\section{SOCIAL SPREADING PHENOMENA}
\label{sec:rumor}

Opinion dynamics deals with the competition between different possible
responses to the same political question/issue.  A key feature is that
the alternatives have the same or at least comparable levels of
plausibility, so that in the interaction between two agents each of
them can in principle influence the other.  
In phenomena like the propagation of rumors or news,
the interaction is intrinsically asymmetric: possible states are very
different in nature.
The flow is only from those who know to those who do not.
The propagation of rumors or news is an instance of the vast class of
social spreading phenomena, that includes the diffusion of fads, the
adoption of  technological innovations, the success of
consumer products mediated by word-of-mouth.
The investigation ot such types of dynamics has a long tradition in
sociology and economics~\cite{bass69, rogers03, bikhchandani92}.
Work along this line has been performed recently by statistical physicists
on the diffusion of innovations~\cite{guardiola02,llas03}, the occurrence of
information cascades in social and economic systems~\cite{watts02b,centola07b},
disaster spreading in infrastructures~\cite{buzna06}, the emergence of
``hits'' in markets~\cite{sinha06,hidalgo06}.
Many of the models introduced for these phenomena assume that a
local threshold in the fraction of active neighbors must be overcome
for the spreading process to occur.
Here we will, somewhat arbitrarily, review only the activity on the
problem of rumor spreading. This phenomenon is modeled without a threshold:
hearing a rumor from a single neighbor is generally enough for a single
agent to start spreading it.
It is then clear that rumor spreading bears a lot of resemblance
with the evolution of an epidemics, with informed people playing the
role of infected agents and uninformed of
susceptible ones~\cite{rapoport53,goffman64}.  Obviously there are
crucial qualitative differences: rumor/news spreading is intentional;
it usually involves an (at least perceived) advantage for the
receiver, etc.  However most of such differences lie in the
interpretation of parameters; the analogy is strong and the field is
usually seen as closer to epidemiology than to opinion dynamics.

When considering rumor spreading some of the relevant questions to
address are similar to those for epidemiology: How many people will
eventually be reached by the news? Is there an 'epidemic threshold'
for the rate of spreading, separating a regime where a finite fraction
of people will be informed from one with the info remaining confined
to a small neighborhood?  What is the detailed temporal evolution?
Other issues, more connected to technological applications, deal with
the cost of the spreading process and its efficiency.
Rumor dynamics has also appealing connections with the search
for robust scalable communication protocols in large distributed
systems~\cite{vogels03,kermarrec03} and ``viral'' strategies in
marketing~\cite{leskovec06}. 

Detailed applications of the common models for epidemics to the
investigation of empirical data on the dissemination of ideas
exist~\cite{goffman66,bettencourt06}, but the most popular model for
rumor spreading, introduced in~\cite{daley64} (DK), has an important
difference.  As in the SIR model for epidemiology~\cite{anderson91},
agents are divided in three classes: ignorants, spreaders and
stiflers, i.e., those who have lost interest in diffusing the
information/rumor.  Their role is exactly the same of the susceptible,
infected, recovered agents of the SIR model. The only difference is
that while for an epidemics infected (I) people become recovered or
removed (R) {\em spontaneously} with a certain rate, typically people
stop propagating a rumor when they realize that those they want to
inform are already informed.  Hence the transition to state R is
proportional to the density of spreaders $s(t)$ in the SIR model,
while it is proportional to $s(t)[s(t)+r(t)]$ in the DK model, where
$r(t)$ is the density of stiflers.

The DK model has been studied analytically in the case of homogeneous
mixing, revealing that there is no threshold: for any rate $\lambda$
of the spreading process a finite fraction $r_\infty$ of people would
be informed~\cite{sudbury85}, given by the solution of
\be
r_{\infty} = 1-e^{-(1+\lambda/\alpha) r_{\infty}},
\ee
where $\alpha$ is the
proportionality constant of the transition rate to state R.
The same result holds for the very similar Maki-Thompson model~\cite{maki73}.
Hence the
nonlinear transition rate removes the threshold of the SIR model.
Clearly when both mechanisms for the damping of the propagation are
present (self-recovery and the nonlinear DK mechanism) a threshold is
recovered, since the linear term prevails for small
$s$~\cite{nekovee07}.

In the context of statistical physics the focus has been on the
behavior of the DK model on complex
networks~\cite{liu03b,moreno04a,moreno04b}.  When scale-free networks
are considered, the fraction $r_\infty$ of people reached decreases
compared to homogeneous nets. This occurs because hubs tend to become
stiflers soon and hence hamper the propagation. However, if one
considers the efficiency $E$ of the spreading process, defined as the
ratio between $r_\infty$ and the total traffic $L$ generated, it is
found that for any value of the parameters scale-free networks are
more efficient than homogeneous ones, and in a broad range of
parameters they are more efficient than the trivial broadcast spreading
mechanism (i.e., each node transmits the message to all its
neighbors).

A remarkable phenomenon occurs when the DK dynamics takes place on the
small-world Watts-Strogatz (WS) network~\cite{zanette01b,zanette02}.
In this case there is an 'epidemic' transition depending on the
rewiring parameter $p$. For $p>p_c$ the rumor propagates to a finite
fraction $r_\infty$ of the network sites.  For $p<p_c$ instead the
rumor remains localized around its origin, so $r_\infty$ vanishes in
the thermodynamic limit.  Notice that $p_c$ is finite for $N \to
\infty$, at odds with the geometric threshold characterizing the
small-world properties of WS networks, that vanishes in the limit of
infinite network.  Hence the transition is dynamic in nature and
cannot be ascribed to a pure geometric effect.

The diffusion of corruption has also been modeled as an epidemic-like
process, with people accepting or practicing a corrupt behavior
corresponding to infected individuals.  The main difference with
respect to usual epidemiological models is that the chance of an
individual to become corrupt is a strongly nonlinear function of the
number of corrupt neighbors. Other modifications include global
coupling terms, modeling the process of people getting corrupt because
of a perceived high prevalence of corruption in the society and the
response of the society as a whole, which is proportional to the
fraction of non-corrupt people.  The resulting phenomenology is quite
rich~\cite{blanchard05}.

It is also worth mentioning the model of Dodds and Watts for social
contagion~\cite{dodds04}, that introduces memory in the basic 
SIR model thus taking into account the effect of repeated exposure
to a rumor. Also information spreading in a population of diffusing
agents~\cite{agliari07} has been studied.

Finally, some activity has been devoted to the related problem of
gossip spreading. While rumors are about some topic of general
interest so that they may potentially extend to all, gossip is the
spreading of a rumor about some person and hence it is by definition a
local phenomenon; it may concern only people close to the subject in
the social network. If only nearest neighbors of the subject can
spread, the fraction of them reached by the gossip exhibits a minimum
as a function of the degree $k$ for some empirical and model social
networks.  Hence there is an ideal number of connections to minimize
the gossip propagation~\cite{lind07,lind07b}.

\section{COEVOLUTION OF STATES AND TOPOLOGY}
\label{sec:coevolution}

All models considered in the previous sections are defined on static
substrates: the interaction pattern is fixed and only opinions, not
connections, are allowed to change.  The opposite case is often
considered in many studies of network formation: vertices are endowed
with quenched attributes and links are formed or removed depending
on such fixed node properties.

In fact, real systems are mostly in between these two extreme cases:
both intrinsic properties of nodes (like opinions) and connections
among them vary in time over comparable temporal scales.  The
interplay of the two evolutions is then a natural issue to be
investigated.  More interestingly, in many cases the two evolutions
are explicitly coupled: if an agent finds that one of his contacts is
too different, he tends to severe the connection and look for other
interaction partners more akin to his own properties.

The investigation of the coevolution of networks and states has
recently attracted interest in several contexts, including
self-organization in Boolean networks~\cite{bornholdt00},
synchronization~\cite{ito02,ito03} and game-theoretic
approaches~\cite{zimmermann01,eguiluz00,zimmermann04,ehrhardt06}.  A
review about these developments is~\cite{gross08}, that also contains
a useful attempt to classify generic patterns of behavior in the
field.  For opinion and cultural dynamics the study of adaptive
networks is still at the beginning, but it promises to be a very
active field in the next years.

In coevolving systems, three ingredients are to be specified.
The dynamics taking place on nodes usually belongs to standard classes
of opinion or cultural dynamics (voter-like, majority-rule,
continuous opinion, etc.). Link dynamics can be of many types.
Examples exist of prescriptions allowing only link deletion
or independent link addition or removal. A more common, simple yet
quite realistic, assumption is that links are rewired: an agent unhappy
about one connection cuts it and forms a new link with
another agent. In this way the average connectivity $\langle k \rangle$
is conserved.
The third ingredient is the relative rate of link or node updates, generally
modeled by a probability $\phi$: a node is chosen at random; if it is
equal to one of its neighbors (or all, depending on the node dynamics)
nothing happens. Otherwise with probability
$\phi$ it rewires one of its connections; with probability $1-\phi$
it updates its state leaving the topology unchanged.

\begin{figure}
\includegraphics[width=\columnwidth]{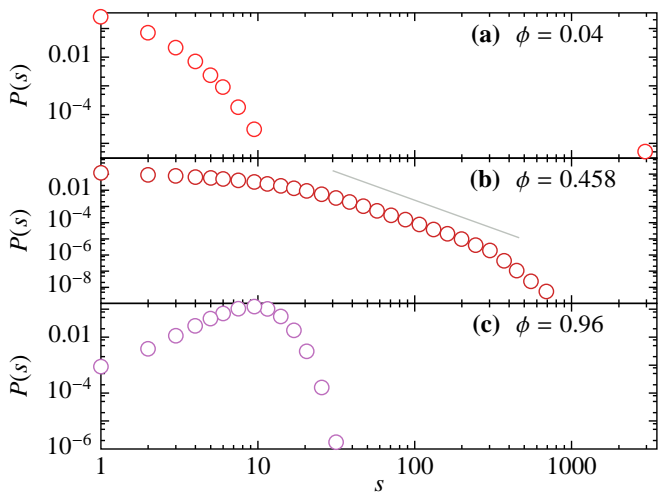}
\caption{Coevolution of opinions and networks. Histograms of cluster
  sizes in the consensus state for values of $\phi$ above, at and
  below the critical point in panels (a), (b) and (c), respectively.
  In panel (b) the distribution appears to follow a power law for part
  of its range with exponent $-3.5\pm 0.3$, as indicated by the solid
  line. From~\cite{holme06}.}
\label{holme06_2}
\end{figure}

If node dynamics leads toward consensus, it is reasonable
to expect that for small $\phi$ there will be asymptotically a
single network with all nodes in the same state.
In the opposite limit $\phi \to 1$ instead, rewiring will quickly
split the network in disconnected components that will separately reach
a different consensus state.
These two limit behaviors are hence separated by a topological
phase transition for some critical value $\phi_c$.
Such type of transition was first reported in~\cite{holme06},
where Potts variables assuming $G$ different values are
defined on the nodes of a network. $G$ is proportional to the number
of vertices, so that $\gamma=G/N$ is constant.  At each time step a
node and a neighbor are selected and with probability $1-\phi$ the
node picks the opinion of the neighbor. With probability $\phi$
instead, the node rewires the link to a new vertex chosen randomly
among those having its same opinion.
Dynamics continues up to complete separation
in components, within which there is full consensus.
For large $\phi$ only rewiring is allowed and the resulting topological
clusters trivially coincide with the
sets of initial holders of individual opinions. The distribution of
their sizes is multinomial. For small $\phi$ practically only
opinion changes are allowed and the final clusters are the
components of the initial graph.  The phase transition occurring
at $\phi_c$ is characterized by a power-law distribution of cluster sizes,
with an exponent about $-3.5$
(Fig.~\ref{holme06_2}), which differs from the one at the threshold of
the giant component formation in random graphs (mean field percolation
class). 
Finite size scaling gives additional exponents that are
universal (do not depend on $\langle k \rangle$ and $\gamma$) and
different from percolation. Critical slowing down is also observed.

A similar dynamics, but for a binary state variable, is investigated
in~\cite{Vazquez08}. A master equation approach gives an absorbing 
phase-transition between an active phase for
$\phi<\phi_c=(\langle k \rangle-2)/(\langle k \rangle-1)$,
and a frozen phase with zero active links for $\phi>\phi_c$.
In finite systems, numerical simulations show that in the absorbing
phase the network is composed of two disconnected components, one
with all spins up and the other with all spins down, while the
active phase corresponds to all nodes belonging to the same giant component.

Nardini et al.~\cite{nardini07} investigate a very similar model, with
only the slight difference that links are rewired to nodes selected
completely at random, with no preference for similar vertices holding
the same opinion. They focus on the approach to consensus in the limit
of small rewiring rate, finding opposite behaviors if the selected
node picks the opinion of the neighbor (direct voter dynamics) or the
neighbor is convinced by the node (reverse voter). In the first case
the consensus time becomes logarithmic in the system size $N$; in the
second it diverges exponentially with $N$. A mean field approach clarifies the
origin of the difference and also allows to understand why the
consensus time remains proportional to $N$ for the AB-model,
independently of the direct or reverse update prescription.

Differently from the models just discussed, in the first papers on
opinion dynamics in adaptive networks~\cite{gil06,zanette06}
links can only be deleted.
In the initial configuration, binary opinions are distributed randomly
on the nodes of a fully connected network.
If two neighbors disagree, one of them is set equal to the other with
probability $p_1$ (voter dynamics).  With probability $(1-p_1)p_2$
instead they get disconnected.
Iterating this dynamics a stationary state is reached depending only on
the combination
$q=p_1/[p_1+(1-p_1)p_2]$.  For small $q$ the system breaks down in two
communities of similar size and opposite opinions, with a large
fraction of internal connections. For large $q$ there are two
possibilities: a single community with the same opinion or one well
connected community with a set of poorly connected smaller
communities.  In correspondence to an intermediate value $q_c$, the
total density of links exhibits a minimum $r_c(q_c)$; both $r_c$ and
$q_c$ vanish for large system size.

Starting from the usual Axelrod model (see Sec.~\ref{sec:axelrod})
in~\cite{centola07} a further step is added: if the overlap $\omega$
between two nodes is exactly zero, the link is removed and one of
the agents connects to another randomly chosen vertex.  In this way
the transition $q_c$ between a monocultural state and fragmentation
is moved to much larger values: coevolution favors consensus.  At
$q_c$ both a cultural and a topological transition take place: the
system becomes separated in cultural groups that also form
topologically disconnected network subsets.  For even higher values of
the variability of the initial state $q=q^*>q_c$ another transition
occurs, involving only the network structure.  For $q>q^*$ the system
remains culturally disordered but a giant component is formed again.
In this regime it is likely that each vertex is completely different
from its neighbors, therefore it continuously breaks links and looks
(unsuccessfully) for new more similar partners. The transition
occurring at $q_c$ can be explained~\cite{vazquez07b} in terms of the
competition between the temporal scales of cultural and topological
evolution.  The topological transition occurring at $q^*$ can be
instead seen as the value of $q$ such that the temporal scale for
reaching a topologically stationary state is maximum.  Another model
of adaptive network coupled to vectorial opinions is
introduced in~\cite{grabowski06}.

Network rewiring has also been considered for the dynamics of the
Deffuant model~\cite{kozma08}.  At each time step with probability
$1-w$ a step of the usual opinion dynamics is performed
with confidence bound $d$, otherwise one
agent breaks one link and reconnects it to a randomly chosen other
node. By changing $w$ it is then possible to go from pure opinion
dynamics in a static environment to fast topological evolution in a
quenched opinion state.  Coevolution has opposite effects on the two
transitions exhibited by the model on static ER networks.  The
confidence bound threshold $d_1$, above which consensus is found,
grows with $w$.  The threshold $d_2$ of the transition between a polarized state (for
$d_2<d<d_1$) and a fragmented one with no macroscopic domains (for
$d<d_2$) goes instead to zero: $d_2(w>0)=0$.  The fragmented state
disappears because, even for small $d$, a node can rewire its
connections and find other agents with whom to reach an agreement.
Another coevolving generalization of Deffuant model
is~\cite{stauffer06a}.

Finally, a Glauber zero-temperature spin dynamics has been recently
considered in~\cite{benczik07}, together with a particular link update
rule, different from all previous cases: at each time step two nodes
carrying equal spins are connected with probability $p$, while nodes
with opposite values of the spin variable are connected with probability $1-p$.
No topological fragmentation occurs in this case. Consensus is always
reached on finite systems but exceedingly long disordered metastable
states exist.

\section{OUTLOOK} 
\label{sec:outlook}

With this review we made a first attempt to summarize the many
activities in the field of the so-called social dynamics. Our point of
view has been that of reviewing what has been done so far in this
young but rapidly evolving area, placing the main emphasis on the
statistical physics approach, i.e., on the contributions the physics
community has been giving to social oriented studies.

Though it is generally very difficult to isolate the contribution of a
given community to an intrinsically interdisciplinary endeavor, it is
nevertheless useful to identify the contribution the physics community
has been giving and the role it could play in the future. In this
perspective it is clear how the statistical physics role in social
dynamics has been mainly focused on modeling, either by introducing
brand new models to capture the main features of a given phenomenology
or performing detailed analysis of already existing models, e.g.,
focusing for instance on the existence of phase transitions,
universality, etc.. An inspiring principle has been provided by the
quest for simplicity.  This has several advantages. It allows for
discovering underlying universalities, i.e., realizing that behind the
details of the different models there could be a level where the
mathematical structure is similar. This implies, in its turn, the
possibility to perform mapping with other known models and exploit the
background of the already acquired knowledge for those models. In
addition physicists placed a great emphasis on the role of scales
(system sizes, timescales, etc.)  as well as on the topology (i.e.,
the network of interactions) underlying the observed phenomenology.

Closely related to modeling is the data analysis activity, both
considering synthetic data coming from simulations and empirical data
gathered from observations of real systems or collected in the
framework of newly devised experiments. Data analysis is very
important not only for the identification of new phenomenologies or
surprising features, but also for the validation of the models against
empirical data.  In this way a positive feedback mechanism could be
triggered between the theoretical and the experimental activities in
order to make the results robust, well understood and concrete.

Methodologically we can identify several important directions the
research in this area should possibly follow. It would be crucial
fostering the interactions across disciplines by promoting scientific
activities with concrete mutual exchanges among social scientists,
physicists, mathematicians and computer scientists. This would help
both in identifying the problems and sharpening the focus, as well as
in devising the most suitable theoretical concepts and tools to
approach the research. 

As for the modeling activity it would highly desirable to identify
general classes of behavior, not based on microscopic definitions, but
rather on large-scale universal characteristics, in order to converge
to a shared theoretical framework based on few fundamental
paradigms. The identification of which phenomena are actually
described by the theoretical models must become a priority. For
instance, the celebrated Axelrod model has not yet been shown to
describe at least semi-quantitatively any concrete situation. Without
applications the intense activity on modeling risks to be only a
conceptual exercise.

In this perspective a crucial factor will be most likely represented
by the availability of large sets of empirical quantitative data. The
research carried out so far only rarely relied on empirical datasets,
often insufficient to discriminate among different modeling
schemes. The joint interdisciplinary activity should then include
systematic campaigns of data gathering as well as the devising of new
experimental setups for a continuous monitoring of social
activities. From this point of view the web may be of great help, both
as a platform to perform controlled online social experiments, and as
a repository of empirical data on large-scale phenomena, like
elections and consumer behavior. It is only in this way that a
virtuous circle involving data collection, data analysis, modeling and
predictions could be triggered, giving rise to an ever more rigorous
and focused approach to socially motivated problems. A successful
example in this perspective is the study of traffic and pedestrian
behaviors, that in the last few years has attained a high level of
maturity, leading to reliable quantitative predictions and
control~\cite{helbing07} (see also Sec.~\ref{sec:collective}).

We conclude this review by highlighting a few interesting directions
that could possibly have a boosting effect on the research in the area
of social dynamics.

\subsubsection{Information dynamics and the Social Web} 

Though only a few years old, the growth of the World Wide Web and its effect
on the society have been astonishing, spreading from the research in
high-energy physics into other scientific disciplines, academe in
general, commerce, entertainment, politics and almost anywhere where
communication serves a purpose. Innovation has widened the
possibilities for communication. Blogs, wikis and social bookmark
tools allow the immediacy of conversation, while the potential of
multimedia and interactivity is vast. The reason for this immediate
success is the fact that no specific skills are needed for
participating.  In the so-called Web 2.0~\cite{oreilly05} users
acquire a completely new role: not only information seekers and
consumers, but information architects, cooperate in shaping the way
in which knowledge is structured and organized, driven by the notion
of meaning and semantics. In this perspective the web is acquiring the
status of a platform for {\em social computing}, able to coordinate
and exploit the cognitive abilities of the users for a given task. One
striking example is given by a series of {\em web
  games}~\cite{vonahn04}, where pairs of players are required to
coordinate the assignment of shared labels to pictures. As a side effect these
games provide a categorization of the images content, an extraordinary
difficult task for artificial vision systems.  More generally, the
idea that the individual, selfish activity of users on the web can
possess very useful side effects, is far more general than the example
cited. The techniques to profit from such an unprecedented opportunity
are, however, far from trivial. Specific technical and theoretical
tools need to be developed in order to take advantage of such a huge
quantity of data and to extract from this noisy source solid and
usable information\cite{huberman04,arrow03}. Such tools should
explicitly consider how users interact on the web, how they manage the
continuous flow of data they receive (see
Sec.~\ref{sec:humandynamics}), and, ultimately, what are the basic
mechanisms involved in their brain activity. In this sense, it is
likely that the new social platforms appearing on the web, could
rapidly become a very interesting laboratory for social sciences. In
particular we expect the web to have a strong impact on the studies of
opinion formation, political and cultural trends, globalization
patterns, consumers behavior, marketing strategies.

\subsubsection{Language and communication systems}

Language dynamics is a promising field which encompasses a broader
range of applications with respect to what described in
Sec.~\ref{sec:language}~\cite{loreto07}. In many biological,
technological and social systems, a crucial problem is that of the
communication among the different components, i.e., the elementary
units of the systems. The agents interact among themselves and with
the environment in a sensorial and non-symbolic way, their
communication system not being predetermined nor fixed from a global
entity. The communication system emerges spontaneously as a result of
the interactions of the agents and it could change continuously due to
the mutations occurring in the agents, in their objectives as well as
in the environment. An important question concerns how conventions are
established, how communication arises, what kind of communication
systems are possible and what are the prerequisites for such an
emergence to occur. In this perspective the emergence of a common
vocabulary only represents a first stage while it is interesting to
investigate the emergence of higher forms of agreement, e.g.,
compositionality, categories, syntactic or grammatical structures. It
is clear how important it would be to cast a theoretical
framework where all these problems could be defined, formalized and
solved. That would be a major input for the comprehension of many
social phenomena as well as for devising new technological
instruments.

\subsubsection{Evolution of social networks}

As real and online social systems grow ever larger, their analysis
becomes more complicated, due to their intrinsic dynamic nature, the
heterogeneity of the individuals, their interests, behavior etc.. In
this perspective, the discovery of communities, i.e., the
identification of more homogeneous groups of individuals, is a major
challenge. In this context, one has to distinguish the communities as
typically intended in social network analysis
(SNA)~\cite{scott00,wasserman94,freeman04} from a broader definition
of communities. In SNA one defines communities over a communication
relationship between the users, e.g. if they regularly exchange
e-mails or talk to each other. In a more general context, for
e.g. providing recommendation strategies, one is more interested in
finding communities of users with homogeneous interests and
behavior. Such homogeneity is independent of contacts between the
users although in most cases there will be at least a partial overlap
between communities defined by the user contacts and those by common
interests and behavior. Two important areas of research can be
identified. On the one hand, there is the question of which observable
features in the available data structures are best suited for
inferring relationships between individuals or users. Selecting
a feature affects the method used to detect communities~\cite{girvan02}, 
which may be different if one operates in the context of recommendation systems 
or in the context of semantic networks. On the other hand important
advances are foreseeable in the domain of coevolution of dynamics and
the underlying social substrates. This topic is still in its infancy,
despite the strong interdependence of dynamics and networks in
virtually all real phenomena. Empirical data on these processes are
becoming available: it is now possible to monitor in detail the
evolution of large scale social systems~\cite{palla07}.

\section*{Acknowledgments}

We wish to thank A.~Baldassarri, A.~Baronchelli, A.~Barrat, R.~Blythe,
E.~Caglioti, C.~Cattuto, L.~Dall'Asta, I.~Dornic, J.P.~Eckmann,
M.~Felici, S.~Galam, G.~Gosti, C.~Hidalgo, P.~Holme, N.F.~Johnson, N.L.~Komarova, 
R.~Lambiotte, J.~Lorenz, A.~McKane, M.~Marsili, J.~Minett, M.~Nowak, 
J.P.~Onnela, F.~Radicchi, J.J.~Ramasco, S.~Redner, M.~San Miguel, 
V.D.P.~Servedio, D.~Stauffer, L.~Steels, D.~Urbig, A.~Vespignani, 
D.~Vilone, T.~Vicsek, W.S.-Y.~Wang. This work was partially supported 
by the EU under contract IST-1940 (ECAgents) and IST-34721 (TAGora). 
The ECAgents and TAGora projects are funded by the Future and 
Emerging Technologies program (IST-FET) of the European Commission.

\bibliographystyle{apsrmp}


\begin{thebibliography}{552}
\expandafter\ifx\csname natexlab\endcsname\relax\def\natexlab#1{#1}\fi
\expandafter\ifx\csname bibnamefont\endcsname\relax
  \def\bibnamefont#1{#1}\fi
\expandafter\ifx\csname bibfnamefont\endcsname\relax
  \def\bibfnamefont#1{#1}\fi
\expandafter\ifx\csname citenamefont\endcsname\relax
  \def\citenamefont#1{#1}\fi
\expandafter\ifx\csname url\endcsname\relax
  \def\url#1{\texttt{#1}}\fi
\expandafter\ifx\csname urlprefix\endcsname\relax\def\urlprefix{URL }\fi
\providecommand{\bibinfo}[2]{#2}
\providecommand{\eprint}[2][]{\url{#2}}

\bibitem[{\citenamefont{Abrams and Strogatz}(2003)}]{abrams03}
\bibinfo{author}{\bibnamefont{Abrams}, \bibfnamefont{D.}}, and
  \bibinfo{author}{\bibfnamefont{S.}~\bibnamefont{Strogatz}},
  \bibinfo{year}{2003}, \bibinfo{journal}{Nature}
  \textbf{\bibinfo{volume}{424}}, \bibinfo{pages}{900}.

\bibitem[{\citenamefont{Agliari} \emph{et~al.}(2006)\citenamefont{Agliari,
  Burioni, Cassi, and Neri}}]{agliari07}
\bibinfo{author}{\bibnamefont{Agliari}, \bibfnamefont{E.}},
  \bibinfo{author}{\bibfnamefont{R.}~\bibnamefont{Burioni}},
  \bibinfo{author}{\bibfnamefont{D.}~\bibnamefont{Cassi}}, and
  \bibinfo{author}{\bibfnamefont{F.~M.} \bibnamefont{Neri}},
  \bibinfo{year}{2006}, \bibinfo{journal}{Phys. Rev. E}
  \textbf{\bibinfo{volume}{73}}(\bibinfo{number}{4}), \bibinfo{eid}{046138}
  (pages~\bibinfo{numpages}{8}).

\bibitem[{\citenamefont{von Ahn and Dabbish}(2004)}]{vonahn04}
\bibinfo{author}{\bibnamefont{von Ahn}, \bibfnamefont{L.}}, and
  \bibinfo{author}{\bibfnamefont{L.}~\bibnamefont{Dabbish}},
  \bibinfo{year}{2004}, in \emph{\bibinfo{booktitle}{CHI '04: Proceedings of
  the SIGCHI conference on Human factors in computing systems}}
  (\bibinfo{publisher}{ACM Press}, \bibinfo{address}{New York, NY, USA}), pp.
  \bibinfo{pages}{319--326}.

\bibitem[{\citenamefont{Aiello and Lu}(2001)}]{aiello01}
\bibinfo{author}{\bibnamefont{Aiello}, \bibfnamefont{F.~C.~W.}}, and
  \bibinfo{author}{\bibfnamefont{L.}~\bibnamefont{Lu}}, \bibinfo{year}{2001},
  \bibinfo{journal}{Exp. Math.} \textbf{\bibinfo{volume}{10}},
  \bibinfo{pages}{53}.

\bibitem[{\citenamefont{Albert and Barab\'asi}(2002)}]{barabasi99b}
\bibinfo{author}{\bibnamefont{Albert}, \bibfnamefont{R.}}, and
  \bibinfo{author}{\bibfnamefont{A.-L.} \bibnamefont{Barab\'asi}},
  \bibinfo{year}{2002}, \bibinfo{journal}{Rev. Mod. Phys.}
  \textbf{\bibinfo{volume}{74}}(\bibinfo{number}{1}), \bibinfo{pages}{47}.

\bibitem[{\citenamefont{{Aldashev} and {Carletti}}(2007)}]{aldashev07}
\bibinfo{author}{\bibnamefont{{Aldashev}}, \bibfnamefont{G.}}, and
  \bibinfo{author}{\bibfnamefont{T.}~\bibnamefont{{Carletti}}},
  \bibinfo{year}{2007}, \eprint{arXiv:0712.1140}.

\bibitem[{\citenamefont{Alder and Wainwright}(1957)}]{alder57}
\bibinfo{author}{\bibnamefont{Alder}, \bibfnamefont{B.~J.}}, and
  \bibinfo{author}{\bibfnamefont{T.~E.} \bibnamefont{Wainwright}},
  \bibinfo{year}{1957}, \bibinfo{journal}{J. Chem. Phys.}
  \textbf{\bibinfo{volume}{27}}, \bibinfo{pages}{1208}.

\bibitem[{\citenamefont{Alder and Wainwright}(1959)}]{alder59}
\bibinfo{author}{\bibnamefont{Alder}, \bibfnamefont{B.~J.}}, and
  \bibinfo{author}{\bibfnamefont{T.~E.} \bibnamefont{Wainwright}},
  \bibinfo{year}{1959}, \bibinfo{journal}{J. Chem. Phys.}
  \textbf{\bibinfo{volume}{31}}, \bibinfo{pages}{459}.

\bibitem[{\citenamefont{{Aleksiejuk}}
  \emph{et~al.}(2002)\citenamefont{{Aleksiejuk}, {Ho{\l}yst}, and
  {Stauffer}}}]{aleksiejuk02}
\bibinfo{author}{\bibnamefont{{Aleksiejuk}}, \bibfnamefont{A.}},
  \bibinfo{author}{\bibfnamefont{J.~A.} \bibnamefont{{Ho{\l}yst}}}, and
  \bibinfo{author}{\bibfnamefont{D.}~\bibnamefont{{Stauffer}}},
  \bibinfo{year}{2002}, \bibinfo{journal}{Physica A}
  \textbf{\bibinfo{volume}{310}}, \bibinfo{pages}{260}.

\bibitem[{\citenamefont{Alfi} \emph{et~al.}(2007)\citenamefont{Alfi, Parisi,
  and Pietronero}}]{alfi07}
\bibinfo{author}{\bibnamefont{Alfi}, \bibfnamefont{V.}},
  \bibinfo{author}{\bibfnamefont{G.}~\bibnamefont{Parisi}}, and
  \bibinfo{author}{\bibfnamefont{L.}~\bibnamefont{Pietronero}},
  \bibinfo{year}{2007}, \bibinfo{journal}{Nat. Phys.}
  \textbf{\bibinfo{volume}{3}}, \bibinfo{pages}{746}.

\bibitem[{\citenamefont{{Allee}}(1942)}]{allee42}
\bibinfo{author}{\bibnamefont{{Allee}}, \bibfnamefont{W.~C.}},
  \bibinfo{year}{1942}, \bibinfo{journal}{Biol. Symp.}
  \textbf{\bibinfo{volume}{8}}, \bibinfo{pages}{139}.

\bibitem[{\citenamefont{Alves} \emph{et~al.}(2002)\citenamefont{Alves, Neto,
  and Martins}}]{alves02}
\bibinfo{author}{\bibnamefont{Alves}, \bibfnamefont{S.}},
  \bibinfo{author}{\bibfnamefont{N.~O.} \bibnamefont{Neto}}, and
  \bibinfo{author}{\bibfnamefont{M.~L.} \bibnamefont{Martins}},
  \bibinfo{year}{2002}, \bibinfo{journal}{Physica A}
  \textbf{\bibinfo{volume}{316}}(\bibinfo{number}{1-4}), \bibinfo{pages}{601}.

\bibitem[{\citenamefont{Amblard and Deffuant}(2004)}]{amblard04}
\bibinfo{author}{\bibnamefont{Amblard}, \bibfnamefont{F.}}, and
  \bibinfo{author}{\bibfnamefont{G.}~\bibnamefont{Deffuant}},
  \bibinfo{year}{2004}, \bibinfo{journal}{Physica A}
  \textbf{\bibinfo{volume}{343}}, \bibinfo{pages}{725}.

\bibitem[{\citenamefont{Anderson and May}(1991)}]{anderson91}
\bibinfo{author}{\bibnamefont{Anderson}, \bibfnamefont{R.}}, and
  \bibinfo{author}{\bibfnamefont{R.}~\bibnamefont{May}}, \bibinfo{year}{1991},
  \emph{\bibinfo{title}{Infectious Diseases of Humans: Dynamics and Control}}
  (\bibinfo{publisher}{Oxford University Press}, \bibinfo{address}{Oxford,
  UK}).

\bibitem[{\citenamefont{{Andresen}}
  \emph{et~al.}(2007)\citenamefont{{Andresen}, {Hansen}, {Hansen},
  {Vasconcelos}, and {Andrade}}}]{andresen07}
\bibinfo{author}{\bibnamefont{{Andresen}}, \bibfnamefont{C.~A.}},
  \bibinfo{author}{\bibfnamefont{H.~F.} \bibnamefont{{Hansen}}},
  \bibinfo{author}{\bibfnamefont{A.}~\bibnamefont{{Hansen}}},
  \bibinfo{author}{\bibfnamefont{G.~L.} \bibnamefont{{Vasconcelos}}}, and
  \bibinfo{author}{\bibfnamefont{J.~S.} \bibnamefont{{Andrade}},
  \bibfnamefont{Jr}}, \bibinfo{year}{2007}, \eprint{arXiv:0712.1945}.

\bibitem[{\citenamefont{Antal} \emph{et~al.}(2006)\citenamefont{Antal, Redner,
  and Sood}}]{antal06}
\bibinfo{author}{\bibnamefont{Antal}, \bibfnamefont{T.}},
  \bibinfo{author}{\bibfnamefont{S.}~\bibnamefont{Redner}}, and
  \bibinfo{author}{\bibfnamefont{V.}~\bibnamefont{Sood}}, \bibinfo{year}{2006},
  \bibinfo{journal}{Phys. Rev. Lett.}
  \textbf{\bibinfo{volume}{96}}(\bibinfo{number}{18}), \bibinfo{pages}{188104}.

\bibitem[{\citenamefont{{Araripe}} \emph{et~al.}(2006)\citenamefont{{Araripe},
  {Costa Filho}, {Herrmann}, and {Andrade}}}]{araripe06}
\bibinfo{author}{\bibnamefont{{Araripe}}, \bibfnamefont{L.~E.}},
  \bibinfo{author}{\bibfnamefont{R.~N.} \bibnamefont{{Costa Filho}}},
  \bibinfo{author}{\bibfnamefont{H.~J.} \bibnamefont{{Herrmann}}}, and
  \bibinfo{author}{\bibfnamefont{J.~S.} \bibnamefont{{Andrade}}},
  \bibinfo{year}{2006}, \bibinfo{journal}{Int. J. Mod. Phys.C}
  \textbf{\bibinfo{volume}{17}}, \bibinfo{pages}{1809}.

\bibitem[{\citenamefont{Arrow}(2003)}]{arrow03}
\bibinfo{author}{\bibnamefont{Arrow}, \bibfnamefont{K.}}, \bibinfo{year}{2003},
  \bibinfo{journal}{Inform. Syst. Front.}
  \textbf{\bibinfo{volume}{5}}(\bibinfo{number}{1}), \bibinfo{pages}{5}.

\bibitem[{\citenamefont{{Ausloos} and {Petroni}}(2007)}]{ausloos07}
\bibinfo{author}{\bibnamefont{{Ausloos}}, \bibfnamefont{M.}}, and
  \bibinfo{author}{\bibfnamefont{F.}~\bibnamefont{{Petroni}}},
  \bibinfo{year}{2007}, \bibinfo{journal}{Europhys. Lett.}
  \textbf{\bibinfo{volume}{77}}, \bibinfo{pages}{38002}.

\bibitem[{\citenamefont{Axelrod}(1997)}]{axelrod97}
\bibinfo{author}{\bibnamefont{Axelrod}, \bibfnamefont{R.}},
  \bibinfo{year}{1997}, \bibinfo{journal}{J. Conflict Resolut.}
  \textbf{\bibinfo{volume}{41}}(\bibinfo{number}{2}), \bibinfo{pages}{203}.

\bibitem[{\citenamefont{Axelrod}(2006)}]{axelrod06b}
\bibinfo{author}{\bibnamefont{Axelrod}, \bibfnamefont{R.}},
  \bibinfo{year}{2006}, in \emph{\bibinfo{booktitle}{Handbook of Computational
  Economics}}, edited by
  \bibinfo{editor}{\bibfnamefont{L.}~\bibnamefont{Tesfatsion}} and
  \bibinfo{editor}{\bibfnamefont{K.~L.} \bibnamefont{Judd}}
  (\bibinfo{publisher}{Elsevier}), volume~\bibinfo{volume}{2}, pp.
  \bibinfo{pages}{1565--1584}.

\bibitem[{\citenamefont{Bagnoli} \emph{et~al.}(2004)\citenamefont{Bagnoli,
  Berrones, and Franci}}]{bagnoli04}
\bibinfo{author}{\bibnamefont{Bagnoli}, \bibfnamefont{F.}},
  \bibinfo{author}{\bibfnamefont{A.}~\bibnamefont{Berrones}}, and
  \bibinfo{author}{\bibfnamefont{F.}~\bibnamefont{Franci}},
  \bibinfo{year}{2004}, \bibinfo{journal}{Physica A}
  \textbf{\bibinfo{volume}{332}}, \bibinfo{pages}{509}.

\bibitem[{\citenamefont{Bagnoli} \emph{et~al.}(2007)\citenamefont{Bagnoli,
  Carletti, Fanelli, Guarino, and Guazzini}}]{bagnoli07}
\bibinfo{author}{\bibnamefont{Bagnoli}, \bibfnamefont{F.}},
  \bibinfo{author}{\bibfnamefont{T.}~\bibnamefont{Carletti}},
  \bibinfo{author}{\bibfnamefont{D.}~\bibnamefont{Fanelli}},
  \bibinfo{author}{\bibfnamefont{A.}~\bibnamefont{Guarino}}, and
  \bibinfo{author}{\bibfnamefont{A.}~\bibnamefont{Guazzini}},
  \bibinfo{year}{2007}, \bibinfo{journal}{Phys. Rev. E}
  \textbf{\bibinfo{volume}{76}}(\bibinfo{number}{6}), \bibinfo{eid}{066105}.

\bibitem[{\citenamefont{Ball}(2004)}]{ball04}
\bibinfo{author}{\bibnamefont{Ball}, \bibfnamefont{P.}}, \bibinfo{year}{2004},
  \emph{\bibinfo{title}{Critical Mass: How One Thing Leads to Another}}
  (\bibinfo{publisher}{{Farrar, Straus and Giroux}}, \bibinfo{address}{London,
  UK}).

\bibitem[{\citenamefont{Barab\'asi}(2005)}]{barabasi05}
\bibinfo{author}{\bibnamefont{Barab\'asi}, \bibfnamefont{A.-L.}},
  \bibinfo{year}{2005}, \bibinfo{journal}{Nature}
  \textbf{\bibinfo{volume}{435}}, \bibinfo{pages}{207}.

\bibitem[{\citenamefont{Barab\'asi and Albert}(1999)}]{barabasi99}
\bibinfo{author}{\bibnamefont{Barab\'asi}, \bibfnamefont{A.-L.}}, and
  \bibinfo{author}{\bibfnamefont{R.}~\bibnamefont{Albert}},
  \bibinfo{year}{1999}, \bibinfo{journal}{Science}
  \textbf{\bibinfo{volume}{286}}, \bibinfo{pages}{509}.

\bibitem[{\citenamefont{Barab\'asi}
  \emph{et~al.}(2002)\citenamefont{Barab\'asi, Jeong, N\'eda, Ravasz, Schubert,
  and Vicsek}}]{barabasi02}
\bibinfo{author}{\bibnamefont{Barab\'asi}, \bibfnamefont{A.-L.}},
  \bibinfo{author}{\bibfnamefont{H.}~\bibnamefont{Jeong}},
  \bibinfo{author}{\bibfnamefont{Z.}~\bibnamefont{N\'eda}},
  \bibinfo{author}{\bibfnamefont{E.}~\bibnamefont{Ravasz}},
  \bibinfo{author}{\bibfnamefont{A.}~\bibnamefont{Schubert}}, and
  \bibinfo{author}{\bibfnamefont{T.}~\bibnamefont{Vicsek}},
  \bibinfo{year}{2002}, \bibinfo{journal}{Physica A}
  \textbf{\bibinfo{volume}{311}}(\bibinfo{number}{3-4}), \bibinfo{pages}{590}.

\bibitem[{\citenamefont{Baronchelli}
  \emph{et~al.}(2006{\natexlab{a}})\citenamefont{Baronchelli, Dall'Asta,
  Barrat, and Loreto}}]{baronchelli06c}
\bibinfo{author}{\bibnamefont{Baronchelli}, \bibfnamefont{A.}},
  \bibinfo{author}{\bibfnamefont{L.}~\bibnamefont{Dall'Asta}},
  \bibinfo{author}{\bibfnamefont{A.}~\bibnamefont{Barrat}}, and
  \bibinfo{author}{\bibfnamefont{V.}~\bibnamefont{Loreto}},
  \bibinfo{year}{2006}{\natexlab{a}}, \bibinfo{journal}{Phys. Rev. E}
  \textbf{\bibinfo{volume}{73}}(\bibinfo{number}{015102}).

\bibitem[{\citenamefont{{Baronchelli}}
  \emph{et~al.}(2007)\citenamefont{{Baronchelli}, {Dall'Asta}, {Barrat}, and
  {Loreto}}}]{baronchelli07}
\bibinfo{author}{\bibnamefont{{Baronchelli}}, \bibfnamefont{A.}},
  \bibinfo{author}{\bibfnamefont{L.}~\bibnamefont{{Dall'Asta}}},
  \bibinfo{author}{\bibfnamefont{A.}~\bibnamefont{{Barrat}}}, and
  \bibinfo{author}{\bibfnamefont{V.}~\bibnamefont{{Loreto}}},
  \bibinfo{year}{2007}, \bibinfo{journal}{Phys. Rev. E}
  \textbf{\bibinfo{volume}{76}}, \bibinfo{pages}{051102}.

\bibitem[{\citenamefont{Baronchelli}
  \emph{et~al.}(2006{\natexlab{b}})\citenamefont{Baronchelli, Felici, Loreto,
  Caglioti, and Steels}}]{baronchelli06a}
\bibinfo{author}{\bibnamefont{Baronchelli}, \bibfnamefont{A.}},
  \bibinfo{author}{\bibfnamefont{M.}~\bibnamefont{Felici}},
  \bibinfo{author}{\bibfnamefont{V.}~\bibnamefont{Loreto}},
  \bibinfo{author}{\bibfnamefont{E.}~\bibnamefont{Caglioti}}, and
  \bibinfo{author}{\bibfnamefont{L.}~\bibnamefont{Steels}},
  \bibinfo{year}{2006}{\natexlab{b}}, \bibinfo{journal}{J. Stat. Mech.}
  \textbf{\bibinfo{volume}{P06014}}.

\bibitem[{\citenamefont{Bartolozzi}
  \emph{et~al.}(2005)\citenamefont{Bartolozzi, Leinweber, and
  Thomas}}]{bartolozzi05}
\bibinfo{author}{\bibnamefont{Bartolozzi}, \bibfnamefont{M.}},
  \bibinfo{author}{\bibfnamefont{D.~B.} \bibnamefont{Leinweber}}, and
  \bibinfo{author}{\bibfnamefont{A.~W.} \bibnamefont{Thomas}},
  \bibinfo{year}{2005}, \bibinfo{journal}{Phys. Rev. E}
  \textbf{\bibinfo{volume}{72}}(\bibinfo{number}{4}), \bibinfo{pages}{046113}.

\bibitem[{\citenamefont{Bass}(1969)}]{bass69}
\bibinfo{author}{\bibnamefont{Bass}, \bibfnamefont{F.~M.}},
  \bibinfo{year}{1969}, \bibinfo{journal}{Management Science}
  \textbf{\bibinfo{volume}{15}}(\bibinfo{number}{5}), \bibinfo{pages}{215}.

\bibitem[{\citenamefont{Baxter} \emph{et~al.}(2006)\citenamefont{Baxter,
  Blythe, Croft, and McKane}}]{baxter06}
\bibinfo{author}{\bibnamefont{Baxter}, \bibfnamefont{G.~J.}},
  \bibinfo{author}{\bibfnamefont{R.~A.} \bibnamefont{Blythe}},
  \bibinfo{author}{\bibfnamefont{W.}~\bibnamefont{Croft}}, and
  \bibinfo{author}{\bibfnamefont{A.~J.} \bibnamefont{McKane}},
  \bibinfo{year}{2006}, \bibinfo{journal}{Phys. Rev. E}
  \textbf{\bibinfo{volume}{73}}(\bibinfo{number}{4}), \bibinfo{pages}{046118}.

\bibitem[{\citenamefont{{Baxter}} \emph{et~al.}(2008)\citenamefont{{Baxter},
  {Blythe}, and {McKane}}}]{baxter08}
\bibinfo{author}{\bibnamefont{{Baxter}}, \bibfnamefont{G.~J.}},
  \bibinfo{author}{\bibfnamefont{R.~A.} \bibnamefont{{Blythe}}}, and
  \bibinfo{author}{\bibfnamefont{A.~J.} \bibnamefont{{McKane}}},
  \bibinfo{year}{2008}, \eprint{arXiv:0801.3083}.

\bibitem[{\citenamefont{Bedogne and Rodgers}(2007)}]{bedogne07}
\bibinfo{author}{\bibnamefont{Bedogne}, \bibfnamefont{C.}}, and
  \bibinfo{author}{\bibfnamefont{G.~J.} \bibnamefont{Rodgers}},
  \bibinfo{year}{2007}, \bibinfo{journal}{Physica A}
  \textbf{\bibinfo{volume}{385}}, \bibinfo{pages}{356}.

\bibitem[{\citenamefont{Behera and Schweitzer}(2003)}]{behera03}
\bibinfo{author}{\bibnamefont{Behera}, \bibfnamefont{L.}}, and
  \bibinfo{author}{\bibfnamefont{F.}~\bibnamefont{Schweitzer}},
  \bibinfo{year}{2003}, \bibinfo{journal}{Int. J. Mod. Phys. C}
  \textbf{\bibinfo{volume}{14}}(\bibinfo{number}{10}), \bibinfo{pages}{1331}.

\bibitem[{\citenamefont{{Ben-Naim}}(2005)}]{bennaim05}
\bibinfo{author}{\bibnamefont{{Ben-Naim}}, \bibfnamefont{E.}},
  \bibinfo{year}{2005}, \bibinfo{journal}{Europhys. Lett.}
  \textbf{\bibinfo{volume}{69}}, \bibinfo{pages}{671}.

\bibitem[{\citenamefont{{Ben-Naim}}
  \emph{et~al.}(2006{\natexlab{a}})\citenamefont{{Ben-Naim}, {Kahng}, and
  {Kim}}}]{bennaim06a}
\bibinfo{author}{\bibnamefont{{Ben-Naim}}, \bibfnamefont{E.}},
  \bibinfo{author}{\bibfnamefont{B.}~\bibnamefont{{Kahng}}}, and
  \bibinfo{author}{\bibfnamefont{J.~S.} \bibnamefont{{Kim}}},
  \bibinfo{year}{2006}{\natexlab{a}}, \bibinfo{journal}{J. Stat. Mech.}
  \textbf{\bibinfo{volume}{P07001}}.

\bibitem[{\citenamefont{Ben-Naim} \emph{et~al.}(2003)\citenamefont{Ben-Naim,
  Krapivsky, and Redner}}]{bennaim03a}
\bibinfo{author}{\bibnamefont{Ben-Naim}, \bibfnamefont{E.}},
  \bibinfo{author}{\bibfnamefont{P.}~\bibnamefont{Krapivsky}}, and
  \bibinfo{author}{\bibfnamefont{S.}~\bibnamefont{Redner}},
  \bibinfo{year}{2003}, \bibinfo{journal}{Physica D}
  \textbf{\bibinfo{volume}{183}}, \bibinfo{pages}{190}.

\bibitem[{\citenamefont{{Ben-Naim} and {Redner}}(2005)}]{bennaim05a}
\bibinfo{author}{\bibnamefont{{Ben-Naim}}, \bibfnamefont{E.}}, and
  \bibinfo{author}{\bibfnamefont{S.}~\bibnamefont{{Redner}}},
  \bibinfo{year}{2005}, \bibinfo{journal}{J. Stat. Mech.}
  \textbf{\bibinfo{volume}{L11002}}.

\bibitem[{\citenamefont{{Ben-Naim}}
  \emph{et~al.}(2006{\natexlab{b}})\citenamefont{{Ben-Naim}, {Vazquez}, and
  {Redner}}}]{bennaim06}
\bibinfo{author}{\bibnamefont{{Ben-Naim}}, \bibfnamefont{E.}},
  \bibinfo{author}{\bibfnamefont{F.}~\bibnamefont{{Vazquez}}}, and
  \bibinfo{author}{\bibfnamefont{S.}~\bibnamefont{{Redner}}},
  \bibinfo{year}{2006}{\natexlab{b}}, \bibinfo{journal}{Eur. Phys. J. B}
  \textbf{\bibinfo{volume}{49}}, \bibinfo{pages}{531}.

\bibitem[{\citenamefont{{Ben-Naim}}
  \emph{et~al.}(2007)\citenamefont{{Ben-Naim}, {Vazquez}, and
  {Redner}}}]{bennaim07}
\bibinfo{author}{\bibnamefont{{Ben-Naim}}, \bibfnamefont{E.}},
  \bibinfo{author}{\bibfnamefont{F.}~\bibnamefont{{Vazquez}}}, and
  \bibinfo{author}{\bibfnamefont{S.}~\bibnamefont{{Redner}}},
  \bibinfo{year}{2007}, \bibinfo{journal}{J. Korean Phys. Soc.}
  \textbf{\bibinfo{volume}{50}}, \bibinfo{pages}{124}.

\bibitem[{\citenamefont{{Benczik}} \emph{et~al.}(2007)\citenamefont{{Benczik},
  {Benczik}, {Schmittmann}, and {Zia}}}]{benczik07}
\bibinfo{author}{\bibnamefont{{Benczik}}, \bibfnamefont{I.~J.}},
  \bibinfo{author}{\bibfnamefont{S.~Z.} \bibnamefont{{Benczik}}},
  \bibinfo{author}{\bibfnamefont{B.}~\bibnamefont{{Schmittmann}}}, and
  \bibinfo{author}{\bibfnamefont{R.~K.~P.} \bibnamefont{{Zia}}},
  \bibinfo{year}{2007}, \eprint{arXiv:0709.4042}.

\bibitem[{\citenamefont{{Bernardes}}
  \emph{et~al.}(2002)\citenamefont{{Bernardes}, {Stauffer}, and
  {Kert{\'e}sz}}}]{bernardes02}
\bibinfo{author}{\bibnamefont{{Bernardes}}, \bibfnamefont{A.~T.}},
  \bibinfo{author}{\bibfnamefont{D.}~\bibnamefont{{Stauffer}}}, and
  \bibinfo{author}{\bibfnamefont{J.}~\bibnamefont{{Kert{\'e}sz}}},
  \bibinfo{year}{2002}, \bibinfo{journal}{Eur. Phys. J. B}
  \textbf{\bibinfo{volume}{25}}, \bibinfo{pages}{123}.

\bibitem[{\citenamefont{Bettencourt}
  \emph{et~al.}(2006)\citenamefont{Bettencourt, Cintr\'on-Arias, Kaiser, and
  Castillo-Ch\'avez}}]{bettencourt06}
\bibinfo{author}{\bibnamefont{Bettencourt}, \bibfnamefont{L.}},
  \bibinfo{author}{\bibfnamefont{A.}~\bibnamefont{Cintr\'on-Arias}},
  \bibinfo{author}{\bibfnamefont{D.~I.} \bibnamefont{Kaiser}}, and
  \bibinfo{author}{\bibfnamefont{C.}~\bibnamefont{Castillo-Ch\'avez}},
  \bibinfo{year}{2006}, \bibinfo{journal}{Physica A}
  \textbf{\bibinfo{volume}{364}}, \bibinfo{pages}{513}.

\bibitem[{\citenamefont{Bikhchandani}
  \emph{et~al.}(1992)\citenamefont{Bikhchandani, Hirshleifer, and
  Welch}}]{bikhchandani92}
\bibinfo{author}{\bibnamefont{Bikhchandani}, \bibfnamefont{S.}},
  \bibinfo{author}{\bibfnamefont{D.}~\bibnamefont{Hirshleifer}}, and
  \bibinfo{author}{\bibfnamefont{I.}~\bibnamefont{Welch}},
  \bibinfo{year}{1992}, \bibinfo{journal}{J. Polit. Econ.}
  \textbf{\bibinfo{volume}{100}}(\bibinfo{number}{5}), \bibinfo{pages}{992}.

\bibitem[{\citenamefont{Binney} \emph{et~al.}(1992)\citenamefont{Binney,
  Dowrick, Fisher, and Newman}}]{binney92}
\bibinfo{author}{\bibnamefont{Binney}, \bibfnamefont{J.}},
  \bibinfo{author}{\bibfnamefont{N.}~\bibnamefont{Dowrick}},
  \bibinfo{author}{\bibfnamefont{A.}~\bibnamefont{Fisher}}, and
  \bibinfo{author}{\bibfnamefont{M.}~\bibnamefont{Newman}},
  \bibinfo{year}{1992}, \emph{\bibinfo{title}{The Theory of Critical Phenomena:
  An Introduction to the Renormalization Group}} (\bibinfo{publisher}{Oxford
  University Press}, \bibinfo{address}{Oxford, UK}).

\bibitem[{\citenamefont{{Blanchard} and {Hongler}}(2006)}]{blanchard06}
\bibinfo{author}{\bibnamefont{{Blanchard}}, \bibfnamefont{P.}}, and
  \bibinfo{author}{\bibfnamefont{M.~.} \bibnamefont{{Hongler}}},
  \bibinfo{year}{2006}, \eprint{cond-mat/0608156}.

\bibitem[{\citenamefont{{Blanchard}}
  \emph{et~al.}(2005)\citenamefont{{Blanchard}, {Krueger}, {Krueger}, and
  {Martin}}}]{blanchard05}
\bibinfo{author}{\bibnamefont{{Blanchard}}, \bibfnamefont{P.}},
  \bibinfo{author}{\bibfnamefont{A.}~\bibnamefont{{Krueger}}},
  \bibinfo{author}{\bibfnamefont{T.}~\bibnamefont{{Krueger}}}, and
  \bibinfo{author}{\bibfnamefont{P.}~\bibnamefont{{Martin}}},
  \bibinfo{year}{2005}, \eprint{arXiv:physics/0505031}.

\bibitem[{\citenamefont{Blue and Adler}(1998)}]{blue98}
\bibinfo{author}{\bibnamefont{Blue}, \bibfnamefont{V.~J.}}, and
  \bibinfo{author}{\bibfnamefont{J.~L.} \bibnamefont{Adler}},
  \bibinfo{year}{1998}, \bibinfo{journal}{Transp. Res. Rec.}
  \textbf{\bibinfo{volume}{1644}}, \bibinfo{pages}{29}.

\bibitem[{\citenamefont{Blue and Adler}(2000)}]{blue00}
\bibinfo{author}{\bibnamefont{Blue}, \bibfnamefont{V.~J.}}, and
  \bibinfo{author}{\bibfnamefont{J.~L.} \bibnamefont{Adler}},
  \bibinfo{year}{2000}, \bibinfo{journal}{Transp. Res. Rec.}
  \textbf{\bibinfo{volume}{1710}}, \bibinfo{pages}{20}.

\bibitem[{\citenamefont{Blythe and McKane}(2007)}]{blythe07}
\bibinfo{author}{\bibnamefont{Blythe}, \bibfnamefont{R.~A.}}, and
  \bibinfo{author}{\bibfnamefont{A.~J.} \bibnamefont{McKane}},
  \bibinfo{year}{2007}, \bibinfo{journal}{J. of Stat. Mech.}
  \textbf{\bibinfo{volume}{P07018}}.

\bibitem[{\citenamefont{Boccaletti}
  \emph{et~al.}(2006)\citenamefont{Boccaletti, Latora, Moreno, Chavez, and
  Hwang}}]{boccaletti06}
\bibinfo{author}{\bibnamefont{Boccaletti}, \bibfnamefont{S.}},
  \bibinfo{author}{\bibfnamefont{V.}~\bibnamefont{Latora}},
  \bibinfo{author}{\bibfnamefont{Y.}~\bibnamefont{Moreno}},
  \bibinfo{author}{\bibfnamefont{M.}~\bibnamefont{Chavez}}, and
  \bibinfo{author}{\bibfnamefont{D.~U.} \bibnamefont{Hwang}},
  \bibinfo{year}{2006}, \bibinfo{journal}{Physics Reports}
  \textbf{\bibinfo{volume}{424}}(\bibinfo{number}{4-5}), \bibinfo{pages}{175},
  \urlprefix\url{http://dx.doi.org/10.1016/j.physrep.2005.10.009}.

\bibitem[{\citenamefont{Bonabeau} \emph{et~al.}(1995)\citenamefont{Bonabeau,
  Theraulaz, and Deneubourg}}]{bonabeau95}
\bibinfo{author}{\bibnamefont{Bonabeau}, \bibfnamefont{E.}},
  \bibinfo{author}{\bibfnamefont{G.}~\bibnamefont{Theraulaz}}, and
  \bibinfo{author}{\bibfnamefont{J.-L.} \bibnamefont{Deneubourg}},
  \bibinfo{year}{1995}, \bibinfo{journal}{Physica A}
  \textbf{\bibinfo{volume}{217}}, \bibinfo{pages}{373}.

\bibitem[{\citenamefont{Bonnekoh}(2003)}]{bonnekoh03}
\bibinfo{author}{\bibnamefont{Bonnekoh}, \bibfnamefont{J.}},
  \bibinfo{year}{2003}, \bibinfo{journal}{Int. J. Mod. Phys. C}
  \textbf{\bibinfo{volume}{14}}(\bibinfo{number}{9}), \bibinfo{pages}{1231}.

\bibitem[{\citenamefont{Bordogna and Albano}(2007)}]{bordogna07}
\bibinfo{author}{\bibnamefont{Bordogna}, \bibfnamefont{C.}}, and
  \bibinfo{author}{\bibfnamefont{E.}~\bibnamefont{Albano}},
  \bibinfo{year}{2007}, \bibinfo{journal}{J. Phys.-Condens Mat.}
  \textbf{\bibinfo{volume}{19}}(\bibinfo{number}{1-2}),
  \bibinfo{pages}{065144}.

\bibitem[{\citenamefont{Borghesi and Galam}(2006)}]{borghesi06}
\bibinfo{author}{\bibnamefont{Borghesi}, \bibfnamefont{C.}}, and
  \bibinfo{author}{\bibfnamefont{S.}~\bibnamefont{Galam}},
  \bibinfo{year}{2006}, \bibinfo{journal}{Phys. Rev. E}
  \textbf{\bibinfo{volume}{73}}(\bibinfo{number}{6}), \bibinfo{pages}{066118}.

\bibitem[{\citenamefont{Bornholdt and Rohlf}(2000)}]{bornholdt00}
\bibinfo{author}{\bibnamefont{Bornholdt}, \bibfnamefont{S.}}, and
  \bibinfo{author}{\bibfnamefont{T.}~\bibnamefont{Rohlf}},
  \bibinfo{year}{2000}, \bibinfo{journal}{Phys. Rev. Lett.}
  \textbf{\bibinfo{volume}{84}}(\bibinfo{number}{26}), \bibinfo{pages}{6114}.

\bibitem[{\citenamefont{Bouchaud and Potters}(2000)}]{bouchaud00}
\bibinfo{author}{\bibnamefont{Bouchaud}, \bibfnamefont{J.-P.}}, and
  \bibinfo{author}{\bibfnamefont{M.}~\bibnamefont{Potters}},
  \bibinfo{year}{2000}, \emph{\bibinfo{title}{Theory of financial risks}}
  (\bibinfo{publisher}{Cambridge University Press},
  \bibinfo{address}{Cambridge, UK}).

\bibitem[{\citenamefont{Boyer and Miramontes}(2003)}]{boyer03}
\bibinfo{author}{\bibnamefont{Boyer}, \bibfnamefont{D.}}, and
  \bibinfo{author}{\bibfnamefont{O.}~\bibnamefont{Miramontes}},
  \bibinfo{year}{2003}, \bibinfo{journal}{Phys. Rev. E}
  \textbf{\bibinfo{volume}{67}}(\bibinfo{number}{3}), \bibinfo{pages}{035102}.

\bibitem[{\citenamefont{Bray}(1994)}]{bray94}
\bibinfo{author}{\bibnamefont{Bray}, \bibfnamefont{A.}}, \bibinfo{year}{1994},
  \bibinfo{journal}{Adv. Phys.}
  \textbf{\bibinfo{volume}{43}}(\bibinfo{number}{3}), \bibinfo{pages}{357}.

\bibitem[{\citenamefont{Brockmann} \emph{et~al.}(2006)\citenamefont{Brockmann,
  Hufnagel, and Geisel}}]{brockmann06}
\bibinfo{author}{\bibnamefont{Brockmann}, \bibfnamefont{D.}},
  \bibinfo{author}{\bibfnamefont{L.}~\bibnamefont{Hufnagel}}, and
  \bibinfo{author}{\bibfnamefont{T.}~\bibnamefont{Geisel}},
  \bibinfo{year}{2006}, \bibinfo{journal}{Nature}
  \textbf{\bibinfo{volume}{439}}, \bibinfo{pages}{462}.

\bibitem[{\citenamefont{Bub} \emph{et~al.}(2002)\citenamefont{Bub, Shrier, and
  Glass}}]{bub02}
\bibinfo{author}{\bibnamefont{Bub}, \bibfnamefont{G.}},
  \bibinfo{author}{\bibfnamefont{A.}~\bibnamefont{Shrier}}, and
  \bibinfo{author}{\bibfnamefont{L.}~\bibnamefont{Glass}},
  \bibinfo{year}{2002}, \bibinfo{journal}{Phys. Rev. Lett.}
  \textbf{\bibinfo{volume}{88}}(\bibinfo{number}{5}), \bibinfo{pages}{058101}.

\bibitem[{\citenamefont{Buchanan}(2007)}]{buchanan07}
\bibinfo{author}{\bibnamefont{Buchanan}, \bibfnamefont{M.}},
  \bibinfo{year}{2007}, \emph{\bibinfo{title}{The social atom}}
  (\bibinfo{publisher}{Bloomsbury}, \bibinfo{address}{New York, NY, USA}).

\bibitem[{\citenamefont{Burstedde} \emph{et~al.}(2001)\citenamefont{Burstedde,
  Klauck, Schadschneider, and Zittartz}}]{burstedde01}
\bibinfo{author}{\bibnamefont{Burstedde}, \bibfnamefont{C.}},
  \bibinfo{author}{\bibfnamefont{A.}~\bibnamefont{Klauck}},
  \bibinfo{author}{\bibfnamefont{A.}~\bibnamefont{Schadschneider}}, and
  \bibinfo{author}{\bibfnamefont{J.}~\bibnamefont{Zittartz}},
  \bibinfo{year}{2001}, \bibinfo{journal}{Physica A}
  \textbf{\bibinfo{volume}{295}}, \bibinfo{pages}{507}.

\bibitem[{\citenamefont{{Buzna}} \emph{et~al.}(2006)\citenamefont{{Buzna},
  {Peters}, and {Helbing}}}]{buzna06}
\bibinfo{author}{\bibnamefont{{Buzna}}, \bibfnamefont{L.}},
  \bibinfo{author}{\bibfnamefont{K.}~\bibnamefont{{Peters}}}, and
  \bibinfo{author}{\bibfnamefont{D.}~\bibnamefont{{Helbing}}},
  \bibinfo{year}{2006}, \bibinfo{journal}{Physica A}
  \textbf{\bibinfo{volume}{363}}, \bibinfo{pages}{132}.

\bibitem[{\citenamefont{Caldarelli}(2007)}]{caldarelli07}
\bibinfo{author}{\bibnamefont{Caldarelli}, \bibfnamefont{G.}},
  \bibinfo{year}{2007}, \emph{\bibinfo{title}{Scale-Free Networks}}
  (\bibinfo{publisher}{Oxford University Press}, \bibinfo{address}{Oxford,
  UK}).

\bibitem[{\citenamefont{Campos} \emph{et~al.}(2003)\citenamefont{Campos,
  de~Oliveira, and Moreira}}]{campos03}
\bibinfo{author}{\bibnamefont{Campos}, \bibfnamefont{P.}},
  \bibinfo{author}{\bibfnamefont{V.}~\bibnamefont{de~Oliveira}}, and
  \bibinfo{author}{\bibfnamefont{F.~G.~B.} \bibnamefont{Moreira}},
  \bibinfo{year}{2003}, \bibinfo{journal}{Phys. Rev. E}
  \textbf{\bibinfo{volume}{67}}(\bibinfo{number}{2}), \bibinfo{pages}{026104}.

\bibitem[{\citenamefont{Carletti} \emph{et~al.}(2006)\citenamefont{Carletti,
  Fanelli, Grolli, and Guarino}}]{carletti06}
\bibinfo{author}{\bibnamefont{Carletti}, \bibfnamefont{T.}},
  \bibinfo{author}{\bibfnamefont{D.}~\bibnamefont{Fanelli}},
  \bibinfo{author}{\bibfnamefont{S.}~\bibnamefont{Grolli}}, and
  \bibinfo{author}{\bibfnamefont{A.}~\bibnamefont{Guarino}},
  \bibinfo{year}{2006}, \bibinfo{journal}{Europhys. Lett.}
  \textbf{\bibinfo{volume}{74}}(\bibinfo{number}{2}), \bibinfo{pages}{222}.

\bibitem[{\citenamefont{{Carletti}}
  \emph{et~al.}(2007)\citenamefont{{Carletti}, {Fanelli}, {Guarino}, {Bagnoli},
  and {Guazzini}}}]{carletti07}
\bibinfo{author}{\bibnamefont{{Carletti}}, \bibfnamefont{T.}},
  \bibinfo{author}{\bibfnamefont{D.}~\bibnamefont{{Fanelli}}},
  \bibinfo{author}{\bibfnamefont{A.}~\bibnamefont{{Guarino}}},
  \bibinfo{author}{\bibfnamefont{F.}~\bibnamefont{{Bagnoli}}}, and
  \bibinfo{author}{\bibfnamefont{A.}~\bibnamefont{{Guazzini}}},
  \bibinfo{year}{2007}, \eprint{arXiv:0712.1024}.

\bibitem[{\citenamefont{{Castellano}}(2005)}]{castellano05}
\bibinfo{author}{\bibnamefont{{Castellano}}, \bibfnamefont{C.}},
  \bibinfo{year}{2005}, in \emph{\bibinfo{booktitle}{AIP Conf. Proc. 779}},
  edited by \bibinfo{editor}{\bibfnamefont{P.}~\bibnamefont{{Garrido}}},
  \bibinfo{editor}{\bibfnamefont{J.}~\bibnamefont{{Marro}}}, and
  \bibinfo{editor}{\bibfnamefont{M.~A.} \bibnamefont{{Mu{\~n}oz}}}, volume
  \bibinfo{volume}{779}, pp. \bibinfo{pages}{114--120}.

\bibitem[{\citenamefont{Castellano}
  \emph{et~al.}(2005)\citenamefont{Castellano, Loreto, Barrat, Cecconi, and
  Parisi}}]{castellano05b}
\bibinfo{author}{\bibnamefont{Castellano}, \bibfnamefont{C.}},
  \bibinfo{author}{\bibfnamefont{V.}~\bibnamefont{Loreto}},
  \bibinfo{author}{\bibfnamefont{A.}~\bibnamefont{Barrat}},
  \bibinfo{author}{\bibfnamefont{F.}~\bibnamefont{Cecconi}}, and
  \bibinfo{author}{\bibfnamefont{D.}~\bibnamefont{Parisi}},
  \bibinfo{year}{2005}, \bibinfo{journal}{Phys. Rev. E}
  \textbf{\bibinfo{volume}{71}}(\bibinfo{number}{6}), \bibinfo{pages}{066107}.

\bibitem[{\citenamefont{Castellano}
  \emph{et~al.}(2000)\citenamefont{Castellano, Marsili, and
  Vespignani}}]{castellano00}
\bibinfo{author}{\bibnamefont{Castellano}, \bibfnamefont{C.}},
  \bibinfo{author}{\bibfnamefont{M.}~\bibnamefont{Marsili}}, and
  \bibinfo{author}{\bibfnamefont{A.}~\bibnamefont{Vespignani}},
  \bibinfo{year}{2000}, \bibinfo{journal}{Phys. Rev. Lett.}
  \textbf{\bibinfo{volume}{85}}(\bibinfo{number}{16}), \bibinfo{pages}{3536}.

\bibitem[{\citenamefont{Castellano}
  \emph{et~al.}(2003)\citenamefont{Castellano, Vilone, and
  Vespignani}}]{castellano03}
\bibinfo{author}{\bibnamefont{Castellano}, \bibfnamefont{C.}},
  \bibinfo{author}{\bibfnamefont{D.}~\bibnamefont{Vilone}}, and
  \bibinfo{author}{\bibfnamefont{A.}~\bibnamefont{Vespignani}},
  \bibinfo{year}{2003}, \bibinfo{journal}{Europhys. Lett.}
  \textbf{\bibinfo{volume}{63}}(\bibinfo{number}{1}), \bibinfo{pages}{153}.

\bibitem[{\citenamefont{Castell\'{o}}
  \emph{et~al.}(2006)\citenamefont{Castell\'{o}, Egu\'{i}luz, and {San
  Miguel}}}]{castello06}
\bibinfo{author}{\bibnamefont{Castell\'{o}}, \bibfnamefont{X.}},
  \bibinfo{author}{\bibfnamefont{V.}~\bibnamefont{Egu\'{i}luz}}, and
  \bibinfo{author}{\bibfnamefont{M.}~\bibnamefont{{San Miguel}}},
  \bibinfo{year}{2006}, \bibinfo{journal}{New J. Phys.}
  \textbf{\bibinfo{volume}{8}}(\bibinfo{number}{12}), \bibinfo{pages}{308}.

\bibitem[{\citenamefont{Castell\'{o}}
  \emph{et~al.}(2007)\citenamefont{Castell\'{o}, Toivonen, Egu\'{i}luz,
  Saram\"{a}ki, aski, and Miguel}}]{castello07}
\bibinfo{author}{\bibnamefont{Castell\'{o}}, \bibfnamefont{X.}},
  \bibinfo{author}{\bibfnamefont{R.}~\bibnamefont{Toivonen}},
  \bibinfo{author}{\bibfnamefont{V.~M.} \bibnamefont{Egu\'{i}luz}},
  \bibinfo{author}{\bibfnamefont{J.}~\bibnamefont{Saram\"{a}ki}},
  \bibinfo{author}{\bibfnamefont{K.~K.} \bibnamefont{aski}}, and
  \bibinfo{author}{\bibfnamefont{M.~S.} \bibnamefont{Miguel}},
  \bibinfo{year}{2007}, \bibinfo{journal}{Europhys. Lett.}
  \textbf{\bibinfo{volume}{79}}(\bibinfo{number}{6}), \bibinfo{pages}{66006
  (6pp)}.

\bibitem[{\citenamefont{Catanzaro} \emph{et~al.}(2005)\citenamefont{Catanzaro,
  Bogu{\~n}\'a, and Pastor-Satorras}}]{catanzaro05}
\bibinfo{author}{\bibnamefont{Catanzaro}, \bibfnamefont{M.}},
  \bibinfo{author}{\bibfnamefont{M.}~\bibnamefont{Bogu{\~n}\'a}}, and
  \bibinfo{author}{\bibfnamefont{R.}~\bibnamefont{Pastor-Satorras}},
  \bibinfo{year}{2005}, \bibinfo{journal}{Phys. Rev. E}
  \textbf{\bibinfo{volume}{71}}(\bibinfo{number}{2}), \bibinfo{pages}{027103}.

\bibitem[{\citenamefont{Cattuto} \emph{et~al.}(2007)\citenamefont{Cattuto,
  Loreto, and Pietronero}}]{cattuto07}
\bibinfo{author}{\bibnamefont{Cattuto}, \bibfnamefont{C.}},
  \bibinfo{author}{\bibfnamefont{V.}~\bibnamefont{Loreto}}, and
  \bibinfo{author}{\bibfnamefont{L.}~\bibnamefont{Pietronero}},
  \bibinfo{year}{2007}, \bibinfo{journal}{Proc. Natl. Acad. Sci. USA}
  \textbf{\bibinfo{volume}{104}}, \bibinfo{pages}{1461}.

\bibitem[{\citenamefont{Cattuto} \emph{et~al.}(2006)\citenamefont{Cattuto,
  Loreto, and Servedio}}]{cattuto06}
\bibinfo{author}{\bibnamefont{Cattuto}, \bibfnamefont{C.}},
  \bibinfo{author}{\bibfnamefont{V.}~\bibnamefont{Loreto}}, and
  \bibinfo{author}{\bibfnamefont{V.}~\bibnamefont{Servedio}},
  \bibinfo{year}{2006}, \bibinfo{journal}{Europhys. Lett.}
  \textbf{\bibinfo{volume}{76}}(\bibinfo{number}{2}), \bibinfo{pages}{208}.

\bibitem[{\citenamefont{Centola} \emph{et~al.}(2007)\citenamefont{Centola,
  Egu\'iluz, and Macy}}]{centola07b}
\bibinfo{author}{\bibnamefont{Centola}, \bibfnamefont{D.}},
  \bibinfo{author}{\bibfnamefont{V.~M.} \bibnamefont{Egu\'iluz}}, and
  \bibinfo{author}{\bibfnamefont{M.}~\bibnamefont{Macy}}, \bibinfo{year}{2007},
  \bibinfo{journal}{Physica A}
  \textbf{\bibinfo{volume}{374}}(\bibinfo{number}{1}), \bibinfo{pages}{449}.

\bibitem[{\citenamefont{{Centola}} \emph{et~al.}(2007)\citenamefont{{Centola},
  {Gonz\'alez-Avella}, {Eguiluz}, and {San Miguel}}}]{centola07}
\bibinfo{author}{\bibnamefont{{Centola}}, \bibfnamefont{D.}},
  \bibinfo{author}{\bibfnamefont{J.~C.} \bibnamefont{{Gonz\'alez-Avella}}},
  \bibinfo{author}{\bibfnamefont{V.~M.} \bibnamefont{{Eguiluz}}}, and
  \bibinfo{author}{\bibfnamefont{M.}~\bibnamefont{{San Miguel}}},
  \bibinfo{year}{2007}, \bibinfo{journal}{J. of Confl. Res.}
  \textbf{\bibinfo{volume}{51}}, \bibinfo{pages}{905}.

\bibitem[{\citenamefont{Chakrabarti}
  \emph{et~al.}(2006)\citenamefont{Chakrabarti, Chakraborti, and
  Chatterjee}}]{chakrabarti06}
\bibinfo{editor}{\bibnamefont{Chakrabarti}, \bibfnamefont{B.~K.}},
  \bibinfo{editor}{\bibfnamefont{A.}~\bibnamefont{Chakraborti}}, and
  \bibinfo{editor}{\bibfnamefont{A.}~\bibnamefont{Chatterjee}} (eds.),
  \bibinfo{year}{2006}, \emph{\bibinfo{title}{Econophysics and Sociophysics:
  Trends and Perspectives}} (\bibinfo{publisher}{Wiley VCH}).

\bibitem[{\citenamefont{Chang}(2001)}]{chang01}
\bibinfo{author}{\bibnamefont{Chang}, \bibfnamefont{I.}}, \bibinfo{year}{2001},
  \bibinfo{journal}{Int. J. Mod. Phys. C}
  \textbf{\bibinfo{volume}{12}}(\bibinfo{number}{10}), \bibinfo{pages}{1509}.

\bibitem[{\citenamefont{Chase}(1980)}]{chase80}
\bibinfo{author}{\bibnamefont{Chase}, \bibfnamefont{I.~D.}},
  \bibinfo{year}{1980}, \bibinfo{journal}{Am. Sociol. Review}
  \textbf{\bibinfo{volume}{45}}, \bibinfo{pages}{905}.

\bibitem[{\citenamefont{Chase}(1982)}]{chase82}
\bibinfo{author}{\bibnamefont{Chase}, \bibfnamefont{I.~D.}},
  \bibinfo{year}{1982}, \bibinfo{journal}{Behavior}
  \textbf{\bibinfo{volume}{80}}, \bibinfo{pages}{218}.

\bibitem[{\citenamefont{Chase} \emph{et~al.}(1994)\citenamefont{Chase,
  Bartolomeo, and Dugatkin}}]{chase94}
\bibinfo{author}{\bibnamefont{Chase}, \bibfnamefont{I.~D.}},
  \bibinfo{author}{\bibfnamefont{C.}~\bibnamefont{Bartolomeo}}, and
  \bibinfo{author}{\bibfnamefont{L.~A.} \bibnamefont{Dugatkin}},
  \bibinfo{year}{1994}, \bibinfo{journal}{Anim. Behav.}
  \textbf{\bibinfo{volume}{48}}, \bibinfo{pages}{393}.

\bibitem[{\citenamefont{Chase} \emph{et~al.}(2002)\citenamefont{Chase, Tovey,
  Spangler-Martin, and Manfredonia}}]{chase02}
\bibinfo{author}{\bibnamefont{Chase}, \bibfnamefont{I.~D.}},
  \bibinfo{author}{\bibfnamefont{C.}~\bibnamefont{Tovey}},
  \bibinfo{author}{\bibfnamefont{D.}~\bibnamefont{Spangler-Martin}}, and
  \bibinfo{author}{\bibfnamefont{M.}~\bibnamefont{Manfredonia}},
  \bibinfo{year}{2002}, \bibinfo{journal}{Proc. Natl. Acad. Sci. USA}
  \textbf{\bibinfo{volume}{99}}(\bibinfo{number}{8}), \bibinfo{pages}{5744}.

\bibitem[{\citenamefont{Chatterjee and Seneta}(1977)}]{chatterjee77}
\bibinfo{author}{\bibnamefont{Chatterjee}, \bibfnamefont{S.}}, and
  \bibinfo{author}{\bibfnamefont{E.}~\bibnamefont{Seneta}},
  \bibinfo{year}{1977}, \bibinfo{journal}{J. Appl. Prob.}
  \textbf{\bibinfo{volume}{14}}, \bibinfo{pages}{89}.

\bibitem[{\citenamefont{{Chen} and {Redner}}(2005)}]{chen05a}
\bibinfo{author}{\bibnamefont{{Chen}}, \bibfnamefont{P.}}, and
  \bibinfo{author}{\bibfnamefont{S.}~\bibnamefont{{Redner}}},
  \bibinfo{year}{2005}, \bibinfo{journal}{J. Phys. A}
  \textbf{\bibinfo{volume}{38}}, \bibinfo{pages}{7239}.

\bibitem[{\citenamefont{Chen and Redner}(2005)}]{chen05}
\bibinfo{author}{\bibnamefont{Chen}, \bibfnamefont{P.}}, and
  \bibinfo{author}{\bibfnamefont{S.}~\bibnamefont{Redner}},
  \bibinfo{year}{2005}, \bibinfo{journal}{Phys. Rev. E}
  \textbf{\bibinfo{volume}{71}}(\bibinfo{number}{3}), \bibinfo{pages}{036101}.

\bibitem[{\citenamefont{Chessa and Murre}(2004)}]{chessa04}
\bibinfo{author}{\bibnamefont{Chessa}, \bibfnamefont{A.}}, and
  \bibinfo{author}{\bibfnamefont{J.}~\bibnamefont{Murre}},
  \bibinfo{year}{2004}, \bibinfo{journal}{Physica A}
  \textbf{\bibinfo{volume}{333}}, \bibinfo{pages}{541}.

\bibitem[{\citenamefont{Chessa and Murre}(2006)}]{chessa06}
\bibinfo{author}{\bibnamefont{Chessa}, \bibfnamefont{A.}}, and
  \bibinfo{author}{\bibfnamefont{J.}~\bibnamefont{Murre}},
  \bibinfo{year}{2006}, \bibinfo{journal}{Physica A}
  \textbf{\bibinfo{volume}{366}}, \bibinfo{pages}{539}.

\bibitem[{\citenamefont{Chomsky}(1965)}]{chomsky1965}
\bibinfo{author}{\bibnamefont{Chomsky}, \bibfnamefont{N.}},
  \bibinfo{year}{1965}, \emph{\bibinfo{title}{Aspects of the Theory of Syntax}}
  (\bibinfo{publisher}{The MIT Press}, \bibinfo{address}{Cambridge, MA, USA}).

\bibitem[{\citenamefont{{Chowdhury}}
  \emph{et~al.}(2000)\citenamefont{{Chowdhury}, {Santen}, and
  {Schadschneider}}}]{chowdhury00}
\bibinfo{author}{\bibnamefont{{Chowdhury}}, \bibfnamefont{D.}},
  \bibinfo{author}{\bibfnamefont{L.}~\bibnamefont{{Santen}}}, and
  \bibinfo{author}{\bibfnamefont{A.}~\bibnamefont{{Schadschneider}}},
  \bibinfo{year}{2000}, \bibinfo{journal}{Phys. Rep.}
  \textbf{\bibinfo{volume}{329}}, \bibinfo{pages}{199}.

\bibitem[{\citenamefont{Clifford and Sudbury}(1973)}]{clifford73}
\bibinfo{author}{\bibnamefont{Clifford}, \bibfnamefont{P.}}, and
  \bibinfo{author}{\bibfnamefont{A.}~\bibnamefont{Sudbury}},
  \bibinfo{year}{1973}, \bibinfo{journal}{Biometrika}
  \textbf{\bibinfo{volume}{60}}(\bibinfo{number}{3}), \bibinfo{pages}{581}.

\bibitem[{\citenamefont{Cobham}(1954)}]{cobham54}
\bibinfo{author}{\bibnamefont{Cobham}, \bibfnamefont{A.}},
  \bibinfo{year}{1954}, \bibinfo{journal}{J. Oper. Res. Soc. Am.}
  \textbf{\bibinfo{volume}{2}}, \bibinfo{pages}{70}.

\bibitem[{\citenamefont{Cohen} \emph{et~al.}(1986)\citenamefont{Cohen, Hajnal,
  and Newman}}]{cohen86}
\bibinfo{author}{\bibnamefont{Cohen}, \bibfnamefont{J.~E.}},
  \bibinfo{author}{\bibfnamefont{J.}~\bibnamefont{Hajnal}}, and
  \bibinfo{author}{\bibfnamefont{C.~M.} \bibnamefont{Newman}},
  \bibinfo{year}{1986}, \bibinfo{journal}{Stoch. Proc. Appl.}
  \textbf{\bibinfo{volume}{22}}, \bibinfo{pages}{315}.

\bibitem[{\citenamefont{Conte} \emph{et~al.}(1997)\citenamefont{Conte,
  Hegselmann, and Terna}}]{conte97}
\bibinfo{author}{\bibnamefont{Conte}, \bibfnamefont{R.}},
  \bibinfo{author}{\bibfnamefont{R.}~\bibnamefont{Hegselmann}}, and
  \bibinfo{author}{\bibfnamefont{P.}~\bibnamefont{Terna}},
  \bibinfo{year}{1997}, \emph{\bibinfo{title}{Simulating Social Phenomena}}
  (\bibinfo{publisher}{Springer Verlag}, \bibinfo{address}{Berlin-Heidelberg,
  Germany}).

\bibitem[{\citenamefont{Costa~Filho}
  \emph{et~al.}(1999)\citenamefont{Costa~Filho, Almeida, Andrade, and
  Moreira}}]{costafilho99}
\bibinfo{author}{\bibnamefont{Costa~Filho}, \bibfnamefont{R.}},
  \bibinfo{author}{\bibfnamefont{M.}~\bibnamefont{Almeida}},
  \bibinfo{author}{\bibfnamefont{J.}~\bibnamefont{Andrade}}, and
  \bibinfo{author}{\bibfnamefont{J.}~\bibnamefont{Moreira}},
  \bibinfo{year}{1999}, \bibinfo{journal}{Phys. Rev. E}
  \textbf{\bibinfo{volume}{60}}(\bibinfo{number}{1}), \bibinfo{pages}{1067}.

\bibitem[{\citenamefont{Cox}(1989)}]{cox89}
\bibinfo{author}{\bibnamefont{Cox}, \bibfnamefont{J.}}, \bibinfo{year}{1989},
  \bibinfo{journal}{Ann. Probab.}
  \textbf{\bibinfo{volume}{17}}(\bibinfo{number}{4}), \bibinfo{pages}{1333}.

\bibitem[{\citenamefont{Cox and Griffeath}(1986)}]{cox86}
\bibinfo{author}{\bibnamefont{Cox}, \bibfnamefont{J.}}, and
  \bibinfo{author}{\bibfnamefont{D.}~\bibnamefont{Griffeath}},
  \bibinfo{year}{1986}, \bibinfo{journal}{Ann. Probab.}
  \textbf{\bibinfo{volume}{14}}(\bibinfo{number}{2}), \bibinfo{pages}{347}.

\bibitem[{\citenamefont{Croft}(2000)}]{croft00}
\bibinfo{author}{\bibnamefont{Croft}, \bibfnamefont{W.}}, \bibinfo{year}{2000},
  \emph{\bibinfo{title}{Explaining Language Change: An Evolutionary Approach}}
  (\bibinfo{publisher}{Longman}).

\bibitem[{\citenamefont{Czir\'ok} \emph{et~al.}(1999)\citenamefont{Czir\'ok,
  Barab\'asi, and Vicsek}}]{czirok99c}
\bibinfo{author}{\bibnamefont{Czir\'ok}, \bibfnamefont{A.}},
  \bibinfo{author}{\bibfnamefont{A.-L.} \bibnamefont{Barab\'asi}}, and
  \bibinfo{author}{\bibfnamefont{T.}~\bibnamefont{Vicsek}},
  \bibinfo{year}{1999}, \bibinfo{journal}{Phys. Rev. Lett.}
  \textbf{\bibinfo{volume}{82}}(\bibinfo{number}{1}), \bibinfo{pages}{209}.

\bibitem[{\citenamefont{{Czir{\'o}k}}
  \emph{et~al.}(1997)\citenamefont{{Czir{\'o}k}, {Stanley}, and
  {Vicsek}}}]{czirok97}
\bibinfo{author}{\bibnamefont{{Czir{\'o}k}}, \bibfnamefont{A.}},
  \bibinfo{author}{\bibfnamefont{H.~E.} \bibnamefont{{Stanley}}}, and
  \bibinfo{author}{\bibfnamefont{T.}~\bibnamefont{{Vicsek}}},
  \bibinfo{year}{1997}, \bibinfo{journal}{J. Phys. A}
  \textbf{\bibinfo{volume}{30}}, \bibinfo{pages}{1375}.

\bibitem[{\citenamefont{{Czir\'ok}}
  \emph{et~al.}(1999)\citenamefont{{Czir\'ok}, {Vicsek}, and
  {Vicsek}}}]{czirok99}
\bibinfo{author}{\bibnamefont{{Czir\'ok}}, \bibfnamefont{A.}},
  \bibinfo{author}{\bibfnamefont{M.}~\bibnamefont{{Vicsek}}}, and
  \bibinfo{author}{\bibfnamefont{T.}~\bibnamefont{{Vicsek}}},
  \bibinfo{year}{1999}, \bibinfo{journal}{Physica A}
  \textbf{\bibinfo{volume}{264}}, \bibinfo{pages}{299}.

\bibitem[{\citenamefont{{Czir{\'o}k} and {Vicsek}}(1999)}]{czirok99a}
\bibinfo{author}{\bibnamefont{{Czir{\'o}k}}, \bibfnamefont{A.}}, and
  \bibinfo{author}{\bibfnamefont{T.}~\bibnamefont{{Vicsek}}},
  \bibinfo{year}{1999}, in \emph{\bibinfo{booktitle}{Lecture Notes in Physics,
  Berlin Springer Verlag}}, edited by
  \bibinfo{editor}{\bibfnamefont{D.}~\bibnamefont{{Reguera}}},
  \bibinfo{editor}{\bibfnamefont{J.~M.~G.} \bibnamefont{{Vilar}}}, and
  \bibinfo{editor}{\bibfnamefont{J.~M.} \bibnamefont{{Rub{\'{\i}}}}}, volume
  \bibinfo{volume}{527}, pp. \bibinfo{pages}{152--+}.

\bibitem[{\citenamefont{{Czir\'ok} and {Vicsek}}(2000)}]{czirok00}
\bibinfo{author}{\bibnamefont{{Czir\'ok}}, \bibfnamefont{A.}}, and
  \bibinfo{author}{\bibfnamefont{T.}~\bibnamefont{{Vicsek}}},
  \bibinfo{year}{2000}, \bibinfo{journal}{Physica A}
  \textbf{\bibinfo{volume}{281}}, \bibinfo{pages}{17}.

\bibitem[{\citenamefont{{da Fontoura Costa}}(2005)}]{costa05}
\bibinfo{author}{\bibnamefont{{da Fontoura Costa}}, \bibfnamefont{L.}},
  \bibinfo{year}{2005}, \bibinfo{journal}{Int. J. Mod. Phys. C}
  \textbf{\bibinfo{volume}{16}}, \bibinfo{pages}{1001}.

\bibitem[{\citenamefont{{Daley} and {Kendall}}(1964)}]{daley64}
\bibinfo{author}{\bibnamefont{{Daley}}, \bibfnamefont{D.~J.}}, and
  \bibinfo{author}{\bibfnamefont{D.~G.} \bibnamefont{{Kendall}}},
  \bibinfo{year}{1964}, \bibinfo{journal}{Nature}
  \textbf{\bibinfo{volume}{204}}, \bibinfo{pages}{1118}.

\bibitem[{\citenamefont{{Dall'Asta} and {Baronchelli}}(2006)}]{dallasta06c}
\bibinfo{author}{\bibnamefont{{Dall'Asta}}, \bibfnamefont{L.}}, and
  \bibinfo{author}{\bibfnamefont{A.}~\bibnamefont{{Baronchelli}}},
  \bibinfo{year}{2006}, \bibinfo{journal}{J. Phys. A}
  \textbf{\bibinfo{volume}{39}}, \bibinfo{pages}{14851}.

\bibitem[{\citenamefont{Dall'Asta}
  \emph{et~al.}(2006{\natexlab{a}})\citenamefont{Dall'Asta, Baronchelli,
  Barrat, and Loreto}}]{dallasta06}
\bibinfo{author}{\bibnamefont{Dall'Asta}, \bibfnamefont{L.}},
  \bibinfo{author}{\bibfnamefont{A.}~\bibnamefont{Baronchelli}},
  \bibinfo{author}{\bibfnamefont{A.}~\bibnamefont{Barrat}}, and
  \bibinfo{author}{\bibfnamefont{V.}~\bibnamefont{Loreto}},
  \bibinfo{year}{2006}{\natexlab{a}}, \bibinfo{journal}{Europhys. Lett.}
  \textbf{\bibinfo{volume}{73}}(\bibinfo{number}{6}), \bibinfo{pages}{969}.

\bibitem[{\citenamefont{Dall'Asta}
  \emph{et~al.}(2006{\natexlab{b}})\citenamefont{Dall'Asta, Baronchelli,
  Barrat, and Loreto}}]{dallasta06b}
\bibinfo{author}{\bibnamefont{Dall'Asta}, \bibfnamefont{L.}},
  \bibinfo{author}{\bibfnamefont{A.}~\bibnamefont{Baronchelli}},
  \bibinfo{author}{\bibfnamefont{A.}~\bibnamefont{Barrat}}, and
  \bibinfo{author}{\bibfnamefont{V.}~\bibnamefont{Loreto}},
  \bibinfo{year}{2006}{\natexlab{b}}, \bibinfo{journal}{Phys. Rev. E}
  \textbf{\bibinfo{volume}{74}}(\bibinfo{number}{3}), \bibinfo{pages}{036105}.

\bibitem[{\citenamefont{Dall'Asta and Castellano}(2007)}]{dallasta07}
\bibinfo{author}{\bibnamefont{Dall'Asta}, \bibfnamefont{L.}}, and
  \bibinfo{author}{\bibfnamefont{C.}~\bibnamefont{Castellano}},
  \bibinfo{year}{2007}, \bibinfo{journal}{Europhys. Lett.}
  \textbf{\bibinfo{volume}{77}}(\bibinfo{number}{6}), \bibinfo{pages}{60005
  (6pp)}.

\bibitem[{\citenamefont{{de la Lama}} \emph{et~al.}(2005)\citenamefont{{de la
  Lama}, {L{\'o}pez}, and {Wio}}}]{delalama05}
\bibinfo{author}{\bibnamefont{{de la Lama}}, \bibfnamefont{M.~S.}},
  \bibinfo{author}{\bibfnamefont{J.~M.} \bibnamefont{{L{\'o}pez}}}, and
  \bibinfo{author}{\bibfnamefont{H.~S.} \bibnamefont{{Wio}}},
  \bibinfo{year}{2005}, \bibinfo{journal}{Europhys. Lett.}
  \textbf{\bibinfo{volume}{72}}, \bibinfo{pages}{851}.

\bibitem[{\citenamefont{{de la Lama}} \emph{et~al.}(2006)\citenamefont{{de la
  Lama}, {Szendro}, {Iglesias}, and {Wio}}}]{delalama06}
\bibinfo{author}{\bibnamefont{{de la Lama}}, \bibfnamefont{M.~S.}},
  \bibinfo{author}{\bibfnamefont{I.~G.} \bibnamefont{{Szendro}}},
  \bibinfo{author}{\bibfnamefont{J.~R.} \bibnamefont{{Iglesias}}}, and
  \bibinfo{author}{\bibfnamefont{H.~S.} \bibnamefont{{Wio}}},
  \bibinfo{year}{2006}, \bibinfo{journal}{Eur. Phys. J. B}
  \textbf{\bibinfo{volume}{51}}, \bibinfo{pages}{435}.

\bibitem[{\citenamefont{{De Sanctis} and Galla}(2007)}]{desanctis07}
\bibinfo{author}{\bibnamefont{{De Sanctis}}, \bibfnamefont{L.}}, and
  \bibinfo{author}{\bibfnamefont{T.}~\bibnamefont{Galla}},
  \bibinfo{year}{2007}, \eprint{arxiv:0707.3428}.

\bibitem[{\citenamefont{{de Saussure}}(1916)}]{saussure1916}
\bibinfo{author}{\bibnamefont{{de Saussure}}, \bibfnamefont{F.}},
  \bibinfo{year}{1916}, \emph{\bibinfo{title}{Cours de linguistique
  g\'en\'erale}} (\bibinfo{publisher}{Bayot}, \bibinfo{address}{Paris}).

\bibitem[{\citenamefont{Deffuant}(2006)}]{deffuant06}
\bibinfo{author}{\bibnamefont{Deffuant}, \bibfnamefont{G.}},
  \bibinfo{year}{2006}, \bibinfo{journal}{JASSS}
  \textbf{\bibinfo{volume}{9}}(\bibinfo{number}{3}).

\bibitem[{\citenamefont{{Deffuant}}
  \emph{et~al.}(2004)\citenamefont{{Deffuant}, {Amblard}, and
  {Weisbuch}}}]{deffuant04}
\bibinfo{author}{\bibnamefont{{Deffuant}}, \bibfnamefont{G.}},
  \bibinfo{author}{\bibfnamefont{F.}~\bibnamefont{{Amblard}}}, and
  \bibinfo{author}{\bibfnamefont{G.}~\bibnamefont{{Weisbuch}}},
  \bibinfo{year}{2004}, \eprint{arxiv:cond-mat/0410199}.

\bibitem[{\citenamefont{Deffuant} \emph{et~al.}(2002)\citenamefont{Deffuant,
  Amblard, Weisbuch, and Faure}}]{deffuant02}
\bibinfo{author}{\bibnamefont{Deffuant}, \bibfnamefont{G.}},
  \bibinfo{author}{\bibfnamefont{F.}~\bibnamefont{Amblard}},
  \bibinfo{author}{\bibfnamefont{G.}~\bibnamefont{Weisbuch}}, and
  \bibinfo{author}{\bibfnamefont{T.}~\bibnamefont{Faure}},
  \bibinfo{year}{2002}, \bibinfo{journal}{JASSS}
  \textbf{\bibinfo{volume}{5}}(\bibinfo{number}{4}).

\bibitem[{\citenamefont{Deffuant} \emph{et~al.}(2000)\citenamefont{Deffuant,
  Neau, Amblard, and Weisbuch}}]{deffuant00}
\bibinfo{author}{\bibnamefont{Deffuant}, \bibfnamefont{G.}},
  \bibinfo{author}{\bibfnamefont{D.}~\bibnamefont{Neau}},
  \bibinfo{author}{\bibfnamefont{F.}~\bibnamefont{Amblard}}, and
  \bibinfo{author}{\bibfnamefont{G.}~\bibnamefont{Weisbuch}},
  \bibinfo{year}{2000}, \bibinfo{journal}{Adv. Compl. Sys.}
  \textbf{\bibinfo{volume}{3}}(\bibinfo{number}{1-4}), \bibinfo{pages}{87}.

\bibitem[{\citenamefont{Dezs\"{o}} \emph{et~al.}(2006)\citenamefont{Dezs\"{o},
  Almaas, Luk\'{a}cs, R\'{a}cz, Szakad\'{a}t, and Barab\'{a}si}}]{dezso06}
\bibinfo{author}{\bibnamefont{Dezs\"{o}}, \bibfnamefont{Z.}},
  \bibinfo{author}{\bibfnamefont{E.}~\bibnamefont{Almaas}},
  \bibinfo{author}{\bibfnamefont{A.}~\bibnamefont{Luk\'{a}cs}},
  \bibinfo{author}{\bibfnamefont{B.}~\bibnamefont{R\'{a}cz}},
  \bibinfo{author}{\bibfnamefont{I.}~\bibnamefont{Szakad\'{a}t}}, and
  \bibinfo{author}{\bibfnamefont{A.-L.} \bibnamefont{Barab\'{a}si}},
  \bibinfo{year}{2006}, \bibinfo{journal}{Phys. Rev. E}
  \textbf{\bibinfo{volume}{73}}(\bibinfo{number}{6}), \bibinfo{pages}{066132}.

\bibitem[{\citenamefont{{Di Mare} and {Latora}}(2007)}]{dimare06}
\bibinfo{author}{\bibnamefont{{Di Mare}}, \bibfnamefont{A.}}, and
  \bibinfo{author}{\bibfnamefont{V.}~\bibnamefont{{Latora}}},
  \bibinfo{year}{2007}, \bibinfo{journal}{Int. J. Mod. Phys. C}
  \textbf{\bibinfo{volume}{18}}, \bibinfo{pages}{1377}.

\bibitem[{\citenamefont{Dickman and Tretyakov}(1995)}]{dickman95}
\bibinfo{author}{\bibnamefont{Dickman}, \bibfnamefont{R.}}, and
  \bibinfo{author}{\bibfnamefont{A.}~\bibnamefont{Tretyakov}},
  \bibinfo{year}{1995}, \bibinfo{journal}{Phys. Rev. E}
  \textbf{\bibinfo{volume}{52}}(\bibinfo{number}{3}), \bibinfo{pages}{3218}.

\bibitem[{\citenamefont{Dodds and Watts}(2004)}]{dodds04}
\bibinfo{author}{\bibnamefont{Dodds}, \bibfnamefont{P.~S.}}, and
  \bibinfo{author}{\bibfnamefont{D.~J.} \bibnamefont{Watts}},
  \bibinfo{year}{2004}, \bibinfo{journal}{Phys. Rev. Lett.}
  \textbf{\bibinfo{volume}{92}}(\bibinfo{number}{21}), \bibinfo{pages}{218701}.

\bibitem[{\citenamefont{Dornic}(1998)}]{dornicthese}
\bibinfo{author}{\bibnamefont{Dornic}, \bibfnamefont{I.}},
  \bibinfo{year}{1998}, \bibinfo{title}{Phd thesis, universit\'e de nice,
  sophia-antipolis}.

\bibitem[{\citenamefont{Dornic} \emph{et~al.}(2001)\citenamefont{Dornic,
  Chat\'e, Chave, and Hinrichsen}}]{dornic01}
\bibinfo{author}{\bibnamefont{Dornic}, \bibfnamefont{I.}},
  \bibinfo{author}{\bibfnamefont{H.}~\bibnamefont{Chat\'e}},
  \bibinfo{author}{\bibfnamefont{J.}~\bibnamefont{Chave}}, and
  \bibinfo{author}{\bibfnamefont{H.}~\bibnamefont{Hinrichsen}},
  \bibinfo{year}{2001}, \bibinfo{journal}{Phys. Rev. Lett.}
  \textbf{\bibinfo{volume}{87}}(\bibinfo{number}{4}), \bibinfo{pages}{045701}.

\bibitem[{\citenamefont{Dorogovtsev and Mendes}(2002)}]{dorogovtsev02}
\bibinfo{author}{\bibnamefont{Dorogovtsev}, \bibfnamefont{S.}}, and
  \bibinfo{author}{\bibfnamefont{J.}~\bibnamefont{Mendes}},
  \bibinfo{year}{2002}, \bibinfo{journal}{Adv. Phys.}
  \textbf{\bibinfo{volume}{51}}, \bibinfo{pages}{1079}.

\bibitem[{\citenamefont{Dorogovtsev and Mendes}(2003)}]{dorogovtsev03}
\bibinfo{author}{\bibnamefont{Dorogovtsev}, \bibfnamefont{S.}}, and
  \bibinfo{author}{\bibfnamefont{J.}~\bibnamefont{Mendes}},
  \bibinfo{year}{2003}, \emph{\bibinfo{title}{Evolution of Networks: From
  Biological Nets to the Internet and WWW}} (\bibinfo{publisher}{Oxford
  University Press}, \bibinfo{address}{Oxford, UK}).

\bibitem[{\citenamefont{{Drouffe} and {Godr{\`e}che}}(1999)}]{drouffe99}
\bibinfo{author}{\bibnamefont{{Drouffe}}, \bibfnamefont{J.-M.}}, and
  \bibinfo{author}{\bibfnamefont{C.}~\bibnamefont{{Godr{\`e}che}}},
  \bibinfo{year}{1999}, \bibinfo{journal}{J. Phys. A}
  \textbf{\bibinfo{volume}{32}}, \bibinfo{pages}{249}.

\bibitem[{\citenamefont{Droz} \emph{et~al.}(2003)\citenamefont{Droz, Ferreira,
  and Lipowski}}]{droz03}
\bibinfo{author}{\bibnamefont{Droz}, \bibfnamefont{M.}},
  \bibinfo{author}{\bibfnamefont{A.}~\bibnamefont{Ferreira}}, and
  \bibinfo{author}{\bibfnamefont{A.}~\bibnamefont{Lipowski}},
  \bibinfo{year}{2003}, \bibinfo{journal}{Phys. Rev. E}
  \textbf{\bibinfo{volume}{67}}(\bibinfo{number}{5}), \bibinfo{pages}{056108}.

\bibitem[{\citenamefont{Dugatkin} \emph{et~al.}(1994)\citenamefont{Dugatkin,
  Alfieri, and Moore}}]{dugatkin94}
\bibinfo{author}{\bibnamefont{Dugatkin}, \bibfnamefont{L.~A.}},
  \bibinfo{author}{\bibfnamefont{M.~S.} \bibnamefont{Alfieri}}, and
  \bibinfo{author}{\bibfnamefont{A.~J.} \bibnamefont{Moore}},
  \bibinfo{year}{1994}, \bibinfo{journal}{Ethology}
  \textbf{\bibinfo{volume}{97}}, \bibinfo{pages}{94}.

\bibitem[{\citenamefont{Ebel} \emph{et~al.}(2002)\citenamefont{Ebel, Mielsch,
  and Bornholdt}}]{ebel02}
\bibinfo{author}{\bibnamefont{Ebel}, \bibfnamefont{H.}},
  \bibinfo{author}{\bibfnamefont{L.-I.} \bibnamefont{Mielsch}}, and
  \bibinfo{author}{\bibfnamefont{S.}~\bibnamefont{Bornholdt}},
  \bibinfo{year}{2002}, \bibinfo{journal}{Phys. Rev. E}
  \textbf{\bibinfo{volume}{66}}(\bibinfo{number}{3}), \bibinfo{pages}{035103}.

\bibitem[{\citenamefont{Eckmann} \emph{et~al.}(2004)\citenamefont{Eckmann,
  Moses, and Sergi}}]{eckmann04}
\bibinfo{author}{\bibnamefont{Eckmann}, \bibfnamefont{J.}},
  \bibinfo{author}{\bibfnamefont{E.}~\bibnamefont{Moses}}, and
  \bibinfo{author}{\bibfnamefont{D.}~\bibnamefont{Sergi}},
  \bibinfo{year}{2004}, \bibinfo{journal}{Proc. Natl. Acad. Sci. USA}
  \textbf{\bibinfo{volume}{101}}(\bibinfo{number}{40}), \bibinfo{pages}{14333}.

\bibitem[{\citenamefont{Egu\'iluz and Zimmermann}(2000)}]{eguiluz00}
\bibinfo{author}{\bibnamefont{Egu\'iluz}, \bibfnamefont{V.~M.}}, and
  \bibinfo{author}{\bibfnamefont{M.~G.} \bibnamefont{Zimmermann}},
  \bibinfo{year}{2000}, \bibinfo{journal}{Phys. Rev. Lett.}
  \textbf{\bibinfo{volume}{85}}(\bibinfo{number}{26}), \bibinfo{pages}{5659}.

\bibitem[{\citenamefont{Ehrhardt} \emph{et~al.}(2006)\citenamefont{Ehrhardt,
  Marsili, and Vega-Redondo}}]{ehrhardt06}
\bibinfo{author}{\bibnamefont{Ehrhardt}, \bibfnamefont{G.~C. M.~A.}},
  \bibinfo{author}{\bibfnamefont{M.}~\bibnamefont{Marsili}}, and
  \bibinfo{author}{\bibfnamefont{F.}~\bibnamefont{Vega-Redondo}},
  \bibinfo{year}{2006}, \bibinfo{journal}{Phys. Rev. E}
  \textbf{\bibinfo{volume}{74}}(\bibinfo{number}{3}), \bibinfo{pages}{036106}.

\bibitem[{\citenamefont{Eigen}(1971)}]{eigen71}
\bibinfo{author}{\bibnamefont{Eigen}, \bibfnamefont{M.}}, \bibinfo{year}{1971},
  \bibinfo{journal}{Naturwissenschaften} \textbf{\bibinfo{volume}{58}},
  \bibinfo{pages}{465}.

\bibitem[{\citenamefont{Eigen and Schuster}(1979)}]{eigen79}
\bibinfo{author}{\bibnamefont{Eigen}, \bibfnamefont{M.}}, and
  \bibinfo{author}{\bibfnamefont{P.}~\bibnamefont{Schuster}},
  \bibinfo{year}{1979}, \emph{\bibinfo{title}{The Hypercycle: A Principle of
  Natural Self-Organization}} (\bibinfo{publisher}{Springer-Verlag},
  \bibinfo{address}{Berlin, Germany}).

\bibitem[{\citenamefont{Elgazzar}(2003)}]{elgazzar03}
\bibinfo{author}{\bibnamefont{Elgazzar}, \bibfnamefont{A.~S.}},
  \bibinfo{year}{2003}, \bibinfo{journal}{Physica A}
  \textbf{\bibinfo{volume}{324}}, \bibinfo{pages}{402}.

\bibitem[{\citenamefont{Epstein and Axtell}(1996)}]{epstein96}
\bibinfo{author}{\bibnamefont{Epstein}, \bibfnamefont{J.}}, and
  \bibinfo{author}{\bibfnamefont{R.}~\bibnamefont{Axtell}},
  \bibinfo{year}{1996}, \emph{\bibinfo{title}{Growing artificial societies:
  social science from the bottom up}} (\bibinfo{publisher}{The Brookings
  Institution}, \bibinfo{address}{Washington, DC, USA}).

\bibitem[{\citenamefont{Erd\"os and R\'enyi}(1959)}]{erdos59}
\bibinfo{author}{\bibnamefont{Erd\"os}, \bibfnamefont{P.}}, and
  \bibinfo{author}{\bibfnamefont{A.}~\bibnamefont{R\'enyi}},
  \bibinfo{year}{1959}, \bibinfo{journal}{Publ. Math. Debrecen}
  \textbf{\bibinfo{volume}{6}}, \bibinfo{pages}{290}.

\bibitem[{\citenamefont{Erd\"os and R\'enyi}(1960)}]{erdos60}
\bibinfo{author}{\bibnamefont{Erd\"os}, \bibfnamefont{P.}}, and
  \bibinfo{author}{\bibfnamefont{A.}~\bibnamefont{R\'enyi}},
  \bibinfo{year}{1960}, \bibinfo{journal}{Publ. Math. Inst. Hung. Acad. Sci.}
  \textbf{\bibinfo{volume}{7}}, \bibinfo{pages}{17}.

\bibitem[{\citenamefont{Evans and Ray}(1993)}]{evans93}
\bibinfo{author}{\bibnamefont{Evans}, \bibfnamefont{J.~W.}}, and
  \bibinfo{author}{\bibfnamefont{T.~R.} \bibnamefont{Ray}},
  \bibinfo{year}{1993}, \bibinfo{journal}{Phys. Rev. E}
  \textbf{\bibinfo{volume}{47}}(\bibinfo{number}{2}), \bibinfo{pages}{1018}.

\bibitem[{\citenamefont{{Farkas}} \emph{et~al.}(2002)\citenamefont{{Farkas},
  {Helbing}, and {Vicsek}}}]{farkas02}
\bibinfo{author}{\bibnamefont{{Farkas}}, \bibfnamefont{I.}},
  \bibinfo{author}{\bibfnamefont{D.}~\bibnamefont{{Helbing}}}, and
  \bibinfo{author}{\bibfnamefont{T.}~\bibnamefont{{Vicsek}}},
  \bibinfo{year}{2002}, \bibinfo{journal}{Nature}
  \textbf{\bibinfo{volume}{419}}, \bibinfo{pages}{131}.

\bibitem[{\citenamefont{Farkas and Vicsek}(2006)}]{farkas06}
\bibinfo{author}{\bibnamefont{Farkas}, \bibfnamefont{I.~J.}}, and
  \bibinfo{author}{\bibfnamefont{T.}~\bibnamefont{Vicsek}},
  \bibinfo{year}{2006}, \bibinfo{journal}{Physica A}
  \textbf{\bibinfo{volume}{369}}, \bibinfo{pages}{830}.

\bibitem[{\citenamefont{Festinger} \emph{et~al.}(1950)\citenamefont{Festinger,
  Schachter, and Back}}]{festinger50}
\bibinfo{author}{\bibnamefont{Festinger}, \bibfnamefont{L.}},
  \bibinfo{author}{\bibfnamefont{S.}~\bibnamefont{Schachter}}, and
  \bibinfo{author}{\bibfnamefont{K.}~\bibnamefont{Back}}, \bibinfo{year}{1950},
  \emph{\bibinfo{title}{Social Pressures in Informal Groups: A Study of Human
  Factors in Housing}} (\bibinfo{publisher}{Harper}, \bibinfo{address}{New
  York, NY, USA}).

\bibitem[{\citenamefont{Filho} \emph{et~al.}(2003)\citenamefont{Filho, Almeida,
  Moreira, and Andrade}}]{costafilho03}
\bibinfo{author}{\bibnamefont{Filho}, \bibfnamefont{R.~C.}},
  \bibinfo{author}{\bibfnamefont{M.}~\bibnamefont{Almeida}},
  \bibinfo{author}{\bibfnamefont{J.}~\bibnamefont{Moreira}}, and
  \bibinfo{author}{\bibfnamefont{J.}~\bibnamefont{Andrade}},
  \bibinfo{year}{2003}, \bibinfo{journal}{Physica A}
  \textbf{\bibinfo{volume}{322}}(\bibinfo{number}{1}), \bibinfo{pages}{698}.

\bibitem[{\citenamefont{{Flache} and {Macy}}(2006)}]{flache06}
\bibinfo{author}{\bibnamefont{{Flache}}, \bibfnamefont{A.}}, and
  \bibinfo{author}{\bibfnamefont{M.~W.} \bibnamefont{{Macy}}},
  \bibinfo{year}{2006}, \eprint{arxiv:physics/0604196}.

\bibitem[{\citenamefont{{Flache} and {Macy}}(2007)}]{flache07}
\bibinfo{author}{\bibnamefont{{Flache}}, \bibfnamefont{A.}}, and
  \bibinfo{author}{\bibfnamefont{M.~W.} \bibnamefont{{Macy}}},
  \bibinfo{year}{2007}, \eprint{arxiv:physics/0701333}.

\bibitem[{\citenamefont{Forster} \emph{et~al.}(1977)\citenamefont{Forster,
  Nelson, and Stephen}}]{forster77}
\bibinfo{author}{\bibnamefont{Forster}, \bibfnamefont{D.}},
  \bibinfo{author}{\bibfnamefont{D.~R.} \bibnamefont{Nelson}}, and
  \bibinfo{author}{\bibfnamefont{M.~J.} \bibnamefont{Stephen}},
  \bibinfo{year}{1977}, \bibinfo{journal}{Phys. Rev. A}
  \textbf{\bibinfo{volume}{16}}(\bibinfo{number}{2}), \bibinfo{pages}{732}.

\bibitem[{\citenamefont{{Fortunato}}(2004)}]{fortunato05a}
\bibinfo{author}{\bibnamefont{{Fortunato}}, \bibfnamefont{S.}},
  \bibinfo{year}{2004}, \bibinfo{journal}{Int. J. Mod. Phys. C}
  \textbf{\bibinfo{volume}{15}}, \bibinfo{pages}{1301}.

\bibitem[{\citenamefont{Fortunato}(2005)}]{fortunato05}
\bibinfo{author}{\bibnamefont{Fortunato}, \bibfnamefont{S.}},
  \bibinfo{year}{2005}, \bibinfo{journal}{Physica A}
  \textbf{\bibinfo{volume}{348}}, \bibinfo{pages}{683}.

\bibitem[{\citenamefont{{Fortunato}}(2005{\natexlab{a}})}]{fortunato05c}
\bibinfo{author}{\bibnamefont{{Fortunato}}, \bibfnamefont{S.}},
  \bibinfo{year}{2005}{\natexlab{a}}, \bibinfo{journal}{Int. J. Mod. Phys. C}
  \textbf{\bibinfo{volume}{16}}, \bibinfo{pages}{259}.

\bibitem[{\citenamefont{{Fortunato}}(2005{\natexlab{b}})}]{fortunato05b}
\bibinfo{author}{\bibnamefont{{Fortunato}}, \bibfnamefont{S.}},
  \bibinfo{year}{2005}{\natexlab{b}}, \bibinfo{journal}{Int. J. Mod. Phys. C}
  \textbf{\bibinfo{volume}{16}}, \bibinfo{pages}{17}.

\bibitem[{\citenamefont{{Fortunato} and {Castellano}}(2007)}]{fortunato07}
\bibinfo{author}{\bibnamefont{{Fortunato}}, \bibfnamefont{S.}}, and
  \bibinfo{author}{\bibfnamefont{C.}~\bibnamefont{{Castellano}}},
  \bibinfo{year}{2007}, \bibinfo{journal}{Phys. Rev. Lett.}
  \textbf{\bibinfo{volume}{99}}, \bibinfo{pages}{138701}.

\bibitem[{\citenamefont{{Fortunato}}
  \emph{et~al.}(2005)\citenamefont{{Fortunato}, {Latora}, {Pluchino}, and
  {Rapisarda}}}]{fortunato05d}
\bibinfo{author}{\bibnamefont{{Fortunato}}, \bibfnamefont{S.}},
  \bibinfo{author}{\bibfnamefont{V.}~\bibnamefont{{Latora}}},
  \bibinfo{author}{\bibfnamefont{A.}~\bibnamefont{{Pluchino}}}, and
  \bibinfo{author}{\bibfnamefont{A.}~\bibnamefont{{Rapisarda}}},
  \bibinfo{year}{2005}, \bibinfo{journal}{Int. J. Mod. Phys. C}
  \textbf{\bibinfo{volume}{16}}, \bibinfo{pages}{1535}.

\bibitem[{\citenamefont{Frachebourg and Krapivsky}(1996)}]{frachebourg96}
\bibinfo{author}{\bibnamefont{Frachebourg}, \bibfnamefont{L.}}, and
  \bibinfo{author}{\bibfnamefont{P.~L.} \bibnamefont{Krapivsky}},
  \bibinfo{year}{1996}, \bibinfo{journal}{Phys. Rev. E}
  \textbf{\bibinfo{volume}{53}}(\bibinfo{number}{4}), \bibinfo{pages}{R3009}.

\bibitem[{\citenamefont{Francis}(1988)}]{francis88}
\bibinfo{author}{\bibnamefont{Francis}, \bibfnamefont{R.~C.}},
  \bibinfo{year}{1988}, \bibinfo{journal}{Ethology}
  \textbf{\bibinfo{volume}{78}}, \bibinfo{pages}{223}.

\bibitem[{\citenamefont{Freeman}(2004)}]{freeman04}
\bibinfo{author}{\bibnamefont{Freeman}, \bibfnamefont{L.}},
  \bibinfo{year}{2004}, \emph{\bibinfo{title}{The Development of Social Network
  Analysis: A Study in the Sociology of Science}}
  (\bibinfo{publisher}{BookSurge Publishing}, \bibinfo{address}{Vancouver,
  Canada}).

\bibitem[{\citenamefont{{Friedman} and {Friedman}}(1984)}]{friedman84}
\bibinfo{author}{\bibnamefont{{Friedman}}, \bibfnamefont{R.~D.}}, and
  \bibinfo{author}{\bibfnamefont{M.}~\bibnamefont{{Friedman}}},
  \bibinfo{year}{1984}, \emph{\bibinfo{title}{The Tyranny of the Status Quo}}
  (\bibinfo{publisher}{Harcourt Brace Company}, \bibinfo{address}{Orlando, FL,
  USA}).

\bibitem[{\citenamefont{Fukui and Ishibashi}(1999)}]{fukui99}
\bibinfo{author}{\bibnamefont{Fukui}, \bibfnamefont{M.}}, and
  \bibinfo{author}{\bibfnamefont{Y.}~\bibnamefont{Ishibashi}},
  \bibinfo{year}{1999}, \bibinfo{journal}{J. Phys. Soc. Jpn.}
  \textbf{\bibinfo{volume}{68}}, \bibinfo{pages}{2861}.

\bibitem[{\citenamefont{Gabrielli and Caldarelli}(2007)}]{gabrielli07}
\bibinfo{author}{\bibnamefont{Gabrielli}, \bibfnamefont{A.}}, and
  \bibinfo{author}{\bibfnamefont{G.}~\bibnamefont{Caldarelli}},
  \bibinfo{year}{2007}, \bibinfo{journal}{Phys. Rev. Lett.}
  \textbf{\bibinfo{volume}{98}}(\bibinfo{number}{20}), \bibinfo{pages}{208701}.

\bibitem[{\citenamefont{{Galam}}(1986)}]{galam86}
\bibinfo{author}{\bibnamefont{{Galam}}, \bibfnamefont{S.}},
  \bibinfo{year}{1986}, \bibinfo{journal}{J. Math. Psychol.}
  \textbf{\bibinfo{volume}{30}}, \bibinfo{pages}{426}.

\bibitem[{\citenamefont{{Galam}}(1990)}]{galam90}
\bibinfo{author}{\bibnamefont{{Galam}}, \bibfnamefont{S.}},
  \bibinfo{year}{1990}, \bibinfo{journal}{J. Stat. Phys.}
  \textbf{\bibinfo{volume}{61}}, \bibinfo{pages}{943}.

\bibitem[{\citenamefont{{Galam}}(1999)}]{galam99}
\bibinfo{author}{\bibnamefont{{Galam}}, \bibfnamefont{S.}},
  \bibinfo{year}{1999}, \bibinfo{journal}{Physica A}
  \textbf{\bibinfo{volume}{274}}, \bibinfo{pages}{132}.

\bibitem[{\citenamefont{{Galam}}(2000)}]{galam00}
\bibinfo{author}{\bibnamefont{{Galam}}, \bibfnamefont{S.}},
  \bibinfo{year}{2000}, \bibinfo{journal}{Physica A}
  \textbf{\bibinfo{volume}{285}}, \bibinfo{pages}{66}.

\bibitem[{\citenamefont{{Galam}}(2002)}]{galam02}
\bibinfo{author}{\bibnamefont{{Galam}}, \bibfnamefont{S.}},
  \bibinfo{year}{2002}, \bibinfo{journal}{Eur. Phys. J. B}
  \textbf{\bibinfo{volume}{25}}(\bibinfo{number}{4}), \bibinfo{pages}{403}.

\bibitem[{\citenamefont{{Galam}}(2004)}]{galam04}
\bibinfo{author}{\bibnamefont{{Galam}}, \bibfnamefont{S.}},
  \bibinfo{year}{2004}, \bibinfo{journal}{Physica A}
  \textbf{\bibinfo{volume}{333}}, \bibinfo{pages}{453}.

\bibitem[{\citenamefont{{Galam}}(2005{\natexlab{a}})}]{galam05}
\bibinfo{author}{\bibnamefont{{Galam}}, \bibfnamefont{S.}},
  \bibinfo{year}{2005}{\natexlab{a}}, \bibinfo{journal}{Phys. Rev. E}
  \textbf{\bibinfo{volume}{71}}(\bibinfo{number}{4}), \bibinfo{pages}{046123}.

\bibitem[{\citenamefont{{Galam}}(2005{\natexlab{b}})}]{galam05b}
\bibinfo{author}{\bibnamefont{{Galam}}, \bibfnamefont{S.}},
  \bibinfo{year}{2005}{\natexlab{b}}, \bibinfo{journal}{Europhys. Lett.}
  \textbf{\bibinfo{volume}{70}}, \bibinfo{pages}{705}.

\bibitem[{\citenamefont{{Galam}} \emph{et~al.}(2002)\citenamefont{{Galam},
  {Chopard}, and {Droz}}}]{galam02a}
\bibinfo{author}{\bibnamefont{{Galam}}, \bibfnamefont{S.}},
  \bibinfo{author}{\bibfnamefont{B.}~\bibnamefont{{Chopard}}}, and
  \bibinfo{author}{\bibfnamefont{M.}~\bibnamefont{{Droz}}},
  \bibinfo{year}{2002}, \bibinfo{journal}{Physica A}
  \textbf{\bibinfo{volume}{314}}, \bibinfo{pages}{256}.

\bibitem[{\citenamefont{{Galam}} \emph{et~al.}(1982)\citenamefont{{Galam},
  {Gefen}, and {Shapir}}}]{galam82}
\bibinfo{author}{\bibnamefont{{Galam}}, \bibfnamefont{S.}},
  \bibinfo{author}{\bibfnamefont{Y.}~\bibnamefont{{Gefen}}}, and
  \bibinfo{author}{\bibfnamefont{Y.}~\bibnamefont{{Shapir}}},
  \bibinfo{year}{1982}, \bibinfo{journal}{J. Math. Sociol.}
  \textbf{\bibinfo{volume}{9}}, \bibinfo{pages}{1}.

\bibitem[{\citenamefont{{Galam} and {Jacobs}}(2007)}]{galam07}
\bibinfo{author}{\bibnamefont{{Galam}}, \bibfnamefont{S.}}, and
  \bibinfo{author}{\bibfnamefont{F.}~\bibnamefont{{Jacobs}}},
  \bibinfo{year}{2007}, \bibinfo{journal}{Physica A}
  \textbf{\bibinfo{volume}{381}}, \bibinfo{pages}{366}.

\bibitem[{\citenamefont{Galam and Moscovici}(1991)}]{galam91}
\bibinfo{author}{\bibnamefont{Galam}, \bibfnamefont{S.}}, and
  \bibinfo{author}{\bibfnamefont{S.}~\bibnamefont{Moscovici}},
  \bibinfo{year}{1991}, \bibinfo{journal}{Eur. J. Soc. Psychol.}
  \textbf{\bibinfo{volume}{21}}, \bibinfo{pages}{49}.

\bibitem[{\citenamefont{{Galea}}(2003)}]{galea03}
\bibinfo{editor}{\bibnamefont{{Galea}}, \bibfnamefont{E.~R.}} (ed.),
  \bibinfo{year}{2003}, \emph{\bibinfo{title}{Pedestrian and Evacuation
  Dynamics}} (\bibinfo{publisher}{CMS Press}, \bibinfo{address}{London, UK}).

\bibitem[{\citenamefont{{Gallos}}(2005)}]{gallos05}
\bibinfo{author}{\bibnamefont{{Gallos}}, \bibfnamefont{L.~K.}},
  \bibinfo{year}{2005}, \bibinfo{journal}{Int. J. Mod. Phys. C}
  \textbf{\bibinfo{volume}{16}}, \bibinfo{pages}{1329}.

\bibitem[{\citenamefont{Galton}(1874)}]{galton1874}
\bibinfo{author}{\bibnamefont{Galton}, \bibfnamefont{F.}},
  \bibinfo{year}{1874}, \emph{\bibinfo{title}{English Men of Science: Their
  Nature and Nurture}} (\bibinfo{publisher}{MacMillan \& Co.},
  \bibinfo{address}{London, UK}).

\bibitem[{\citenamefont{Gammaitoni}
  \emph{et~al.}(1998)\citenamefont{Gammaitoni, H\"anggi, Jung, and
  Marchesoni}}]{gammaitoni98}
\bibinfo{author}{\bibnamefont{Gammaitoni}, \bibfnamefont{L.}},
  \bibinfo{author}{\bibfnamefont{P.}~\bibnamefont{H\"anggi}},
  \bibinfo{author}{\bibfnamefont{P.}~\bibnamefont{Jung}}, and
  \bibinfo{author}{\bibfnamefont{F.}~\bibnamefont{Marchesoni}},
  \bibinfo{year}{1998}, \bibinfo{journal}{Rev. Mod. Phys.}
  \textbf{\bibinfo{volume}{70}}(\bibinfo{number}{1}), \bibinfo{pages}{223}.

\bibitem[{\citenamefont{{Gekle}} \emph{et~al.}(2005)\citenamefont{{Gekle},
  {Peliti}, and {Galam}}}]{gekle05}
\bibinfo{author}{\bibnamefont{{Gekle}}, \bibfnamefont{S.}},
  \bibinfo{author}{\bibfnamefont{L.}~\bibnamefont{{Peliti}}}, and
  \bibinfo{author}{\bibfnamefont{S.}~\bibnamefont{{Galam}}},
  \bibinfo{year}{2005}, \bibinfo{journal}{Eur. Phys. J. B}
  \textbf{\bibinfo{volume}{45}}, \bibinfo{pages}{569}.

\bibitem[{\citenamefont{{Gil} and {Zanette}}(2006)}]{gil06}
\bibinfo{author}{\bibnamefont{{Gil}}, \bibfnamefont{S.}}, and
  \bibinfo{author}{\bibfnamefont{D.~H.} \bibnamefont{{Zanette}}},
  \bibinfo{year}{2006}, \bibinfo{journal}{Phys. Lett. A}
  \textbf{\bibinfo{volume}{356}}, \bibinfo{pages}{89}.

\bibitem[{\citenamefont{Girvan and Newman}(2002)}]{girvan02}
\bibinfo{author}{\bibnamefont{Girvan}, \bibfnamefont{M.}}, and
  \bibinfo{author}{\bibfnamefont{M.~E.} \bibnamefont{Newman}},
  \bibinfo{year}{2002}, \bibinfo{journal}{Proc. Natl. Acad. Sci. USA}
  \textbf{\bibinfo{volume}{99}}(\bibinfo{number}{12}), \bibinfo{pages}{7821}.

\bibitem[{\citenamefont{Glauber}(1963)}]{glauber63}
\bibinfo{author}{\bibnamefont{Glauber}, \bibfnamefont{R.~J.}},
  \bibinfo{year}{1963}, \bibinfo{journal}{J. Math. Phys.}
  \textbf{\bibinfo{volume}{4}}(\bibinfo{number}{2}), \bibinfo{pages}{294}.

\bibitem[{\citenamefont{{Goffman}}(1966)}]{goffman66}
\bibinfo{author}{\bibnamefont{{Goffman}}, \bibfnamefont{W.}},
  \bibinfo{year}{1966}, \bibinfo{journal}{Nature}
  \textbf{\bibinfo{volume}{212}}, \bibinfo{pages}{449}.

\bibitem[{\citenamefont{{Goffman} and Newill}(1964)}]{goffman64}
\bibinfo{author}{\bibnamefont{{Goffman}}, \bibfnamefont{W.}}, and
  \bibinfo{author}{\bibfnamefont{V.~A.} \bibnamefont{Newill}},
  \bibinfo{year}{1964}, \bibinfo{journal}{Nature}
  \textbf{\bibinfo{volume}{204}}, \bibinfo{pages}{225}.

\bibitem[{\citenamefont{Goh} \emph{et~al.}(2001)\citenamefont{Goh, Kahng, and
  Kim}}]{goh01}
\bibinfo{author}{\bibnamefont{Goh}, \bibfnamefont{K.-I.}},
  \bibinfo{author}{\bibfnamefont{B.}~\bibnamefont{Kahng}}, and
  \bibinfo{author}{\bibfnamefont{D.}~\bibnamefont{Kim}}, \bibinfo{year}{2001},
  \bibinfo{journal}{Phys. Rev. Lett.}
  \textbf{\bibinfo{volume}{87}}(\bibinfo{number}{27}), \bibinfo{pages}{278701}.

\bibitem[{\citenamefont{Golder and Huberman}(2006)}]{golder06}
\bibinfo{author}{\bibnamefont{Golder}, \bibfnamefont{S.}}, and
  \bibinfo{author}{\bibfnamefont{B.~A.} \bibnamefont{Huberman}},
  \bibinfo{year}{2006}, \bibinfo{journal}{J. Inf. Sys.}
  \textbf{\bibinfo{volume}{32}}(\bibinfo{number}{2}), \bibinfo{pages}{198}.

\bibitem[{\citenamefont{Gonzalez} \emph{et~al.}(2008)\citenamefont{Gonzalez,
  Hidalgo, and Barabasi}}]{gonzalez08}
\bibinfo{author}{\bibnamefont{Gonzalez}, \bibfnamefont{M.~C.}},
  \bibinfo{author}{\bibfnamefont{C.~A.} \bibnamefont{Hidalgo}}, and
  \bibinfo{author}{\bibfnamefont{A.-L.} \bibnamefont{Barabasi}},
  \bibinfo{year}{2008}, \bibinfo{journal}{Nature}
  \textbf{\bibinfo{volume}{453}}(\bibinfo{number}{7196}), \bibinfo{pages}{779}.

\bibitem[{\citenamefont{{Gonz{\'a}lez}}
  \emph{et~al.}(2004)\citenamefont{{Gonz{\'a}lez}, {Sousa}, and
  {Herrmann}}}]{gonzalez04}
\bibinfo{author}{\bibnamefont{{Gonz{\'a}lez}}, \bibfnamefont{M.~C.}},
  \bibinfo{author}{\bibfnamefont{A.~O.} \bibnamefont{{Sousa}}}, and
  \bibinfo{author}{\bibfnamefont{H.~J.} \bibnamefont{{Herrmann}}},
  \bibinfo{year}{2004}, \bibinfo{journal}{Int. J. Mod. Phys. C}
  \textbf{\bibinfo{volume}{15}}(\bibinfo{number}{1}), \bibinfo{pages}{45}.

\bibitem[{\citenamefont{{Gonz{\'a}lez}}
  \emph{et~al.}(2006)\citenamefont{{Gonz{\'a}lez}, {Sousa}, and
  {Herrmann}}}]{gonzalez06}
\bibinfo{author}{\bibnamefont{{Gonz{\'a}lez}}, \bibfnamefont{M.~C.}},
  \bibinfo{author}{\bibfnamefont{A.~O.} \bibnamefont{{Sousa}}}, and
  \bibinfo{author}{\bibfnamefont{H.~J.} \bibnamefont{{Herrmann}}},
  \bibinfo{year}{2006}, \bibinfo{journal}{Eur. Phys. J. B}
  \textbf{\bibinfo{volume}{49}}, \bibinfo{pages}{253}.

\bibitem[{\citenamefont{Gonz\'{a}lez-Avella}
  \emph{et~al.}(2007)\citenamefont{Gonz\'{a}lez-Avella, Cosenza, Klemm,
  Egu\'{i}luz, and {San Miguel}}}]{gonzalez-avella07}
\bibinfo{author}{\bibnamefont{Gonz\'{a}lez-Avella}, \bibfnamefont{J.}},
  \bibinfo{author}{\bibfnamefont{M.}~\bibnamefont{Cosenza}},
  \bibinfo{author}{\bibfnamefont{K.}~\bibnamefont{Klemm}},
  \bibinfo{author}{\bibfnamefont{V.}~\bibnamefont{Egu\'{i}luz}}, and
  \bibinfo{author}{\bibfnamefont{M.}~\bibnamefont{{San Miguel}}},
  \bibinfo{year}{2007}, \bibinfo{journal}{JASSS}
  \textbf{\bibinfo{volume}{10}}(\bibinfo{number}{3}), \bibinfo{pages}{9}.

\bibitem[{\citenamefont{Gonz\'alez-Avella}
  \emph{et~al.}(2005)\citenamefont{Gonz\'alez-Avella, Cosenza, and
  Tucci}}]{gonzalez-avella05}
\bibinfo{author}{\bibnamefont{Gonz\'alez-Avella}, \bibfnamefont{J.~C.}},
  \bibinfo{author}{\bibfnamefont{M.~G.} \bibnamefont{Cosenza}}, and
  \bibinfo{author}{\bibfnamefont{K.}~\bibnamefont{Tucci}},
  \bibinfo{year}{2005}, \bibinfo{journal}{Phys. Rev. E}
  \textbf{\bibinfo{volume}{72}}(\bibinfo{number}{6}), \bibinfo{pages}{065102}.

\bibitem[{\citenamefont{Gonz\'alez-Avella}
  \emph{et~al.}(2006)\citenamefont{Gonz\'alez-Avella, Egu\'iluz, Cosenza,
  Klemm, Herrera, and {San Miguel}}}]{gonzalez-avella06}
\bibinfo{author}{\bibnamefont{Gonz\'alez-Avella}, \bibfnamefont{J.~C.}},
  \bibinfo{author}{\bibfnamefont{V.~M.} \bibnamefont{Egu\'iluz}},
  \bibinfo{author}{\bibfnamefont{M.~G.} \bibnamefont{Cosenza}},
  \bibinfo{author}{\bibfnamefont{K.}~\bibnamefont{Klemm}},
  \bibinfo{author}{\bibfnamefont{J.~L.} \bibnamefont{Herrera}}, and
  \bibinfo{author}{\bibfnamefont{M.}~\bibnamefont{{San Miguel}}},
  \bibinfo{year}{2006}, \bibinfo{journal}{Phys. Rev. E}
  \textbf{\bibinfo{volume}{73}}(\bibinfo{number}{4}), \bibinfo{pages}{046119}.

\bibitem[{\citenamefont{Grabowski and
  Kosi\'nski}(2006{\natexlab{a}})}]{grabowski06}
\bibinfo{author}{\bibnamefont{Grabowski}, \bibfnamefont{A.}}, and
  \bibinfo{author}{\bibfnamefont{R.~A.} \bibnamefont{Kosi\'nski}},
  \bibinfo{year}{2006}{\natexlab{a}}, \bibinfo{journal}{Phys. Rev. E}
  \textbf{\bibinfo{volume}{73}}(\bibinfo{number}{1}), \bibinfo{pages}{016135}.

\bibitem[{\citenamefont{Grabowski and
  Kosi\'nski}(2006{\natexlab{b}})}]{grabowski06a}
\bibinfo{author}{\bibnamefont{Grabowski}, \bibfnamefont{A.}}, and
  \bibinfo{author}{\bibfnamefont{R.~A.} \bibnamefont{Kosi\'nski}},
  \bibinfo{year}{2006}{\natexlab{b}}, \bibinfo{journal}{Physica A}
  \textbf{\bibinfo{volume}{361}}, \bibinfo{pages}{651}.

\bibitem[{\citenamefont{Granovetter}(1973)}]{granovetter73}
\bibinfo{author}{\bibnamefont{Granovetter}, \bibfnamefont{M.}},
  \bibinfo{year}{1973}, \bibinfo{journal}{Am. J. Sociol.}
  \textbf{\bibinfo{volume}{78}}(\bibinfo{number}{6}), \bibinfo{pages}{1360}.

\bibitem[{\citenamefont{Granovetter}(1983)}]{granovetter83}
\bibinfo{author}{\bibnamefont{Granovetter}, \bibfnamefont{M.}},
  \bibinfo{year}{1983}, \bibinfo{journal}{Sociol. Theory}
  \textbf{\bibinfo{volume}{1}}, \bibinfo{pages}{201}.

\bibitem[{\citenamefont{Granovsky and Madras}(1995)}]{granovsky95}
\bibinfo{author}{\bibnamefont{Granovsky}, \bibfnamefont{B.}}, and
  \bibinfo{author}{\bibfnamefont{N.}~\bibnamefont{Madras}},
  \bibinfo{year}{1995}, \bibinfo{journal}{Stoch. Proc. Appl.}
  \textbf{\bibinfo{volume}{55}}, \bibinfo{pages}{23}.

\bibitem[{\citenamefont{Greenberg and Hastings}(1978)}]{greenberg78}
\bibinfo{author}{\bibnamefont{Greenberg}, \bibfnamefont{J.~M.}}, and
  \bibinfo{author}{\bibfnamefont{S.~P.} \bibnamefont{Hastings}},
  \bibinfo{year}{1978}, \bibinfo{journal}{SIAM J. Appl. Math.}
  \textbf{\bibinfo{volume}{34}}(\bibinfo{number}{3}), \bibinfo{pages}{515}.

\bibitem[{\citenamefont{Grinstein and Linsker}(2006)}]{grinstein06}
\bibinfo{author}{\bibnamefont{Grinstein}, \bibfnamefont{G.}}, and
  \bibinfo{author}{\bibfnamefont{R.}~\bibnamefont{Linsker}},
  \bibinfo{year}{2006}, \bibinfo{journal}{Phys. Rev. Lett.}
  \textbf{\bibinfo{volume}{97}}(\bibinfo{number}{13}), \bibinfo{pages}{130201}.

\bibitem[{\citenamefont{{Gross} and {Blasius}}(2008)}]{gross08}
\bibinfo{author}{\bibnamefont{{Gross}}, \bibfnamefont{T.}}, and
  \bibinfo{author}{\bibfnamefont{B.}~\bibnamefont{{Blasius}}},
  \bibinfo{year}{2008}, \bibinfo{journal}{J. R. Soc. Interface}
  \textbf{\bibinfo{volume}{5}}, \bibinfo{pages}{259}.

\bibitem[{\citenamefont{Guardiola} \emph{et~al.}(2002)\citenamefont{Guardiola,
  D\'iaz-Guilera, P\'erez, Arenas, and Llas}}]{guardiola02}
\bibinfo{author}{\bibnamefont{Guardiola}, \bibfnamefont{X.}},
  \bibinfo{author}{\bibfnamefont{A.}~\bibnamefont{D\'iaz-Guilera}},
  \bibinfo{author}{\bibfnamefont{C.~J.} \bibnamefont{P\'erez}},
  \bibinfo{author}{\bibfnamefont{A.}~\bibnamefont{Arenas}}, and
  \bibinfo{author}{\bibfnamefont{M.}~\bibnamefont{Llas}}, \bibinfo{year}{2002},
  \bibinfo{journal}{Phys. Rev. E}
  \textbf{\bibinfo{volume}{66}}(\bibinfo{number}{2}), \bibinfo{pages}{026121}.

\bibitem[{\citenamefont{{Guhl}}(1968)}]{guhl68}
\bibinfo{author}{\bibnamefont{{Guhl}}, \bibfnamefont{A.~M.}},
  \bibinfo{year}{1968}, \bibinfo{journal}{Anim. Behav.}
  \textbf{\bibinfo{volume}{16}}, \bibinfo{pages}{219}.

\bibitem[{\citenamefont{Haken}(1978)}]{haken77}
\bibinfo{author}{\bibnamefont{Haken}, \bibfnamefont{H.}}, \bibinfo{year}{1978},
  \emph{\bibinfo{title}{Synergetics; An Introduction. Non-equilibrium Phase
  Transitions and Self-Organization in Physics, Chemistry and Biology}}
  (\bibinfo{publisher}{Springer Verlag}, \bibinfo{address}{Berlin-Heidelberg,
  Germany}).

\bibitem[{\citenamefont{Hammal} \emph{et~al.}(2005)\citenamefont{Hammal, Chate,
  Dornic, and Mu{\~n}oz}}]{alhammal05}
\bibinfo{author}{\bibnamefont{Hammal}, \bibfnamefont{O.~A.}},
  \bibinfo{author}{\bibfnamefont{H.}~\bibnamefont{Chate}},
  \bibinfo{author}{\bibfnamefont{I.}~\bibnamefont{Dornic}}, and
  \bibinfo{author}{\bibfnamefont{M.~A.} \bibnamefont{Mu{\~n}oz}},
  \bibinfo{year}{2005}, \bibinfo{journal}{Phys. Rev. Lett.}
  \textbf{\bibinfo{volume}{94}}(\bibinfo{number}{23}), \bibinfo{pages}{230601}.

\bibitem[{\citenamefont{Hankin and Wright}(1958)}]{hankin58}
\bibinfo{author}{\bibnamefont{Hankin}, \bibfnamefont{B.~D.}}, and
  \bibinfo{author}{\bibfnamefont{R.~A.} \bibnamefont{Wright}},
  \bibinfo{year}{1958}, \bibinfo{journal}{Opns. Res. Quart.}
  \textbf{\bibinfo{volume}{9}}, \bibinfo{pages}{81}.

\bibitem[{\citenamefont{Hauert} \emph{et~al.}(2005)\citenamefont{Hauert, Nowak,
  and Lieberman}}]{lieberman05}
\bibinfo{author}{\bibnamefont{Hauert}, \bibfnamefont{J.}},
  \bibinfo{author}{\bibfnamefont{M.}~\bibnamefont{Nowak}}, and
  \bibinfo{author}{\bibfnamefont{E.}~\bibnamefont{Lieberman}},
  \bibinfo{year}{2005}, \bibinfo{journal}{Nature}
  \textbf{\bibinfo{volume}{433}}, \bibinfo{pages}{312}.

\bibitem[{\citenamefont{{He}} \emph{et~al.}(2004)\citenamefont{{He}, {Li}, and
  {Luo}}}]{he04}
\bibinfo{author}{\bibnamefont{{He}}, \bibfnamefont{M.}},
  \bibinfo{author}{\bibfnamefont{B.}~\bibnamefont{{Li}}}, and
  \bibinfo{author}{\bibfnamefont{L.}~\bibnamefont{{Luo}}},
  \bibinfo{year}{2004}, \bibinfo{journal}{Int. J. Mod. Phys. C}
  \textbf{\bibinfo{volume}{15}}, \bibinfo{pages}{997}.

\bibitem[{\citenamefont{Hegselmann and Krause}(2002)}]{hegselmann02}
\bibinfo{author}{\bibnamefont{Hegselmann}, \bibfnamefont{R.}}, and
  \bibinfo{author}{\bibfnamefont{U.}~\bibnamefont{Krause}},
  \bibinfo{year}{2002}, \bibinfo{journal}{JASSS}
  \textbf{\bibinfo{volume}{5}}(\bibinfo{number}{3}).

\bibitem[{\citenamefont{Hegselmann and Krause}(2005)}]{hegselmann05}
\bibinfo{author}{\bibnamefont{Hegselmann}, \bibfnamefont{R.}}, and
  \bibinfo{author}{\bibfnamefont{U.}~\bibnamefont{Krause}},
  \bibinfo{year}{2005}, \bibinfo{journal}{Comput. Econ.}
  \textbf{\bibinfo{volume}{25}}(\bibinfo{number}{4}), \bibinfo{pages}{381}.

\bibitem[{\citenamefont{Helbing}(1991)}]{helbing91}
\bibinfo{author}{\bibnamefont{Helbing}, \bibfnamefont{D.}},
  \bibinfo{year}{1991}, \emph{\bibinfo{title}{Quantitative sociodynamics}}
  (\bibinfo{publisher}{Springer}, \bibinfo{address}{Heidelberg, Germany}).

\bibitem[{\citenamefont{Helbing}(1993{\natexlab{a}})}]{helbing93a}
\bibinfo{author}{\bibnamefont{Helbing}, \bibfnamefont{D.}},
  \bibinfo{year}{1993}{\natexlab{a}}, \bibinfo{journal}{Physica A}
  \textbf{\bibinfo{volume}{196}}(\bibinfo{number}{2}), \bibinfo{pages}{546}.

\bibitem[{\citenamefont{Helbing}(1993{\natexlab{b}})}]{helbing93}
\bibinfo{author}{\bibnamefont{Helbing}, \bibfnamefont{D.}},
  \bibinfo{year}{1993}{\natexlab{b}}, \bibinfo{journal}{Physica A}
  \textbf{\bibinfo{volume}{193}}(\bibinfo{number}{2}), \bibinfo{pages}{241}.

\bibitem[{\citenamefont{{Helbing}}(1994)}]{helbing94}
\bibinfo{author}{\bibnamefont{{Helbing}}, \bibfnamefont{D.}},
  \bibinfo{year}{1994}, \bibinfo{journal}{J. Math. Sociol.}
  \textbf{\bibinfo{volume}{19}}(\bibinfo{number}{3}), \bibinfo{pages}{189}.

\bibitem[{\citenamefont{Helbing}(2001)}]{helbing01}
\bibinfo{author}{\bibnamefont{Helbing}, \bibfnamefont{D.}},
  \bibinfo{year}{2001}, \bibinfo{journal}{Rev. Mod. Phys.}
  \textbf{\bibinfo{volume}{73}}(\bibinfo{number}{4}), \bibinfo{pages}{1067}.

\bibitem[{\citenamefont{{Helbing}} \emph{et~al.}(2002)\citenamefont{{Helbing},
  Farkas, Moln{\'a}r, and Vicsek}}]{helbing02}
\bibinfo{author}{\bibnamefont{{Helbing}}, \bibfnamefont{D.}},
  \bibinfo{author}{\bibfnamefont{I.}~\bibnamefont{Farkas}},
  \bibinfo{author}{\bibfnamefont{P.}~\bibnamefont{Moln{\'a}r}}, and
  \bibinfo{author}{\bibfnamefont{T.}~\bibnamefont{Vicsek}},
  \bibinfo{year}{2002}, in \emph{\bibinfo{booktitle}{Pedestrian and Evacuation
  Dynamics}}, edited by
  \bibinfo{editor}{\bibfnamefont{M.}~\bibnamefont{Schreckenberg}} and
  \bibinfo{editor}{\bibfnamefont{S.~D.} \bibnamefont{Sharma}}
  (\bibinfo{publisher}{Springer Verlag}, \bibinfo{address}{Berlin, Germany}),
  pp. \bibinfo{pages}{19--58}.

\bibitem[{\citenamefont{{Helbing}}
  \emph{et~al.}(2000{\natexlab{a}})\citenamefont{{Helbing}, {Farkas}, and
  {Vicsek}}}]{helbing00a}
\bibinfo{author}{\bibnamefont{{Helbing}}, \bibfnamefont{D.}},
  \bibinfo{author}{\bibfnamefont{I.}~\bibnamefont{{Farkas}}}, and
  \bibinfo{author}{\bibfnamefont{T.}~\bibnamefont{{Vicsek}}},
  \bibinfo{year}{2000}{\natexlab{a}}, \bibinfo{journal}{Nature}
  \textbf{\bibinfo{volume}{407}}, \bibinfo{pages}{487}.

\bibitem[{\citenamefont{{Helbing}}
  \emph{et~al.}(2000{\natexlab{b}})\citenamefont{{Helbing}, {Farkas}, and
  {Vicsek}}}]{helbing00}
\bibinfo{author}{\bibnamefont{{Helbing}}, \bibfnamefont{D.}},
  \bibinfo{author}{\bibfnamefont{I.~J.} \bibnamefont{{Farkas}}}, and
  \bibinfo{author}{\bibfnamefont{T.}~\bibnamefont{{Vicsek}}},
  \bibinfo{year}{2000}{\natexlab{b}}, \bibinfo{journal}{Phys. Rev. Lett.}
  \textbf{\bibinfo{volume}{84}}, \bibinfo{pages}{1240}.

\bibitem[{\citenamefont{Helbing} \emph{et~al.}(2007)\citenamefont{Helbing,
  Johansson, and Al-Abideen}}]{helbing07}
\bibinfo{author}{\bibnamefont{Helbing}, \bibfnamefont{D.}},
  \bibinfo{author}{\bibfnamefont{A.}~\bibnamefont{Johansson}}, and
  \bibinfo{author}{\bibfnamefont{H.~Z.} \bibnamefont{Al-Abideen}},
  \bibinfo{year}{2007}, \bibinfo{journal}{Phys. Rev. E}
  \textbf{\bibinfo{volume}{75}}(\bibinfo{number}{4}), \bibinfo{pages}{046109}.

\bibitem[{\citenamefont{{Helbing} and {Moln{\'a}r}}(1995)}]{helbing95}
\bibinfo{author}{\bibnamefont{{Helbing}}, \bibfnamefont{D.}}, and
  \bibinfo{author}{\bibfnamefont{P.}~\bibnamefont{{Moln{\'a}r}}},
  \bibinfo{year}{1995}, \bibinfo{journal}{Phys. Rev. E}
  \textbf{\bibinfo{volume}{51}}, \bibinfo{pages}{4282}.

\bibitem[{\citenamefont{{Henderson}}(1971)}]{henderson71}
\bibinfo{author}{\bibnamefont{{Henderson}}, \bibfnamefont{L.~F.}},
  \bibinfo{year}{1971}, \bibinfo{journal}{Nature}
  \textbf{\bibinfo{volume}{229}}, \bibinfo{pages}{381}.

\bibitem[{\citenamefont{{Hendrickx}}(2007)}]{hendrickx07}
\bibinfo{author}{\bibnamefont{{Hendrickx}}, \bibfnamefont{J.~M.}},
  \bibinfo{year}{2007}, \eprint{arxiv:0708.4343}.

\bibitem[{\citenamefont{Hewitt}(1970)}]{hewitt70}
\bibinfo{author}{\bibnamefont{Hewitt}, \bibfnamefont{C.}},
  \bibinfo{year}{1970}, \emph{\bibinfo{title}{PLANNER: A Language for
  Manipulating Models and Proving Theorems in a Robot}}
  (\bibinfo{publisher}{dspace.mit.edu}).

\bibitem[{\citenamefont{{Hidalgo}}(2006)}]{hidalgo06bis}
\bibinfo{author}{\bibnamefont{{Hidalgo}}, \bibfnamefont{C.~A.}},
  \bibinfo{year}{2006}, \bibinfo{journal}{Physica A}
  \textbf{\bibinfo{volume}{369}}, \bibinfo{pages}{877}.

\bibitem[{\citenamefont{{Hidalgo}} \emph{et~al.}(2006)\citenamefont{{Hidalgo},
  {Castro}, and {Rodriguez-Sickert}}}]{hidalgo06}
\bibinfo{author}{\bibnamefont{{Hidalgo}}, \bibfnamefont{C.~A.}},
  \bibinfo{author}{\bibfnamefont{A.}~\bibnamefont{{Castro}}}, and
  \bibinfo{author}{\bibfnamefont{C.}~\bibnamefont{{Rodriguez-Sickert}}},
  \bibinfo{year}{2006}, \bibinfo{journal}{New J. Phys.}
  \textbf{\bibinfo{volume}{8}}, \bibinfo{pages}{52}.

\bibitem[{\citenamefont{Hofbauer and Sigmund}(1998)}]{hofbauer98}
\bibinfo{author}{\bibnamefont{Hofbauer}, \bibfnamefont{J.}}, and
  \bibinfo{author}{\bibfnamefont{K.}~\bibnamefont{Sigmund}},
  \bibinfo{year}{1998}, \emph{\bibinfo{title}{Evolutionary Games and Population
  Dynamics}} (\bibinfo{publisher}{{Cambridge University Press}},
  \bibinfo{address}{Cambridge, UK}).

\bibitem[{\citenamefont{Hogeweg and Hesper}(1983)}]{hogeweg83}
\bibinfo{author}{\bibnamefont{Hogeweg}, \bibfnamefont{P.}}, and
  \bibinfo{author}{\bibfnamefont{B.}~\bibnamefont{Hesper}},
  \bibinfo{year}{1983}, \bibinfo{journal}{Behav. Ecol. Sociobiol.}
  \textbf{\bibinfo{volume}{12}}, \bibinfo{pages}{271}.

\bibitem[{\citenamefont{Holley and Liggett}(1975)}]{holley75}
\bibinfo{author}{\bibnamefont{Holley}, \bibfnamefont{R.}}, and
  \bibinfo{author}{\bibfnamefont{T.}~\bibnamefont{Liggett}},
  \bibinfo{year}{1975}, \bibinfo{journal}{Ann. Probab.}
  \textbf{\bibinfo{volume}{3}}(\bibinfo{number}{4}), \bibinfo{pages}{643}.

\bibitem[{\citenamefont{Holme}(2003)}]{holme03}
\bibinfo{author}{\bibnamefont{Holme}, \bibfnamefont{P.}}, \bibinfo{year}{2003},
  \bibinfo{journal}{Europhys. Lett.}
  \textbf{\bibinfo{volume}{64}}(\bibinfo{number}{3}), \bibinfo{pages}{427}.

\bibitem[{\citenamefont{Holme} \emph{et~al.}(2004)\citenamefont{Holme, Edling,
  and Liljeros}}]{holme04}
\bibinfo{author}{\bibnamefont{Holme}, \bibfnamefont{P.}},
  \bibinfo{author}{\bibfnamefont{C.~R.} \bibnamefont{Edling}}, and
  \bibinfo{author}{\bibfnamefont{F.}~\bibnamefont{Liljeros}},
  \bibinfo{year}{2004}, \bibinfo{journal}{Soc. Networks}
  \textbf{\bibinfo{volume}{26}}, \bibinfo{pages}{155}.

\bibitem[{\citenamefont{Holme and Newman}(2006)}]{holme06}
\bibinfo{author}{\bibnamefont{Holme}, \bibfnamefont{P.}}, and
  \bibinfo{author}{\bibfnamefont{M.~E.~J.} \bibnamefont{Newman}},
  \bibinfo{year}{2006}, \bibinfo{journal}{Phys. Rev. E}
  \textbf{\bibinfo{volume}{74}}(\bibinfo{number}{5}), \bibinfo{pages}{056108}.

\bibitem[{\citenamefont{{Ho{\l}yst}}
  \emph{et~al.}(2000)\citenamefont{{Ho{\l}yst}, {Kacperski}, and
  {Schweitzer}}}]{holyst00}
\bibinfo{author}{\bibnamefont{{Ho{\l}yst}}, \bibfnamefont{J.~A.}},
  \bibinfo{author}{\bibfnamefont{K.}~\bibnamefont{{Kacperski}}}, and
  \bibinfo{author}{\bibfnamefont{F.}~\bibnamefont{{Schweitzer}}},
  \bibinfo{year}{2000}, \bibinfo{journal}{Physica A}
  \textbf{\bibinfo{volume}{285}}, \bibinfo{pages}{199}.

\bibitem[{\citenamefont{{Ho{\l}yst}}
  \emph{et~al.}(2001)\citenamefont{{Ho{\l}yst}, {Kacperski}, and
  {Schweitzer}}}]{holyst01}
\bibinfo{author}{\bibnamefont{{Ho{\l}yst}}, \bibfnamefont{J.~A.}},
  \bibinfo{author}{\bibfnamefont{K.}~\bibnamefont{{Kacperski}}}, and
  \bibinfo{author}{\bibfnamefont{F.}~\bibnamefont{{Schweitzer}}},
  \bibinfo{year}{2001}, \bibinfo{journal}{Ann. Rev. Comp. Phys.}
  \textbf{\bibinfo{volume}{9}}, \bibinfo{pages}{253}.

\bibitem[{\citenamefont{Huang}(1987)}]{huang87}
\bibinfo{author}{\bibnamefont{Huang}, \bibfnamefont{K.}}, \bibinfo{year}{1987},
  \emph{\bibinfo{title}{Statistical mechanics}} (\bibinfo{publisher}{Wiley},
  \bibinfo{address}{New York, NY, USA}).

\bibitem[{\citenamefont{{Huberman} and {Adamic}}(2004)}]{huberman04}
\bibinfo{author}{\bibnamefont{{Huberman}}, \bibfnamefont{B.~A.}}, and
  \bibinfo{author}{\bibfnamefont{L.~A.} \bibnamefont{{Adamic}}},
  \bibinfo{year}{2004}, in \emph{\bibinfo{booktitle}{LNP Vol. 650: Complex
  Networks}}, edited by
  \bibinfo{editor}{\bibfnamefont{E.}~\bibnamefont{{Ben-Naim}}},
  \bibinfo{editor}{\bibfnamefont{H.}~\bibnamefont{{Frauenfelder}}}, and
  \bibinfo{editor}{\bibfnamefont{Z.}~\bibnamefont{{Toroczkai}}}, pp.
  \bibinfo{pages}{371--398}.

\bibitem[{\citenamefont{Hurford}(1989)}]{hurford89}
\bibinfo{author}{\bibnamefont{Hurford}, \bibfnamefont{J.}},
  \bibinfo{year}{1989}, \bibinfo{journal}{Lingua}
  \textbf{\bibinfo{volume}{77}}(\bibinfo{number}{2}), \bibinfo{pages}{187}.

\bibitem[{\citenamefont{Indekeu}(2004)}]{indekeu04}
\bibinfo{author}{\bibnamefont{Indekeu}, \bibfnamefont{J.~O.}},
  \bibinfo{year}{2004}, \bibinfo{journal}{Physica A}
  \textbf{\bibinfo{volume}{333}}, \bibinfo{pages}{461}.

\bibitem[{\citenamefont{Ito and Kaneko}(2001)}]{ito02}
\bibinfo{author}{\bibnamefont{Ito}, \bibfnamefont{J.}}, and
  \bibinfo{author}{\bibfnamefont{K.}~\bibnamefont{Kaneko}},
  \bibinfo{year}{2001}, \bibinfo{journal}{Phys. Rev. Lett.}
  \textbf{\bibinfo{volume}{88}}(\bibinfo{number}{2}), \bibinfo{pages}{028701}.

\bibitem[{\citenamefont{Ito and Kaneko}(2003)}]{ito03}
\bibinfo{author}{\bibnamefont{Ito}, \bibfnamefont{J.}}, and
  \bibinfo{author}{\bibfnamefont{K.}~\bibnamefont{Kaneko}},
  \bibinfo{year}{2003}, \bibinfo{journal}{Phys. Rev. E}
  \textbf{\bibinfo{volume}{67}}(\bibinfo{number}{4}), \bibinfo{pages}{046226}.

\bibitem[{\citenamefont{{Jiang}} \emph{et~al.}(2007)\citenamefont{{Jiang},
  {Hua}, and {Chen}}}]{jiang07}
\bibinfo{author}{\bibnamefont{{Jiang}}, \bibfnamefont{L.-l.}},
  \bibinfo{author}{\bibfnamefont{D.-y.} \bibnamefont{{Hua}}}, and
  \bibinfo{author}{\bibfnamefont{T.}~\bibnamefont{{Chen}}},
  \bibinfo{year}{2007}, \bibinfo{journal}{J. Phys. A}
  \textbf{\bibinfo{volume}{40}}, \bibinfo{pages}{11271}.

\bibitem[{\citenamefont{{Jiang}} \emph{et~al.}(2008)\citenamefont{{Jiang},
  {Hua}, {Zhu}, {Wang}, and {Zhou}}}]{jiang08}
\bibinfo{author}{\bibnamefont{{Jiang}}, \bibfnamefont{L.-L.}},
  \bibinfo{author}{\bibfnamefont{D.-Y.} \bibnamefont{{Hua}}},
  \bibinfo{author}{\bibfnamefont{J.-F.} \bibnamefont{{Zhu}}},
  \bibinfo{author}{\bibfnamefont{B.-H.} \bibnamefont{{Wang}}}, and
  \bibinfo{author}{\bibfnamefont{T.}~\bibnamefont{{Zhou}}},
  \bibinfo{year}{2008}, \eprint{arXiv:0801.1896}.

\bibitem[{\citenamefont{Jin} \emph{et~al.}(2001)\citenamefont{Jin, Girvan, and
  Newman}}]{jin01}
\bibinfo{author}{\bibnamefont{Jin}, \bibfnamefont{E.~M.}},
  \bibinfo{author}{\bibfnamefont{M.}~\bibnamefont{Girvan}}, and
  \bibinfo{author}{\bibfnamefont{M.~E.~J.} \bibnamefont{Newman}},
  \bibinfo{year}{2001}, \bibinfo{journal}{Phys. Rev. E}
  \textbf{\bibinfo{volume}{64}}(\bibinfo{number}{4}), \bibinfo{pages}{046132}.

\bibitem[{\citenamefont{Johansen}(2001)}]{johansen01}
\bibinfo{author}{\bibnamefont{Johansen}, \bibfnamefont{A.}},
  \bibinfo{year}{2001}, \bibinfo{journal}{Physica A}
  \textbf{\bibinfo{volume}{3}}, \bibinfo{pages}{539}.

\bibitem[{\citenamefont{Johansen}(2004)}]{johansen04}
\bibinfo{author}{\bibnamefont{Johansen}, \bibfnamefont{A.}},
  \bibinfo{year}{2004}, \bibinfo{journal}{Physica A}
  \textbf{\bibinfo{volume}{1}}, \bibinfo{pages}{286}.

\bibitem[{\citenamefont{Johansen and Sornette}(2000)}]{johansen00}
\bibinfo{author}{\bibnamefont{Johansen}, \bibfnamefont{A.}}, and
  \bibinfo{author}{\bibfnamefont{D.}~\bibnamefont{Sornette}},
  \bibinfo{year}{2000}, \bibinfo{journal}{Physica A}
  \textbf{\bibinfo{volume}{276}}, \bibinfo{pages}{338}.

\bibitem[{\citenamefont{Kacperski and Ho{\l}yst}(1996)}]{kacperski96}
\bibinfo{author}{\bibnamefont{Kacperski}, \bibfnamefont{K.}}, and
  \bibinfo{author}{\bibfnamefont{J.~A.} \bibnamefont{Ho{\l}yst}},
  \bibinfo{year}{1996}, \bibinfo{journal}{J. Stat. Phys.}
  \textbf{\bibinfo{volume}{84}}(\bibinfo{number}{1-2}), \bibinfo{pages}{169}.

\bibitem[{\citenamefont{Kacperski and Ho{\l}yst}(1997)}]{kacperski97}
\bibinfo{author}{\bibnamefont{Kacperski}, \bibfnamefont{K.}}, and
  \bibinfo{author}{\bibfnamefont{J.~A.} \bibnamefont{Ho{\l}yst}},
  \bibinfo{year}{1997}, in \emph{\bibinfo{booktitle}{Self Organization of
  Complex Structures}}, edited by
  \bibinfo{editor}{\bibfnamefont{F.}~\bibnamefont{Schweitzer}}
  (\bibinfo{publisher}{Gordon and Breach Publ.}, \bibinfo{address}{Amsterdam,
  The Netherlands}), volume~\bibinfo{volume}{II}, pp.
  \bibinfo{pages}{367--378}.

\bibitem[{\citenamefont{Kacperski and Ho{\l}yst}(1999)}]{kacperski99}
\bibinfo{author}{\bibnamefont{Kacperski}, \bibfnamefont{K.}}, and
  \bibinfo{author}{\bibfnamefont{J.~A.} \bibnamefont{Ho{\l}yst}},
  \bibinfo{year}{1999}, \bibinfo{journal}{Physica A}
  \textbf{\bibinfo{volume}{269}}(\bibinfo{number}{2}), \bibinfo{pages}{511}.

\bibitem[{\citenamefont{Kacperski and Ho{\l}yst}(2000)}]{kacperski00}
\bibinfo{author}{\bibnamefont{Kacperski}, \bibfnamefont{K.}}, and
  \bibinfo{author}{\bibfnamefont{J.~A.} \bibnamefont{Ho{\l}yst}},
  \bibinfo{year}{2000}, \bibinfo{journal}{Physica A}
  \textbf{\bibinfo{volume}{287}}(\bibinfo{number}{3}), \bibinfo{pages}{631}.

\bibitem[{\citenamefont{Ke} \emph{et~al.}(2008)\citenamefont{Ke, Gong, and
  Wang}}]{ke04-->08}
\bibinfo{author}{\bibnamefont{Ke}, \bibfnamefont{J.}},
  \bibinfo{author}{\bibfnamefont{T.}~\bibnamefont{Gong}}, and
  \bibinfo{author}{\bibfnamefont{W.-Y.} \bibnamefont{Wang}},
  \bibinfo{year}{2008}, \bibinfo{journal}{Comm. Comput. Phys.}
  \textbf{\bibinfo{volume}{3}}(\bibinfo{number}{4}), \bibinfo{pages}{935},
  \bibinfo{note}{originally presented at the 5th Conference on Language
  evolution, Leipzig, Germany, March 2004.}

\bibitem[{\citenamefont{Ke} \emph{et~al.}(2002)\citenamefont{Ke, Minett, Au,
  and Wang}}]{ke02a}
\bibinfo{author}{\bibnamefont{Ke}, \bibfnamefont{J.}},
  \bibinfo{author}{\bibfnamefont{J.}~\bibnamefont{Minett}},
  \bibinfo{author}{\bibfnamefont{C.-P.} \bibnamefont{Au}}, and
  \bibinfo{author}{\bibfnamefont{W.}~\bibnamefont{Wang}}, \bibinfo{year}{2002},
  \bibinfo{journal}{Complexity}
  \textbf{\bibinfo{volume}{7}}(\bibinfo{number}{3}), \bibinfo{pages}{41}.

\bibitem[{\citenamefont{Kermarrec} \emph{et~al.}(2003)\citenamefont{Kermarrec,
  Massoulie, and Ganesh}}]{kermarrec03}
\bibinfo{author}{\bibnamefont{Kermarrec}, \bibfnamefont{A.-M.}},
  \bibinfo{author}{\bibfnamefont{L.}~\bibnamefont{Massoulie}}, and
  \bibinfo{author}{\bibfnamefont{A.}~\bibnamefont{Ganesh}},
  \bibinfo{year}{2003}, \bibinfo{journal}{IEEE T. Parall. Distr.}
  \textbf{\bibinfo{volume}{14}}(\bibinfo{number}{3}), \bibinfo{pages}{248}.

\bibitem[{\citenamefont{Kerner}(2004)}]{kerner04}
\bibinfo{author}{\bibnamefont{Kerner}, \bibfnamefont{B.}},
  \bibinfo{year}{2004}, \emph{\bibinfo{title}{The physics of traffic}}
  (\bibinfo{publisher}{Springer Verlag}, \bibinfo{address}{Berlin, Germany}).

\bibitem[{\citenamefont{Kirby}(2001)}]{kirby01}
\bibinfo{author}{\bibnamefont{Kirby}, \bibfnamefont{S.}}, \bibinfo{year}{2001},
  \bibinfo{journal}{IEEE Trans. Evol. Comp.}
  \textbf{\bibinfo{volume}{5}}(\bibinfo{number}{2}), \bibinfo{pages}{102}.

\bibitem[{\citenamefont{Kleinberg}(2002)}]{kleinberg02}
\bibinfo{author}{\bibnamefont{Kleinberg}, \bibfnamefont{J.}},
  \bibinfo{year}{2002}, \bibinfo{title}{Bursty and hierarchical structure in
  streams}.

\bibitem[{\citenamefont{Klemm}
  \emph{et~al.}(2003{\natexlab{a}})\citenamefont{Klemm, Egu\'iluz, Toral, and
  {San Miguel}}}]{klemm03}
\bibinfo{author}{\bibnamefont{Klemm}, \bibfnamefont{K.}},
  \bibinfo{author}{\bibfnamefont{V.~M.} \bibnamefont{Egu\'iluz}},
  \bibinfo{author}{\bibfnamefont{R.}~\bibnamefont{Toral}}, and
  \bibinfo{author}{\bibfnamefont{M.}~\bibnamefont{{San Miguel}}},
  \bibinfo{year}{2003}{\natexlab{a}}, \bibinfo{journal}{Phys. Rev. E}
  \textbf{\bibinfo{volume}{67}}(\bibinfo{number}{4}),
  \bibinfo{pages}{045101(R)}.

\bibitem[{\citenamefont{Klemm}
  \emph{et~al.}(2003{\natexlab{b}})\citenamefont{Klemm, Egu\'iluz, Toral, and
  {San Miguel}}}]{klemm03b}
\bibinfo{author}{\bibnamefont{Klemm}, \bibfnamefont{K.}},
  \bibinfo{author}{\bibfnamefont{V.~M.} \bibnamefont{Egu\'iluz}},
  \bibinfo{author}{\bibfnamefont{R.}~\bibnamefont{Toral}}, and
  \bibinfo{author}{\bibfnamefont{M.}~\bibnamefont{{San Miguel}}},
  \bibinfo{year}{2003}{\natexlab{b}}, \bibinfo{journal}{Phys. Rev. E}
  \textbf{\bibinfo{volume}{67}}(\bibinfo{number}{2}), \bibinfo{pages}{026120}.

\bibitem[{\citenamefont{Klemm}
  \emph{et~al.}(2003{\natexlab{c}})\citenamefont{Klemm, Egu\'iluz, Toral, and
  {San Miguel}}}]{klemm03c}
\bibinfo{author}{\bibnamefont{Klemm}, \bibfnamefont{K.}},
  \bibinfo{author}{\bibfnamefont{V.~M.} \bibnamefont{Egu\'iluz}},
  \bibinfo{author}{\bibfnamefont{R.}~\bibnamefont{Toral}}, and
  \bibinfo{author}{\bibfnamefont{M.}~\bibnamefont{{San Miguel}}},
  \bibinfo{year}{2003}{\natexlab{c}}, \bibinfo{journal}{Physica A}
  \textbf{\bibinfo{volume}{327}}(\bibinfo{number}{1-2}), \bibinfo{pages}{1}.

\bibitem[{\citenamefont{Klemm} \emph{et~al.}(2005)\citenamefont{Klemm,
  Egu{\'i}luz, Toral, and {San Miguel}}}]{klemm05}
\bibinfo{author}{\bibnamefont{Klemm}, \bibfnamefont{K.}},
  \bibinfo{author}{\bibfnamefont{V.~M.} \bibnamefont{Egu{\'i}luz}},
  \bibinfo{author}{\bibfnamefont{R.}~\bibnamefont{Toral}}, and
  \bibinfo{author}{\bibfnamefont{M.}~\bibnamefont{{San Miguel}}},
  \bibinfo{year}{2005}, \bibinfo{journal}{J. Econ. Dyn. Control}
  \textbf{\bibinfo{volume}{29}}(\bibinfo{number}{1-2}), \bibinfo{pages}{321}.

\bibitem[{\citenamefont{{Klietsch}}(2005)}]{klietsch05}
\bibinfo{author}{\bibnamefont{{Klietsch}}, \bibfnamefont{N.}},
  \bibinfo{year}{2005}, \bibinfo{journal}{Int. J. Mod. Phys. C}
  \textbf{\bibinfo{volume}{16}}, \bibinfo{pages}{577}.

\bibitem[{\citenamefont{{Klimek}} \emph{et~al.}(2007)\citenamefont{{Klimek},
  {Lambiotte}, and {Thurner}}}]{klimek07}
\bibinfo{author}{\bibnamefont{{Klimek}}, \bibfnamefont{P.}},
  \bibinfo{author}{\bibfnamefont{R.}~\bibnamefont{{Lambiotte}}}, and
  \bibinfo{author}{\bibfnamefont{S.}~\bibnamefont{{Thurner}}},
  \bibinfo{year}{2007}, \eprint{arxiv:0706.4058}.

\bibitem[{\citenamefont{{Kohring}}(1996)}]{kohring96}
\bibinfo{author}{\bibnamefont{{Kohring}}, \bibfnamefont{G.~A.}},
  \bibinfo{year}{1996}, \bibinfo{journal}{J. de Phys. I}
  \textbf{\bibinfo{volume}{6}}, \bibinfo{pages}{301}.

\bibitem[{\citenamefont{Komarova} \emph{et~al.}(2001)\citenamefont{Komarova,
  Niyogi, and Nowak}}]{komarova01}
\bibinfo{author}{\bibnamefont{Komarova}, \bibfnamefont{N.}},
  \bibinfo{author}{\bibfnamefont{P.}~\bibnamefont{Niyogi}}, and
  \bibinfo{author}{\bibfnamefont{M.}~\bibnamefont{Nowak}},
  \bibinfo{year}{2001}, \bibinfo{journal}{J. Theor. Bio.}
  \textbf{\bibinfo{volume}{209}}(\bibinfo{number}{1}), \bibinfo{pages}{43}.

\bibitem[{\citenamefont{Komarova and Nowak}(2001)}]{komarova01b}
\bibinfo{author}{\bibnamefont{Komarova}, \bibfnamefont{N.}}, and
  \bibinfo{author}{\bibfnamefont{M.}~\bibnamefont{Nowak}},
  \bibinfo{year}{2001}, \bibinfo{journal}{Bull. Math. Bio.}
  \textbf{\bibinfo{volume}{63}}(\bibinfo{number}{3}), \bibinfo{pages}{451}.

\bibitem[{\citenamefont{Komarova and Niyogi}(2004)}]{komarova04b}
\bibinfo{author}{\bibnamefont{Komarova}, \bibfnamefont{N.~L.}}, and
  \bibinfo{author}{\bibfnamefont{P.}~\bibnamefont{Niyogi}},
  \bibinfo{year}{2004}, \bibinfo{journal}{Artif. Intell.}
  \textbf{\bibinfo{volume}{154}}(\bibinfo{number}{1-2}), \bibinfo{pages}{1}.

\bibitem[{\citenamefont{Kondrat and Sznajd-Weron}(2008)}]{kondrat08}
\bibinfo{author}{\bibnamefont{Kondrat}, \bibfnamefont{G.}}, and
  \bibinfo{author}{\bibfnamefont{K.}~\bibnamefont{Sznajd-Weron}},
  \bibinfo{year}{2008}, \bibinfo{journal}{Phys. Rev. E}
  \textbf{\bibinfo{volume}{77}}(\bibinfo{number}{2}), \bibinfo{eid}{021127}
  (pages~\bibinfo{numpages}{8}).

\bibitem[{\citenamefont{{Kosmidis}}
  \emph{et~al.}(2005)\citenamefont{{Kosmidis}, {Halley}, and
  {Argyrakis}}}]{kosmidis05}
\bibinfo{author}{\bibnamefont{{Kosmidis}}, \bibfnamefont{K.}},
  \bibinfo{author}{\bibfnamefont{J.~M.} \bibnamefont{{Halley}}}, and
  \bibinfo{author}{\bibfnamefont{P.}~\bibnamefont{{Argyrakis}}},
  \bibinfo{year}{2005}, \bibinfo{journal}{Physica A}
  \textbf{\bibinfo{volume}{353}}, \bibinfo{pages}{595}.

\bibitem[{\citenamefont{Kosterlitz and Thouless}(1973)}]{kosterlitz73}
\bibinfo{author}{\bibnamefont{Kosterlitz}, \bibfnamefont{J.~M.}}, and
  \bibinfo{author}{\bibfnamefont{D.~J.} \bibnamefont{Thouless}},
  \bibinfo{year}{1973}, \bibinfo{journal}{J. Phys. C}
  \textbf{\bibinfo{volume}{6}}, \bibinfo{pages}{1181}.

\bibitem[{\citenamefont{Kozma and Barrat}(2008)}]{kozma08}
\bibinfo{author}{\bibnamefont{Kozma}, \bibfnamefont{B.}}, and
  \bibinfo{author}{\bibfnamefont{A.}~\bibnamefont{Barrat}},
  \bibinfo{year}{2008}, \bibinfo{journal}{Phys. Rev. E}
  \textbf{\bibinfo{volume}{77}}(\bibinfo{number}{1}), \bibinfo{eid}{016102}
  (pages~\bibinfo{numpages}{10}).

\bibitem[{\citenamefont{Krapivsky}(1992)}]{krapivsky92}
\bibinfo{author}{\bibnamefont{Krapivsky}, \bibfnamefont{P.~L.}},
  \bibinfo{year}{1992}, \bibinfo{journal}{Phys. Rev. A}
  \textbf{\bibinfo{volume}{45}}(\bibinfo{number}{2}), \bibinfo{pages}{1067}.

\bibitem[{\citenamefont{Krapivsky and Redner}(2003)}]{krapivsky03}
\bibinfo{author}{\bibnamefont{Krapivsky}, \bibfnamefont{P.~L.}}, and
  \bibinfo{author}{\bibfnamefont{S.}~\bibnamefont{Redner}},
  \bibinfo{year}{2003}, \bibinfo{journal}{Phys. Rev. Lett.}
  \textbf{\bibinfo{volume}{90}}(\bibinfo{number}{23}), \bibinfo{pages}{238701}.

\bibitem[{\citenamefont{{Krauss}}(1992)}]{krauss92}
\bibinfo{author}{\bibnamefont{{Krauss}}, \bibfnamefont{M.}},
  \bibinfo{year}{1992}, \bibinfo{journal}{Language}
  \textbf{\bibinfo{volume}{68}}, \bibinfo{pages}{4}.

\bibitem[{\citenamefont{Kubo} \emph{et~al.}(1985)\citenamefont{Kubo, Toda, and
  Hashitsume}}]{kubo85}
\bibinfo{author}{\bibnamefont{Kubo}, \bibfnamefont{R.}},
  \bibinfo{author}{\bibfnamefont{M.}~\bibnamefont{Toda}}, and
  \bibinfo{author}{\bibfnamefont{N.}~\bibnamefont{Hashitsume}},
  \bibinfo{year}{1985}, \emph{\bibinfo{title}{Statistical Physics}}
  (\bibinfo{publisher}{Springer Verlag}, \bibinfo{address}{Berlin, Germany}).

\bibitem[{\citenamefont{Kuperman}(2006)}]{kuperman06}
\bibinfo{author}{\bibnamefont{Kuperman}, \bibfnamefont{M.~N.}},
  \bibinfo{year}{2006}, \bibinfo{journal}{Phys. Rev. E}
  \textbf{\bibinfo{volume}{73}}(\bibinfo{number}{4}), \bibinfo{pages}{046139}.

\bibitem[{\citenamefont{Kuperman and Zanette}(2002)}]{kuperman02}
\bibinfo{author}{\bibnamefont{Kuperman}, \bibfnamefont{M.~N.}}, and
  \bibinfo{author}{\bibfnamefont{D.}~\bibnamefont{Zanette}},
  \bibinfo{year}{2002}, \bibinfo{journal}{Eur. Phys. J. B}
  \textbf{\bibinfo{volume}{26}}, \bibinfo{pages}{387}.

\bibitem[{\citenamefont{{Kuramoto}}(1975)}]{kuramoto75}
\bibinfo{author}{\bibnamefont{{Kuramoto}}, \bibfnamefont{Y.}},
  \bibinfo{year}{1975}, in \emph{\bibinfo{booktitle}{Mathematical Problems in
  Theoretical Physics}}, edited by
  \bibinfo{editor}{\bibfnamefont{H.}~\bibnamefont{{Araki}}}
  (\bibinfo{publisher}{Springer Verlag}, \bibinfo{address}{Berlin, Germany}),
  volume~\bibinfo{volume}{39} of \emph{\bibinfo{series}{Lecture Notes in
  Physics}}, pp. \bibinfo{pages}{420--422}.

\bibitem[{\citenamefont{{Lacasa} and {Luque}}(2006)}]{lacasa06}
\bibinfo{author}{\bibnamefont{{Lacasa}}, \bibfnamefont{L.}}, and
  \bibinfo{author}{\bibfnamefont{B.}~\bibnamefont{{Luque}}},
  \bibinfo{year}{2006}, \bibinfo{journal}{Physica A}
  \textbf{\bibinfo{volume}{366}}, \bibinfo{pages}{472}.

\bibitem[{\citenamefont{{Laguna}} \emph{et~al.}(2003)\citenamefont{{Laguna},
  {Abramson}, and {Zanette}}}]{laguna03}
\bibinfo{author}{\bibnamefont{{Laguna}}, \bibfnamefont{M.~F.}},
  \bibinfo{author}{\bibfnamefont{G.}~\bibnamefont{{Abramson}}}, and
  \bibinfo{author}{\bibfnamefont{D.~H.} \bibnamefont{{Zanette}}},
  \bibinfo{year}{2003}, \bibinfo{journal}{Physica A}
  \textbf{\bibinfo{volume}{329}}, \bibinfo{pages}{459}.

\bibitem[{\citenamefont{Laguna} \emph{et~al.}(2004)\citenamefont{Laguna,
  Abramson, and Zanette}}]{laguna04}
\bibinfo{author}{\bibnamefont{Laguna}, \bibfnamefont{M.~F.}},
  \bibinfo{author}{\bibfnamefont{G.}~\bibnamefont{Abramson}}, and
  \bibinfo{author}{\bibfnamefont{D.~H.} \bibnamefont{Zanette}},
  \bibinfo{year}{2004}, \bibinfo{journal}{Complexity Archive}
  \textbf{\bibinfo{volume}{9}}(\bibinfo{number}{4}), \bibinfo{pages}{31}.

\bibitem[{\citenamefont{{Laguna}} \emph{et~al.}(2005)\citenamefont{{Laguna},
  {Risau Gusman}, {Abramson}, {Gon{\c c}alves}, and {Iglesias}}}]{laguna05}
\bibinfo{author}{\bibnamefont{{Laguna}}, \bibfnamefont{M.~F.}},
  \bibinfo{author}{\bibfnamefont{S.}~\bibnamefont{{Risau Gusman}}},
  \bibinfo{author}{\bibfnamefont{G.}~\bibnamefont{{Abramson}}},
  \bibinfo{author}{\bibfnamefont{S.}~\bibnamefont{{Gon{\c c}alves}}}, and
  \bibinfo{author}{\bibfnamefont{J.~R.} \bibnamefont{{Iglesias}}},
  \bibinfo{year}{2005}, \bibinfo{journal}{Physica A}
  \textbf{\bibinfo{volume}{351}}, \bibinfo{pages}{580}.

\bibitem[{\citenamefont{{Lambiotte}}(2007)}]{lambiotte07}
\bibinfo{author}{\bibnamefont{{Lambiotte}}, \bibfnamefont{R.}},
  \bibinfo{year}{2007}, \bibinfo{journal}{Europhys. Lett.}
  \textbf{\bibinfo{volume}{78}}, \bibinfo{pages}{68002}.

\bibitem[{\citenamefont{Lambiotte and Ausloos}(2007)}]{lambiotte07a}
\bibinfo{author}{\bibnamefont{Lambiotte}, \bibfnamefont{R.}}, and
  \bibinfo{author}{\bibfnamefont{M.}~\bibnamefont{Ausloos}},
  \bibinfo{year}{2007}, \bibinfo{journal}{J. Stat. Mech.}
  \textbf{\bibinfo{volume}{P08026}}.

\bibitem[{\citenamefont{Lambiotte}
  \emph{et~al.}(2007{\natexlab{a}})\citenamefont{Lambiotte, Ausloos, and
  Ho{\l}yst}}]{lambiotte06}
\bibinfo{author}{\bibnamefont{Lambiotte}, \bibfnamefont{R.}},
  \bibinfo{author}{\bibfnamefont{M.}~\bibnamefont{Ausloos}}, and
  \bibinfo{author}{\bibfnamefont{J.~A.} \bibnamefont{Ho{\l}yst}},
  \bibinfo{year}{2007}{\natexlab{a}}, \bibinfo{journal}{Phys. Rev. E}
  \textbf{\bibinfo{volume}{75}}(\bibinfo{number}{3}), \bibinfo{pages}{030101}.

\bibitem[{\citenamefont{{Lambiotte} and {Redner}}(2007)}]{lambiotte07c}
\bibinfo{author}{\bibnamefont{{Lambiotte}}, \bibfnamefont{R.}}, and
  \bibinfo{author}{\bibfnamefont{S.}~\bibnamefont{{Redner}}},
  \bibinfo{year}{2007}, \eprint{arXiv:0712.0364}.

\bibitem[{\citenamefont{Lambiotte and Redner}(2007)}]{lambiotte07b}
\bibinfo{author}{\bibnamefont{Lambiotte}, \bibfnamefont{R.}}, and
  \bibinfo{author}{\bibfnamefont{S.}~\bibnamefont{Redner}},
  \bibinfo{year}{2007}, \bibinfo{journal}{J. Stat. Mech.}
  \textbf{\bibinfo{volume}{L10001}}.

\bibitem[{\citenamefont{Lambiotte}
  \emph{et~al.}(2007{\natexlab{b}})\citenamefont{Lambiotte, Thurner, and
  Hanel}}]{lambiotte07d}
\bibinfo{author}{\bibnamefont{Lambiotte}, \bibfnamefont{R.}},
  \bibinfo{author}{\bibfnamefont{S.}~\bibnamefont{Thurner}}, and
  \bibinfo{author}{\bibfnamefont{R.}~\bibnamefont{Hanel}},
  \bibinfo{year}{2007}{\natexlab{b}}, \bibinfo{journal}{Phys. Rev. E}
  \textbf{\bibinfo{volume}{76}}(\bibinfo{number}{4}), \bibinfo{eid}{046101}
  (pages~\bibinfo{numpages}{8}).

\bibitem[{\citenamefont{{Landau} and {Binder}}(2005)}]{landau05}
\bibinfo{author}{\bibnamefont{{Landau}}, \bibfnamefont{D.}}, and
  \bibinfo{author}{\bibfnamefont{K.}~\bibnamefont{{Binder}}},
  \bibinfo{year}{2005}, \emph{\bibinfo{title}{A Guide to Monte Carlo
  Simulations in Statistical Physics}} (\bibinfo{publisher}{Cambridge
  University Press}, \bibinfo{address}{Cambridge, UK}).

\bibitem[{\citenamefont{Landau}(1951{\natexlab{a}})}]{landau51}
\bibinfo{author}{\bibnamefont{Landau}, \bibfnamefont{H.~G.}},
  \bibinfo{year}{1951}{\natexlab{a}}, \bibinfo{journal}{Bull. Math. Biophys.}
  \textbf{\bibinfo{volume}{13}}, \bibinfo{pages}{1}.

\bibitem[{\citenamefont{Landau}(1951{\natexlab{b}})}]{landau51a}
\bibinfo{author}{\bibnamefont{Landau}, \bibfnamefont{H.~G.}},
  \bibinfo{year}{1951}{\natexlab{b}}, \bibinfo{journal}{Bull. Math. Biophys.}
  \textbf{\bibinfo{volume}{13}}, \bibinfo{pages}{245}.

\bibitem[{\citenamefont{Langton}(1996)}]{langton96}
\bibinfo{author}{\bibnamefont{Langton}, \bibfnamefont{C.}},
  \bibinfo{year}{1996}, \emph{\bibinfo{title}{Artificial Life}}
  (\bibinfo{publisher}{Addison-Wesley Longman Publishing Co., Inc.},
  \bibinfo{address}{Boston, MA, USA}).

\bibitem[{\citenamefont{{Latan\'e}}(1981)}]{latane81}
\bibinfo{author}{\bibnamefont{{Latan\'e}}, \bibfnamefont{B.}},
  \bibinfo{year}{1981}, \bibinfo{journal}{Am. Psychol.}
  \textbf{\bibinfo{volume}{36}}, \bibinfo{pages}{343}.

\bibitem[{\citenamefont{Lenaerts} \emph{et~al.}(2005)\citenamefont{Lenaerts,
  Jansen, Tuyls, and de~Vylder}}]{lenaerts05}
\bibinfo{author}{\bibnamefont{Lenaerts}, \bibfnamefont{T.}},
  \bibinfo{author}{\bibfnamefont{B.}~\bibnamefont{Jansen}},
  \bibinfo{author}{\bibfnamefont{K.}~\bibnamefont{Tuyls}}, and
  \bibinfo{author}{\bibfnamefont{B.}~\bibnamefont{de~Vylder}},
  \bibinfo{year}{2005}, \bibinfo{journal}{J. Theor. Bio.}
  \textbf{\bibinfo{volume}{235}}(\bibinfo{number}{4}), \bibinfo{pages}{566}.

\bibitem[{\citenamefont{{Leone}} \emph{et~al.}(2002)\citenamefont{{Leone},
  {V{\'a}zquez}, {Vespignani}, and {Zecchina}}}]{leone02}
\bibinfo{author}{\bibnamefont{{Leone}}, \bibfnamefont{M.}},
  \bibinfo{author}{\bibfnamefont{A.}~\bibnamefont{{V{\'a}zquez}}},
  \bibinfo{author}{\bibfnamefont{A.}~\bibnamefont{{Vespignani}}}, and
  \bibinfo{author}{\bibfnamefont{R.}~\bibnamefont{{Zecchina}}},
  \bibinfo{year}{2002}, \bibinfo{journal}{Eur. Phys. J. B}
  \textbf{\bibinfo{volume}{28}}, \bibinfo{pages}{191}.

\bibitem[{\citenamefont{Leskovec} \emph{et~al.}(2006)\citenamefont{Leskovec,
  Adamic, and Huberman}}]{leskovec06}
\bibinfo{author}{\bibnamefont{Leskovec}, \bibfnamefont{J.}},
  \bibinfo{author}{\bibfnamefont{L.}~\bibnamefont{Adamic}}, and
  \bibinfo{author}{\bibfnamefont{B.}~\bibnamefont{Huberman}},
  \bibinfo{year}{2006}, in \emph{\bibinfo{booktitle}{EC 06: Proceedings of the
  7th ACM conference on Electronic commerce}} (\bibinfo{publisher}{ACM Press},
  \bibinfo{address}{New York, NY, USA}), pp. \bibinfo{pages}{228--237}.

\bibitem[{\citenamefont{{Lewenstein}}
  \emph{et~al.}(1992)\citenamefont{{Lewenstein}, {Nowak}, and
  {Latan{\'e}}}}]{lewenstein92}
\bibinfo{author}{\bibnamefont{{Lewenstein}}, \bibfnamefont{M.}},
  \bibinfo{author}{\bibfnamefont{A.}~\bibnamefont{{Nowak}}}, and
  \bibinfo{author}{\bibfnamefont{B.}~\bibnamefont{{Latan{\'e}}}},
  \bibinfo{year}{1992}, \bibinfo{journal}{Phys. Rev. A}
  \textbf{\bibinfo{volume}{45}}, \bibinfo{pages}{763}.

\bibitem[{\citenamefont{Lewis}(1969)}]{lewis69}
\bibinfo{author}{\bibnamefont{Lewis}, \bibfnamefont{D.}}, \bibinfo{year}{1969},
  \emph{\bibinfo{title}{Convention: a philosophical study}}
  (\bibinfo{publisher}{Harvard University Press}, \bibinfo{address}{Cambridge,
  MA, USA}).

\bibitem[{\citenamefont{Li} \emph{et~al.}(2006)\citenamefont{Li, Zheng, and
  Hui}}]{li06}
\bibinfo{author}{\bibnamefont{Li}, \bibfnamefont{P.-P.}},
  \bibinfo{author}{\bibfnamefont{D.-F.} \bibnamefont{Zheng}}, and
  \bibinfo{author}{\bibfnamefont{P.~M.} \bibnamefont{Hui}},
  \bibinfo{year}{2006}, \bibinfo{journal}{Phys. Rev. E}
  \textbf{\bibinfo{volume}{73}}(\bibinfo{number}{5}), \bibinfo{pages}{056128}.

\bibitem[{\citenamefont{Lieberman} \emph{et~al.}(2007)\citenamefont{Lieberman,
  Michel, Jackson, Tang, and Nowak}}]{lieberman07}
\bibinfo{author}{\bibnamefont{Lieberman}, \bibfnamefont{E.}},
  \bibinfo{author}{\bibfnamefont{J.-B.} \bibnamefont{Michel}},
  \bibinfo{author}{\bibfnamefont{J.}~\bibnamefont{Jackson}},
  \bibinfo{author}{\bibfnamefont{T.}~\bibnamefont{Tang}}, and
  \bibinfo{author}{\bibfnamefont{M.~A.} \bibnamefont{Nowak}},
  \bibinfo{year}{2007}, \bibinfo{journal}{Nature}
  \textbf{\bibinfo{volume}{449}}(\bibinfo{number}{7163}), \bibinfo{pages}{713}.

\bibitem[{\citenamefont{Liggett}(1985)}]{liggett85}
\bibinfo{author}{\bibnamefont{Liggett}, \bibfnamefont{T.~M.}},
  \bibinfo{year}{1985}, \emph{\bibinfo{title}{Interacting particle systems}}
  (\bibinfo{publisher}{Springer Verlag}, \bibinfo{address}{New York, NY, USA}).

\bibitem[{\citenamefont{Liggett}(1999)}]{liggett99}
\bibinfo{author}{\bibnamefont{Liggett}, \bibfnamefont{T.~M.}},
  \bibinfo{year}{1999}, \emph{\bibinfo{title}{Stochastic interacting systems:
  contact, voter and exclusion processes}} (\bibinfo{publisher}{Springer
  Verlag}, \bibinfo{address}{Berlin, Germany}).

\bibitem[{\citenamefont{Liljeros} \emph{et~al.}(2001)\citenamefont{Liljeros,
  Edling, L, Stanley, and {\AA}berg}}]{lilijeros01}
\bibinfo{author}{\bibnamefont{Liljeros}, \bibfnamefont{F.}},
  \bibinfo{author}{\bibfnamefont{C.}~\bibnamefont{Edling}},
  \bibinfo{author}{\bibfnamefont{A.}~\bibnamefont{L}},
  \bibinfo{author}{\bibfnamefont{E.}~\bibnamefont{Stanley}}, and
  \bibinfo{author}{\bibfnamefont{Y.}~\bibnamefont{{\AA}berg}},
  \bibinfo{year}{2001}, \bibinfo{journal}{Nature}
  \textbf{\bibinfo{volume}{411}}, \bibinfo{pages}{907}.

\bibitem[{\citenamefont{{Lima}} \emph{et~al.}(2008)\citenamefont{{Lima},
  {Sousa}, and {Sumuor}}}]{lima08}
\bibinfo{author}{\bibnamefont{{Lima}}, \bibfnamefont{F.~W.~S.}},
  \bibinfo{author}{\bibfnamefont{A.~O.} \bibnamefont{{Sousa}}}, and
  \bibinfo{author}{\bibfnamefont{M.~A.} \bibnamefont{{Sumuor}}},
  \bibinfo{year}{2008}, \eprint{arXiv:0801.4250}.

\bibitem[{\citenamefont{Lind}
  \emph{et~al.}(2007{\natexlab{a}})\citenamefont{Lind, da~Silva, jr., and
  Herrmann}}]{lind07}
\bibinfo{author}{\bibnamefont{Lind}, \bibfnamefont{P.}},
  \bibinfo{author}{\bibfnamefont{L.}~\bibnamefont{da~Silva}},
  \bibinfo{author}{\bibfnamefont{J.~A.} \bibnamefont{jr.}}, and
  \bibinfo{author}{\bibfnamefont{H.}~\bibnamefont{Herrmann}},
  \bibinfo{year}{2007}{\natexlab{a}}, \bibinfo{journal}{Europhys. Lett.}
  \textbf{\bibinfo{volume}{78}}(\bibinfo{number}{6}), \bibinfo{pages}{68005
  (5pp)}.

\bibitem[{\citenamefont{Lind}
  \emph{et~al.}(2007{\natexlab{b}})\citenamefont{Lind, da~Silva, J.~S.~Andrade,
  and Herrmann}}]{lind07b}
\bibinfo{author}{\bibnamefont{Lind}, \bibfnamefont{P.~G.}},
  \bibinfo{author}{\bibfnamefont{L.~R.} \bibnamefont{da~Silva}},
  \bibinfo{author}{\bibfnamefont{J.}~\bibnamefont{J.~S.~Andrade}}, and
  \bibinfo{author}{\bibfnamefont{H.~J.} \bibnamefont{Herrmann}},
  \bibinfo{year}{2007}{\natexlab{b}}, \bibinfo{journal}{Phys. Rev. E}
  \textbf{\bibinfo{volume}{76}}(\bibinfo{number}{3}), \bibinfo{eid}{036117}
  (pages~\bibinfo{numpages}{10}).

\bibitem[{\citenamefont{{Lipowski} and {Lipowska}}(2008)}]{lipowski08}
\bibinfo{author}{\bibnamefont{{Lipowski}}, \bibfnamefont{A.}}, and
  \bibinfo{author}{\bibfnamefont{D.}~\bibnamefont{{Lipowska}}},
  \bibinfo{year}{2008}, \eprint{arXiv:0801.1658}.

\bibitem[{\citenamefont{Liu} \emph{et~al.}(2003)\citenamefont{Liu, Lai, and
  Ye}}]{liu03b}
\bibinfo{author}{\bibnamefont{Liu}, \bibfnamefont{Z.}},
  \bibinfo{author}{\bibfnamefont{Y.-C.} \bibnamefont{Lai}}, and
  \bibinfo{author}{\bibfnamefont{N.}~\bibnamefont{Ye}}, \bibinfo{year}{2003},
  \bibinfo{journal}{Phys. Rev. E}
  \textbf{\bibinfo{volume}{67}}(\bibinfo{number}{3}), \bibinfo{pages}{031911}.

\bibitem[{\citenamefont{Llas} \emph{et~al.}(2003)\citenamefont{Llas, Gleiser,
  L\'opez, and D\'iaz-Guilera}}]{llas03}
\bibinfo{author}{\bibnamefont{Llas}, \bibfnamefont{M.}},
  \bibinfo{author}{\bibfnamefont{P.~M.} \bibnamefont{Gleiser}},
  \bibinfo{author}{\bibfnamefont{J.~M.} \bibnamefont{L\'opez}}, and
  \bibinfo{author}{\bibfnamefont{A.}~\bibnamefont{D\'iaz-Guilera}},
  \bibinfo{year}{2003}, \bibinfo{journal}{Phys. Rev. E}
  \textbf{\bibinfo{volume}{68}}(\bibinfo{number}{6}), \bibinfo{pages}{066101}.

\bibitem[{\citenamefont{Lloyd and May}(2001)}]{lloyd01}
\bibinfo{author}{\bibnamefont{Lloyd}, \bibfnamefont{A.}}, and
  \bibinfo{author}{\bibfnamefont{R.}~\bibnamefont{May}}, \bibinfo{year}{2001},
  \bibinfo{journal}{Science}
  \textbf{\bibinfo{volume}{292}}(\bibinfo{number}{5520}),
  \bibinfo{pages}{1316}.

\bibitem[{\citenamefont{{Lorenz}}(2005{\natexlab{a}})}]{lorenz05}
\bibinfo{author}{\bibnamefont{{Lorenz}}, \bibfnamefont{J.}},
  \bibinfo{year}{2005}{\natexlab{a}}, \bibinfo{journal}{Physica A}
  \textbf{\bibinfo{volume}{355}}, \bibinfo{pages}{217}.

\bibitem[{\citenamefont{{Lorenz}}(2005{\natexlab{b}})}]{lorenz07c}
\bibinfo{author}{\bibnamefont{{Lorenz}}, \bibfnamefont{J.}},
  \bibinfo{year}{2005}{\natexlab{b}}, in \emph{\bibinfo{booktitle}{Proc. IASTED
  Conf. MSO 2005}}, edited by
  \bibinfo{editor}{\bibfnamefont{G.}~\bibnamefont{{Tonella}}},
  \eprint{arxiv:0708.3293}.

\bibitem[{\citenamefont{Lorenz}(2006)}]{lorenz06}
\bibinfo{author}{\bibnamefont{Lorenz}, \bibfnamefont{J.}},
  \bibinfo{year}{2006}, \bibinfo{journal}{JASSS}
  \textbf{\bibinfo{volume}{9}}(\bibinfo{number}{1}).

\bibitem[{\citenamefont{{Lorenz}}(2006)}]{lorenz07a}
\bibinfo{author}{\bibnamefont{{Lorenz}}, \bibfnamefont{J.}},
  \bibinfo{year}{2006}, \bibinfo{journal}{Eur. J. Econ. Soc. Syst.}
  \textbf{\bibinfo{volume}{19}}, \bibinfo{pages}{213}.

\bibitem[{\citenamefont{{Lorenz}}(2007)}]{lorenz07}
\bibinfo{author}{\bibnamefont{{Lorenz}}, \bibfnamefont{J.}},
  \bibinfo{year}{2007}, \bibinfo{journal}{Int. J. Mod. Phys. C}
  \textbf{\bibinfo{volume}{18}}(\bibinfo{number}{12}), \bibinfo{pages}{1819}.

\bibitem[{\citenamefont{{Lorenz}}(2008{\natexlab{a}})}]{lorenz08a}
\bibinfo{author}{\bibnamefont{{Lorenz}}, \bibfnamefont{J.}},
  \bibinfo{year}{2008}{\natexlab{a}}, in \emph{\bibinfo{booktitle}{Managing
  Complexity: Insights, Concepts, Applications}}, edited by
  \bibinfo{editor}{\bibfnamefont{D.}~\bibnamefont{Helbing}}
  (\bibinfo{publisher}{Springer}), \eprint{arxiv:0708.3172}.

\bibitem[{\citenamefont{{Lorenz}}(2008{\natexlab{b}})}]{lorenz08}
\bibinfo{author}{\bibnamefont{{Lorenz}}, \bibfnamefont{J.}},
  \bibinfo{year}{2008}{\natexlab{b}}, \eprint{arXiv:0801.1399}.

\bibitem[{\citenamefont{{Lorenz} and {Urbig}}(2007)}]{lorenz07d}
\bibinfo{author}{\bibnamefont{{Lorenz}}, \bibfnamefont{J.}}, and
  \bibinfo{author}{\bibfnamefont{D.}~\bibnamefont{{Urbig}}},
  \bibinfo{year}{2007}, \bibinfo{journal}{Adv. Compl. Syst.}
  \textbf{\bibinfo{volume}{10}}(\bibinfo{number}{2}), \bibinfo{pages}{251}.

\bibitem[{\citenamefont{Loreto and Steels}(2007)}]{loreto07}
\bibinfo{author}{\bibnamefont{Loreto}, \bibfnamefont{V.}}, and
  \bibinfo{author}{\bibfnamefont{L.}~\bibnamefont{Steels}},
  \bibinfo{year}{2007}, \bibinfo{journal}{Nat. Phys.}
  \textbf{\bibinfo{volume}{3}}, \bibinfo{pages}{758}.

\bibitem[{\citenamefont{Lu} \emph{et~al.}(2008)\citenamefont{Lu, Korniss, and
  Szymanski}}]{lu08}
\bibinfo{author}{\bibnamefont{Lu}, \bibfnamefont{Q.}},
  \bibinfo{author}{\bibfnamefont{G.}~\bibnamefont{Korniss}}, and
  \bibinfo{author}{\bibfnamefont{B.~K.} \bibnamefont{Szymanski}},
  \bibinfo{year}{2008}, \bibinfo{journal}{Phys. Rev. E}
  \textbf{\bibinfo{volume}{77}}(\bibinfo{number}{1}), \bibinfo{eid}{016111}
  (pages~\bibinfo{numpages}{10}).

\bibitem[{\citenamefont{Lux}(2007)}]{lux07}
\bibinfo{author}{\bibnamefont{Lux}, \bibfnamefont{T.}}, \bibinfo{year}{2007},
  \urlprefix\url{http://www2.warwick.ac.uk/fac/soc/wbs/research/wfri/wpaperser%
ies/wf06-256.pdf}.

\bibitem[{\citenamefont{Lyra} \emph{et~al.}(2003)\citenamefont{Lyra, Costa,
  Filho, and Jr.}}]{lyra03}
\bibinfo{author}{\bibnamefont{Lyra}, \bibfnamefont{M.~L.}},
  \bibinfo{author}{\bibfnamefont{U.~M.~S.} \bibnamefont{Costa}},
  \bibinfo{author}{\bibfnamefont{R.~N.~C.} \bibnamefont{Filho}}, and
  \bibinfo{author}{\bibfnamefont{J.~S.~A.} \bibnamefont{Jr.}},
  \bibinfo{year}{2003}, \bibinfo{journal}{Europhys. Lett.}
  \textbf{\bibinfo{volume}{62}}(\bibinfo{number}{1}), \bibinfo{pages}{131}.

\bibitem[{\citenamefont{Macy and Willer}(2002)}]{macy02}
\bibinfo{author}{\bibnamefont{Macy}, \bibfnamefont{M.}}, and
  \bibinfo{author}{\bibfnamefont{R.}~\bibnamefont{Willer}},
  \bibinfo{year}{2002}, \bibinfo{journal}{Ann. Rev. Sociol.}
  \textbf{\bibinfo{volume}{28}}, \bibinfo{pages}{143}.

\bibitem[{\citenamefont{Maes}(1991)}]{maes91}
\bibinfo{author}{\bibnamefont{Maes}, \bibfnamefont{P.}}, \bibinfo{year}{1991},
  \emph{\bibinfo{title}{Designing Autonomous Agents: Theory and Practice from
  Biology to Engineering and Back}} (\bibinfo{publisher}{MIT Press},
  \bibinfo{address}{Cambridge, MA, USA}).

\bibitem[{\citenamefont{Majorana}(1942)}]{majorana42}
\bibinfo{author}{\bibnamefont{Majorana}, \bibfnamefont{E.}},
  \bibinfo{year}{1942}, \bibinfo{journal}{Scientia}
  \textbf{\bibinfo{volume}{Feb-Mar}}, \bibinfo{pages}{58}.

\bibitem[{\citenamefont{Majorana}(2005)}]{majorana42eng}
\bibinfo{author}{\bibnamefont{Majorana}, \bibfnamefont{E.}},
  \bibinfo{year}{2005}, \bibinfo{journal}{J. Quantit. Fin.}
  \textbf{\bibinfo{volume}{5}}, \bibinfo{pages}{133}.

\bibitem[{\citenamefont{Maki and Thompson}(1973)}]{maki73}
\bibinfo{author}{\bibnamefont{Maki}, \bibfnamefont{D.~P.}}, and
  \bibinfo{author}{\bibfnamefont{M.}~\bibnamefont{Thompson}},
  \bibinfo{year}{1973}, \emph{\bibinfo{title}{Mathematical Models and
  Applications, with Emphasis on the Social, Life, and Management Sciences}}
  (\bibinfo{publisher}{Prentice-Hall}, \bibinfo{address}{Englewood Cliffs,
  NJ}).

\bibitem[{\citenamefont{{Malarz}} \emph{et~al.}(2006)\citenamefont{{Malarz},
  {Stauffer}, and {Ku{\l}akowski}}}]{malarz06b}
\bibinfo{author}{\bibnamefont{{Malarz}}, \bibfnamefont{K.}},
  \bibinfo{author}{\bibfnamefont{D.}~\bibnamefont{{Stauffer}}}, and
  \bibinfo{author}{\bibfnamefont{K.}~\bibnamefont{{Ku{\l}akowski}}},
  \bibinfo{year}{2006}, \bibinfo{journal}{Eur. Phys. J. B}
  \textbf{\bibinfo{volume}{50}}, \bibinfo{pages}{195}.

\bibitem[{\citenamefont{Maniccam}(2003)}]{maniccam03}
\bibinfo{author}{\bibnamefont{Maniccam}, \bibfnamefont{S.}},
  \bibinfo{year}{2003}, \bibinfo{journal}{Physica A}
  \textbf{\bibinfo{volume}{321}}, \bibinfo{pages}{653}.

\bibitem[{\citenamefont{{Mantegna} and {Stanley}}(1999)}]{mantegna99}
\bibinfo{author}{\bibnamefont{{Mantegna}}, \bibfnamefont{R.~N.}}, and
  \bibinfo{author}{\bibfnamefont{H.~E.} \bibnamefont{{Stanley}}},
  \bibinfo{year}{1999}, \emph{\bibinfo{title}{{Introduction to Econophysics}}}
  (\bibinfo{publisher}{Cambridge University Press},
  \bibinfo{address}{Cambridge, UK}).

\bibitem[{\citenamefont{Marsili}(1994)}]{marsili94}
\bibinfo{author}{\bibnamefont{Marsili}, \bibfnamefont{M.}},
  \bibinfo{year}{1994}, \bibinfo{journal}{J. Stat. Phys.}
  \textbf{\bibinfo{volume}{V77}}(\bibinfo{number}{3}), \bibinfo{pages}{733}.

\bibitem[{\citenamefont{{Martins}}(2007)}]{martins07}
\bibinfo{author}{\bibnamefont{{Martins}}, \bibfnamefont{A.~C.~R.}},
  \bibinfo{year}{2007}, \eprint{arXiv:0711.1199}.

\bibitem[{\citenamefont{{Martins}}(2008)}]{martins08}
\bibinfo{author}{\bibnamefont{{Martins}}, \bibfnamefont{A.~C.~R.}},
  \bibinfo{year}{2008}, \eprint{arXiv:0801.2411}.

\bibitem[{\citenamefont{Matsuda} \emph{et~al.}(1992)\citenamefont{Matsuda,
  Ogita, Sasaki, and Sato}}]{matsuda92}
\bibinfo{author}{\bibnamefont{Matsuda}, \bibfnamefont{H.}},
  \bibinfo{author}{\bibfnamefont{N.}~\bibnamefont{Ogita}},
  \bibinfo{author}{\bibfnamefont{A.}~\bibnamefont{Sasaki}}, and
  \bibinfo{author}{\bibfnamefont{K.}~\bibnamefont{Sato}}, \bibinfo{year}{1992},
  \bibinfo{journal}{Prog. Theor. Phys.}
  \textbf{\bibinfo{volume}{88}}(\bibinfo{number}{6}), \bibinfo{pages}{1035}.

\bibitem[{\citenamefont{May}(2006)}]{may06}
\bibinfo{author}{\bibnamefont{May}, \bibfnamefont{R.}}, \bibinfo{year}{2006},
  \bibinfo{journal}{Trends Ecol. Evol.} \textbf{\bibinfo{volume}{21}},
  \bibinfo{pages}{394}.

\bibitem[{\citenamefont{{Mazzitello}}
  \emph{et~al.}(2007)\citenamefont{{Mazzitello}, {Candia}, and
  {Dossetti}}}]{mazzitello07}
\bibinfo{author}{\bibnamefont{{Mazzitello}}, \bibfnamefont{K.~I.}},
  \bibinfo{author}{\bibfnamefont{J.}~\bibnamefont{{Candia}}}, and
  \bibinfo{author}{\bibfnamefont{V.}~\bibnamefont{{Dossetti}}},
  \bibinfo{year}{2007}, \bibinfo{journal}{Int. J. Mod. Phys. C}
  \textbf{\bibinfo{volume}{18}}(\bibinfo{number}{9}), \bibinfo{pages}{1475}.

\bibitem[{\citenamefont{McCarthy}(1959)}]{mccarthy59}
\bibinfo{author}{\bibnamefont{McCarthy}, \bibfnamefont{J.}},
  \bibinfo{year}{1959}, in \emph{\bibinfo{booktitle}{Proceedings of the
  {T}eddington Conference on the Mechanization of Thought Processes}}
  (\bibinfo{publisher}{Her Majesty's Stationary Office},
  \bibinfo{address}{London, UK}), pp. \bibinfo{pages}{75--91}.

\bibitem[{\citenamefont{Meakin and Scalapino}(1987)}]{meakin87}
\bibinfo{author}{\bibnamefont{Meakin}, \bibfnamefont{P.}}, and
  \bibinfo{author}{\bibfnamefont{D.~J.} \bibnamefont{Scalapino}},
  \bibinfo{year}{1987}, \bibinfo{journal}{J. Chem. Phys.}
  \textbf{\bibinfo{volume}{87}}(\bibinfo{number}{1}), \bibinfo{pages}{731}.

\bibitem[{\citenamefont{Mermin and Wagner}(1966)}]{mermin66}
\bibinfo{author}{\bibnamefont{Mermin}, \bibfnamefont{N.~D.}}, and
  \bibinfo{author}{\bibfnamefont{H.}~\bibnamefont{Wagner}},
  \bibinfo{year}{1966}, \bibinfo{journal}{Phys. Rev. Lett.}
  \textbf{\bibinfo{volume}{17}}(\bibinfo{number}{22}), \bibinfo{pages}{1133}.

\bibitem[{\citenamefont{Metropolis}
  \emph{et~al.}(1953)\citenamefont{Metropolis, Rosenbluth, Rosenbluth, Teller,
  and Teller}}]{metropolis53}
\bibinfo{author}{\bibnamefont{Metropolis}, \bibfnamefont{N.}},
  \bibinfo{author}{\bibfnamefont{A.}~\bibnamefont{Rosenbluth}},
  \bibinfo{author}{\bibfnamefont{M.}~\bibnamefont{Rosenbluth}},
  \bibinfo{author}{\bibfnamefont{A.}~\bibnamefont{Teller}}, and
  \bibinfo{author}{\bibfnamefont{E.}~\bibnamefont{Teller}},
  \bibinfo{year}{1953}, \bibinfo{journal}{J. Chem. Phys.}
  \textbf{\bibinfo{volume}{21}}, \bibinfo{pages}{1087}.

\bibitem[{\citenamefont{Meyer and Wilson}(1990)}]{meyer90}
\bibinfo{editor}{\bibnamefont{Meyer}, \bibfnamefont{J.-A.}}, and
  \bibinfo{editor}{\bibfnamefont{S.~W.} \bibnamefont{Wilson}} (eds.),
  \bibinfo{year}{1990}, \emph{\bibinfo{title}{From animals to animats:
  proceedings of the First International Conference on Simulation of Adaptive
  Behavior}} (\bibinfo{publisher}{MIT Press}, \bibinfo{address}{Cambridge, MA,
  USA}).

\bibitem[{\citenamefont{{Mezard}} \emph{et~al.}(1987)\citenamefont{{Mezard},
  {Parisi}, and {Virasoro}}}]{mezard87}
\bibinfo{author}{\bibnamefont{{Mezard}}, \bibfnamefont{M.}},
  \bibinfo{author}{\bibfnamefont{G.}~\bibnamefont{{Parisi}}}, and
  \bibinfo{author}{\bibfnamefont{M.}~\bibnamefont{{Virasoro}}},
  \bibinfo{year}{1987}, \emph{\bibinfo{title}{Spin Glass Theory and Beyond}}
  (\bibinfo{publisher}{World Scientific Publishing},
  \bibinfo{address}{Singapore}).

\bibitem[{\citenamefont{{Michard} and {Bouchaud}}(2005)}]{michard05}
\bibinfo{author}{\bibnamefont{{Michard}}, \bibfnamefont{Q.}}, and
  \bibinfo{author}{\bibfnamefont{J.-P.} \bibnamefont{{Bouchaud}}},
  \bibinfo{year}{2005}, \bibinfo{journal}{Eur. Phys. J. B}
  \textbf{\bibinfo{volume}{47}}, \bibinfo{pages}{151}.

\bibitem[{\citenamefont{Milgram}(1967)}]{milgram67}
\bibinfo{author}{\bibnamefont{Milgram}, \bibfnamefont{S.}},
  \bibinfo{year}{1967}, \bibinfo{journal}{Psychol. Today}
  \textbf{\bibinfo{volume}{1}}, \bibinfo{pages}{60}.

\bibitem[{\citenamefont{Minett and Wang}(2008)}]{minett04->08}
\bibinfo{author}{\bibnamefont{Minett}, \bibfnamefont{J.~W.}}, and
  \bibinfo{author}{\bibfnamefont{W.~S.~Y.} \bibnamefont{Wang}},
  \bibinfo{year}{2008}, \bibinfo{journal}{Lingua}
  \textbf{\bibinfo{volume}{118}}(\bibinfo{number}{1}), \bibinfo{pages}{19},
  \bibinfo{note}{preprint 2004}.

\bibitem[{\citenamefont{Minsky}(1961)}]{minsky61}
\bibinfo{author}{\bibnamefont{Minsky}, \bibfnamefont{M.}},
  \bibinfo{year}{1961}, \bibinfo{journal}{Proc. Inst. Rad. Eng.}
  \textbf{\bibinfo{volume}{49}}, \bibinfo{pages}{8}.

\bibitem[{\citenamefont{{Mira} and {Paredes}}(2005)}]{mira05}
\bibinfo{author}{\bibnamefont{{Mira}}, \bibfnamefont{J.}}, and
  \bibinfo{author}{\bibfnamefont{{\'A}.}~\bibnamefont{{Paredes}}},
  \bibinfo{year}{2005}, \bibinfo{journal}{Europhys. Lett.}
  \textbf{\bibinfo{volume}{69}}, \bibinfo{pages}{1031}.

\bibitem[{\citenamefont{Mitzenmacher}(2003)}]{mitzenmacher03}
\bibinfo{author}{\bibnamefont{Mitzenmacher}, \bibfnamefont{M.}},
  \bibinfo{year}{2003}, \bibinfo{journal}{Internet Math.}
  \textbf{\bibinfo{volume}{1}}(\bibinfo{number}{2}).

\bibitem[{\citenamefont{Mobilia}(2003)}]{mobilia03}
\bibinfo{author}{\bibnamefont{Mobilia}, \bibfnamefont{M.}},
  \bibinfo{year}{2003}, \bibinfo{journal}{Phys. Rev. Lett.}
  \textbf{\bibinfo{volume}{91}}(\bibinfo{number}{2}), \bibinfo{pages}{028701}.

\bibitem[{\citenamefont{Mobilia and Georgiev}(2005)}]{mobilia05}
\bibinfo{author}{\bibnamefont{Mobilia}, \bibfnamefont{M.}}, and
  \bibinfo{author}{\bibfnamefont{I.~T.} \bibnamefont{Georgiev}},
  \bibinfo{year}{2005}, \bibinfo{journal}{Phys. Rev. E}
  \textbf{\bibinfo{volume}{71}}(\bibinfo{number}{4}), \bibinfo{pages}{046102}.

\bibitem[{\citenamefont{Mobilia} \emph{et~al.}(2007)\citenamefont{Mobilia,
  Petersen, and Redner}}]{mobilia07}
\bibinfo{author}{\bibnamefont{Mobilia}, \bibfnamefont{M.}},
  \bibinfo{author}{\bibfnamefont{A.}~\bibnamefont{Petersen}}, and
  \bibinfo{author}{\bibfnamefont{S.}~\bibnamefont{Redner}},
  \bibinfo{year}{2007}, \bibinfo{journal}{J. Stat. Mech.}
  \textbf{\bibinfo{volume}{P08029}}.

\bibitem[{\citenamefont{Mobilia and Redner}(2003)}]{mobilia03b}
\bibinfo{author}{\bibnamefont{Mobilia}, \bibfnamefont{M.}}, and
  \bibinfo{author}{\bibfnamefont{S.}~\bibnamefont{Redner}},
  \bibinfo{year}{2003}, \bibinfo{journal}{Phys. Rev. E}
  \textbf{\bibinfo{volume}{68}}(\bibinfo{number}{4}), \bibinfo{pages}{046106}.

\bibitem[{\citenamefont{Molloy and Reed}(1995)}]{molloy_reed95}
\bibinfo{author}{\bibnamefont{Molloy}, \bibfnamefont{M.}}, and
  \bibinfo{author}{\bibfnamefont{B.}~\bibnamefont{Reed}}, \bibinfo{year}{1995},
  \bibinfo{journal}{Random Struct. Algor.} \textbf{\bibinfo{volume}{6}},
  \bibinfo{pages}{161}.

\bibitem[{\citenamefont{Molloy and Reed}(1998)}]{molloy_reed98}
\bibinfo{author}{\bibnamefont{Molloy}, \bibfnamefont{M.}}, and
  \bibinfo{author}{\bibfnamefont{B.}~\bibnamefont{Reed}}, \bibinfo{year}{1998},
  \bibinfo{journal}{Comb. Probab. Comput.} \textbf{\bibinfo{volume}{7}},
  \bibinfo{pages}{295}.

\bibitem[{\citenamefont{{Morales-Matamoros}}
  \emph{et~al.}(2006)\citenamefont{{Morales-Matamoros}, {Mart\'inez-Cruz}, and
  {Tejeida-Padilla}}}]{morales06}
\bibinfo{author}{\bibnamefont{{Morales-Matamoros}}, \bibfnamefont{O.}},
  \bibinfo{author}{\bibfnamefont{M.~A.} \bibnamefont{{Mart\'inez-Cruz}}}, and
  \bibinfo{author}{\bibfnamefont{R.}~\bibnamefont{{Tejeida-Padilla}}},
  \bibinfo{year}{2006}, in \emph{\bibinfo{booktitle}{50th Annual Meeting of the
  ISSS}}.

\bibitem[{\citenamefont{{Moreira}} \emph{et~al.}(2001)\citenamefont{{Moreira},
  {Andrade}, and {Stauffer}}}]{moreira01}
\bibinfo{author}{\bibnamefont{{Moreira}}, \bibfnamefont{A.~A.}},
  \bibinfo{author}{\bibfnamefont{J.~S.} \bibnamefont{{Andrade}}}, and
  \bibinfo{author}{\bibfnamefont{D.}~\bibnamefont{{Stauffer}}},
  \bibinfo{year}{2001}, \bibinfo{journal}{Int. J. Mod. Phys. C}
  \textbf{\bibinfo{volume}{12}}, \bibinfo{pages}{39}.

\bibitem[{\citenamefont{Moreno}(1934)}]{moreno34}
\bibinfo{author}{\bibnamefont{Moreno}, \bibfnamefont{J.}},
  \bibinfo{year}{1934}, \emph{\bibinfo{title}{Who shall survive? : a new
  approach to the problem of Human Interrelations}},
  volume~\bibinfo{volume}{58} of \emph{\bibinfo{series}{Nervous and mental
  disease monograph series}} (\bibinfo{publisher}{Nervous and Mental Disease
  Publ.}, \bibinfo{address}{Washington, USA}).

\bibitem[{\citenamefont{Moreno}
  \emph{et~al.}(2004{\natexlab{a}})\citenamefont{Moreno, Nekovee, and
  Pacheco}}]{moreno04b}
\bibinfo{author}{\bibnamefont{Moreno}, \bibfnamefont{Y.}},
  \bibinfo{author}{\bibfnamefont{M.}~\bibnamefont{Nekovee}}, and
  \bibinfo{author}{\bibfnamefont{A.}~\bibnamefont{Pacheco}},
  \bibinfo{year}{2004}{\natexlab{a}}, \bibinfo{journal}{Phys. Rev. E}
  \textbf{\bibinfo{volume}{69}}(\bibinfo{number}{6}), \bibinfo{pages}{066130}.

\bibitem[{\citenamefont{Moreno}
  \emph{et~al.}(2004{\natexlab{b}})\citenamefont{Moreno, Nekovee, and
  Vespignani}}]{moreno04a}
\bibinfo{author}{\bibnamefont{Moreno}, \bibfnamefont{Y.}},
  \bibinfo{author}{\bibfnamefont{M.}~\bibnamefont{Nekovee}}, and
  \bibinfo{author}{\bibfnamefont{A.}~\bibnamefont{Vespignani}},
  \bibinfo{year}{2004}{\natexlab{b}}, \bibinfo{journal}{Phys. Rev. E}
  \textbf{\bibinfo{volume}{69}}(\bibinfo{number}{5}), \bibinfo{pages}{055101}.

\bibitem[{\citenamefont{Muramatsu} \emph{et~al.}(1999)\citenamefont{Muramatsu,
  Irie, and Nagatani}}]{muramatsu99}
\bibinfo{author}{\bibnamefont{Muramatsu}, \bibfnamefont{M.}},
  \bibinfo{author}{\bibfnamefont{T.}~\bibnamefont{Irie}}, and
  \bibinfo{author}{\bibfnamefont{T.}~\bibnamefont{Nagatani}},
  \bibinfo{year}{1999}, \bibinfo{journal}{Physica A}
  \textbf{\bibinfo{volume}{267}}, \bibinfo{pages}{487}.

\bibitem[{\citenamefont{Muramatsu and Nagatani}(2000)}]{muramatsu00a}
\bibinfo{author}{\bibnamefont{Muramatsu}, \bibfnamefont{M.}}, and
  \bibinfo{author}{\bibfnamefont{T.}~\bibnamefont{Nagatani}},
  \bibinfo{year}{2000}, \bibinfo{journal}{Physica A}
  \textbf{\bibinfo{volume}{286}}, \bibinfo{pages}{377}.

\bibitem[{\citenamefont{{Nagatani}}(2002)}]{nagatani02}
\bibinfo{author}{\bibnamefont{{Nagatani}}, \bibfnamefont{T.}},
  \bibinfo{year}{2002}, \bibinfo{journal}{Rep. Prog. in Phys.}
  \textbf{\bibinfo{volume}{65}}, \bibinfo{pages}{1331}.

\bibitem[{\citenamefont{Nakayama and Nakamura}(2004)}]{nakayama04}
\bibinfo{author}{\bibnamefont{Nakayama}, \bibfnamefont{S.}}, and
  \bibinfo{author}{\bibfnamefont{Y.}~\bibnamefont{Nakamura}},
  \bibinfo{year}{2004}, \bibinfo{journal}{Physica A}
  \textbf{\bibinfo{volume}{337}}, \bibinfo{pages}{625}.

\bibitem[{\citenamefont{{Nardini}} \emph{et~al.}(2007)\citenamefont{{Nardini},
  {Kozma}, and {Barrat}}}]{nardini07}
\bibinfo{author}{\bibnamefont{{Nardini}}, \bibfnamefont{C.}},
  \bibinfo{author}{\bibfnamefont{B.}~\bibnamefont{{Kozma}}}, and
  \bibinfo{author}{\bibfnamefont{A.}~\bibnamefont{{Barrat}}},
  \bibinfo{year}{2007}, \eprint{arXiv:0711.1261}.

\bibitem[{\citenamefont{Nash}(1950)}]{nash50}
\bibinfo{author}{\bibnamefont{Nash}, \bibfnamefont{J.}}, \bibinfo{year}{1950},
  \bibinfo{journal}{Proc. Natl. Acad. Sci. USA} \textbf{\bibinfo{volume}{36}},
  \bibinfo{pages}{48}.

\bibitem[{\citenamefont{N\'eda} \emph{et~al.}(2003)\citenamefont{N\'eda,
  Nikitin, and Vicsek}}]{neda03}
\bibinfo{author}{\bibnamefont{N\'eda}, \bibfnamefont{Z.}},
  \bibinfo{author}{\bibfnamefont{A.}~\bibnamefont{Nikitin}}, and
  \bibinfo{author}{\bibfnamefont{T.}~\bibnamefont{Vicsek}},
  \bibinfo{year}{2003}, \bibinfo{journal}{Physica A}
  \textbf{\bibinfo{volume}{321}}(\bibinfo{number}{1-2}), \bibinfo{pages}{238}.

\bibitem[{\citenamefont{{N{\'e}da}}
  \emph{et~al.}(2000)\citenamefont{{N{\'e}da}, {Ravasz}, {Brechet}, {Vicsek},
  and {Barab{\'a}si}}}]{neda00a}
\bibinfo{author}{\bibnamefont{{N{\'e}da}}, \bibfnamefont{Z.}},
  \bibinfo{author}{\bibfnamefont{E.}~\bibnamefont{{Ravasz}}},
  \bibinfo{author}{\bibfnamefont{Y.}~\bibnamefont{{Brechet}}},
  \bibinfo{author}{\bibfnamefont{T.}~\bibnamefont{{Vicsek}}}, and
  \bibinfo{author}{\bibfnamefont{A.-L.} \bibnamefont{{Barab{\'a}si}}},
  \bibinfo{year}{2000}, \bibinfo{journal}{Nature}
  \textbf{\bibinfo{volume}{403}}, \bibinfo{pages}{849}.

\bibitem[{\citenamefont{N\'eda} \emph{et~al.}(2000)\citenamefont{N\'eda,
  Ravasz, Vicsek, Brechet, and Barab\'asi}}]{neda00}
\bibinfo{author}{\bibnamefont{N\'eda}, \bibfnamefont{Z.}},
  \bibinfo{author}{\bibfnamefont{E.}~\bibnamefont{Ravasz}},
  \bibinfo{author}{\bibfnamefont{T.}~\bibnamefont{Vicsek}},
  \bibinfo{author}{\bibfnamefont{Y.}~\bibnamefont{Brechet}}, and
  \bibinfo{author}{\bibfnamefont{A.~L.} \bibnamefont{Barab\'asi}},
  \bibinfo{year}{2000}, \bibinfo{journal}{Phys. Rev. E}
  \textbf{\bibinfo{volume}{61}}(\bibinfo{number}{6}), \bibinfo{pages}{6987}.

\bibitem[{\citenamefont{Nekovee} \emph{et~al.}(2007)\citenamefont{Nekovee,
  Moreno, Bianconi, and Marsili}}]{nekovee07}
\bibinfo{author}{\bibnamefont{Nekovee}, \bibfnamefont{M.}},
  \bibinfo{author}{\bibfnamefont{Y.}~\bibnamefont{Moreno}},
  \bibinfo{author}{\bibfnamefont{G.}~\bibnamefont{Bianconi}}, and
  \bibinfo{author}{\bibfnamefont{M.}~\bibnamefont{Marsili}},
  \bibinfo{year}{2007}, \bibinfo{journal}{Physica A}
  \textbf{\bibinfo{volume}{374}}, \bibinfo{pages}{457}.

\bibitem[{\citenamefont{{Nettle}}(1999{\natexlab{a}})}]{nettle99a}
\bibinfo{author}{\bibnamefont{{Nettle}}, \bibfnamefont{D.}},
  \bibinfo{year}{1999}{\natexlab{a}}, \bibinfo{journal}{Lingua}
  \textbf{\bibinfo{volume}{108}}, \bibinfo{pages}{95}.

\bibitem[{\citenamefont{{Nettle}}(1999{\natexlab{b}})}]{nettle99b}
\bibinfo{author}{\bibnamefont{{Nettle}}, \bibfnamefont{D.}},
  \bibinfo{year}{1999}{\natexlab{b}}, \bibinfo{journal}{Lingua}
  \textbf{\bibinfo{volume}{108}}, \bibinfo{pages}{119}.

\bibitem[{\citenamefont{von Neumann and Morgenstern}(1947)}]{vonneumann47}
\bibinfo{author}{\bibnamefont{von Neumann}, \bibfnamefont{J.}}, and
  \bibinfo{author}{\bibfnamefont{O.}~\bibnamefont{Morgenstern}},
  \bibinfo{year}{1947}, \emph{\bibinfo{title}{Theory of games and economic
  behavior}} (\bibinfo{publisher}{Princeton University Press},
  \bibinfo{address}{Princeton, USA}).

\bibitem[{\citenamefont{Neumann}(1966)}]{vonneumann66}
\bibinfo{author}{\bibnamefont{Neumann}, \bibfnamefont{J.~V.}},
  \bibinfo{year}{1966}, \emph{\bibinfo{title}{Theory of Self-Reproducing
  Automata}} (\bibinfo{publisher}{University of Illinois Press},
  \bibinfo{address}{Champaign, IL, USA}).

\bibitem[{\citenamefont{Newman}(2001{\natexlab{a}})}]{newman01}
\bibinfo{author}{\bibnamefont{Newman}, \bibfnamefont{M.}},
  \bibinfo{year}{2001}{\natexlab{a}}, \bibinfo{journal}{Phys. Rev. E}
  \textbf{\bibinfo{volume}{64}}(\bibinfo{number}{1}), \bibinfo{pages}{016131}.

\bibitem[{\citenamefont{Newman}(2001{\natexlab{b}})}]{newman01b}
\bibinfo{author}{\bibnamefont{Newman}, \bibfnamefont{M.}},
  \bibinfo{year}{2001}{\natexlab{b}}, \bibinfo{journal}{Phys. Rev. E}
  \textbf{\bibinfo{volume}{64}}(\bibinfo{number}{1}), \bibinfo{pages}{016132}.

\bibitem[{\citenamefont{Newman}(2003{\natexlab{a}})}]{newman03b}
\bibinfo{author}{\bibnamefont{Newman}, \bibfnamefont{M.}},
  \bibinfo{year}{2003}{\natexlab{a}}, \bibinfo{journal}{SIAM Review}
  \textbf{\bibinfo{volume}{45}}(\bibinfo{number}{2}), \bibinfo{pages}{167}.

\bibitem[{\citenamefont{Newman}(2004)}]{newman04}
\bibinfo{author}{\bibnamefont{Newman}, \bibfnamefont{M.}},
  \bibinfo{year}{2004}, \bibinfo{journal}{Proc. Natl. Acad. Sci. USA}
  \textbf{\bibinfo{volume}{101}}, \bibinfo{pages}{5200}.

\bibitem[{\citenamefont{Newman} \emph{et~al.}(2002)\citenamefont{Newman,
  Forrest, and Balthrop}}]{newman02}
\bibinfo{author}{\bibnamefont{Newman}, \bibfnamefont{M.}},
  \bibinfo{author}{\bibfnamefont{S.}~\bibnamefont{Forrest}}, and
  \bibinfo{author}{\bibfnamefont{J.}~\bibnamefont{Balthrop}},
  \bibinfo{year}{2002}, \bibinfo{journal}{Phys. Rev. E}
  \textbf{\bibinfo{volume}{66}}(\bibinfo{number}{3}), \bibinfo{pages}{035101}.

\bibitem[{\citenamefont{Newman and Park}(2003)}]{newman03c}
\bibinfo{author}{\bibnamefont{Newman}, \bibfnamefont{M.}}, and
  \bibinfo{author}{\bibfnamefont{J.}~\bibnamefont{Park}}, \bibinfo{year}{2003},
  \bibinfo{journal}{Phys. Rev. E} \textbf{\bibinfo{volume}{68}},
  \bibinfo{pages}{036122}.

\bibitem[{\citenamefont{Newman and Watts}(1999)}]{newman99}
\bibinfo{author}{\bibnamefont{Newman}, \bibfnamefont{M.}}, and
  \bibinfo{author}{\bibfnamefont{D.}~\bibnamefont{Watts}},
  \bibinfo{year}{1999}, \bibinfo{journal}{Phys. Rev. E}
  \textbf{\bibinfo{volume}{60}}(\bibinfo{number}{6}), \bibinfo{pages}{7332}.

\bibitem[{\citenamefont{Newman}(2003{\natexlab{b}})}]{newman03}
\bibinfo{author}{\bibnamefont{Newman}, \bibfnamefont{M.~E.~J.}},
  \bibinfo{year}{2003}{\natexlab{b}}, \bibinfo{journal}{Phys. Rev. E}
  \textbf{\bibinfo{volume}{67}}(\bibinfo{number}{2}), \bibinfo{pages}{026126}.

\bibitem[{\citenamefont{Newman}(2005)}]{newman05}
\bibinfo{author}{\bibnamefont{Newman}, \bibfnamefont{M.~E.~J.}},
  \bibinfo{year}{2005}, \bibinfo{journal}{Contemp. Phys.}
  \textbf{\bibinfo{volume}{46}}, \bibinfo{pages}{323}.

\bibitem[{\citenamefont{Niyogi}(2006)}]{niyogi06}
\bibinfo{author}{\bibnamefont{Niyogi}, \bibfnamefont{P.}},
  \bibinfo{year}{2006}, \emph{\bibinfo{title}{The Computational Nature of
  Language Learning and Evolution}} (\bibinfo{publisher}{MIT Press},
  \bibinfo{address}{Cambridge, MA, USA}).

\bibitem[{\citenamefont{Nowak} \emph{et~al.}(1990)\citenamefont{Nowak, Szamrej,
  and Latan\'e}}]{nowak90}
\bibinfo{author}{\bibnamefont{Nowak}, \bibfnamefont{A.}},
  \bibinfo{author}{\bibfnamefont{J.}~\bibnamefont{Szamrej}}, and
  \bibinfo{author}{\bibfnamefont{B.}~\bibnamefont{Latan\'e}},
  \bibinfo{year}{1990}, \bibinfo{journal}{Psychol. Rev.}
  \textbf{\bibinfo{volume}{97}}(\bibinfo{number}{3}), \bibinfo{pages}{362}.

\bibitem[{\citenamefont{{Nowak}}(2006)}]{nowak06}
\bibinfo{author}{\bibnamefont{{Nowak}}, \bibfnamefont{M.}},
  \bibinfo{year}{2006}, \emph{\bibinfo{title}{Evolutionary Dynamics: Exploring
  the Equations of Life}} (\bibinfo{publisher}{Harvard University Press},
  \bibinfo{address}{Cambridge, MA, USA}).

\bibitem[{\citenamefont{Nowak and Krakauer}(1999)}]{nowak99}
\bibinfo{author}{\bibnamefont{Nowak}, \bibfnamefont{M.}}, and
  \bibinfo{author}{\bibfnamefont{D.}~\bibnamefont{Krakauer}},
  \bibinfo{year}{1999}, \bibinfo{journal}{Proc. Natl. Acad. Sci. USA}
  \textbf{\bibinfo{volume}{96}}(\bibinfo{number}{14}), \bibinfo{pages}{8028}.

\bibitem[{\citenamefont{Nowak}
  \emph{et~al.}(1999{\natexlab{a}})\citenamefont{Nowak, Krakauer, and
  Dress}}]{nowak99c}
\bibinfo{author}{\bibnamefont{Nowak}, \bibfnamefont{M.}},
  \bibinfo{author}{\bibfnamefont{D.}~\bibnamefont{Krakauer}}, and
  \bibinfo{author}{\bibfnamefont{A.}~\bibnamefont{Dress}},
  \bibinfo{year}{1999}{\natexlab{a}}, \bibinfo{journal}{Proc. R. Soc. London}
  \textbf{\bibinfo{volume}{266}}(\bibinfo{number}{1433}),
  \bibinfo{pages}{2131}.

\bibitem[{\citenamefont{Nowak}
  \emph{et~al.}(1999{\natexlab{b}})\citenamefont{Nowak, Plotkin, and
  Krakauer}}]{nowak99b}
\bibinfo{author}{\bibnamefont{Nowak}, \bibfnamefont{M.}},
  \bibinfo{author}{\bibfnamefont{J.}~\bibnamefont{Plotkin}}, and
  \bibinfo{author}{\bibfnamefont{D.}~\bibnamefont{Krakauer}},
  \bibinfo{year}{1999}{\natexlab{b}}, \bibinfo{journal}{J. Theor. Bio.}
  \textbf{\bibinfo{volume}{200}}(\bibinfo{number}{2}), \bibinfo{pages}{147}.

\bibitem[{\citenamefont{{Ochrombel}}(2001)}]{ochrombel01}
\bibinfo{author}{\bibnamefont{{Ochrombel}}, \bibfnamefont{R.}},
  \bibinfo{year}{2001}, \bibinfo{journal}{Int. J. Mod. Phys. C}
  \textbf{\bibinfo{volume}{12}}, \bibinfo{pages}{1091}.

\bibitem[{\citenamefont{Odagaki and Tsujiguchi}(2006)}]{odagaki06}
\bibinfo{author}{\bibnamefont{Odagaki}, \bibfnamefont{T.}}, and
  \bibinfo{author}{\bibfnamefont{M.}~\bibnamefont{Tsujiguchi}},
  \bibinfo{year}{2006}, \bibinfo{journal}{Physica A}
  \textbf{\bibinfo{volume}{367}}, \bibinfo{pages}{435}.

\bibitem[{\citenamefont{Oliphant}(1997)}]{oliphant97a}
\bibinfo{author}{\bibnamefont{Oliphant}, \bibfnamefont{M.}},
  \bibinfo{year}{1997}, \emph{\bibinfo{title}{Formal approaches to innate and
  learned communicaton: laying the foundation for language}}, Ph.D. thesis,
  \bibinfo{school}{University of California, San Diego}.

\bibitem[{\citenamefont{Oliphant and Batali}(1996)}]{oliphant96b}
\bibinfo{author}{\bibnamefont{Oliphant}, \bibfnamefont{M.}}, and
  \bibinfo{author}{\bibfnamefont{J.}~\bibnamefont{Batali}},
  \bibinfo{year}{1996}, \bibinfo{journal}{The newsletter of the Center for
  research on language} \textbf{\bibinfo{volume}{11}}(\bibinfo{number}{1}).

\bibitem[{\citenamefont{Oliveira and Barab\'asi}(2005)}]{oliveira05}
\bibinfo{author}{\bibnamefont{Oliveira}, \bibfnamefont{J.~G.}}, and
  \bibinfo{author}{\bibfnamefont{A.~L.} \bibnamefont{Barab\'asi}},
  \bibinfo{year}{2005}, \bibinfo{journal}{Nature}
  \textbf{\bibinfo{volume}{437}}, \bibinfo{pages}{1251}.

\bibitem[{\citenamefont{de~Oliveira}(2003)}]{deoliveira03}
\bibinfo{author}{\bibnamefont{de~Oliveira}, \bibfnamefont{M.}},
  \bibinfo{year}{2003}, \bibinfo{journal}{Phys. Rev. E}
  \textbf{\bibinfo{volume}{67}}(\bibinfo{number}{6}), \bibinfo{pages}{066101}.

\bibitem[{\citenamefont{de~Oliveira}
  \emph{et~al.}(1993)\citenamefont{de~Oliveira, Mendes, and
  Santos}}]{deoliveira93}
\bibinfo{author}{\bibnamefont{de~Oliveira}, \bibfnamefont{M.}},
  \bibinfo{author}{\bibfnamefont{J.}~\bibnamefont{Mendes}}, and
  \bibinfo{author}{\bibfnamefont{M.}~\bibnamefont{Santos}},
  \bibinfo{year}{1993}, \bibinfo{journal}{J. Phys. A}
  \textbf{\bibinfo{volume}{26}}(\bibinfo{number}{10}), \bibinfo{pages}{2317}.

\bibitem[{\citenamefont{de~Oliveira}(1992)}]{oliveira92}
\bibinfo{author}{\bibnamefont{de~Oliveira}, \bibfnamefont{M.~J.}},
  \bibinfo{year}{1992}, \bibinfo{journal}{J. Stat. Phys.}
  \textbf{\bibinfo{volume}{66}}, \bibinfo{pages}{273}.

\bibitem[{\citenamefont{de~Oliveira}
  \emph{et~al.}(2007)\citenamefont{de~Oliveira, Stauffer, Lima, Sousa, Schulze,
  and de~Oliveira}}]{oliveira07}
\bibinfo{author}{\bibnamefont{de~Oliveira}, \bibfnamefont{P.}},
  \bibinfo{author}{\bibfnamefont{D.}~\bibnamefont{Stauffer}},
  \bibinfo{author}{\bibfnamefont{F.}~\bibnamefont{Lima}},
  \bibinfo{author}{\bibfnamefont{A.}~\bibnamefont{Sousa}},
  \bibinfo{author}{\bibfnamefont{C.}~\bibnamefont{Schulze}}, and
  \bibinfo{author}{\bibfnamefont{S.~M.} \bibnamefont{de~Oliveira}},
  \bibinfo{year}{2007}, \bibinfo{journal}{Physica A}
  \textbf{\bibinfo{volume}{376}}, \bibinfo{pages}{609}.

\bibitem[{\citenamefont{de~Oliveira}
  \emph{et~al.}(2006{\natexlab{a}})\citenamefont{de~Oliveira, Campos, Gomes,
  and Tsang}}]{viviane06a}
\bibinfo{author}{\bibnamefont{de~Oliveira}, \bibfnamefont{V.~M.}},
  \bibinfo{author}{\bibfnamefont{P.~R.} \bibnamefont{Campos}},
  \bibinfo{author}{\bibfnamefont{M.}~\bibnamefont{Gomes}}, and
  \bibinfo{author}{\bibfnamefont{I.}~\bibnamefont{Tsang}},
  \bibinfo{year}{2006}{\natexlab{a}}, \bibinfo{journal}{Physica A}
  \textbf{\bibinfo{volume}{368}}, \bibinfo{pages}{257}.

\bibitem[{\citenamefont{de~Oliveira}
  \emph{et~al.}(2006{\natexlab{b}})\citenamefont{de~Oliveira, Gomes, and
  Tsang}}]{viviane06}
\bibinfo{author}{\bibnamefont{de~Oliveira}, \bibfnamefont{V.~M.}},
  \bibinfo{author}{\bibfnamefont{M.}~\bibnamefont{Gomes}}, and
  \bibinfo{author}{\bibfnamefont{I.}~\bibnamefont{Tsang}},
  \bibinfo{year}{2006}{\natexlab{b}}, \bibinfo{journal}{Physica A}
  \textbf{\bibinfo{volume}{361}}, \bibinfo{pages}{361}.

\bibitem[{\citenamefont{Onnela} \emph{et~al.}(2007)\citenamefont{Onnela,
  Saram{\"a}ki, Hyv{\"o}nen, Szab\'o, Lazer, Kaski, Kert\'esz, and
  Barab\'asi}}]{onnela07}
\bibinfo{author}{\bibnamefont{Onnela}, \bibfnamefont{J.}},
  \bibinfo{author}{\bibfnamefont{J.}~\bibnamefont{Saram{\"a}ki}},
  \bibinfo{author}{\bibfnamefont{J.}~\bibnamefont{Hyv{\"o}nen}},
  \bibinfo{author}{\bibfnamefont{G.}~\bibnamefont{Szab\'o}},
  \bibinfo{author}{\bibfnamefont{D.}~\bibnamefont{Lazer}},
  \bibinfo{author}{\bibfnamefont{K.}~\bibnamefont{Kaski}},
  \bibinfo{author}{\bibfnamefont{J.}~\bibnamefont{Kert\'esz}}, and
  \bibinfo{author}{\bibfnamefont{A.~L.} \bibnamefont{Barab\'asi}},
  \bibinfo{year}{2007}, \bibinfo{journal}{Proc. Natl. Acad. Sci. USA}
  \textbf{\bibinfo{volume}{104}}(\bibinfo{number}{18}), \bibinfo{pages}{7332}.

\bibitem[{\citenamefont{O'Reilly}(2005)}]{oreilly05}
\bibinfo{author}{\bibnamefont{O'Reilly}, \bibfnamefont{T.}},
  \bibinfo{year}{2005}, \urlprefix\url{http://www.oreillynet.com/lpt/a/6228}.

\bibitem[{\citenamefont{Osborne and Rubinstein}(1994)}]{osborne94}
\bibinfo{author}{\bibnamefont{Osborne}, \bibfnamefont{M.}}, and
  \bibinfo{author}{\bibfnamefont{A.}~\bibnamefont{Rubinstein}},
  \bibinfo{year}{1994}, \emph{\bibinfo{title}{A Course in Game Theory}}
  (\bibinfo{publisher}{MIT Press}, \bibinfo{address}{Cambridge, MA, USA}).

\bibitem[{\citenamefont{Palla} \emph{et~al.}(2007)\citenamefont{Palla,
  Barab\'asi, and Vicsek}}]{palla07}
\bibinfo{author}{\bibnamefont{Palla}, \bibfnamefont{G.}},
  \bibinfo{author}{\bibfnamefont{A.-L.} \bibnamefont{Barab\'asi}}, and
  \bibinfo{author}{\bibfnamefont{T.}~\bibnamefont{Vicsek}},
  \bibinfo{year}{2007}, \bibinfo{journal}{Nature}
  \textbf{\bibinfo{volume}{446}}(\bibinfo{number}{7136}), \bibinfo{pages}{664}.

\bibitem[{\citenamefont{{Parravano}}
  \emph{et~al.}(2006)\citenamefont{{Parravano}, {Rivera-Ram{\'i}rez}, and
  {Cosenza}}}]{parravano06}
\bibinfo{author}{\bibnamefont{{Parravano}}, \bibfnamefont{A.}},
  \bibinfo{author}{\bibfnamefont{H.}~\bibnamefont{{Rivera-Ram{\'i}rez}}}, and
  \bibinfo{author}{\bibfnamefont{M.~G.} \bibnamefont{{Cosenza}}},
  \bibinfo{year}{2006}, \eprint{arXiv:nlin/0612030}.

\bibitem[{\citenamefont{{Parrish} and {Hamner}}(1997)}]{parrish97}
\bibinfo{author}{\bibnamefont{{Parrish}}, \bibfnamefont{J.~K.}}, and
  \bibinfo{author}{\bibfnamefont{W.~H.} \bibnamefont{{Hamner}}},
  \bibinfo{year}{1997}, \emph{\bibinfo{title}{Animal Groups in Three
  Dimensions}} (\bibinfo{publisher}{Cambridge University Press},
  \bibinfo{address}{New York, NY, USA}).

\bibitem[{\citenamefont{Pastor-Satorras}
  \emph{et~al.}(2001)\citenamefont{Pastor-Satorras, V\'azquez, and
  Vespignani}}]{pastor-satorras01}
\bibinfo{author}{\bibnamefont{Pastor-Satorras}, \bibfnamefont{R.}},
  \bibinfo{author}{\bibfnamefont{A.}~\bibnamefont{V\'azquez}}, and
  \bibinfo{author}{\bibfnamefont{A.}~\bibnamefont{Vespignani}},
  \bibinfo{year}{2001}, \bibinfo{journal}{Phys. Rev. Lett.}
  \textbf{\bibinfo{volume}{87}}(\bibinfo{number}{25}), \bibinfo{pages}{258701}.

\bibitem[{\citenamefont{Pastor-Satorras and
  Vespignani}(2001)}]{pastor-satorras01b}
\bibinfo{author}{\bibnamefont{Pastor-Satorras}, \bibfnamefont{R.}}, and
  \bibinfo{author}{\bibfnamefont{A.}~\bibnamefont{Vespignani}},
  \bibinfo{year}{2001}, \bibinfo{journal}{Phys. Rev. Lett.}
  \textbf{\bibinfo{volume}{86}}(\bibinfo{number}{14}), \bibinfo{pages}{3200}.

\bibitem[{\citenamefont{Pastor-Satorras and
  Vespignani}(2004)}]{pastor-satorras04}
\bibinfo{author}{\bibnamefont{Pastor-Satorras}, \bibfnamefont{R.}}, and
  \bibinfo{author}{\bibfnamefont{A.}~\bibnamefont{Vespignani}},
  \bibinfo{year}{2004}, \emph{\bibinfo{title}{Evolution and Structure of the
  Internet: A Statistical Physics Approach}} (\bibinfo{publisher}{Cambridge
  University Press}, \bibinfo{address}{New York, NY, USA}).

\bibitem[{\citenamefont{Patriarca and Lepp\"anen}(2004)}]{patriarca04}
\bibinfo{author}{\bibnamefont{Patriarca}, \bibfnamefont{M.}}, and
  \bibinfo{author}{\bibfnamefont{T.}~\bibnamefont{Lepp\"anen}},
  \bibinfo{year}{2004}, \bibinfo{journal}{Physica A}
  \textbf{\bibinfo{volume}{338}}, \bibinfo{pages}{296}.

\bibitem[{\citenamefont{{Pereira} and {Moreira}}(2005)}]{pereira05}
\bibinfo{author}{\bibnamefont{{Pereira}}, \bibfnamefont{L.~F.}}, and
  \bibinfo{author}{\bibfnamefont{F.~G.} \bibnamefont{{Moreira}}},
  \bibinfo{year}{2005}, \bibinfo{journal}{Phys. Rev. E}
  \textbf{\bibinfo{volume}{71}}(\bibinfo{number}{1}), \bibinfo{pages}{016123}.

\bibitem[{\citenamefont{Pinasco and Romanelli}(2006)}]{pinasco06}
\bibinfo{author}{\bibnamefont{Pinasco}, \bibfnamefont{J.}}, and
  \bibinfo{author}{\bibfnamefont{L.}~\bibnamefont{Romanelli}},
  \bibinfo{year}{2006}, \bibinfo{journal}{Physica A}
  \textbf{\bibinfo{volume}{361}}, \bibinfo{pages}{355}.

\bibitem[{\citenamefont{Pinker and Bloom}(1990)}]{pinker90}
\bibinfo{author}{\bibnamefont{Pinker}, \bibfnamefont{S.}}, and
  \bibinfo{author}{\bibfnamefont{P.}~\bibnamefont{Bloom}},
  \bibinfo{year}{1990}, \bibinfo{journal}{Behav. Brain Sci.}
  \textbf{\bibinfo{volume}{13}}(\bibinfo{number}{4}), \bibinfo{pages}{707}.

\bibitem[{\citenamefont{Plotkin and Nowak}(2000)}]{plotkin00}
\bibinfo{author}{\bibnamefont{Plotkin}, \bibfnamefont{J.}}, and
  \bibinfo{author}{\bibfnamefont{M.}~\bibnamefont{Nowak}},
  \bibinfo{year}{2000}, \bibinfo{journal}{J. Theor. Bio.}
  \textbf{\bibinfo{volume}{205}}(\bibinfo{number}{1}), \bibinfo{pages}{147}.

\bibitem[{\citenamefont{Pluchino} \emph{et~al.}(2006)\citenamefont{Pluchino,
  Boccaletti, Latora, and Rapisarda}}]{pluchino06}
\bibinfo{author}{\bibnamefont{Pluchino}, \bibfnamefont{A.}},
  \bibinfo{author}{\bibfnamefont{S.}~\bibnamefont{Boccaletti}},
  \bibinfo{author}{\bibfnamefont{V.}~\bibnamefont{Latora}}, and
  \bibinfo{author}{\bibfnamefont{A.}~\bibnamefont{Rapisarda}},
  \bibinfo{year}{2006}, \bibinfo{journal}{Physica A}
  \textbf{\bibinfo{volume}{372}}(\bibinfo{number}{2}), \bibinfo{pages}{316}.

\bibitem[{\citenamefont{{Pluchino}}
  \emph{et~al.}(2005)\citenamefont{{Pluchino}, {Latora}, and
  {Rapisarda}}}]{pluchino05}
\bibinfo{author}{\bibnamefont{{Pluchino}}, \bibfnamefont{A.}},
  \bibinfo{author}{\bibfnamefont{V.}~\bibnamefont{{Latora}}}, and
  \bibinfo{author}{\bibfnamefont{A.}~\bibnamefont{{Rapisarda}}},
  \bibinfo{year}{2005}, \bibinfo{journal}{Int. J. Mod. Phys. C}
  \textbf{\bibinfo{volume}{16}}, \bibinfo{pages}{515}.

\bibitem[{\citenamefont{{Porfiri}} \emph{et~al.}(2007)\citenamefont{{Porfiri},
  {Bollt}, and {Stilwell}}}]{porfiri07}
\bibinfo{author}{\bibnamefont{{Porfiri}}, \bibfnamefont{M.}},
  \bibinfo{author}{\bibfnamefont{E.~M.} \bibnamefont{{Bollt}}}, and
  \bibinfo{author}{\bibfnamefont{D.~J.} \bibnamefont{{Stilwell}}},
  \bibinfo{year}{2007}, \bibinfo{journal}{Eur. Phys. J. B}
  \textbf{\bibinfo{volume}{57}}, \bibinfo{pages}{481}.

\bibitem[{\citenamefont{Rapoport}(1953)}]{rapoport53}
\bibinfo{author}{\bibnamefont{Rapoport}, \bibfnamefont{A.}},
  \bibinfo{year}{1953}, \bibinfo{journal}{Bull. Math. Biophys.}
  \textbf{\bibinfo{volume}{15}}, \bibinfo{pages}{523}.

\bibitem[{\citenamefont{Redner}(2001)}]{redner01}
\bibinfo{author}{\bibnamefont{Redner}, \bibfnamefont{S.}},
  \bibinfo{year}{2001}, \emph{\bibinfo{title}{A guide to first-passage
  processes}} (\bibinfo{publisher}{Cambridge University Press},
  \bibinfo{address}{Cambridge, UK}).

\bibitem[{\citenamefont{Ridley}(2003)}]{ridley03}
\bibinfo{author}{\bibnamefont{Ridley}, \bibfnamefont{M.}},
  \bibinfo{year}{2003}, \emph{\bibinfo{title}{Nature Via Nurture: Genes,
  Experience, and What Makes Us Human}} (\bibinfo{publisher}{Fourth Estate}).

\bibitem[{\citenamefont{{Rodrigues} and {da F.~Costa}}(2005)}]{rodrigues05}
\bibinfo{author}{\bibnamefont{{Rodrigues}}, \bibfnamefont{F.~A.}}, and
  \bibinfo{author}{\bibfnamefont{L.}~\bibnamefont{{da F.~Costa}}},
  \bibinfo{year}{2005}, \bibinfo{journal}{Int. J. Mod. Phys. C}
  \textbf{\bibinfo{volume}{16}}, \bibinfo{pages}{1785}.

\bibitem[{\citenamefont{Roehner}(2007)}]{roehner07}
\bibinfo{author}{\bibnamefont{Roehner}, \bibfnamefont{B.}},
  \bibinfo{year}{2007}, \emph{\bibinfo{title}{Driving Forces in Physical,
  Biological and Socio-economic Phenomena}} (\bibinfo{publisher}{Cambridge
  University Press}, \bibinfo{address}{Cambridge, UK}).

\bibitem[{\citenamefont{{Roehner}} \emph{et~al.}(2004)\citenamefont{{Roehner},
  {Sornette}, and {Andersen}}}]{roehner04}
\bibinfo{author}{\bibnamefont{{Roehner}}, \bibfnamefont{B.~M.}},
  \bibinfo{author}{\bibfnamefont{D.}~\bibnamefont{{Sornette}}}, and
  \bibinfo{author}{\bibfnamefont{J.~V.} \bibnamefont{{Andersen}}},
  \bibinfo{year}{2004}, \bibinfo{journal}{Int. J. Mod. Phys. C}
  \textbf{\bibinfo{volume}{15}}, \bibinfo{pages}{809}.

\bibitem[{\citenamefont{Rogers}(2003)}]{rogers03}
\bibinfo{author}{\bibnamefont{Rogers}, \bibfnamefont{E.~M.}},
  \bibinfo{year}{2003}, \emph{\bibinfo{title}{Diffusion of Innovations}}
  (\bibinfo{publisher}{Free Press}, \bibinfo{address}{New York}),
  \bibinfo{edition}{5th ed.} edition, ISBN \bibinfo{isbn}{0743222091},
  \urlprefix\url{http://www.loc.gov/catdir/enhancements/fy0641/2003049022-s.ht%
ml}.

\bibitem[{\citenamefont{{Sabatelli} and {Richmond}}(2003)}]{sabatelli03}
\bibinfo{author}{\bibnamefont{{Sabatelli}}, \bibfnamefont{L.}}, and
  \bibinfo{author}{\bibfnamefont{P.}~\bibnamefont{{Richmond}}},
  \bibinfo{year}{2003}, \bibinfo{journal}{Int. J. Mod. Phys. C}
  \textbf{\bibinfo{volume}{14}}, \bibinfo{pages}{1223}.

\bibitem[{\citenamefont{Sabatelli and Richmond}(2004)}]{sabatelli04}
\bibinfo{author}{\bibnamefont{Sabatelli}, \bibfnamefont{L.}}, and
  \bibinfo{author}{\bibfnamefont{P.}~\bibnamefont{Richmond}},
  \bibinfo{year}{2004}, \bibinfo{journal}{Physica A}
  \textbf{\bibinfo{volume}{334}}, \bibinfo{pages}{274}.

\bibitem[{\citenamefont{Saloma} \emph{et~al.}(2003)\citenamefont{Saloma, Perez,
  Tapang, Lim, and Palmes-Saloma}}]{saloma03}
\bibinfo{author}{\bibnamefont{Saloma}, \bibfnamefont{C.}},
  \bibinfo{author}{\bibfnamefont{G.~J.} \bibnamefont{Perez}},
  \bibinfo{author}{\bibfnamefont{G.}~\bibnamefont{Tapang}},
  \bibinfo{author}{\bibfnamefont{M.}~\bibnamefont{Lim}}, and
  \bibinfo{author}{\bibfnamefont{C.}~\bibnamefont{Palmes-Saloma}},
  \bibinfo{year}{2003}, \bibinfo{journal}{Proc. Natl. Acad. Sci. USA}
  \textbf{\bibinfo{volume}{100}}, \bibinfo{pages}{11947}.

\bibitem[{\citenamefont{S\'anchez} \emph{et~al.}(2002)\citenamefont{S\'anchez,
  L\'opez, and Rodr\'iguez}}]{sanchez02}
\bibinfo{author}{\bibnamefont{S\'anchez}, \bibfnamefont{A.~D.}},
  \bibinfo{author}{\bibfnamefont{J.~M.} \bibnamefont{L\'opez}}, and
  \bibinfo{author}{\bibfnamefont{M.~A.} \bibnamefont{Rodr\'iguez}},
  \bibinfo{year}{2002}, \bibinfo{journal}{Phys. Rev. Lett.}
  \textbf{\bibinfo{volume}{88}}(\bibinfo{number}{4}), \bibinfo{pages}{048701}.

\bibitem[{\citenamefont{{Schadschneider}}
  \emph{et~al.}(2008)\citenamefont{{Schadschneider}, {Klingsch}, {Kluepfel},
  {Kretz}, {Rogsch}, and {Seyfried}}}]{schadschneider08}
\bibinfo{author}{\bibnamefont{{Schadschneider}}, \bibfnamefont{A.}},
  \bibinfo{author}{\bibfnamefont{W.}~\bibnamefont{{Klingsch}}},
  \bibinfo{author}{\bibfnamefont{H.}~\bibnamefont{{Kluepfel}}},
  \bibinfo{author}{\bibfnamefont{T.}~\bibnamefont{{Kretz}}},
  \bibinfo{author}{\bibfnamefont{C.}~\bibnamefont{{Rogsch}}}, and
  \bibinfo{author}{\bibfnamefont{A.}~\bibnamefont{{Seyfried}}},
  \bibinfo{year}{2008}, \eprint{arXiv:0802.1620}.

\bibitem[{\citenamefont{{Schelling}}(1971)}]{schelling71}
\bibinfo{author}{\bibnamefont{{Schelling}}, \bibfnamefont{T.~C.}},
  \bibinfo{year}{1971}, \bibinfo{journal}{J. Math. Sociol.}
  \textbf{\bibinfo{volume}{1}}(\bibinfo{number}{2}), \bibinfo{pages}{143}.

\bibitem[{\citenamefont{Scheucher and Spohn}(1988)}]{scheucher89}
\bibinfo{author}{\bibnamefont{Scheucher}, \bibfnamefont{M.}}, and
  \bibinfo{author}{\bibfnamefont{H.}~\bibnamefont{Spohn}},
  \bibinfo{year}{1988}, \bibinfo{journal}{J. Stat. Phys.}
  \textbf{\bibinfo{volume}{53}}(\bibinfo{number}{1}), \bibinfo{pages}{279}.

\bibitem[{\citenamefont{{Schneider}}(2004)}]{schneider04}
\bibinfo{author}{\bibnamefont{{Schneider}}, \bibfnamefont{J.~J.}},
  \bibinfo{year}{2004}, \bibinfo{journal}{Int. J. Mod. Phys. C}
  \textbf{\bibinfo{volume}{15}}, \bibinfo{pages}{659}.

\bibitem[{\citenamefont{{Schneider} and
  {Hirtreiter}}(2005{\natexlab{a}})}]{schneider05}
\bibinfo{author}{\bibnamefont{{Schneider}}, \bibfnamefont{J.~J.}}, and
  \bibinfo{author}{\bibfnamefont{C.}~\bibnamefont{{Hirtreiter}}},
  \bibinfo{year}{2005}{\natexlab{a}}, \bibinfo{journal}{Int. J. Mod. Phys. C}
  \textbf{\bibinfo{volume}{16}}, \bibinfo{pages}{157}.

\bibitem[{\citenamefont{{Schneider} and
  {Hirtreiter}}(2005{\natexlab{b}})}]{schneider05a}
\bibinfo{author}{\bibnamefont{{Schneider}}, \bibfnamefont{J.~J.}}, and
  \bibinfo{author}{\bibfnamefont{C.}~\bibnamefont{{Hirtreiter}}},
  \bibinfo{year}{2005}{\natexlab{b}}, \bibinfo{journal}{Physica A}
  \textbf{\bibinfo{volume}{353}}, \bibinfo{pages}{539}.

\bibitem[{\citenamefont{{Schneider} and
  {Hirtreiter}}(2005{\natexlab{c}})}]{schneider05b}
\bibinfo{author}{\bibnamefont{{Schneider}}, \bibfnamefont{J.~J.}}, and
  \bibinfo{author}{\bibfnamefont{C.}~\bibnamefont{{Hirtreiter}}},
  \bibinfo{year}{2005}{\natexlab{c}}, \bibinfo{journal}{Int. J. Mod. Phys. C}
  \textbf{\bibinfo{volume}{16}}, \bibinfo{pages}{1165}.

\bibitem[{\citenamefont{{Schreckenberg} and {Sharma}}(2001)}]{schreckenberg01}
\bibinfo{editor}{\bibnamefont{{Schreckenberg}}, \bibfnamefont{M.}}, and
  \bibinfo{editor}{\bibfnamefont{S.~D.} \bibnamefont{{Sharma}}} (eds.),
  \bibinfo{year}{2001}, \emph{\bibinfo{title}{Pedestrian and Evacuation
  Dynamics}} (\bibinfo{publisher}{Springer-Verlag},
  \bibinfo{address}{Berlin-Heidelberg, Germany}).

\bibitem[{\citenamefont{Schulze}(2003{\natexlab{a}})}]{schulze03}
\bibinfo{author}{\bibnamefont{Schulze}, \bibfnamefont{C.}},
  \bibinfo{year}{2003}{\natexlab{a}}, \bibinfo{journal}{Int. J. Mod. Phys. C}
  \textbf{\bibinfo{volume}{14}}(\bibinfo{number}{1}), \bibinfo{pages}{95}.

\bibitem[{\citenamefont{Schulze}(2003{\natexlab{b}})}]{schulze03a}
\bibinfo{author}{\bibnamefont{Schulze}, \bibfnamefont{C.}},
  \bibinfo{year}{2003}{\natexlab{b}}, \bibinfo{journal}{Physica A}
  \textbf{\bibinfo{volume}{324}}(\bibinfo{number}{3-4}), \bibinfo{pages}{717}.

\bibitem[{\citenamefont{{Schulze}}(2004)}]{schulze04}
\bibinfo{author}{\bibnamefont{{Schulze}}, \bibfnamefont{C.}},
  \bibinfo{year}{2004}, \bibinfo{journal}{Int. J. Mod. Phys. C}
  \textbf{\bibinfo{volume}{15}}, \bibinfo{pages}{569}.

\bibitem[{\citenamefont{{Schulze} and
  {Stauffer}}(2005{\natexlab{a}})}]{schulze05b}
\bibinfo{author}{\bibnamefont{{Schulze}}, \bibfnamefont{C.}}, and
  \bibinfo{author}{\bibfnamefont{D.}~\bibnamefont{{Stauffer}}},
  \bibinfo{year}{2005}{\natexlab{a}}, \bibinfo{journal}{Int. J. Mod. Phys. C}
  \textbf{\bibinfo{volume}{16}}, \bibinfo{pages}{781}.

\bibitem[{\citenamefont{{Schulze} and
  {Stauffer}}(2005{\natexlab{b}})}]{schulze05a}
\bibinfo{author}{\bibnamefont{{Schulze}}, \bibfnamefont{C.}}, and
  \bibinfo{author}{\bibfnamefont{D.}~\bibnamefont{{Stauffer}}},
  \bibinfo{year}{2005}{\natexlab{b}}, in \emph{\bibinfo{booktitle}{AIP Conf.
  Proc. 779: Modeling Cooperative Behavior in the Social Sciences}}, edited by
  \bibinfo{editor}{\bibfnamefont{P.}~\bibnamefont{{Garrido}}},
  \bibinfo{editor}{\bibfnamefont{J.}~\bibnamefont{{Marro}}}, and
  \bibinfo{editor}{\bibfnamefont{M.~A.} \bibnamefont{{Mu{\~n}oz}}}, pp.
  \bibinfo{pages}{49--55}.

\bibitem[{\citenamefont{{Schulze} and {Stauffer}}(2006)}]{schulze06}
\bibinfo{author}{\bibnamefont{{Schulze}}, \bibfnamefont{C.}}, and
  \bibinfo{author}{\bibfnamefont{D.}~\bibnamefont{{Stauffer}}},
  \bibinfo{year}{2006}, \bibinfo{journal}{Adv. Compl. Syst.}
  \textbf{\bibinfo{volume}{9}}(\bibinfo{number}{3}), \bibinfo{pages}{183}.

\bibitem[{\citenamefont{{Schulze}} \emph{et~al.}(2008)\citenamefont{{Schulze},
  {Stauffer}, and {Wichmann}}}]{schulze08}
\bibinfo{author}{\bibnamefont{{Schulze}}, \bibfnamefont{C.}},
  \bibinfo{author}{\bibfnamefont{D.}~\bibnamefont{{Stauffer}}}, and
  \bibinfo{author}{\bibfnamefont{S.}~\bibnamefont{{Wichmann}}},
  \bibinfo{year}{2008}, \bibinfo{journal}{Comm. Comput. Phys.}
  \textbf{\bibinfo{volume}{3}}(\bibinfo{number}{2}), \bibinfo{pages}{271}.

\bibitem[{\citenamefont{{Schw{\"a}mmle}}(2005)}]{schwammle05}
\bibinfo{author}{\bibnamefont{{Schw{\"a}mmle}}, \bibfnamefont{V.}},
  \bibinfo{year}{2005}, \bibinfo{journal}{Int. J. Mod. Phys. C}
  \textbf{\bibinfo{volume}{16}}, \bibinfo{pages}{1519}.

\bibitem[{\citenamefont{{Schweitzer}}(2003)}]{schweitzer03}
\bibinfo{author}{\bibnamefont{{Schweitzer}}, \bibfnamefont{F.}},
  \bibinfo{year}{2003}, \emph{\bibinfo{title}{Brownian Agents and Active
  Particles}} (\bibinfo{publisher}{Springer Verlag},
  \bibinfo{address}{Berlin-Heidelberg, Germany}).

\bibitem[{\citenamefont{{Schweitzer} and {Ho{\l}yst}}(2000)}]{schweitzer00}
\bibinfo{author}{\bibnamefont{{Schweitzer}}, \bibfnamefont{F.}}, and
  \bibinfo{author}{\bibfnamefont{J.~A.} \bibnamefont{{Ho{\l}yst}}},
  \bibinfo{year}{2000}, \bibinfo{journal}{Eur. Phys. J. B}
  \textbf{\bibinfo{volume}{15}}, \bibinfo{pages}{723}.

\bibitem[{\citenamefont{Scott}(2000)}]{scott00}
\bibinfo{author}{\bibnamefont{Scott}, \bibfnamefont{J.}}, \bibinfo{year}{2000},
  \emph{\bibinfo{title}{Social Network Analysis: A Handbook}}
  (\bibinfo{publisher}{SAGE Publications}, \bibinfo{address}{London, UK}).

\bibitem[{\citenamefont{{Sethna}} \emph{et~al.}(2001)\citenamefont{{Sethna},
  {Dahmen}, and {Myers}}}]{sethna01}
\bibinfo{author}{\bibnamefont{{Sethna}}, \bibfnamefont{J.~P.}},
  \bibinfo{author}{\bibfnamefont{K.~A.} \bibnamefont{{Dahmen}}}, and
  \bibinfo{author}{\bibfnamefont{C.~R.} \bibnamefont{{Myers}}},
  \bibinfo{year}{2001}, \bibinfo{journal}{Nature}
  \textbf{\bibinfo{volume}{410}}, \bibinfo{pages}{242}.

\bibitem[{\citenamefont{Shibanai} \emph{et~al.}(2001)\citenamefont{Shibanai,
  Yasuno, and Ishiguro}}]{shibanai01}
\bibinfo{author}{\bibnamefont{Shibanai}, \bibfnamefont{Y.}},
  \bibinfo{author}{\bibfnamefont{S.}~\bibnamefont{Yasuno}}, and
  \bibinfo{author}{\bibfnamefont{I.}~\bibnamefont{Ishiguro}},
  \bibinfo{year}{2001}, \bibinfo{journal}{J. Conflict Resolut.}
  \textbf{\bibinfo{volume}{45}}(\bibinfo{number}{1}), \bibinfo{pages}{80}.

\bibitem[{\citenamefont{{Sinha} and {Pan}}(2006)}]{sinha06}
\bibinfo{author}{\bibnamefont{{Sinha}}, \bibfnamefont{S.}}, and
  \bibinfo{author}{\bibfnamefont{R.~K.} \bibnamefont{{Pan}}},
  \bibinfo{year}{2006}, in \emph{\bibinfo{booktitle}{Econophysics \&
  Sociophysics: Trends \& Perspectives}}, edited by
  \bibinfo{editor}{\bibfnamefont{B.~K.} \bibnamefont{Chakrabarti}},
  \bibinfo{editor}{\bibfnamefont{A.}~\bibnamefont{Chakraborti}}, and
  \bibinfo{editor}{\bibfnamefont{A.}~\bibnamefont{Chatterjee}}
  (\bibinfo{publisher}{Wiley VCH}), p. \bibinfo{pages}{417}.

\bibitem[{\citenamefont{Sire and Majumdar}(1995)}]{sire95}
\bibinfo{author}{\bibnamefont{Sire}, \bibfnamefont{C.}}, and
  \bibinfo{author}{\bibfnamefont{S.~N.} \bibnamefont{Majumdar}},
  \bibinfo{year}{1995}, \bibinfo{journal}{Phys. Rev. E}
  \textbf{\bibinfo{volume}{52}}(\bibinfo{number}{1}), \bibinfo{pages}{244}.

\bibitem[{\citenamefont{{Situngkir}}(2004)}]{situngkir04}
\bibinfo{author}{\bibnamefont{{Situngkir}}, \bibfnamefont{H.}},
  \bibinfo{year}{2004}, \eprint{arxiv:nlin/0405002}.

\bibitem[{\citenamefont{Slanina and Lavi{\v c}ka}(2003)}]{slanina03}
\bibinfo{author}{\bibnamefont{Slanina}, \bibfnamefont{F.}}, and
  \bibinfo{author}{\bibfnamefont{H.}~\bibnamefont{Lavi{\v c}ka}},
  \bibinfo{year}{2003}, \bibinfo{journal}{Eur. Phys. J. B}
  \textbf{\bibinfo{volume}{35}}(\bibinfo{number}{2}), \bibinfo{pages}{279}.

\bibitem[{\citenamefont{{Slanina}} \emph{et~al.}(2007)\citenamefont{{Slanina},
  {Sznajd-Weron}, and {Przybyla}}}]{slanina07}
\bibinfo{author}{\bibnamefont{{Slanina}}, \bibfnamefont{F.}},
  \bibinfo{author}{\bibfnamefont{K.}~\bibnamefont{{Sznajd-Weron}}}, and
  \bibinfo{author}{\bibfnamefont{P.}~\bibnamefont{{Przybyla}}},
  \bibinfo{year}{2007}, \eprint{arXiv:0712.2035}.

\bibitem[{\citenamefont{Smith}(1982)}]{smith82}
\bibinfo{author}{\bibnamefont{Smith}, \bibfnamefont{J.~M.}},
  \bibinfo{year}{1982}, \emph{\bibinfo{title}{Evolution and the Theory of
  Games}} (\bibinfo{publisher}{Cambridge University Press},
  \bibinfo{address}{Cambridge, UK}).

\bibitem[{\citenamefont{Smith} \emph{et~al.}(2003)\citenamefont{Smith,
  Brighton, and Kirby}}]{smith03}
\bibinfo{author}{\bibnamefont{Smith}, \bibfnamefont{K.}},
  \bibinfo{author}{\bibfnamefont{H.}~\bibnamefont{Brighton}}, and
  \bibinfo{author}{\bibfnamefont{S.}~\bibnamefont{Kirby}},
  \bibinfo{year}{2003}, \bibinfo{journal}{Adv. Compl. Sys.}
  \textbf{\bibinfo{volume}{6}}(\bibinfo{number}{4}), \bibinfo{pages}{537}.

\bibitem[{\citenamefont{{Sood}} \emph{et~al.}(2007)\citenamefont{{Sood},
  {Antal}, and {Redner}}}]{sood07}
\bibinfo{author}{\bibnamefont{{Sood}}, \bibfnamefont{V.}},
  \bibinfo{author}{\bibfnamefont{T.}~\bibnamefont{{Antal}}}, and
  \bibinfo{author}{\bibfnamefont{S.}~\bibnamefont{{Redner}}},
  \bibinfo{year}{2007}, \eprint{arXiv:0712.4288}.

\bibitem[{\citenamefont{Sood and Redner}(2005)}]{sood05}
\bibinfo{author}{\bibnamefont{Sood}, \bibfnamefont{V.}}, and
  \bibinfo{author}{\bibfnamefont{S.}~\bibnamefont{Redner}},
  \bibinfo{year}{2005}, \bibinfo{journal}{Phys. Rev. Lett.}
  \textbf{\bibinfo{volume}{94}}(\bibinfo{number}{17}), \bibinfo{pages}{178701}.

\bibitem[{\citenamefont{Sousa}(2005)}]{sousa05}
\bibinfo{author}{\bibnamefont{Sousa}, \bibfnamefont{A.}}, \bibinfo{year}{2005},
  \bibinfo{journal}{Physica A} \textbf{\bibinfo{volume}{348}},
  \bibinfo{pages}{701}.

\bibitem[{\citenamefont{Sousa and S\'anchez}(2006)}]{sousa06}
\bibinfo{author}{\bibnamefont{Sousa}, \bibfnamefont{A.}}, and
  \bibinfo{author}{\bibfnamefont{J.}~\bibnamefont{S\'anchez}},
  \bibinfo{year}{2006}, \bibinfo{journal}{Physica A}
  \textbf{\bibinfo{volume}{361}}, \bibinfo{pages}{319}.

\bibitem[{\citenamefont{{Sousa}}(2004)}]{sousa04}
\bibinfo{author}{\bibnamefont{{Sousa}}, \bibfnamefont{A.~O.}},
  \bibinfo{year}{2004}, \eprint{arxiv:cond-mat/0406766}.

\bibitem[{\citenamefont{{Sousa} and {Stauffer}}(2000)}]{sousa00}
\bibinfo{author}{\bibnamefont{{Sousa}}, \bibfnamefont{A.~O.}}, and
  \bibinfo{author}{\bibfnamefont{D.}~\bibnamefont{{Stauffer}}},
  \bibinfo{year}{2000}, \bibinfo{journal}{Int. J. Mod. Phys. C}
  \textbf{\bibinfo{volume}{11}}, \bibinfo{pages}{1063}.

\bibitem[{\citenamefont{{Stark}} \emph{et~al.}(2007)\citenamefont{{Stark},
  {Tessone}, and {Schweitzer}}}]{stark07}
\bibinfo{author}{\bibnamefont{{Stark}}, \bibfnamefont{H.-U.}},
  \bibinfo{author}{\bibfnamefont{C.~J.} \bibnamefont{{Tessone}}}, and
  \bibinfo{author}{\bibfnamefont{F.}~\bibnamefont{{Schweitzer}}},
  \bibinfo{year}{2007}, \eprint{arXiv:0711.1133}.

\bibitem[{\citenamefont{Stauffer}(2002{\natexlab{a}})}]{stauffer02b}
\bibinfo{author}{\bibnamefont{Stauffer}, \bibfnamefont{D.}},
  \bibinfo{year}{2002}{\natexlab{a}}, \bibinfo{journal}{Int. J. Mod. Phys. C}
  \textbf{\bibinfo{volume}{13}}(\bibinfo{number}{7}), \bibinfo{pages}{975}.

\bibitem[{\citenamefont{Stauffer}(2002{\natexlab{b}})}]{stauffer02a}
\bibinfo{author}{\bibnamefont{Stauffer}, \bibfnamefont{D.}},
  \bibinfo{year}{2002}{\natexlab{b}}, \bibinfo{journal}{Int. J. Mod. Phys. C}
  \textbf{\bibinfo{volume}{13}}(\bibinfo{number}{3}), \bibinfo{pages}{315}.

\bibitem[{\citenamefont{Stauffer}(2003{\natexlab{a}})}]{stauffer03a}
\bibinfo{author}{\bibnamefont{Stauffer}, \bibfnamefont{D.}},
  \bibinfo{year}{2003}{\natexlab{a}}, in \emph{\bibinfo{booktitle}{AIP Conf.
  Proc. 690: The Monte Carlo Method in the Physical Sciences}}, edited by
  \bibinfo{editor}{\bibfnamefont{J.~E.} \bibnamefont{Gubernatis}}
  (\bibinfo{publisher}{American Institute of Physics}), pp.
  \bibinfo{pages}{147--155}.

\bibitem[{\citenamefont{Stauffer}(2003{\natexlab{b}})}]{stauffer03}
\bibinfo{author}{\bibnamefont{Stauffer}, \bibfnamefont{D.}},
  \bibinfo{year}{2003}{\natexlab{b}}, \bibinfo{journal}{Int. J. Mod. Phys. C}
  \textbf{\bibinfo{volume}{14}}(\bibinfo{number}{2}), \bibinfo{pages}{237}.

\bibitem[{\citenamefont{{Stauffer}}(2004)}]{stauffer02d}
\bibinfo{author}{\bibnamefont{{Stauffer}}, \bibfnamefont{D.}},
  \bibinfo{year}{2004}, \bibinfo{journal}{J. Math. Sociol.}
  \textbf{\bibinfo{volume}{28}}(\bibinfo{number}{1}), \bibinfo{pages}{28}.

\bibitem[{\citenamefont{Stauffer} \emph{et~al.}(2007)\citenamefont{Stauffer,
  Castell\'o, Egu\'iluz, and {San Miguel}}}]{stauffer07}
\bibinfo{author}{\bibnamefont{Stauffer}, \bibfnamefont{D.}},
  \bibinfo{author}{\bibfnamefont{X.}~\bibnamefont{Castell\'o}},
  \bibinfo{author}{\bibfnamefont{V.~M.} \bibnamefont{Egu\'iluz}}, and
  \bibinfo{author}{\bibfnamefont{M.}~\bibnamefont{{San Miguel}}},
  \bibinfo{year}{2007}, \bibinfo{journal}{Physica A}
  \textbf{\bibinfo{volume}{374}}, \bibinfo{pages}{835}.

\bibitem[{\citenamefont{Stauffer}
  \emph{et~al.}(2006{\natexlab{a}})\citenamefont{Stauffer, Hohnisch, and
  Pittnauer}}]{stauffer06a}
\bibinfo{author}{\bibnamefont{Stauffer}, \bibfnamefont{D.}},
  \bibinfo{author}{\bibfnamefont{M.}~\bibnamefont{Hohnisch}}, and
  \bibinfo{author}{\bibfnamefont{S.}~\bibnamefont{Pittnauer}},
  \bibinfo{year}{2006}{\natexlab{a}}, \bibinfo{journal}{Physica A}
  \textbf{\bibinfo{volume}{370}}, \bibinfo{pages}{734}.

\bibitem[{\citenamefont{Stauffer and Meyer-Ortmanns}(2004)}]{stauffer04d}
\bibinfo{author}{\bibnamefont{Stauffer}, \bibfnamefont{D.}}, and
  \bibinfo{author}{\bibfnamefont{H.}~\bibnamefont{Meyer-Ortmanns}},
  \bibinfo{year}{2004}, \bibinfo{journal}{Int. J. Mod. Phys. C}
  \textbf{\bibinfo{volume}{15}}(\bibinfo{number}{2}), \bibinfo{pages}{241}.

\bibitem[{\citenamefont{Stauffer}
  \emph{et~al.}(2006{\natexlab{b}})\citenamefont{Stauffer, {Moss de Oliveira},
  {de Oliveira}, and {S\'a Martins}}}]{stauffer06e}
\bibinfo{author}{\bibnamefont{Stauffer}, \bibfnamefont{D.}},
  \bibinfo{author}{\bibfnamefont{S.}~\bibnamefont{{Moss de Oliveira}}},
  \bibinfo{author}{\bibfnamefont{P.}~\bibnamefont{{de Oliveira}}}, and
  \bibinfo{author}{\bibfnamefont{J.}~\bibnamefont{{S\'a Martins}}},
  \bibinfo{year}{2006}{\natexlab{b}}, \emph{\bibinfo{title}{Biology, sociology,
  geology by computational physicists}} (\bibinfo{publisher}{Elsevier},
  \bibinfo{address}{Amsterdam}).

\bibitem[{\citenamefont{Stauffer and de~Oliveira}(2002)}]{stauffer02c}
\bibinfo{author}{\bibnamefont{Stauffer}, \bibfnamefont{D.}}, and
  \bibinfo{author}{\bibfnamefont{P.~M.~C.} \bibnamefont{de~Oliveira}},
  \bibinfo{year}{2002}, \eprint{arxiv:cond-mat/0208296}.

\bibitem[{\citenamefont{{Stauffer} and {S\'a Martins}}(2003)}]{stauffer03c}
\bibinfo{author}{\bibnamefont{{Stauffer}}, \bibfnamefont{D.}}, and
  \bibinfo{author}{\bibfnamefont{J.~S.} \bibnamefont{{S\'a Martins}}},
  \bibinfo{year}{2003}, \bibinfo{journal}{Adv. Compl. Sys.}
  \textbf{\bibinfo{volume}{6}}, \bibinfo{pages}{559}.

\bibitem[{\citenamefont{{Stauffer} and {S\'a Martins}}(2004)}]{stauffer04a}
\bibinfo{author}{\bibnamefont{{Stauffer}}, \bibfnamefont{D.}}, and
  \bibinfo{author}{\bibfnamefont{J.~S.} \bibnamefont{{S\'a Martins}}},
  \bibinfo{year}{2004}, \bibinfo{journal}{Physica A}
  \textbf{\bibinfo{volume}{334}}, \bibinfo{pages}{558}.

\bibitem[{\citenamefont{Stauffer}
  \emph{et~al.}(2006{\natexlab{c}})\citenamefont{Stauffer, Schulze, Lima,
  Wichmann, and Solomon}}]{stauffer06b}
\bibinfo{author}{\bibnamefont{Stauffer}, \bibfnamefont{D.}},
  \bibinfo{author}{\bibfnamefont{C.}~\bibnamefont{Schulze}},
  \bibinfo{author}{\bibfnamefont{F.}~\bibnamefont{Lima}},
  \bibinfo{author}{\bibfnamefont{S.}~\bibnamefont{Wichmann}}, and
  \bibinfo{author}{\bibfnamefont{S.}~\bibnamefont{Solomon}},
  \bibinfo{year}{2006}{\natexlab{c}}, \bibinfo{journal}{Physica A}
  \textbf{\bibinfo{volume}{371}}, \bibinfo{pages}{719}.

\bibitem[{\citenamefont{Stauffer} \emph{et~al.}(2004)\citenamefont{Stauffer,
  Sousa, and Schulze}}]{stauffer04}
\bibinfo{author}{\bibnamefont{Stauffer}, \bibfnamefont{D.}},
  \bibinfo{author}{\bibfnamefont{A.}~\bibnamefont{Sousa}}, and
  \bibinfo{author}{\bibfnamefont{C.}~\bibnamefont{Schulze}},
  \bibinfo{year}{2004}, \bibinfo{journal}{JASSS}
  \textbf{\bibinfo{volume}{7}}(\bibinfo{number}{3}).

\bibitem[{\citenamefont{{Stauffer}}
  \emph{et~al.}(2000)\citenamefont{{Stauffer}, {Sousa}, and {de
  Oliveira}}}]{stauffer00}
\bibinfo{author}{\bibnamefont{{Stauffer}}, \bibfnamefont{D.}},
  \bibinfo{author}{\bibfnamefont{A.~O.} \bibnamefont{{Sousa}}}, and
  \bibinfo{author}{\bibfnamefont{S.~M.} \bibnamefont{{de Oliveira}}},
  \bibinfo{year}{2000}, \bibinfo{journal}{Int. J. Mod. Phys. C}
  \textbf{\bibinfo{volume}{11}}, \bibinfo{pages}{1239}.

\bibitem[{\citenamefont{Steels}(1995)}]{steels95}
\bibinfo{author}{\bibnamefont{Steels}, \bibfnamefont{L.}},
  \bibinfo{year}{1995}, \bibinfo{journal}{Artif. Life}
  \textbf{\bibinfo{volume}{2}}(\bibinfo{number}{3}), \bibinfo{pages}{319}.

\bibitem[{\citenamefont{Steels}(1996)}]{steels96d}
\bibinfo{author}{\bibnamefont{Steels}, \bibfnamefont{L.}},
  \bibinfo{year}{1996}, in \emph{\bibinfo{booktitle}{Artificial Life V:
  Proceeding of the Fifth International Workshop on the Synthesis and
  Simulation of Living Systems}}, edited by
  \bibinfo{editor}{\bibfnamefont{C.}~\bibnamefont{Langton}} and
  \bibinfo{editor}{\bibfnamefont{T.}~\bibnamefont{Shimohara}}
  (\bibinfo{publisher}{The MIT Press}, \bibinfo{address}{Cambridge, MA, USA}),
  pp. \bibinfo{pages}{179--184}.

\bibitem[{\citenamefont{Steels}(2005)}]{steels05h}
\bibinfo{author}{\bibnamefont{Steels}, \bibfnamefont{L.}},
  \bibinfo{year}{2005}, \bibinfo{journal}{Connect. Sci.}
  \textbf{\bibinfo{volume}{17}}(\bibinfo{number}{3-4}), \bibinfo{pages}{213}.

\bibitem[{\citenamefont{Steels and Baillie}(2000)}]{steels03b}
\bibinfo{author}{\bibnamefont{Steels}, \bibfnamefont{L.}}, and
  \bibinfo{author}{\bibfnamefont{J.-C.} \bibnamefont{Baillie}},
  \bibinfo{year}{2000}, \bibinfo{journal}{Trends Cogn. Sci.}
  \textbf{\bibinfo{volume}{7}}(\bibinfo{number}{7}), \bibinfo{pages}{308}.

\bibitem[{\citenamefont{Stone}(1961)}]{stone61}
\bibinfo{author}{\bibnamefont{Stone}, \bibfnamefont{M.}}, \bibinfo{year}{1961},
  \bibinfo{journal}{Ann. Math. Stat.} \textbf{\bibinfo{volume}{32}},
  \bibinfo{pages}{1339}.

\bibitem[{\citenamefont{Strogatz and Stewart}(1993)}]{strogatz93}
\bibinfo{author}{\bibnamefont{Strogatz}, \bibfnamefont{S.}}, and
  \bibinfo{author}{\bibfnamefont{I.}~\bibnamefont{Stewart}},
  \bibinfo{year}{1993}, \bibinfo{journal}{Sci. Am.}
  \textbf{\bibinfo{volume}{267}}, \bibinfo{pages}{102}.

\bibitem[{\citenamefont{Strogatz}(1994)}]{strogatz94}
\bibinfo{author}{\bibnamefont{Strogatz}, \bibfnamefont{S.~H.}},
  \bibinfo{year}{1994}, in \emph{\bibinfo{booktitle}{Frontiers in Mathematical
  Biology}}, edited by \bibinfo{editor}{\bibfnamefont{S.~A.}
  \bibnamefont{{Levin}}}, volume \bibinfo{volume}{100} of
  \emph{\bibinfo{series}{Lecture Notes in Biomathematics}}, pp.
  \bibinfo{pages}{122--+}.

\bibitem[{\citenamefont{Suchecki}
  \emph{et~al.}(2005{\natexlab{a}})\citenamefont{Suchecki, Egu\'{i}luz, and
  {San Miguel}}}]{suchecki05}
\bibinfo{author}{\bibnamefont{Suchecki}, \bibfnamefont{K.}},
  \bibinfo{author}{\bibfnamefont{V.~M.} \bibnamefont{Egu\'{i}luz}}, and
  \bibinfo{author}{\bibfnamefont{M.}~\bibnamefont{{San Miguel}}},
  \bibinfo{year}{2005}{\natexlab{a}}, \bibinfo{journal}{Europhys. Lett.}
  \textbf{\bibinfo{volume}{69}}(\bibinfo{number}{2}), \bibinfo{pages}{228}.

\bibitem[{\citenamefont{Suchecki}
  \emph{et~al.}(2005{\natexlab{b}})\citenamefont{Suchecki, Egu\'{i}luz, and
  {San Miguel}}}]{suchecki05b}
\bibinfo{author}{\bibnamefont{Suchecki}, \bibfnamefont{K.}},
  \bibinfo{author}{\bibfnamefont{V.~M.} \bibnamefont{Egu\'{i}luz}}, and
  \bibinfo{author}{\bibfnamefont{M.}~\bibnamefont{{San Miguel}}},
  \bibinfo{year}{2005}{\natexlab{b}}, \bibinfo{journal}{Phys. Rev. E}
  \textbf{\bibinfo{volume}{72}}(\bibinfo{number}{3}), \bibinfo{pages}{036132}.

\bibitem[{\citenamefont{Sudbury}(1985)}]{sudbury85}
\bibinfo{author}{\bibnamefont{Sudbury}, \bibfnamefont{A.}},
  \bibinfo{year}{1985}, \bibinfo{journal}{J. Appl. Prob.}
  \textbf{\bibinfo{volume}{22}}(\bibinfo{number}{2}), \bibinfo{pages}{443}.

\bibitem[{\citenamefont{Sutherland}(2003)}]{sutherland03}
\bibinfo{author}{\bibnamefont{Sutherland}, \bibfnamefont{W.~J.}},
  \bibinfo{year}{2003}, \bibinfo{journal}{Nature}
  \textbf{\bibinfo{volume}{423}}, \bibinfo{pages}{276}.

\bibitem[{\citenamefont{{Szab\'o} and {F{\'a}th}}(2007)}]{szabo07}
\bibinfo{author}{\bibnamefont{{Szab\'o}}, \bibfnamefont{G.}}, and
  \bibinfo{author}{\bibfnamefont{G.}~\bibnamefont{{F{\'a}th}}},
  \bibinfo{year}{2007}, \bibinfo{journal}{Phys. Rep.}
  \textbf{\bibinfo{volume}{446}}(\bibinfo{number}{4-6}), \bibinfo{pages}{97}.

\bibitem[{\citenamefont{{Sznajd-Weron}}(2002)}]{sznajd02a}
\bibinfo{author}{\bibnamefont{{Sznajd-Weron}}, \bibfnamefont{K.}},
  \bibinfo{year}{2002}, \bibinfo{journal}{Phys. Rev. E}
  \textbf{\bibinfo{volume}{66}}(\bibinfo{number}{4}), \bibinfo{pages}{046131}.

\bibitem[{\citenamefont{{Sznajd-Weron}}(2004)}]{sznajd04}
\bibinfo{author}{\bibnamefont{{Sznajd-Weron}}, \bibfnamefont{K.}},
  \bibinfo{year}{2004}, \bibinfo{journal}{Phys. Rev. E}
  \textbf{\bibinfo{volume}{70}}(\bibinfo{number}{3}), \bibinfo{pages}{037104}.

\bibitem[{\citenamefont{{Sznajd-Weron}}(2005{\natexlab{a}})}]{sznajd05b}
\bibinfo{author}{\bibnamefont{{Sznajd-Weron}}, \bibfnamefont{K.}},
  \bibinfo{year}{2005}{\natexlab{a}}, \bibinfo{journal}{Phys. Rev. E}
  \textbf{\bibinfo{volume}{71}}(\bibinfo{number}{4}), \bibinfo{pages}{046110}.

\bibitem[{\citenamefont{{Sznajd-Weron}}(2005{\natexlab{b}})}]{sznajd05a}
\bibinfo{author}{\bibnamefont{{Sznajd-Weron}}, \bibfnamefont{K.}},
  \bibinfo{year}{2005}{\natexlab{b}}, \bibinfo{journal}{Acta Phys. Pol. B}
  \textbf{\bibinfo{volume}{36}}, \bibinfo{pages}{2537}.

\bibitem[{\citenamefont{{Sznajd-Weron} and {Sznajd}}(2000)}]{sznajd00}
\bibinfo{author}{\bibnamefont{{Sznajd-Weron}}, \bibfnamefont{K.}}, and
  \bibinfo{author}{\bibfnamefont{J.}~\bibnamefont{{Sznajd}}},
  \bibinfo{year}{2000}, \bibinfo{journal}{Int. J. Mod. Phys. C}
  \textbf{\bibinfo{volume}{11}}, \bibinfo{pages}{1157}.

\bibitem[{\citenamefont{Sznajd-Weron and Sznajd}(2005)}]{sznajd05}
\bibinfo{author}{\bibnamefont{Sznajd-Weron}, \bibfnamefont{K.}}, and
  \bibinfo{author}{\bibfnamefont{J.}~\bibnamefont{Sznajd}},
  \bibinfo{year}{2005}, \bibinfo{journal}{Physica A}
  \textbf{\bibinfo{volume}{351}}(\bibinfo{number}{2-4}), \bibinfo{pages}{593}.

\bibitem[{\citenamefont{Sznajd-Weron and Weron}(2002)}]{sznajd02}
\bibinfo{author}{\bibnamefont{Sznajd-Weron}, \bibfnamefont{K.}}, and
  \bibinfo{author}{\bibfnamefont{R.}~\bibnamefont{Weron}},
  \bibinfo{year}{2002}, \bibinfo{journal}{Int. J. Mod. Phys. C}
  \textbf{\bibinfo{volume}{13}}(\bibinfo{number}{1}), \bibinfo{pages}{115}.

\bibitem[{\citenamefont{Sznajd-Weron and Weron}(2003)}]{sznajd03}
\bibinfo{author}{\bibnamefont{Sznajd-Weron}, \bibfnamefont{K.}}, and
  \bibinfo{author}{\bibfnamefont{R.}~\bibnamefont{Weron}},
  \bibinfo{year}{2003}, \bibinfo{journal}{Physica A}
  \textbf{\bibinfo{volume}{324}}(\bibinfo{number}{1-2}), \bibinfo{pages}{437}.

\bibitem[{\citenamefont{Tavares} \emph{et~al.}(2007)\citenamefont{Tavares,
  da~Gama, and Nunes}}]{tavares07}
\bibinfo{author}{\bibnamefont{Tavares}, \bibfnamefont{J.~M.}},
  \bibinfo{author}{\bibfnamefont{M.~M.~T.} \bibnamefont{da~Gama}}, and
  \bibinfo{author}{\bibfnamefont{A.}~\bibnamefont{Nunes}},
  \bibinfo{year}{2007}, \eprint{arXiv/0712.4265}.

\bibitem[{\citenamefont{{Tessone} and {Toral}}(2005)}]{tessone05}
\bibinfo{author}{\bibnamefont{{Tessone}}, \bibfnamefont{C.~J.}}, and
  \bibinfo{author}{\bibfnamefont{R.}~\bibnamefont{{Toral}}},
  \bibinfo{year}{2005}, \bibinfo{journal}{Physica A}
  \textbf{\bibinfo{volume}{351}}, \bibinfo{pages}{106}.

\bibitem[{\citenamefont{{Tessone}} \emph{et~al.}(2004)\citenamefont{{Tessone},
  {Toral}, {Amengual}, {Wio}, and {San Miguel}}}]{tessone04}
\bibinfo{author}{\bibnamefont{{Tessone}}, \bibfnamefont{C.~J.}},
  \bibinfo{author}{\bibfnamefont{R.}~\bibnamefont{{Toral}}},
  \bibinfo{author}{\bibfnamefont{P.}~\bibnamefont{{Amengual}}},
  \bibinfo{author}{\bibfnamefont{H.~S.} \bibnamefont{{Wio}}}, and
  \bibinfo{author}{\bibfnamefont{M.}~\bibnamefont{{San Miguel}}},
  \bibinfo{year}{2004}, \bibinfo{journal}{Eur. Phys. J. B}
  \textbf{\bibinfo{volume}{39}}, \bibinfo{pages}{535}.

\bibitem[{\citenamefont{Theraulaz} \emph{et~al.}(1995)\citenamefont{Theraulaz,
  Bonabeau, and Deneubourg}}]{theraulaz95}
\bibinfo{author}{\bibnamefont{Theraulaz}, \bibfnamefont{G.}},
  \bibinfo{author}{\bibfnamefont{E.}~\bibnamefont{Bonabeau}}, and
  \bibinfo{author}{\bibfnamefont{J.-L.} \bibnamefont{Deneubourg}},
  \bibinfo{year}{1995}, \bibinfo{journal}{J. Theor. Biol.}
  \textbf{\bibinfo{volume}{174}}, \bibinfo{pages}{313}.

\bibitem[{\citenamefont{{Toner} and {Tu}}(1995)}]{toner95}
\bibinfo{author}{\bibnamefont{{Toner}}, \bibfnamefont{J.}}, and
  \bibinfo{author}{\bibfnamefont{Y.}~\bibnamefont{{Tu}}}, \bibinfo{year}{1995},
  \bibinfo{journal}{Phys. Rev. Lett.} \textbf{\bibinfo{volume}{75}},
  \bibinfo{pages}{4326}.

\bibitem[{\citenamefont{{Toner} and {Tu}}(1998)}]{toner98}
\bibinfo{author}{\bibnamefont{{Toner}}, \bibfnamefont{J.}}, and
  \bibinfo{author}{\bibfnamefont{Y.}~\bibnamefont{{Tu}}}, \bibinfo{year}{1998},
  \bibinfo{journal}{Phys. Rev. E.} \textbf{\bibinfo{volume}{58}},
  \bibinfo{pages}{4828}.

\bibitem[{\citenamefont{Toral and Tessone}(2007)}]{toral07}
\bibinfo{author}{\bibnamefont{Toral}, \bibfnamefont{R.}}, and
  \bibinfo{author}{\bibfnamefont{C.~J.} \bibnamefont{Tessone}},
  \bibinfo{year}{2007}, \bibinfo{journal}{Comm. Comput. Phys.}
  \textbf{\bibinfo{volume}{2}}(\bibinfo{number}{2}), \bibinfo{pages}{177}.

\bibitem[{\citenamefont{{Toscani}}(2006)}]{toscani06}
\bibinfo{author}{\bibnamefont{{Toscani}}, \bibfnamefont{G.}},
  \bibinfo{year}{2006}, \bibinfo{journal}{Commun. Math. Sci.}
  \textbf{\bibinfo{volume}{4}}(\bibinfo{number}{3}), \bibinfo{pages}{481}.

\bibitem[{\citenamefont{{Traulsen}}
  \emph{et~al.}(2005)\citenamefont{{Traulsen}, {Claussen}, and
  {Hauert}}}]{traulsen05}
\bibinfo{author}{\bibnamefont{{Traulsen}}, \bibfnamefont{A.}},
  \bibinfo{author}{\bibfnamefont{J.~C.} \bibnamefont{{Claussen}}}, and
  \bibinfo{author}{\bibfnamefont{C.}~\bibnamefont{{Hauert}}},
  \bibinfo{year}{2005}, \bibinfo{journal}{Phys. Rev. Lett.}
  \textbf{\bibinfo{volume}{95}}, \bibinfo{pages}{238701}.

\bibitem[{\citenamefont{Travieso and da~Fontoura~Costa}(2006)}]{travieso06}
\bibinfo{author}{\bibnamefont{Travieso}, \bibfnamefont{G.}}, and
  \bibinfo{author}{\bibfnamefont{L.}~\bibnamefont{da~Fontoura~Costa}},
  \bibinfo{year}{2006}, \bibinfo{journal}{Phys. Rev. E}
  \textbf{\bibinfo{volume}{74}}(\bibinfo{number}{3}), \bibinfo{pages}{036112}.

\bibitem[{\citenamefont{Tsujiguchi and Odagaki}(2007)}]{tsujiguchi07}
\bibinfo{author}{\bibnamefont{Tsujiguchi}, \bibfnamefont{M.}}, and
  \bibinfo{author}{\bibfnamefont{T.}~\bibnamefont{Odagaki}},
  \bibinfo{year}{2007}, \bibinfo{journal}{Physica A}
  \textbf{\bibinfo{volume}{375}}(\bibinfo{number}{1}), \bibinfo{pages}{317}.

\bibitem[{\citenamefont{{Tu}} \emph{et~al.}(2005)\citenamefont{{Tu}, {Sousa},
  {Kong}, and {Liu}}}]{yu-song05}
\bibinfo{author}{\bibnamefont{{Tu}}, \bibfnamefont{Y.-S.}},
  \bibinfo{author}{\bibfnamefont{A.~O.} \bibnamefont{{Sousa}}},
  \bibinfo{author}{\bibfnamefont{L.-J.} \bibnamefont{{Kong}}}, and
  \bibinfo{author}{\bibfnamefont{M.-R.} \bibnamefont{{Liu}}},
  \bibinfo{year}{2005}, \bibinfo{journal}{Int. J. Mod. Phys. C}
  \textbf{\bibinfo{volume}{16}}, \bibinfo{pages}{1149}.

\bibitem[{\citenamefont{Ulam}(1960)}]{ulam60}
\bibinfo{author}{\bibnamefont{Ulam}, \bibfnamefont{S.}}, \bibinfo{year}{1960},
  \emph{\bibinfo{title}{A Collection of Mathematical Problems}}
  (\bibinfo{publisher}{Interscience}, \bibinfo{address}{New York, NY, USA}).

\bibitem[{\citenamefont{{Urbig} and {Lorenz}}(2004)}]{urbig07}
\bibinfo{author}{\bibnamefont{{Urbig}}, \bibfnamefont{D.}}, and
  \bibinfo{author}{\bibfnamefont{J.}~\bibnamefont{{Lorenz}}},
  \bibinfo{year}{2004}, in \emph{\bibinfo{booktitle}{Proceedings of the Second
  Conference of the European Social Simulation Association (ESSA)}},
  \eprint{arxiv:0708.3334}.

\bibitem[{\citenamefont{Varela and Bourgine}(1992)}]{varela92}
\bibinfo{editor}{\bibnamefont{Varela}, \bibfnamefont{F.~J.}}, and
  \bibinfo{editor}{\bibfnamefont{P.}~\bibnamefont{Bourgine}} (eds.),
  \bibinfo{year}{1992}, \emph{\bibinfo{title}{Toward a Practice of Autonomous
  Systems: Proceedings of the First European Conference on Artificial Life}}
  (\bibinfo{publisher}{MIT Press}, \bibinfo{address}{Cambridge, MA, USA}).

\bibitem[{\citenamefont{V\'azquez}(2005)}]{vazquez05}
\bibinfo{author}{\bibnamefont{V\'azquez}, \bibfnamefont{A.}},
  \bibinfo{year}{2005}, \bibinfo{journal}{Phys. Rev. Lett.}
  \textbf{\bibinfo{volume}{95}}, \bibinfo{pages}{248701}.

\bibitem[{\citenamefont{V\'azquez} \emph{et~al.}(2006)\citenamefont{V\'azquez,
  Oliveira, Dezs{\"o}, Goh, Kondor, and Barab\'asi}}]{vazquez06}
\bibinfo{author}{\bibnamefont{V\'azquez}, \bibfnamefont{A.}},
  \bibinfo{author}{\bibfnamefont{J.~G.} \bibnamefont{Oliveira}},
  \bibinfo{author}{\bibfnamefont{Z.}~\bibnamefont{Dezs{\"o}}},
  \bibinfo{author}{\bibfnamefont{K.-I.} \bibnamefont{Goh}},
  \bibinfo{author}{\bibfnamefont{I.}~\bibnamefont{Kondor}}, and
  \bibinfo{author}{\bibfnamefont{A.-L.} \bibnamefont{Barab\'asi}},
  \bibinfo{year}{2006}, \bibinfo{journal}{Phys. Rev. E}
  \textbf{\bibinfo{volume}{73}}(\bibinfo{number}{3}), \bibinfo{pages}{036127}.

\bibitem[{\citenamefont{Vazquez} \emph{et~al.}(2008)\citenamefont{Vazquez,
  Egu\'{\i}luz, and Miguel}}]{Vazquez08}
\bibinfo{author}{\bibnamefont{Vazquez}, \bibfnamefont{F.}},
  \bibinfo{author}{\bibfnamefont{V.~M.} \bibnamefont{Egu\'{\i}luz}}, and
  \bibinfo{author}{\bibfnamefont{M.~S.} \bibnamefont{Miguel}},
  \bibinfo{year}{2008}, \bibinfo{journal}{Phys. Rev. Lett.}
  \textbf{\bibinfo{volume}{100}}(\bibinfo{number}{10}), \bibinfo{eid}{108702}
  (pages~\bibinfo{numpages}{4}).

\bibitem[{\citenamefont{{Vazquez}} \emph{et~al.}(2007)\citenamefont{{Vazquez},
  {Gonzalez-Avella}, {Eguiluz}, and {San Miguel}}}]{vazquez07b}
\bibinfo{author}{\bibnamefont{{Vazquez}}, \bibfnamefont{F.}},
  \bibinfo{author}{\bibfnamefont{J.~C.} \bibnamefont{{Gonzalez-Avella}}},
  \bibinfo{author}{\bibfnamefont{V.~M.} \bibnamefont{{Eguiluz}}}, and
  \bibinfo{author}{\bibfnamefont{M.}~\bibnamefont{{San Miguel}}},
  \bibinfo{year}{2007}, \bibinfo{journal}{Phys. Rev. E}
  \textbf{\bibinfo{volume}{76}}, \bibinfo{pages}{046120}.

\bibitem[{\citenamefont{Vazquez} \emph{et~al.}(2003)\citenamefont{Vazquez,
  Krapivsky, and Redner}}]{vazquez03}
\bibinfo{author}{\bibnamefont{Vazquez}, \bibfnamefont{F.}},
  \bibinfo{author}{\bibfnamefont{P.~L.} \bibnamefont{Krapivsky}}, and
  \bibinfo{author}{\bibfnamefont{S.}~\bibnamefont{Redner}},
  \bibinfo{year}{2003}, \bibinfo{journal}{J. Phys. A}
  \textbf{\bibinfo{volume}{36}}(\bibinfo{number}{3}), \bibinfo{pages}{L61}.

\bibitem[{\citenamefont{Vazquez and Redner}(2004)}]{vazquez04}
\bibinfo{author}{\bibnamefont{Vazquez}, \bibfnamefont{F.}}, and
  \bibinfo{author}{\bibfnamefont{S.}~\bibnamefont{Redner}},
  \bibinfo{year}{2004}, \bibinfo{journal}{J. Phys. A}
  \textbf{\bibinfo{volume}{37}}(\bibinfo{number}{35}), \bibinfo{pages}{8479}.

\bibitem[{\citenamefont{Vazquez and Redner}(2007)}]{vazquez07}
\bibinfo{author}{\bibnamefont{Vazquez}, \bibfnamefont{F.}}, and
  \bibinfo{author}{\bibfnamefont{S.}~\bibnamefont{Redner}},
  \bibinfo{year}{2007}, \bibinfo{journal}{Europhys. Lett.}
  \textbf{\bibinfo{volume}{78}}(\bibinfo{number}{1}), \bibinfo{pages}{18002
  (5pp)}.

\bibitem[{\citenamefont{{Vicsek}} \emph{et~al.}(1995)\citenamefont{{Vicsek},
  {Czir{\'o}k}, {Ben-Jacob}, {Cohen}, and {Shochet}}}]{vicsek95}
\bibinfo{author}{\bibnamefont{{Vicsek}}, \bibfnamefont{T.}},
  \bibinfo{author}{\bibfnamefont{A.}~\bibnamefont{{Czir{\'o}k}}},
  \bibinfo{author}{\bibfnamefont{E.}~\bibnamefont{{Ben-Jacob}}},
  \bibinfo{author}{\bibfnamefont{I.}~\bibnamefont{{Cohen}}}, and
  \bibinfo{author}{\bibfnamefont{O.}~\bibnamefont{{Shochet}}},
  \bibinfo{year}{1995}, \bibinfo{journal}{Phys. Rev. Lett.}
  \textbf{\bibinfo{volume}{75}}, \bibinfo{pages}{1226}.

\bibitem[{\citenamefont{Vilone and Castellano}(2004)}]{vilone04}
\bibinfo{author}{\bibnamefont{Vilone}, \bibfnamefont{D.}}, and
  \bibinfo{author}{\bibfnamefont{C.}~\bibnamefont{Castellano}},
  \bibinfo{year}{2004}, \bibinfo{journal}{Phys. Rev. E}
  \textbf{\bibinfo{volume}{69}}(\bibinfo{number}{1}), \bibinfo{pages}{016109}.

\bibitem[{\citenamefont{{Vilone}} \emph{et~al.}(2002)\citenamefont{{Vilone},
  {Vespignani}, and {Castellano}}}]{vilone03}
\bibinfo{author}{\bibnamefont{{Vilone}}, \bibfnamefont{D.}},
  \bibinfo{author}{\bibfnamefont{A.}~\bibnamefont{{Vespignani}}}, and
  \bibinfo{author}{\bibfnamefont{C.}~\bibnamefont{{Castellano}}},
  \bibinfo{year}{2002}, \bibinfo{journal}{Eur. Phys. J. B}
  \textbf{\bibinfo{volume}{30}}, \bibinfo{pages}{399}.

\bibitem[{\citenamefont{Vogels} \emph{et~al.}(2003)\citenamefont{Vogels, van
  Renesse, and Birman}}]{vogels03}
\bibinfo{author}{\bibnamefont{Vogels}, \bibfnamefont{W.}},
  \bibinfo{author}{\bibfnamefont{R.}~\bibnamefont{van Renesse}}, and
  \bibinfo{author}{\bibfnamefont{K.}~\bibnamefont{Birman}},
  \bibinfo{year}{2003}, \bibinfo{journal}{SIGCOMM Comput. Commun. Rev.}
  \textbf{\bibinfo{volume}{33}}(\bibinfo{number}{1}), \bibinfo{pages}{131}.

\bibitem[{\citenamefont{Wang and Minett}(2005)}]{wang05}
\bibinfo{author}{\bibnamefont{Wang}, \bibfnamefont{W.~S.-Y.}}, and
  \bibinfo{author}{\bibfnamefont{J.~W.} \bibnamefont{Minett}},
  \bibinfo{year}{2005}, \bibinfo{journal}{Trends Ecol. Evol.}
  \textbf{\bibinfo{volume}{20}}(\bibinfo{number}{5}), \bibinfo{pages}{263}.

\bibitem[{\citenamefont{Wasserman and Faust}(1994)}]{wasserman94}
\bibinfo{author}{\bibnamefont{Wasserman}, \bibfnamefont{S.}}, and
  \bibinfo{author}{\bibfnamefont{K.}~\bibnamefont{Faust}},
  \bibinfo{year}{1994}, \emph{\bibinfo{title}{Social network analysis}}
  (\bibinfo{publisher}{Cambridge University Press},
  \bibinfo{address}{Cambridge, UK}).

\bibitem[{\citenamefont{Watts and Strogatz}(1998)}]{watts98}
\bibinfo{author}{\bibnamefont{Watts}, \bibfnamefont{D.}}, and
  \bibinfo{author}{\bibfnamefont{S.}~\bibnamefont{Strogatz}},
  \bibinfo{year}{1998}, \bibinfo{journal}{Nature}
  \textbf{\bibinfo{volume}{393}}, \bibinfo{pages}{440}.

\bibitem[{\citenamefont{Watts}(2002)}]{watts02b}
\bibinfo{author}{\bibnamefont{Watts}, \bibfnamefont{D.~J.}},
  \bibinfo{year}{2002}, \bibinfo{journal}{Proc. Natl. Acad. Sci. USA}
  \textbf{\bibinfo{volume}{99}}(\bibinfo{number}{9}), \bibinfo{pages}{5766}.

\bibitem[{\citenamefont{Weidlich}(1971)}]{weidlich71}
\bibinfo{author}{\bibnamefont{Weidlich}, \bibfnamefont{W.}},
  \bibinfo{year}{1971}, \bibinfo{journal}{Br. J. Math. Statist. Psychol.}
  \textbf{\bibinfo{volume}{24}}, \bibinfo{pages}{251}.

\bibitem[{\citenamefont{Weidlich}(1991)}]{weidlich91}
\bibinfo{author}{\bibnamefont{Weidlich}, \bibfnamefont{W.}},
  \bibinfo{year}{1991}, \bibinfo{journal}{Phys. Rep.}
  \textbf{\bibinfo{volume}{204}}, \bibinfo{pages}{1}.

\bibitem[{\citenamefont{{Weidlich}}(2002)}]{weidlich02}
\bibinfo{author}{\bibnamefont{{Weidlich}}, \bibfnamefont{W.}},
  \bibinfo{year}{2002}, \emph{\bibinfo{title}{Sociodynamics: A Systematic
  Approach to Mathematical Modelling in Social Sciences}}
  (\bibinfo{publisher}{Taylor and Francis}, \bibinfo{address}{London, UK}).

\bibitem[{\citenamefont{{Weisbuch}}(2004)}]{weisbuch04}
\bibinfo{author}{\bibnamefont{{Weisbuch}}, \bibfnamefont{G.}},
  \bibinfo{year}{2004}, \bibinfo{journal}{Eur. Phys. J. B}
  \textbf{\bibinfo{volume}{38}}, \bibinfo{pages}{339}.

\bibitem[{\citenamefont{Weisbuch} \emph{et~al.}(2005)\citenamefont{Weisbuch,
  Deffuant, and Amblard}}]{weisbuch05}
\bibinfo{author}{\bibnamefont{Weisbuch}, \bibfnamefont{G.}},
  \bibinfo{author}{\bibfnamefont{G.}~\bibnamefont{Deffuant}}, and
  \bibinfo{author}{\bibfnamefont{F.}~\bibnamefont{Amblard}},
  \bibinfo{year}{2005}, \bibinfo{journal}{Physica A}
  \textbf{\bibinfo{volume}{353}}, \bibinfo{pages}{555}.

\bibitem[{\citenamefont{Weisbuch} \emph{et~al.}(2002)\citenamefont{Weisbuch,
  Deffuant, Amblard, and Nadal}}]{weisbuch02}
\bibinfo{author}{\bibnamefont{Weisbuch}, \bibfnamefont{G.}},
  \bibinfo{author}{\bibfnamefont{G.}~\bibnamefont{Deffuant}},
  \bibinfo{author}{\bibfnamefont{F.}~\bibnamefont{Amblard}}, and
  \bibinfo{author}{\bibfnamefont{J.~P.} \bibnamefont{Nadal}},
  \bibinfo{year}{2002}, in \emph{\bibinfo{booktitle}{Lecture Notes in Economics
  and Mathematical Systems}}, edited by
  \bibinfo{editor}{\bibfnamefont{G.}~\bibnamefont{Fandel}} and
  \bibinfo{editor}{\bibfnamefont{W.}~\bibnamefont{Trockel}}
  (\bibinfo{publisher}{Springer Verlag}, \bibinfo{address}{Berlin-Heidelberg,
  Germany}), volume \bibinfo{volume}{521}, pp. \bibinfo{pages}{225--242}.

\bibitem[{\citenamefont{Weiss}(1999)}]{weiss99}
\bibinfo{editor}{\bibnamefont{Weiss}, \bibfnamefont{G.}} (ed.),
  \bibinfo{year}{1999}, \emph{\bibinfo{title}{Multiagent systems: a modern
  approach to distributed artificial intelligence}} (\bibinfo{publisher}{MIT
  Press}, \bibinfo{address}{Cambridge, MA, USA}).

\bibitem[{\citenamefont{Wichmann}(2008)}]{wichmann08}
\bibinfo{author}{\bibnamefont{Wichmann}, \bibfnamefont{S.}},
  \bibinfo{year}{2008}, \eprint{arXiv:0801.1415}.

\bibitem[{\citenamefont{{Wichmann}}
  \emph{et~al.}(2006)\citenamefont{{Wichmann}, {Stauffer}, {Lima}, and
  {Schulze}}}]{wichmann07}
\bibinfo{author}{\bibnamefont{{Wichmann}}, \bibfnamefont{S.}},
  \bibinfo{author}{\bibfnamefont{D.}~\bibnamefont{{Stauffer}}},
  \bibinfo{author}{\bibfnamefont{F.~W.~S.} \bibnamefont{{Lima}}}, and
  \bibinfo{author}{\bibfnamefont{C.}~\bibnamefont{{Schulze}}},
  \bibinfo{year}{2006}, \bibinfo{journal}{Trans. Philol. Soc.}
  \textbf{\bibinfo{volume}{105}}(\bibinfo{number}{2}), \bibinfo{pages}{126}.

\bibitem[{\citenamefont{Wilkinson and Willemsen}(1983)}]{wilkinson83}
\bibinfo{author}{\bibnamefont{Wilkinson}, \bibfnamefont{D.}}, and
  \bibinfo{author}{\bibfnamefont{J.~F.} \bibnamefont{Willemsen}},
  \bibinfo{year}{1983}, \bibinfo{journal}{J. Phys. A}
  \textbf{\bibinfo{volume}{16}}(\bibinfo{number}{14}), \bibinfo{pages}{3365}.

\bibitem[{\citenamefont{{Wilson}}(1971)}]{wilson71}
\bibinfo{author}{\bibnamefont{{Wilson}}, \bibfnamefont{E.~O.}},
  \bibinfo{year}{1971}, \emph{\bibinfo{title}{The Insect Societies}}
  (\bibinfo{publisher}{Harvard Univ. Press}, \bibinfo{address}{Cambridge, MA,
  USA}).

\bibitem[{\citenamefont{Wio} \emph{et~al.}(2006)\citenamefont{Wio, de~la Lama,
  and L\'opez}}]{wio06}
\bibinfo{author}{\bibnamefont{Wio}, \bibfnamefont{H.~S.}},
  \bibinfo{author}{\bibfnamefont{M.~S.} \bibnamefont{de~la Lama}}, and
  \bibinfo{author}{\bibfnamefont{J.~M.} \bibnamefont{L\'opez}},
  \bibinfo{year}{2006}, \bibinfo{journal}{Physica A}
  \textbf{\bibinfo{volume}{371}}, \bibinfo{pages}{108}.

\bibitem[{\citenamefont{Wittgenstein}(1953{\natexlab{a}})}]{wittgenstein53engl%
ish}
\bibinfo{author}{\bibnamefont{Wittgenstein}, \bibfnamefont{L.}},
  \bibinfo{year}{1953}{\natexlab{a}}, \emph{\bibinfo{title}{Philosophical
  Investigations. (Translated by Anscombe, G.E.M.)}} (\bibinfo{publisher}{Basil
  Blackwell}, \bibinfo{address}{Oxford, UK}).

\bibitem[{\citenamefont{Wittgenstein}(1953{\natexlab{b}})}]{wittgenstein53germ%
an}
\bibinfo{author}{\bibnamefont{Wittgenstein}, \bibfnamefont{L.}},
  \bibinfo{year}{1953}{\natexlab{b}}, \emph{\bibinfo{title}{Philosophische
  Untersuchungen}} (\bibinfo{publisher}{Suhrkamp Verlag},
  \bibinfo{address}{Frankfurt am Main, Germany}).

\bibitem[{\citenamefont{Wooldridge}(2002)}]{wooldridge02}
\bibinfo{author}{\bibnamefont{Wooldridge}, \bibfnamefont{M.}},
  \bibinfo{year}{2002}, \emph{\bibinfo{title}{Introduction to MultiAgent
  Systems}} (\bibinfo{publisher}{John Wiley and Sons},
  \bibinfo{address}{Chichester, UK}).

\bibitem[{\citenamefont{Wu}(1982)}]{wu82}
\bibinfo{author}{\bibnamefont{Wu}, \bibfnamefont{F.}}, \bibinfo{year}{1982},
  \bibinfo{journal}{Rev. Mod. Phys.}
  \textbf{\bibinfo{volume}{54}}(\bibinfo{number}{1}), \bibinfo{pages}{235}.

\bibitem[{\citenamefont{Wu} \emph{et~al.}(2004)\citenamefont{Wu, Huberman,
  Adamic, and Tyler}}]{wu04}
\bibinfo{author}{\bibnamefont{Wu}, \bibfnamefont{F.}},
  \bibinfo{author}{\bibfnamefont{B.}~\bibnamefont{Huberman}},
  \bibinfo{author}{\bibfnamefont{L.}~\bibnamefont{Adamic}}, and
  \bibinfo{author}{\bibfnamefont{J.}~\bibnamefont{Tyler}},
  \bibinfo{year}{2004}, \bibinfo{journal}{Physica A}
  \textbf{\bibinfo{volume}{337}}(\bibinfo{number}{1-2}), \bibinfo{pages}{327}.

\bibitem[{\citenamefont{Wu and Huberman}(2007)}]{wu07}
\bibinfo{author}{\bibnamefont{Wu}, \bibfnamefont{F.}}, and
  \bibinfo{author}{\bibfnamefont{B.~A.} \bibnamefont{Huberman}},
  \bibinfo{year}{2007}, \bibinfo{journal}{Proc. Natl. Acad. Sci. USA}
  \textbf{\bibinfo{volume}{104}}(\bibinfo{number}{45}), \bibinfo{pages}{17599}.

\bibitem[{\citenamefont{Zanette and Montemurro}(2005)}]{zanette05}
\bibinfo{author}{\bibnamefont{Zanette}, \bibfnamefont{D.}}, and
  \bibinfo{author}{\bibfnamefont{M.}~\bibnamefont{Montemurro}},
  \bibinfo{year}{2005}, \bibinfo{journal}{J. Quant. Linguist.}
  \textbf{\bibinfo{volume}{12}}(\bibinfo{number}{1}), \bibinfo{pages}{29}.

\bibitem[{\citenamefont{Zanette}(2001)}]{zanette01b}
\bibinfo{author}{\bibnamefont{Zanette}, \bibfnamefont{D.~H.}},
  \bibinfo{year}{2001}, \bibinfo{journal}{Phys. Rev. E}
  \textbf{\bibinfo{volume}{64}}(\bibinfo{number}{5}), \bibinfo{pages}{050901}.

\bibitem[{\citenamefont{Zanette}(2002)}]{zanette02}
\bibinfo{author}{\bibnamefont{Zanette}, \bibfnamefont{D.~H.}},
  \bibinfo{year}{2002}, \bibinfo{journal}{Phys. Rev. E}
  \textbf{\bibinfo{volume}{65}}(\bibinfo{number}{4}), \bibinfo{pages}{041908}.

\bibitem[{\citenamefont{{Zanette}}(2007{\natexlab{a}})}]{zanette07a}
\bibinfo{author}{\bibnamefont{{Zanette}}, \bibfnamefont{D.~H.}},
  \bibinfo{year}{2007}{\natexlab{a}}, \eprint{arXiv:0711.1360}.

\bibitem[{\citenamefont{{Zanette}}(2007{\natexlab{b}})}]{zanette07}
\bibinfo{author}{\bibnamefont{{Zanette}}, \bibfnamefont{D.~H.}},
  \bibinfo{year}{2007}{\natexlab{b}}, \eprint{arXiv:0710.1511}.

\bibitem[{\citenamefont{{Zanette} and {Gil}}(2006)}]{zanette06}
\bibinfo{author}{\bibnamefont{{Zanette}}, \bibfnamefont{D.~H.}}, and
  \bibinfo{author}{\bibfnamefont{S.}~\bibnamefont{{Gil}}},
  \bibinfo{year}{2006}, \bibinfo{journal}{Physica D}
  \textbf{\bibinfo{volume}{224}}, \bibinfo{pages}{156}.

\bibitem[{\citenamefont{Zillio} \emph{et~al.}(2005)\citenamefont{Zillio,
  Volkov, Banavar, Hubbell, and Maritan}}]{zillio05}
\bibinfo{author}{\bibnamefont{Zillio}, \bibfnamefont{T.}},
  \bibinfo{author}{\bibfnamefont{I.}~\bibnamefont{Volkov}},
  \bibinfo{author}{\bibfnamefont{J.~R.} \bibnamefont{Banavar}},
  \bibinfo{author}{\bibfnamefont{S.~P.} \bibnamefont{Hubbell}}, and
  \bibinfo{author}{\bibfnamefont{A.}~\bibnamefont{Maritan}},
  \bibinfo{year}{2005}, \bibinfo{journal}{Phys. Rev. Lett.}
  \textbf{\bibinfo{volume}{95}}(\bibinfo{number}{9}), \bibinfo{pages}{098101}.

\bibitem[{\citenamefont{{Zimmermann}}
  \emph{et~al.}(2001)\citenamefont{{Zimmermann}, {Eguiluz}, and {San
  Miguel}}}]{zimmermann01}
\bibinfo{author}{\bibnamefont{{Zimmermann}}, \bibfnamefont{M.~G.}},
  \bibinfo{author}{\bibfnamefont{V.~M.} \bibnamefont{{Eguiluz}}}, and
  \bibinfo{author}{\bibfnamefont{M.}~\bibnamefont{{San Miguel}}},
  \bibinfo{year}{2001}, \bibinfo{journal}{Lect. Notes Econ. Math.}
  \textbf{\bibinfo{volume}{503}}, \bibinfo{pages}{73}.

\bibitem[{\citenamefont{Zimmermann}
  \emph{et~al.}(2004)\citenamefont{Zimmermann, Egu{\'i}luz, and {San
  Miguel}}}]{zimmermann04}
\bibinfo{author}{\bibnamefont{Zimmermann}, \bibfnamefont{M.~G.}},
  \bibinfo{author}{\bibfnamefont{V.~M.} \bibnamefont{Egu{\'i}luz}}, and
  \bibinfo{author}{\bibfnamefont{M.}~\bibnamefont{{San Miguel}}},
  \bibinfo{year}{2004}, \bibinfo{journal}{Phys. Rev. E}
  \textbf{\bibinfo{volume}{69}}(\bibinfo{number}{6}), \bibinfo{pages}{065102}.

\bibitem[{\citenamefont{Zipf}(1949)}]{zipf49}
\bibinfo{author}{\bibnamefont{Zipf}, \bibfnamefont{G.~K.}},
  \bibinfo{year}{1949}, \emph{\bibinfo{title}{Human Behavior and the Principle
  of Least Effort}} (\bibinfo{publisher}{Addison-Wesley, Reading MA (USA)}).

\end{thebibliography}



\end{document}